\UseRawInputEncoding 

\documentclass[10pt]{article}
\usepackage[letterpaper, left=1in, top=1in, right=1in, bottom=1in, verbose, ignoremp]{geometry}

\usepackage{url}\RequirePackage[colorlinks,citecolor=blue, linkcolor=blue,urlcolor = blue]{hyperref}
\usepackage{latexsym,amssymb,amsmath,amsfonts,graphicx,color,fancyvrb,amsthm,enumerate,subcaption,mathrsfs}
\usepackage[longnamesfirst,authoryear,round]{natbib}
\usepackage[dvipsnames]{xcolor}
\usepackage{xy}\xyoption{all} \xyoption{poly} \xyoption{knot}
\usepackage{float}
\usepackage{bm}
\usepackage{bbm}
\usepackage{multicol,multirow}
\usepackage{array}
\usepackage{relsize}
\usepackage{chngcntr}
\usepackage{etoolbox}
\usepackage{caption}
\usepackage{tikz}
\usetikzlibrary{decorations.pathreplacing,calc}
\usetikzlibrary{shapes,backgrounds}
\usetikzlibrary{patterns}
\usetikzlibrary{cd}
\usepackage{amsopn}
\usepackage{tabularx}
\usepackage{calc}
\usepackage{algorithm}
\usepackage[noend]{algpseudocode}
\usepackage{changepage}
\usepackage{hyperref}
\usepackage{mathtools}

\usepackage{tikz-cd} 
\usepackage{amscd} 
\usepackage{comment}
\usepackage[capitalize,nameinlink,compress]{cleveref}
\crefname{figure}{Figure}{Figures} 
\crefname{equation}{}{} 
\crefname{assumption}{Assumption}{Assumptions}
\crefname{subsection}{Subsection}{Subsections}
\usepackage{array}
\usepackage{booktabs}
\usepackage[dvipsnames]{xcolor}
\usepackage{tikz}
\usepackage{adjustbox}

\newcolumntype{M}[1]{>{\centering\arraybackslash}m{#1}}

\usepackage{adjustbox}
\usepackage{graphicx}
\usepackage{hyperref}

\definecolor{intdrei}{RGB}{0,85,212}
\definecolor{intvier}{RGB}{0,51,128}
\definecolor{intfunf}{RGB}{0,34,85}
\newcounter{cdrow}

\newtheorem*{theorem*}{Theorem}

\newtheorem*{claim*}{Claim}

\theoremstyle{definition}

\newtheorem*{definition*}{Definition}

\theoremstyle{remark}

\newtheorem*{example*}{Example}

\def\log{{\rm log}}

\def\rn{\mathbb{R}^n}
\def\r0{r_0}
\def\r1{r_1}
\def\r2{r_2}
\def\g1{g_1}
\def\g2{g_2}
\def\g3{g_3}

\makeatletter
\newcommand*{\op}{%
  \DOTSB
  \mathop{\vphantom{\bigoplus}\mathpalette\matt@op\relax}%
  \slimits@
}
\newcommand\matt@op[2]{%
  \vcenter{\m@th\hbox{\resizebox{\widthof{$#1\bigoplus$}}{!}{$\boxplus$}}}%
}
\makeatother

\makeatletter
\def\@biblabel#1{}
\makeatother

\makeatletter
\patchcmd{\NAT@citex}
  {\@citea\NAT@hyper@{%
     \NAT@nmfmt{\NAT@nm}%
     \hyper@natlinkbreak{\NAT@aysep\NAT@spacechar}{\@citeb\@extra@b@citeb}%
     \NAT@date}}
  {\@citea\NAT@nmfmt{\NAT@nm}%
   \NAT@aysep\NAT@spacechar\NAT@hyper@{\NAT@date}}{}{}

\patchcmd{\NAT@citex}
  {\@citea\NAT@hyper@{%
     \NAT@nmfmt{\NAT@nm}%
     \hyper@natlinkbreak{\NAT@spacechar\NAT@@open\if*#1*\else#1\NAT@spacechar\fi}%
       {\@citeb\@extra@b@citeb}%
     \NAT@date}}
  {\@citea\NAT@nmfmt{\NAT@nm}%
   \NAT@spacechar\NAT@@open\if*#1*\else#1\NAT@spacechar\fi\NAT@hyper@{\NAT@date}}
  {}{}

\makeatother

\begin{document}
\def\spacingset#1{\renewcommand{\baselinestretch}%
{#1}\small\normalsize} \spacingset{1}

\begin{flushleft}
{\Large{\textbf{A Topological Gaussian Mixture Model for Bone Marrow Morphology in Leukaemia}}}
\newline
\\
Qiquan Wang$^{1,\dagger}$, Anna Song$^{1,2}$, Antoniana Batsivari$^{2}$, Dominique Bonnet$^{2}$, and Anthea Monod$^{1}$
\\
\bigskip
\bf{1} Department of Mathematics, Imperial College London, UK \\
\bf{2} Haematopoietic Stem Cell Laboratory, The Francis Crick Institute, UK
\\
\bigskip
$\dagger$ Corresponding e-mail: qiquan.wang17@imperial.ac.uk
\end{flushleft}


\begin{abstract}
    Acute myeloid leukaemia (AML) is a type of blood and bone marrow cancer characterized by the proliferation of abnormal clonal haematopoietic cells in the bone marrow leading to bone marrow failure.  Over the course of the disease, angiogenic factors released by leukaemic cells drastically alter the bone marrow vascular niches resulting in observable structural abnormalities. We use a technique from topological data analysis---\emph{persistent homology}---to quantify the images and infer on the disease through the imaged morphological features on a proprietary high-resolution image dataset. We find that persistent homology uncovers succinct dissimilarities between the control, early, and late stages of AML development. We then integrate persistent homology into stage-dependent Gaussian mixture models, proposing a new class of models which are applicable to persistent homology summaries and able to both infer patterns in morphological changes between different stages of progression as well as provide a basis for prediction.
\end{abstract}

\paragraph{Keywords:} Confocal imaging; Classification; Gaussian mixture models; Leukaemia; Persistent homology; Prediction. 


\section{Introduction}
\label{sec:intro}

Acute myeloid leukaemia (AML) is a cancer of the bone marrow and blood which can also infiltrate other haematopoietic organs such as the spleen and other tissues \citep{Behrmann2018-iy}. It is characterized by the proliferation of abnormally differentiated and nonfunctional haematopoietic blast cells, which hijack and alter the bone marrow niches; they suppress normal haematopoiesis in favor of their own growth \citep{Kumar17}.  AML cells are known to intervene with the normal microenvironment equilibrium and remodel the bone marrow \citep{passaro,DUARTE20181507}, giving rise to prominent morphological changes such as increases in the vessel density and vessel leakiness. Overall, the regularity in the bone marrow vasculature degrades, leading to poorer fluid circulation which in turn further exacerbates the disease.  Data on AML are difficult and very costly to obtain and analyze, which contributes to the challenging nature of the disease study.

In this paper, we both quantify and model the progressive alteration of the vessels in the bone marrow microenvironment, captured using 3D confocal microscopy at high resolutions in a proprietary dataset, as AML develops. We utilize a central methodology in topological data analysis---\emph{persistent homology}---to quantitatively summarize both the local and global geometric information within the imaging data in an interpretable manner, as scatter plots in the plane known as \emph{persistence diagrams}. This allows the geometric information to be modeled from the observed patterns and clusters in the persistence diagrams. We then propose and implement a new class of stage-dependent Gaussian mixture models to model the deformation of the bone marrow architecture with the progression of AML across its stages.

The remainder of this paper is organized as follows.  We close this section with an overview of related work quantifying confocal AML imaging and prior use of persistent homology in cancer imaging.  In Section \ref{sec:data_description}, we describe the femur bone microscopy image dataset as our study of interest. In Section \ref{sec:preliminaries}, we then introduce persistent homology and the variant \emph{signed distance persistent homology} (SDPH) used in our application. Applying the techniques to our dataset, in Section \ref{sec:analysis} we present our first contribution: the results of our analysis using SDPH, where we distinguish different stages of progression from the imaging data using both local and global analysis. We then turn to our second contribution, which is to construct a novel class of Gaussian mixture models (GMMs) in Section \ref{sec:modelling} to model the stages of progression in terms of features captured by SDPH. In Section \ref{sec:disucussion}, we conclude with a discussion of our work and future directions for statistical modeling with SDPH.\\

\noindent
\textbf{Related Work.} Current understanding of the morphological progression is far from complete, despite numerous prior studies.  Among those using images to understand the disease, as is our aim in this work, \cite{gomariz, Coutu2017, Coutu2018} quantifiably analyzed the spatial distribution and interactions of different cells and features within the bone marrow niches in high-resolution microscopy images. \cite{SHIH20093161, mri} conducted similar analyses on the leukaemic-induced changes using magnetic resonance images. More broadly, a recent survey by \cite{mohammed2024statisticalanalysisquantitativecancer} summarized some of the modern techniques for quantitative cancer imaging analysis. 

Persistent homology has been successfully implemented to study the shapes within medical imaging data in oncology. Notable examples include the classification of hepatic tumors from magnetic resonance images \citep{oyama2019hepatic}; classification of acute lymphoblastic leukaemia from microscopy images \cite{shah2024allclassification}; tumor segmentation from whole slide images \citep{QAISER2016119}; computer-aided diagnosis \citep{yu2016predicting}; and prognosis \citep{crawford}. More recent work by \cite{pritchard23} and \cite{moon_sedt} demonstrated the use of signed Euclidean distance transform on images to quantify bone microstructure and tumor morphology for classification and prognostics. On a finer level, persistent homology has also been used to summarize cellular architecture in applications to breast cancer \citep{histo, aukerman2022persistent} and prostate cancer \citep{lawson2019persistent}.


\section{Data Description: Confocal Imaging}
\label{sec:data_description}

We study the morphological changes to bone marrow vascular niches caused by AML under control environments by engrafting AML and non-AML cells into immunodeficient mice. AML, human primary samples, and leukaemic cell lines of different types and clonalities were injected intravenously, and the mice were sacrificed at different stages as the disease developed. From each mouse, we retained one of the femur bones for immunostaining and 3D confocal imaging, while other bones were crushed and used to approximate the percentage of engraftment. These data are very costly to obtain in terms of time consumption as well as equipment and resources required: they are among the first to study the structure of the bone marrow vascular niches at such a fine resolution, crucial for the novel investigation into differences between early and late stages of AML as most research has focused on identifying changes and causes in late stages.

\begin{figure}
    \centering
    \includegraphics[width=.8\linewidth]{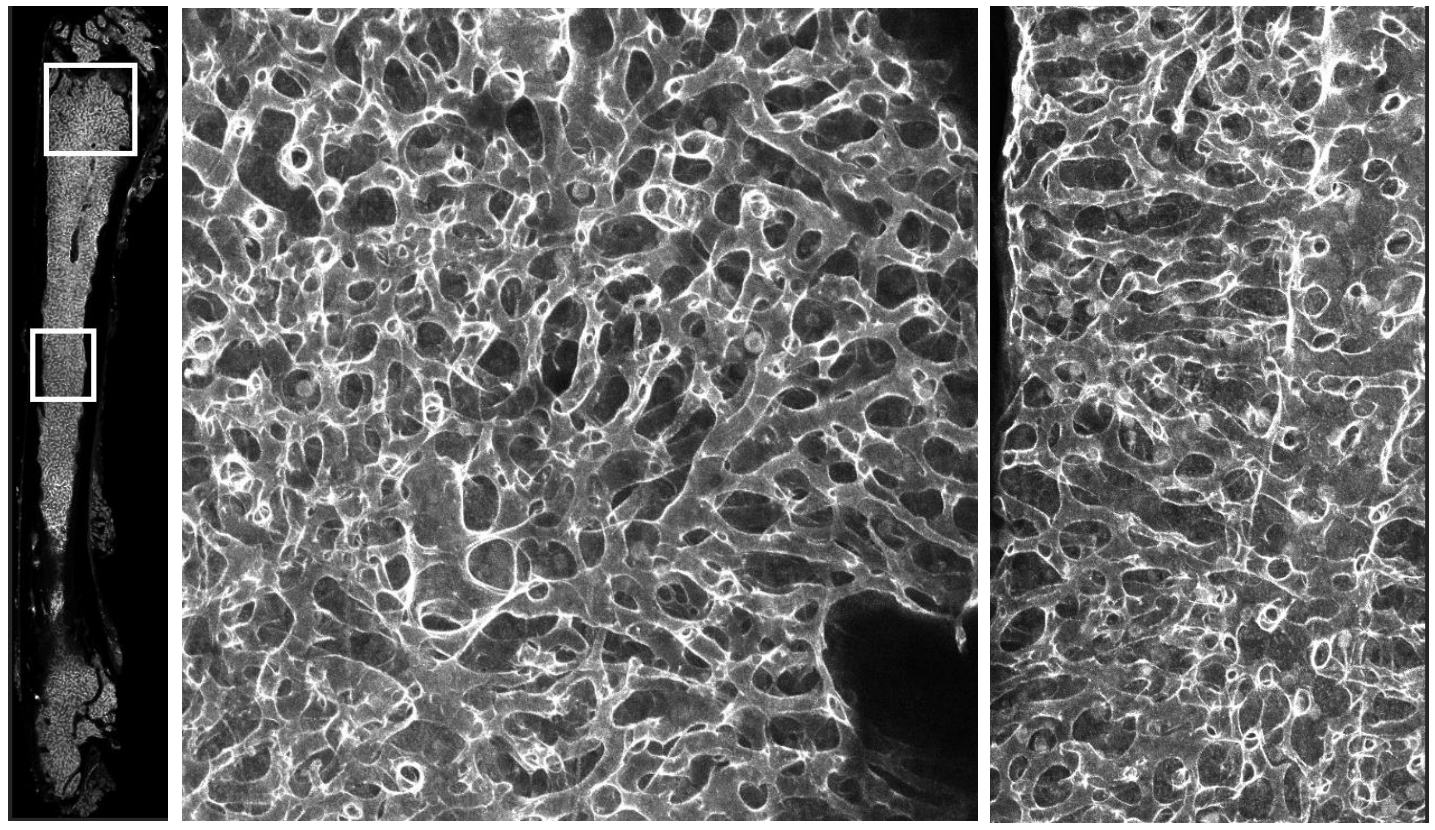}
    \caption{Confocal images of the femur bone of a healthy mouse showing the spongy vasculature within the bone marrow. Central and right images show zoomed-in views of the image on the left, where the white borders in the image indicate the region where the enlarged views are taken from.}
    \label{fig:boneimage}
\end{figure}

Figure \ref{fig:boneimage} shows the dense network of vessels within a healthy bone marrow visualized through confocal microscopy. This network consists of sinusoidal and other capillaries which facilitate the transport of oxygen, nutrients, waste products, as well as blood cells around the bone marrow. The complexities of the spatial arrangement of the vessels and of the interchanges within this network are inherently optimized to enable the essential functions of the bone marrow. Such a network varies from region to region along the femur; additionally, the diverse branching pattern in the capillaries poses a significant challenge in the task of quantifying the vessel structure. Additional factors, such as age, size, and physiological conditions also affect the structure of the vessels. 

\begin{figure}
    \centering
    \includegraphics[width=.5\linewidth]{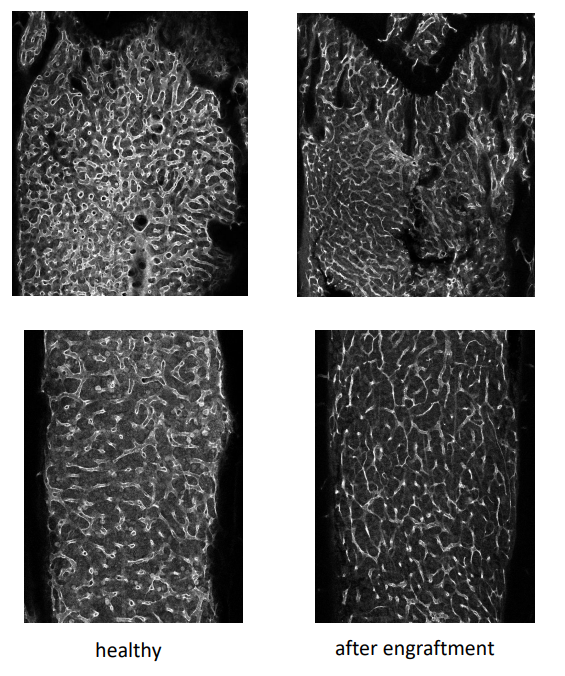}
    \caption{Figures show how healthy bone marrow vasculature (left) is disrupted by AML (right). AML increases the permeability of the vessels and induces greater irregularity and abnormal behavior.}
    \label{fig:healthy_engrafted}
\end{figure}

When AML cells are engrafted into the bone marrow, they release cytokines---growth factors and other molecules which alter the bone marrow niche into one which favors the proliferation of AML cells. Observable alterations include increases in vessel density, proportion of endothelial cells, vascular leakiness, as well as vessel compression, hypoxia, and altered perfusion \citep{passaro}. Figure \ref{fig:healthy_engrafted} shows a comparison between the vessel structure over corresponding regions in a healthy bone marrow and a bone marrow engrafted with AML. We observe that within the healthy bone marrow, the vessels form a smooth, homogeneous, and regular network, whereas in the bone marrow after engraftment, the vessels become thinner, inhomogeneous, more permeable, and irregular.

\begin{figure}
    \centering
    \includegraphics[width=.8\linewidth]{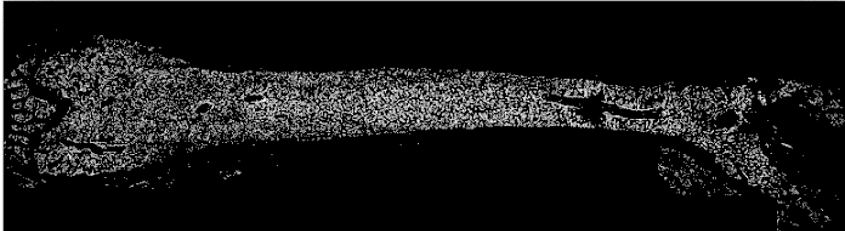}
    \caption{An example slice of the 3D bone marrow vasculature from a mouse femur imaged under the confocal microscope. In this study, we focus on the knee region, located on the left side of the image; the remaining portion of the femur is referred to as the long region.}
    \label{fig:confocalimage}
\end{figure}

Figure \ref{fig:confocalimage} shows a 2D cross-section of a bone under the confocal microscope; in particular, we focus on imaging the endothelial cells lining the bone marrow vessels, which give a complete view of the vasculature. A collection of 27 femur bones were selected for this study to determine how the microenvironment of the vessels in the bone marrow is remodeled by the introduction of various types of leukaemic and non-leukaemic cells at different stages of engraftment. The bone samples are outlined as follows:

\begin{itemize}
    \item 4 CTRL samples: Control samples with no cell injection, hence 0\% engraftment.
    \item 4 MNC samples: Samples engrafted with mononuclear cord blood cells extracted from human umbilical cord at 53\%, 67\%, 75\%, 86\% engraftment.
    \item 7 U937 samples: Samples injected with an immortalized cell line at 1\%, 7\%, 8\%, 10\%, 10\%, 10\% engraftment.
    \item 3 HL60 samples: Samples injected with an immortalized cell line from a different subtype to U937 at 23\%, 25\%, 25\% engraftment.
    \item 6 P1 samples: Samples injected with ``patient 1's'' AML cells at 10\%, 40\%, 44\%, 51\%, 60\%, 76\% engraftment (``patient 1'' is a pseudo ID).
    \item 3 P2 samples: Samples injected with ``patient 2's'' AML cells at 59\%, 88\%, 90\% engraftment (``patient 2'' is a pseudo ID).
\end{itemize}

Biologically, we expect the CTRL samples to display normal murine vessel structure within the bone marrow since these samples were not subjected to any alterations. The MNC cells were harvested from blood sampled from the umbilical cord of human newborns, consisting of a variety of blood cells with a single nucleus, such as lymphocytes, monocytes, eosinophils, basophils, and various stem/progenitor cells. These samples are designed to play the role of ``fake'' control samples for when healthy human cells are engrafted into the murine bone marrow. U937 and HL60 bone samples were injected with different types of cultured immortalized cell lines from human extracts, meaning that the injected population of cells had particular mutations which allowed them to evade normal cellular senescence and proliferate indefinitely. These cells were cultured to retain certain interesting phenotypes and functions before being injected into mice. Notably, these cell lines are monoclonal, meaning that the whole population of cells descends from the same ancestor cell. In contrast, P1 and P2 bone samples were injected with polyclonal cells extracted from human patients, which means that the collections of cells have much more complex mutation landscapes and there is no control over the types of clonal cells that were injected. There are several possible scenarios for the engraftment of these cells in the bone marrow. For example, some clonal subpopulations may be more aggressive and grow more rapidly than others. Alternatively, different clonal subpopulations may cohabitate without singular or group dominance in the whole population. The spatial distribution of these subpopulations is unknown.\\

\noindent
\textbf{Preprocessing and Image Segmentation.} Confocal microscopy of the bone samples generates multiple 2D slices along the depth of the samples which can be stacked together to recreate the 3D images.
To preprocess the stack of confocal microscopy images embedded with noise and distortions, we pass a median filter and apply Niblack local thresholding \citep{niblack1985introduction} to the images to approximate the positions of the vessels and cavities in the image slices. We reconstruct the 3D structures, ensuring linkage between the bone marrow vasculature within each slice using a Willmore regularization method \citep{bretin,Song2022}. Details of the image segmentation and reconstruction process can be found in \cite{annathesis}.


\section{Background and Preliminaries: Persistent Homology}
\label{sec:preliminaries}

In this section, we provide a brief overview of the techniques from topological data analysis that we will use to quantify the confocal images described in the previous section towards our goal of infering on disease progression.

\subsection{Persistent Homology}
\label{sec:ph}

Persistent homology is a methodology that adapts the concept of homology from algebraic topology to the computational setting, notably for real data analysis. Classical homology algebraically counts the number and types of holes of sets and spaces, as a means to characterize them and distinguish between sets and spaces. Here, ``holes'' correspond to connected components, loops, voids, and higher dimensional analogues; specifically, we are interested in those holes which are invariant to continuous deformations, such as stretching and compressing. The \emph{degree} of homology corresponds to different types of holes: degree 0 homology captures the connected components; degree 1 homology captures the loops; degree 2 homology captures the voids or ``bubbles'', and so on. In the context of our application, we are interested in the first 3 degrees of homology, where intuitively, degree 0 homology should provide information corresponding to the network of interconnected vessels running throughout the bone, while degrees 1 and 2 homology capture loops and empty nodules created within such a network.

Computing homology is a difficult task, so computational topology techniques largely rely on \emph{simplicial} homology, where the set or space of interest is discretized into a skeletal structure, where there exist tractable and efficient algorithms to compute homology. The discretization of the set or space of interest represents it as a simplicial complex, which is composed of simplices---vertices, edges, triangles, and higher dimensional structures---assembled in a combinatorial manner, where every face of a simplex in the complex must also belong to that complex.  We can then compute the homology of the simplicial complex combinatorially; we use $H_p$ to denote the $p$th homology group consisting of $p$-simplices.  

In topological data analysis, the approach to discretization of our set or space is important: typically, this is done by sampling a number of points from the set or space, then connecting them to create the simplicial complex by a predetermined connection rule.  A popular and intuitive choice for computation is the \emph{Vietoris--Rips complex}, where the connection rule is that two points are joined by an edge if two balls with the points as centers overlap; higher dimensional structures are created in a similar manner by considering the overlapping of the balls for more points.  A crucial question in topological data analysis, then, is to determine the ``correct'' radius of the balls in order to construct the simplicial complex to then compute homology.  The answer that persistent homology gives is to consider a spectrum of radii, rather than one fixed radius, in order to produce a multiresolutional summary of the homology.  The growth of the radii are dictated by a continuous, real-valued \emph{filtration function}, which produces a \emph{filtration}---an increasing collection of nested subcomplexes to the simplicial complex representing the discretized set or space of interest---according to the predetermined connection rule.

\begin{figure}
    \centering
    \includegraphics[width=\linewidth]{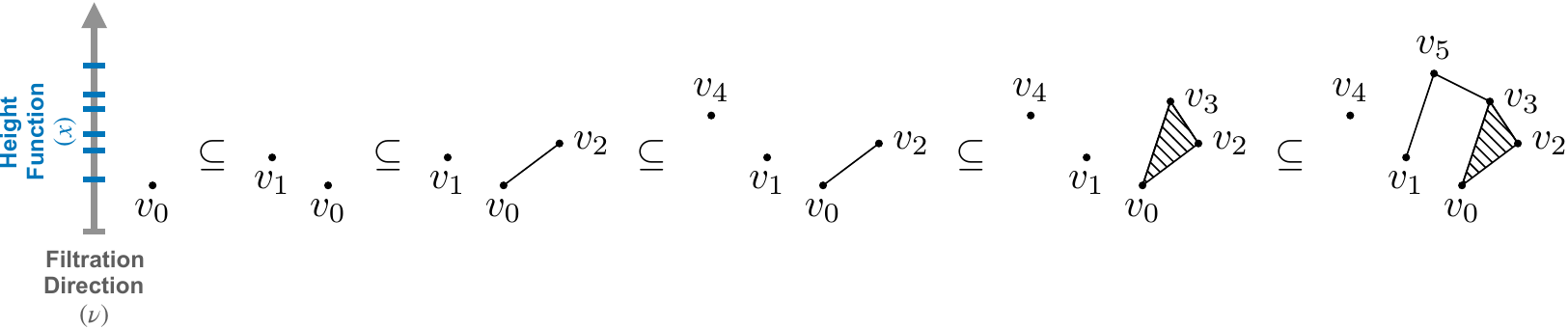}
    \caption{Illustrating a sublevel set filtration by height function \citep{crawford}.  For the vertical direction $\nu$, the value of the height function is tracked by $x$, where as $x$ increases, more topological structure of the simplicial complex $K$ in terms of vertices, edges, and triangles is revealed.}
    \label{fig:sublevel_ex}
\end{figure}

When the set or space of interest is a shape, as in our work, a typical approach to the multiresolutional computation of homology is by considering the \emph{height function} to construct the filtration. Let $f:K\rightarrow \mathbb{R}$ be such a continuous real-valued function on the simplicial complex $K$ representing the topological space. Then the sequence of subcomplexes ${K_x}$ are defined as
$$
K_r= f^{-1}(-\infty, r],
$$
for $r\in \mathbb{R}$. This is referred to as the \emph{sublevel set filtration}, with the property that $K_a \subseteq K_b$ for all $a\leq b \in \mathbb{R}$. An illustration of a sublevel set filtration dictated by the height function is given in Figure \ref{fig:sublevel_ex}, originally appearing in \cite{crawford}.

Persistent homology computes the homology over this sequence of subcomplexes and allows us to track how the topological features of the data, summarized by the number and types of simplices, evolve over the different resolutions. Each feature is tracked by an interval, where the left endpoint (usually referred to as the \emph{birth} time) indicates when the feature appeared, and the right endpoint (usually referred to as the \emph{death} time) indicates when that feature disappears (by absorption into a larger structure), lending interpretability to the procedure: longer intervals usually refer to ``more important'' features while shorter intervals are usually considered to be topological ``noise.''  Formally, this interpretability is given by the \emph{structure theorem for persistent homology} \citep{crawleyboevey2014decomposition}, where the geometric and topological information encoded by the homology groups together with the mapping induced by the inclusions between the sublevel sets generate a \emph{persistence module} $\text{PH}_p(K_\bullet)$ for a fixed degree $p$, which decomposes uniquely (up to isomoprhism) into a direct sum of irreducible interval modules.  The collection of these intervals is the output of persistent homology; it is typically plotted as a scatterplot in the upper half plane, together with the \emph{diagonal} (i.e., the set of zero-length intervals where birth is equal to death), in a \emph{persistence diagram}.

\subsection{Signed Distance Persistent Homology}
\label{sec:sdph}

For our application goal of quantifying bone marrow vasculature imaging and inferencing on AML, we specifically focus on computing persistent homology on filtrations constructed using the \emph{signed distance function}. We refer to this procedure as \emph{signed distance persistent homology} (SDPH). 

For any $A\subset\mathbb{R}^n$, we first define the distance function from any point $x$ to $A$ as
\begin{equation}
\label{eq:dist}
\text{dist}(x, A) = \inf_{y\in A} \|x - y \|.
\end{equation}
This is the distance from $x$ to the closest point in $A$ to $x$. Then, to distinguish whether a point is inside or outside $A$, we apply the notion of sign: the \emph{signed distance function} $d$ at any point $x \in \rn$ is defined 
\begin{equation}
\label{eq:signed_dist}
d(x) = \text{dist}(x, A) - \text{dist}(x, (A^c)^o),
\end{equation}
where $(A^c)^o=A^c \cap (\partial A)^c$ and $\partial A$ is the boundary. This means that any point inside of $A$ will have a negative signed distance, where the absolute value of this distance is equal to the distance to its closest point on the boundary; any point outside of $A$ will have a positive signed distance, indicating the shortest distance to the same boundary. In our application of AML bone marrow vasculature imaging, we assume $A$ to be a nonempty region in $\mathbb{R}^3$ bounded by the vessel walls and impose $C^k$ regularization on the boundary $\partial A$ matching the natural setting of the vessel walls. Then $(A^c)^o$ refers to exactly the extravascular space outside the vessels.

\cite{song2023generalized} developed a generalized Morse theory of the signed distance function in the context of persistent homology; Morse theory is a branch of differential topology that studies topological features using differentiable functions, specifically, in terms of critical points.  By Corollary 5 of \cite{song2023generalized}, for a generically embedded surface in $\mathbb{R}^3$, the signed distance function is a topological Morse function with finitely many critical points. In this setting, there exists a direct correspondence between the non-degenerate critical values of the signed distance function and the interval decomposition of the corresponding persistence module $\text{PH}_p(K_\bullet)$ for fixed degree $p$, in particular, a correspondence to either the birth or death times of topological features captured by homology.


\section{Quantifying AML with Signed Distance Persistent Homology}
\label{sec:analysis}

In this section, we present our first contributions, which is a quantification and analysis of AML bone marrow vasculature imaging using SDPH.  In particular, we show that SDPH is able to rigorously capture the quantitative information within confocal bone marrow vasculature imaging in a way that is interpretable and indicative of different disease stages, given the data.

\subsection{Applying SDPH and Data Interpretation}
\label{sec:interpretation}

\cite{song2023generalized} show that there are six types of non-degenerate critical points of the signed distance function, each with its individual geometric interpretation and corresponding to either a birth or death time of a homological component. In their original notation (see \cite{song2023generalized}), the six types of critical points were referred to as the $0^-$, $1^-$, $2^-$, $1^+$, $2^+$, and $3^+$ critical points, corresponding to their Morse index and whether the points occur within or outside the surface. In this paper for convenience and ease of interpretation, we will alter the notation and rename the critical points as the  $r_0$, $r_1$, $r_2$, $g_1$, $g_2$, and $g_3$ sizes, respectively, denoting the absolute values in the original notation. This relabeling is done with the intuition that the $r_0$, $r_1$, and $r_2$ sizes measure different behaviors in \textit{radius} or thickness of vessels, and that the $g_1$, $g_2$, and $g_3$ measure behaviors in the in interstitial \textit{gaps} between vessels in our imaging application. With this relabeled notation, the morphological interpretation of the types of critical points as measured inside the bone marrow architecture are given as follows.
\begin{itemize}
    \item $r_0$ points characterize the vessel thicknesses at bumps or widening in the vessels
    \item $r_1$ points characterize the vessel thicknesses at narrowing or bottlenecks in the vessels
    \item $r_2$ points characterize the vessel thicknesses where there are dimples or hollows in the vessels
    \item $g_1$ points characterize the spaces between vessels where there are narrowings in vascular loops
    \item $g_2$ points characterize the spaces between vessels where there are widenings in vascular loops
    \item $g_3$ points characterize pockets of space surrounded by vessels
\end{itemize}

\begin{figure}[tb]
    \centering
    \begin{subfigure}[b]{0.3\linewidth}
        \includegraphics[width=\linewidth]{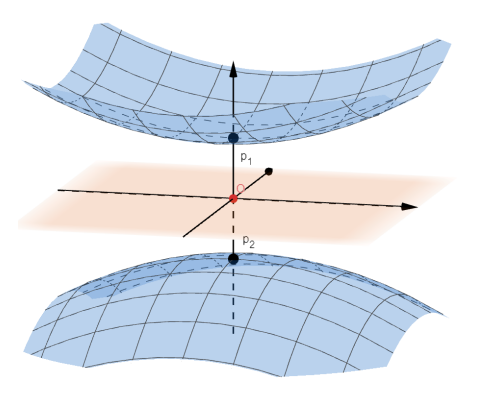}
        \caption{$g_1$ or $r_2$ critical point}
        \label{fig:ex1}
    \end{subfigure}
    \begin{subfigure}[b]{0.3\linewidth}
        \includegraphics[width=\linewidth]{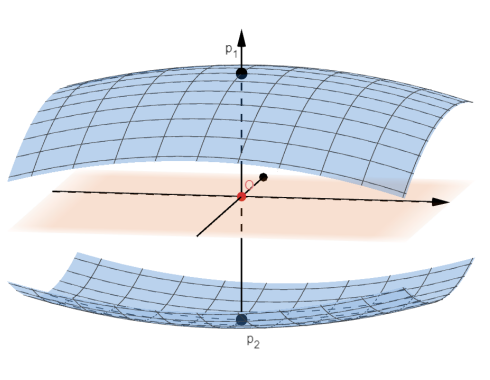}
        \caption{$g_3$ or $r_0$ critical point}
        \label{fig:ex2}
    \end{subfigure}
    \begin{subfigure}[b]{0.3\linewidth}
        \includegraphics[width=\linewidth]{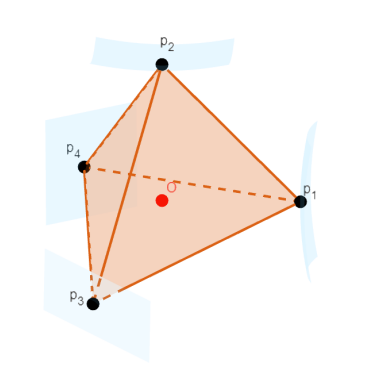}
        \caption{$g_3$ or $r_0$ critical point}
        \label{fig:ex3}
    \end{subfigure}
    \caption{Examples of local morphology of non-degenerate critical points from \cite{song2023generalized}.}\label{fig:example_critical_sizes}
\end{figure}

Figures from \cite{song2023generalized} illustrate the local behavior found near different types of critical points; a selection is presented in Figure \ref{fig:example_critical_sizes}. In these illustrations, the red dots represent the non-degenerate critical points and the black dots their closest points on the surrounding surfaces respectively. In Figure \ref{fig:ex1}, we observe the behavior around either a $g_1$ or $r_2$ critical point. The difference arises from whether we consider the critical point to be inside or outside the lumen. If the point is inside the lumen, then the construction mimics a critical point of type $r_2$ found within a biconcavity of a dimple on a blood vessel. Alternatively, if the point is outside the lumen, then the two pieces of surface in the construction represent the outside linings of two vessels and a critical point of type $g_1$ found between them in the narrowing formed. Both Figures \ref{fig:ex2} and \ref{fig:ex3} illustrate the possible local morphologies around a type $g_3$ or $r_0$ critical point. In Figure \ref{fig:ex2}, the critical point is found within two concave pieces of surface, which may either be the vessel walls around a $r_0$ critical point or the inner lumen of a concavity surrounded by vessels. Figure \ref{fig:ex3} more clearly shows the construction of a critical point found within an open cavity.

\begin{figure}
    \centering
    \includegraphics[width=\linewidth]{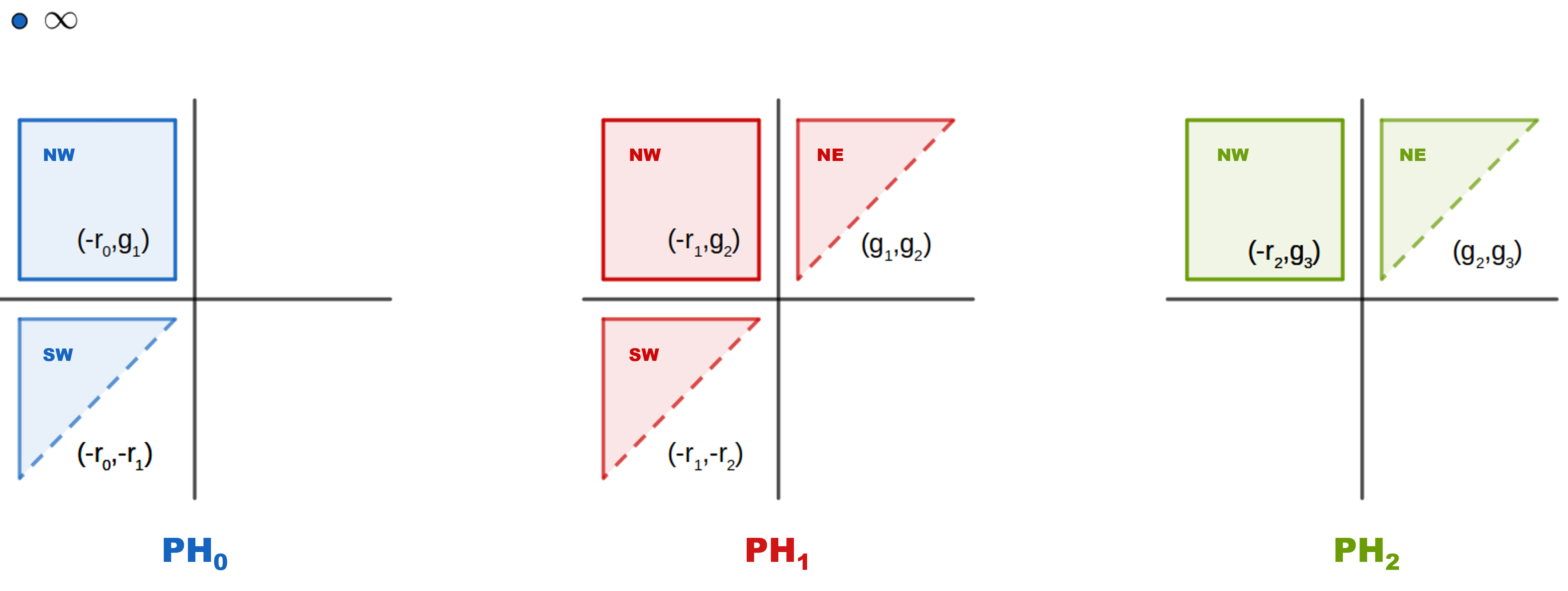}
    \caption{Seven types of pairings of the critical points in each of the quadrants of the persistence diagrams from \cite{song2023generalized}.}
    \label{fig:seven}
\end{figure}

Among the different types of critical points, we identify seven types of pairings arising from the location of the pairs of critical points as birth or death points within quadrants of the persistence diagrams as shown in Figure \ref{fig:seven}. The seven types of pairings are the following:
$$
(-r_0, -r_1), \, (-r_0,g_1), \, (-r_1,-r_2), \, (-r_1,g_2), \, (g_1,g_2), \, (-r_2,g_3), \, (g_2,g_3).
$$
They occur within different quadrants of the persistence diagrams.
We refer to the 0-, 1-, and 2-degree persistence diagrams as $\text{PH}_0$, $\text{PH}_1$, and $\text{PH}_2$. Here, we use cardinal directions to indicate quadrants of the persistence diagrams; recall that since we are using a signed distance filtration function, we have negative birth and death values occurring on the persistence diagrams. 

The pairings of critical points within different quadrants present further geometric interpretations. For example, within the PH$_1$ north-west (NW) quadrant, the number of birth--death points corresponds to the number of loops in the sample, while the individual birth and death times measure the thicknesses and sizes of the loops up to a difference in sign. We also note that the pairing of $-r_2$ and $g_3$ critical sizes should rarely occur for vessels as this alludes to a bubble of extravascular space fully trapped inside the lumen, which intuitively should not normally appear.

Moreover, within PH$_0$ SW, the paired $(-r_0, -r_1)$ critical sizes reveal the fluctuations in the thickness of the vessels. We define a measure for such behavior, which we call the \emph{undulation aspect ratio}, computed by $1-r_1/r_0$, where a value of $0$ represents no fluctuations in vessel thickness. Similarly, in PH$_1$ NW, on the pairing of $(-r_1, g_2)$ critical sizes, we define a \emph{loop aspect ratio} between the thickness of the vessel and the size of the loop formed by the vessel, computed by $g_2/r_1$. A final measure that we introduce is the \emph{waviness aspect ratio}, measured in PH$_1$ NE by $1-g_1/g_2$, which gives a comparison of the extravascular space in different regions where loops are formed by the vessels.

We utilize the critical sizes in analysis of the bone marrow imaging data in two different ways: one is a \emph{``local'' texture analysis} and the other is a \emph{``global'' texture analysis}. In the local analysis, we compute the morphology (i.e., the SDPH diagrams and its summary of critical points) over each image segmented into numerous small regions with the goal of distilling the heterogeneous textures into mixtures of similar and more homogeneous textures through clustering. Then, based on this decomposition into homogeneous textures for each image, we further group the images into what we refer to as \emph{phases}, which correlate with the level of engraftment. For the global analysis, we instead compute the morphology over the whole images and using the density of points on the SDPH diagrams, we find that the clustering behaviors of the approximate densities are in concordance with the results of the local analysis.

It is important to note that there are distinctions between the following terms used in this paper: The \emph{actual} stage of engraftment of the AML or injected cells within the mice, which varies from region to region and cannot be directly measured, which differs from the \emph{estimated} stage of engraftment, which is a global estimate of the amount of engraftment through the process involving crushing all the remaining bones of the mice other than imaged femur bone imaged.  We refer to the estimated stage of engraftment in this paper as simply the level of engraftment. We refer to clusters of similar morphology evaluated with local analysis as \emph{textures} and we observe the imaged samples as being composed of different textures.

\subsection{Spatial Texture Decomposition via Local Texture Analysis}
\label{sec:local_analysis}

We first perform a local texture analysis using the measures and critical sizes as defined in Section \ref{sec:interpretation}, we seek to analyze the localized textures by computing ``local'' SDPH diagrams at equidistant locations throughout the 3D image sample for each bone and determining how the distribution of textures on the bone differ with respect to the progression of the disease. 

We focus our analysis on the textural composition of the knee region in femoral bones by extracting local features within 3D balls centered on a regular grid. Each ball has a radius of 40 voxels in the horizontal plane and 1/5 of the sample thickness in the vertical plane, and ball centres are spaced 5 voxels apart. This yields over 22 million overlapping regions across the full dataset of 27 bone samples. The choice of a 40-voxel radius (approximately 80$\mu$m, since each voxel measures $\sim$2$\mu$m) balances biological relevance and computational feasibility. At this scale, each ball typically captures multiple sinusoidal capillaries (10--20$\mu$m in diameter), along with surrounding tissue and cavities, allowing for informative characterization of local morphology. Smaller radii risk producing sparse or uninformative topological features, while larger radii may over-smooth and obscure important heterogeneity.

To compute SDPH diagrams, we first divide each bone image into 8 coarse subregions over the knee, which support distributed processing and spatial stratification. Within each subregion, we extract SDPH diagrams for each individual ball by including only those topological features whose birth and death points $(b_i, d_i)$ lie entirely within the ball. Each ball-level diagram provides a localized summary of vessel structure. We remove points with persistence less than 0.5 (roughly 0.25$\mu$m), as these are likely to correspond to imaging noise, given the scale of the vessels of interest (10–20$\mu$m).

We summarize each local SDPH diagram using the following 15 features:
\begin{itemize}
    \item means and standard deviations of $r_0$, $r_1$, $g_2$, and $g_3$ critical sizes;
    \item means and standard deviations of undulation aspect ratios, loop aspect ratios, and waviness aspect ratios;
    \item standard deviation of the collection of $(r_1,g_2)$ pairs as an additional measure of spread in PH$_1$ NW, capturing the variation in loop sizes and thicknesses across the network.
\end{itemize}
After normalizing the localized feature vectors, we further group these feature vectors into clusters comprising similar textures. Our dominant approach utilizes the $k$-means algorithm \citep{macqueen1967some} in clustering the locations into groups. The $k$-means algorithm is an iterative method to partition the dataset into $k$ clusters of similar features. By initializing on some random centroids, then iteratively assigning data points to their nearest centroid, then updating the centroids with the means of the assigned data points, we converge to stable centroids and cluster assignments for the data.  We also considered other approaches using alternative clustering techniques, such as Gaussian mixture model clustering \citep{gmm} and clustering for large applications (CLARA, \cite{clara}) which yielded similar results. These results can be found in Section \ref{app:clustering_exploration} of the Supplementary Material.

Figure \ref{fig:knee-texture} shows the distributions of groups of different textures in the knee regions of the bone for the different samples when the number of clusters is set to $k = 3$, represented in the figures by the different colors projected back onto their locations in the bone image. The projected distribution for the long regions of the bone is available in the Supplementary Material. From Figure \ref{fig:knee-texture}, we observe that the CTRL samples consist of largely the white texture, which we arbitrarily assign to be texture A, whereas other samples with higher levels of engraftment contain more of the other two dark and light blue colored-textures, which we denoted as textures B and C.  We also identify a transition in the ratio of different textures across the samples. Figure \ref{fig:texture} shows the varying percentage decomposition into A, B, C textures for the knee region of each bone sample. 

\begin{figure}
    \centering
    \includegraphics[width=\linewidth]{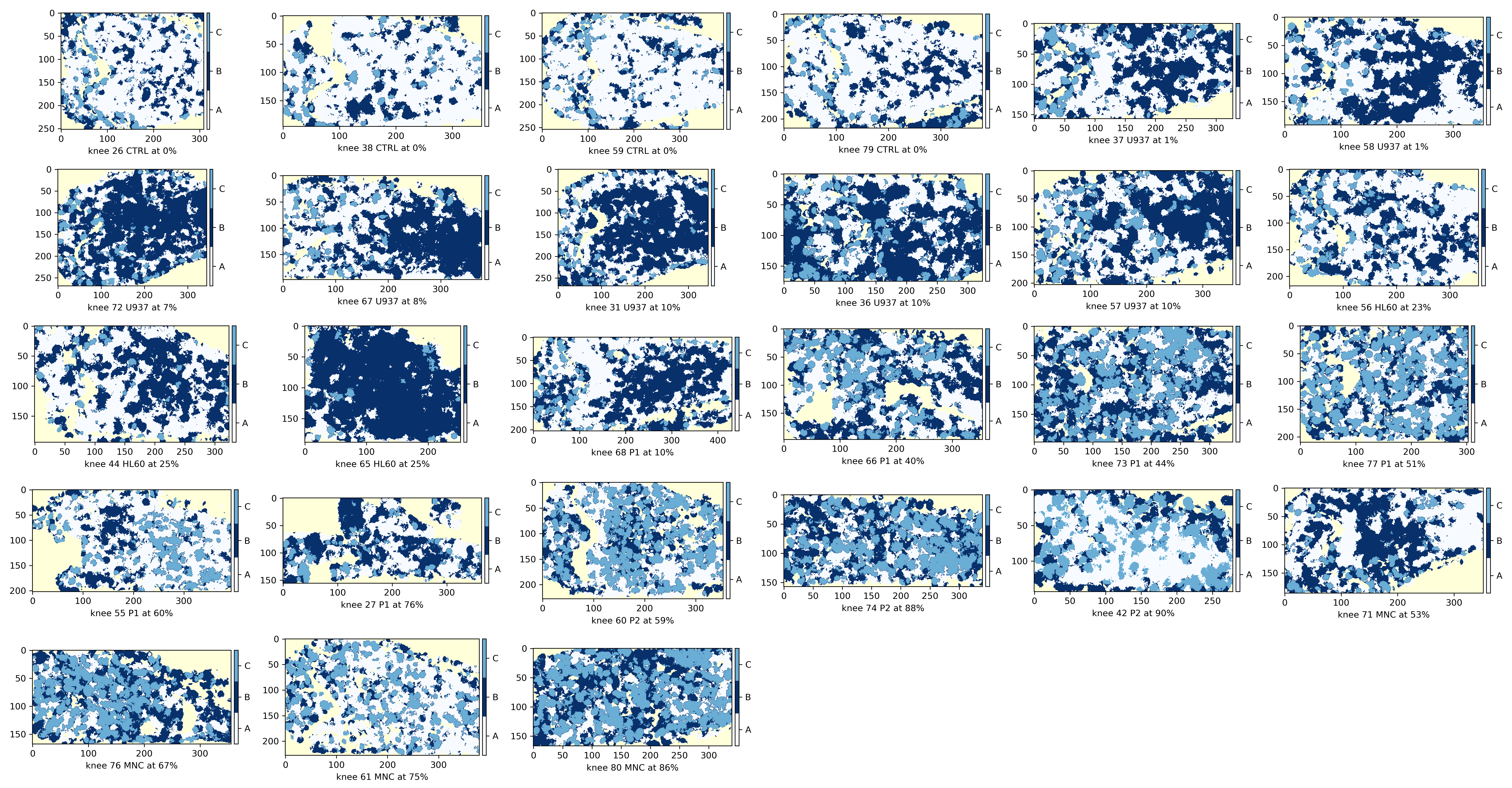}
    \caption{Distribution of 3 groups of textures colored in white, dark blue, and light blue as textures A, B, and C \citep{annathesis}.}
    \label{fig:knee-texture}
\end{figure}

\begin{figure}
    \centering
    \includegraphics[width=.8\linewidth]{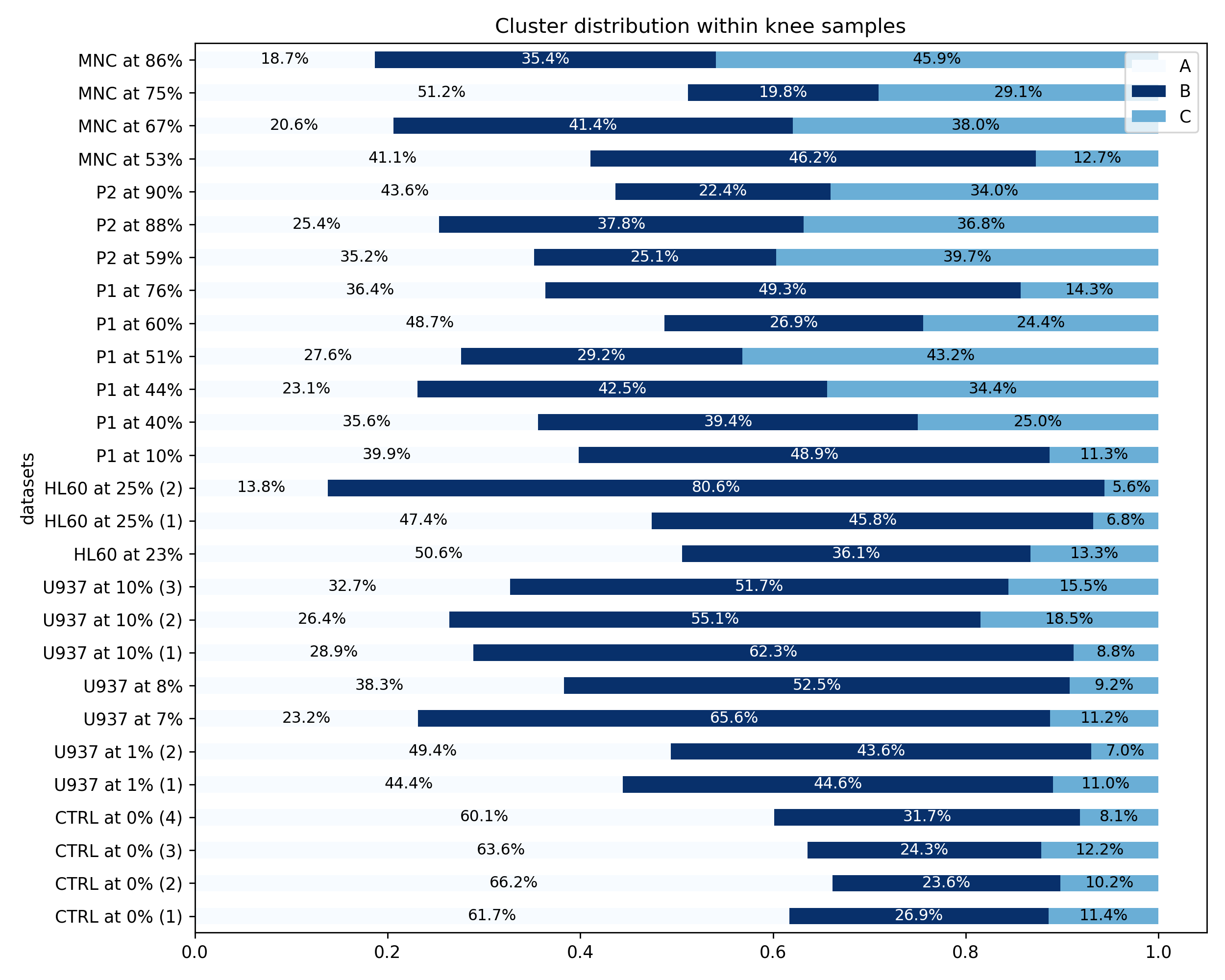}
    \caption{Texture percentage decomposition for the knee region of sample considering 3 clusters as in Figure \ref{fig:knee-texture} \citep{annathesis}.}
    \label{fig:texture}    
\end{figure}

\begin{figure}
    \centering
    \includegraphics[width=.8\linewidth]{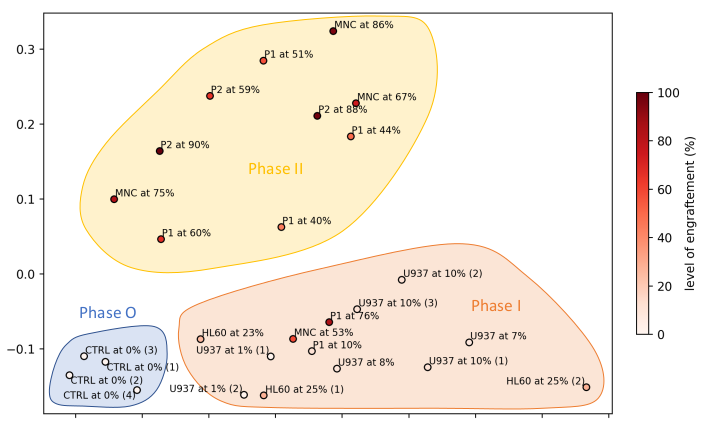}
    \caption{Visualizing the texture decomposition of Figure \ref{fig:texture} in 2D using PCA and highlighting the phase groupings \citep{annathesis}.}
    \label{fig:pca}
\end{figure}

Figure \ref{fig:pca} shows the percentage decomposition of the 3 textures projected onto two dimensions for visualization using principal component analysis (PCA), a dimensionality reduction technique which seeks to emphasize the significant features of the data while retaining as much variation of the data as possible. Based on this, we can approximately divide the samples into 3 groups, which we can see highlighted in the figure in blue, orange, and yellow. It is important to note that this grouping is not unique to this specific combination of $k=3$ and $k$-means clustering algorithm; we find that across a range of values of $k$ and both the GMM and CLARA clustering algorithms, we repeatedly observe clear divisions between the outlined Phase I and II samples. On the other hand, Phase O samples may appear distinct to the other two groups or merged with the other groups for some combinations of $k$ and algorithms, however the occurrence of the latter is the rarer of the two. Certain samples may violate the grouping, for example, for certain combinations, P1 sample at 40\% may appear closer to Phase I samples than Phase II samples. Further supporting evidence is shown in Section \ref{app:clustering_exploration} of the Supplementary Material. 

Overall, based on the percentage texture decomposition, we allocate the knee samples into 3 morphological stages---Phases O, I, and II---as follows:
\begin{itemize}
    \item Phase O consists of all CTRL samples;
    \item Phase I consists of all U937 and HL60 samples, with a MNC sample at 53\%, and P1 samples at 10\% and 76\%;
    \item Phase II consists of all P1 samples except at 10\% and 76\%, all P2 samples, and all MNC samples except at 53\%.
\end{itemize}

Returning to Figure \ref{fig:texture}, we find that, as previously noted, the dominant texture for Phase O samples is texture A, constituting on average over 60\% of the texture decomposition---a considerably higher proportion than that found within the other phases. In contrast, for Phase I samples, the dominance shifts from texture A to texture B, with the samples containing on average between 45\% to 55\% of texture B. Lastly, for Phase II samples, we observe on average a more balanced distribution between the textures, with a similar proportion for each texture. Morphologically, texture A is characterized by thicker vessels compared to textures B and C, whereas texture B features lower $r_1$ values than the other textures and smaller interstitial gaps between the vessels. In contrast, texture C features much larger interstitial gaps between the vessels than found in the other two textures. Further analysis of individual textures can be found in \cite{annathesis}.

Furthermore, we find a correspondence between the phase allocations and the levels of engraftment. By assignment, we imposed that Phase O be solely associated with the CTRL samples (in the ``control'' stage with no engraftment), while Phase I is associated with samples with low percentages of engraftment, mostly varying between 1\% and 25\% of engraftment; we describe this as the early stages of engraftment. In comparison, Phase II contains samples with higher percentages of engraftment, varying between 40\% and 90\%; we associate this with late stages of engraftment. The exceptions to this trend are the two samples MNC at 53\% and P1 at 67\%, whose estimated levels of engraftment may differ to the actual levels of engraftment due to the estimation process described previously.

\subsection{Spatial Texture via Global Texture Analysis}
\label{sec:global_analysis}

\begin{figure}
    \centering
    \includegraphics[width=.8\linewidth]{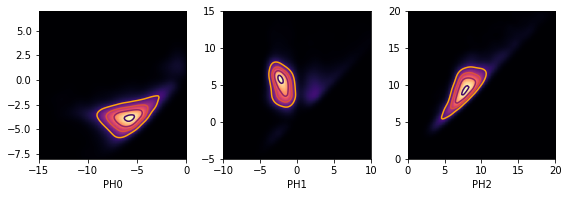}
    \caption{Plots of the persistence-weighted Gaussian kernel estimate for the SDPH diagrams of the knee region of the first CTRL sample with 0\% engraftment. The plots are generated by magma colormap in Python where the darker colours represent regions of low density and lighter colours represent regions of high density.}
    \label{fig:example_heatmap}
\end{figure}

In parallel, we performed a global analysis of the textures by computing SDPH on the entire knee regions of the bone samples and observed commonalities in clustering behavior with the phase groupings from the local texture analysis. To visualize the large collection of birth--death points in persistence diagrams computed for each sample, we compute an approximate density in the form of a Gaussian kernel density estimate. In particular, on each $(\text{birth}, \text{death})$ point in the diagram we applied the Gaussian kernel, $g(x)= \frac{1}{2\pi \sigma^2}\exp(-\frac{\|x\|^2}{2\sigma^2})$ with $\sigma = 0.5$, weighted by their persistence, that is, the difference between the birth and death times, similar to the approaches in \cite{maroulas_non_parametric} and \cite{moon_sedt}. Figure \ref{fig:example_heatmap} shows an example of such Gaussian kernel density estimates (KDEs) computed for the persistence diagrams of the first CTRL sample plotted as contoured heatmaps.

We observe several different types of transitions in density in the different quadrants of the persistence diagrams as the samples vary from different types and levels of engraftments. In particular, Table \ref{tab:transitions} shows a selection of persistence-weighted heatmaps of the density estimates for the PH$_1$ NW quadrants of the SDPH diagrams of the knee samples. We observe increasing variation in $g_2$ values for higher levels of engraftment and from Phase I samples to Phase II samples. This variation suggests that the vascular loops dilate towards later stages of engraftment. Also, moving from Phase O to Phase II samples, we observe a shift in the range of $r_1$ values towards lower values of $r_1$, suggesting thinner vessels and perhaps thinner vessel walls and greater permeability related to later stages of AML. 

Another interesting feature is the spread of the $(g_2,g_3)$ points in the PH$_2$ NE quadrant of the persistence diagrams. We note from Table \ref{tab:transitionsph2ne} that in general, the samples with higher levels of engraftment show greater variation in the distribution of the points in the direction parallel to the diagonal. Other changes can be similarly found in the other quadrants transitioning through the engraftment levels and phases. The full tables of density heatmaps for the knee and long regions are available within Section \ref{app:heatmap} of the Supplementary Material.

\begin{table}
\caption{A selection of heatmaps of each phase with contouring showing the Gaussian kernel density estimates in the PH$_1$ NW quadrant of the persistence diagrams corresponding to the pairing of the $-r_1$ and $g_2$ sizes.}
\label{tab:transitions}
\begin{adjustwidth}{-2cm}{}
\begin{tabular}{cc|cc|cc}
\multicolumn{2}{c|}{Phase O} & \multicolumn{2}{c|}{Phase I} & \multicolumn{2}{c}{Phase II} \\ \hline
\includegraphics[width=8em]{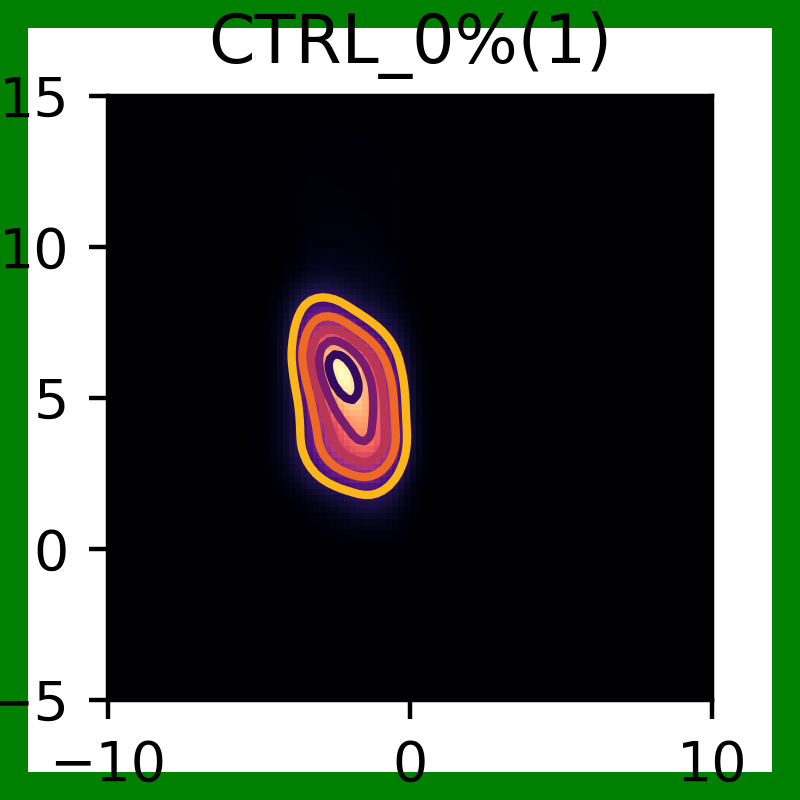} &      
\includegraphics[width=8em]{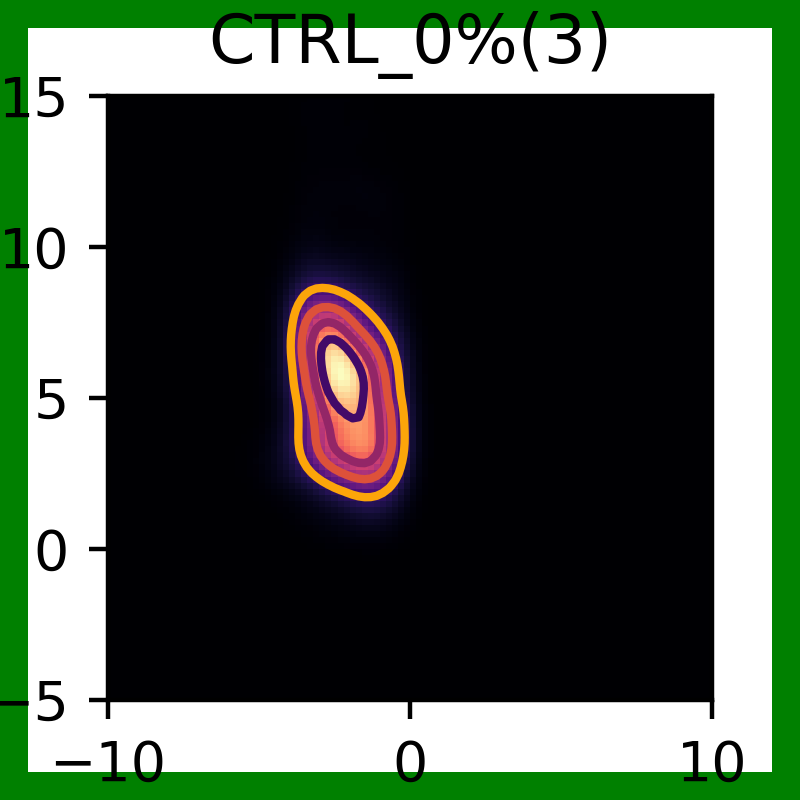} &
\includegraphics[width=8em]{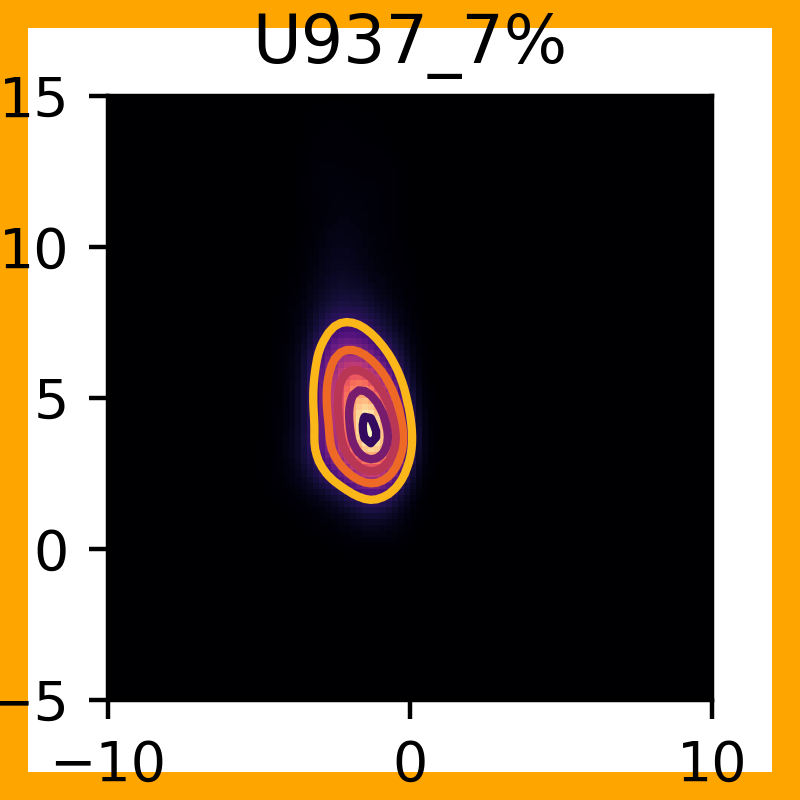} &
\includegraphics[width=8em]{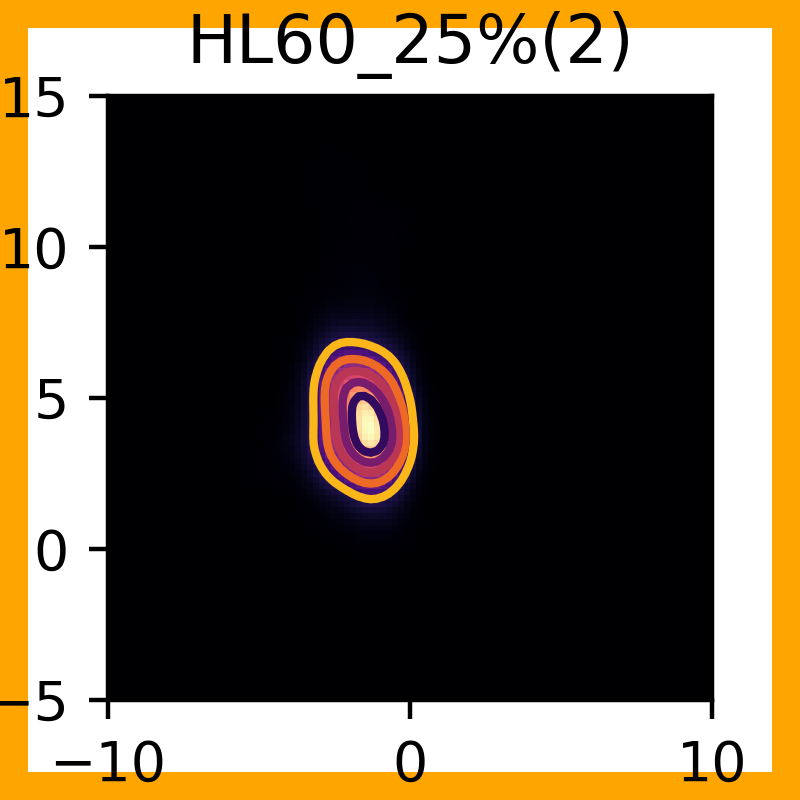} & 
\includegraphics[width=8em]{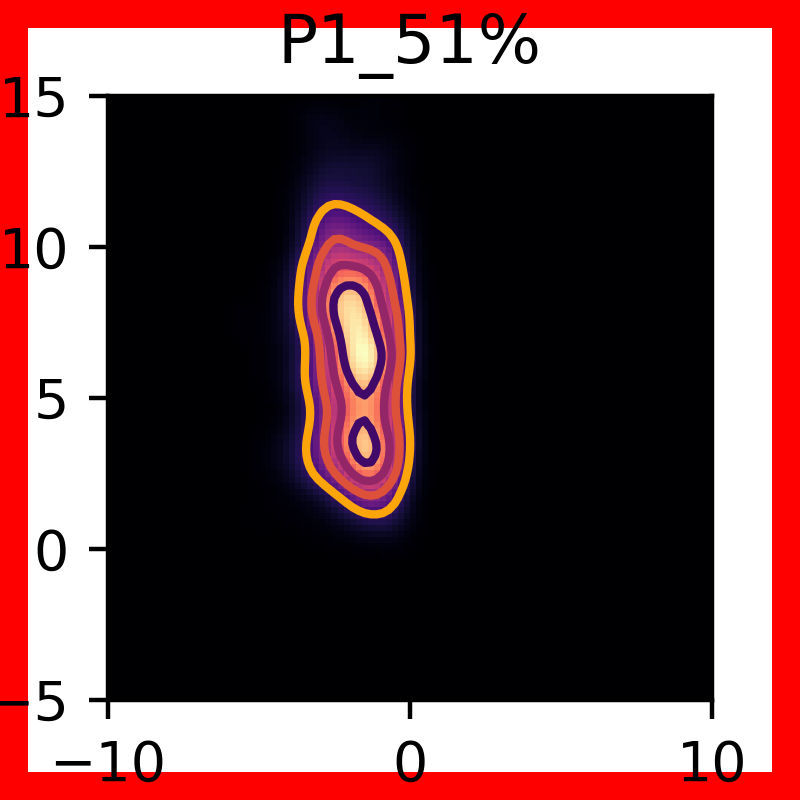} &
\includegraphics[width=8em]{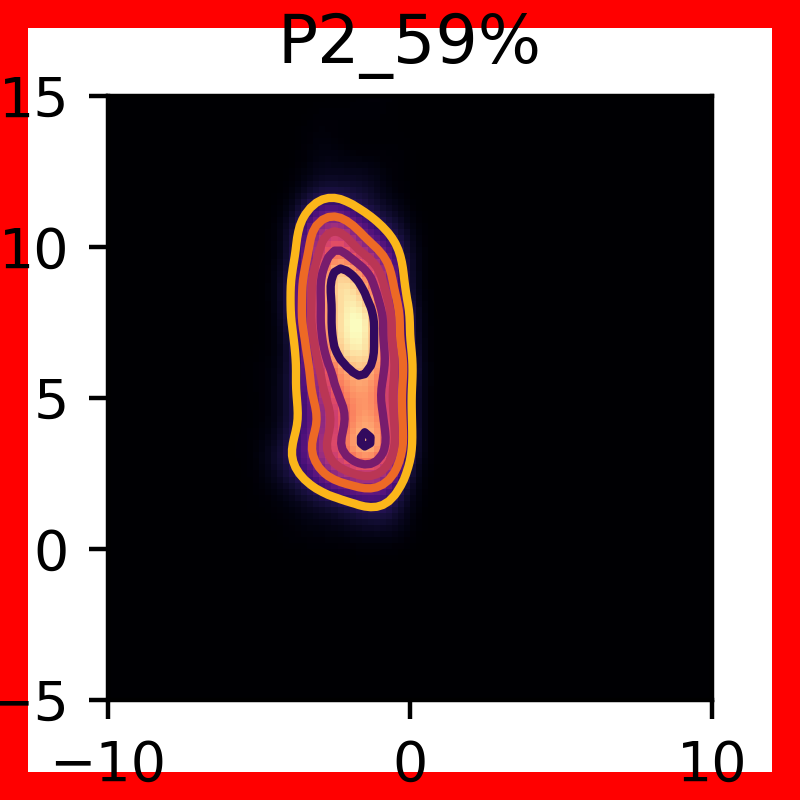}\\ 
\includegraphics[width=8em]{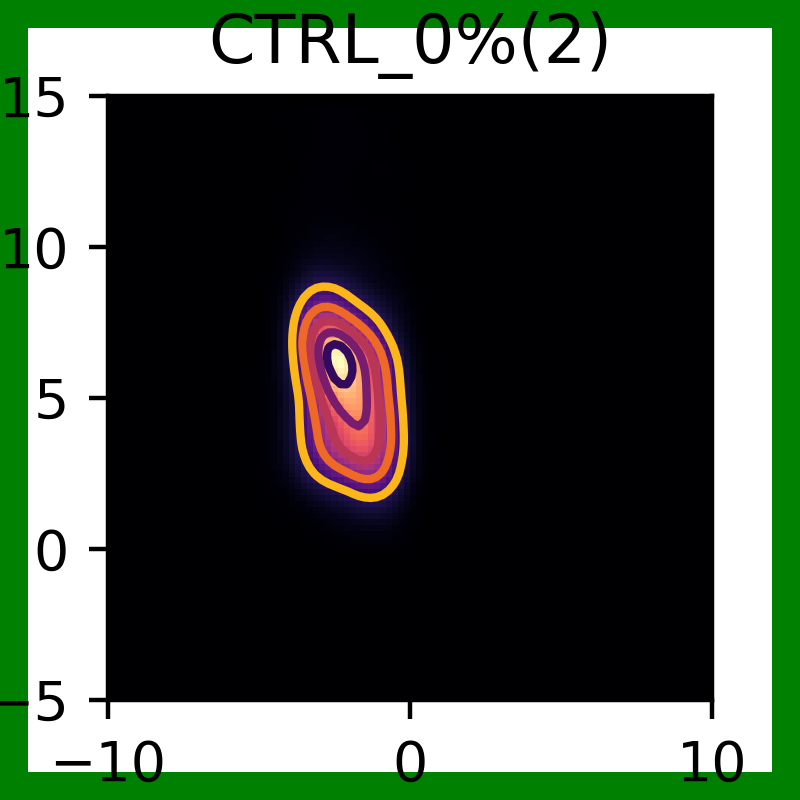} &      
\includegraphics[width=8em]{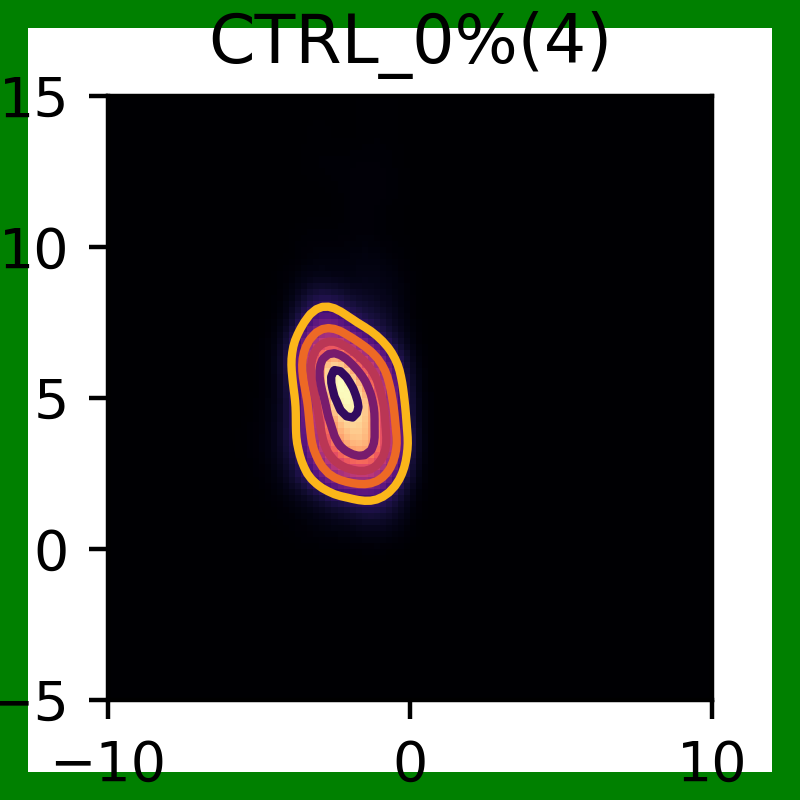} &
\includegraphics[width=8em]{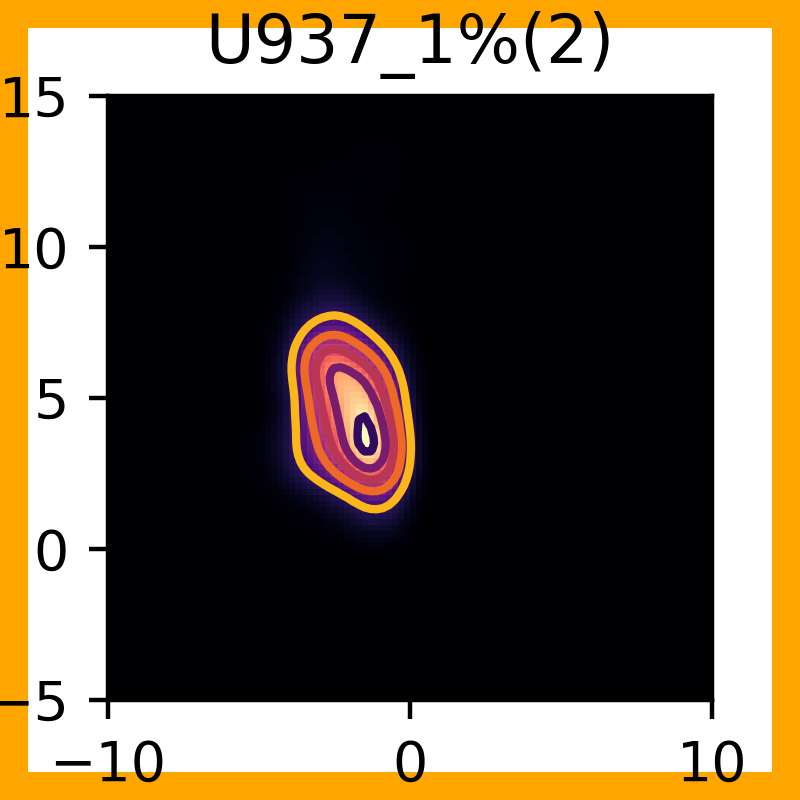} &
\includegraphics[width=8em]{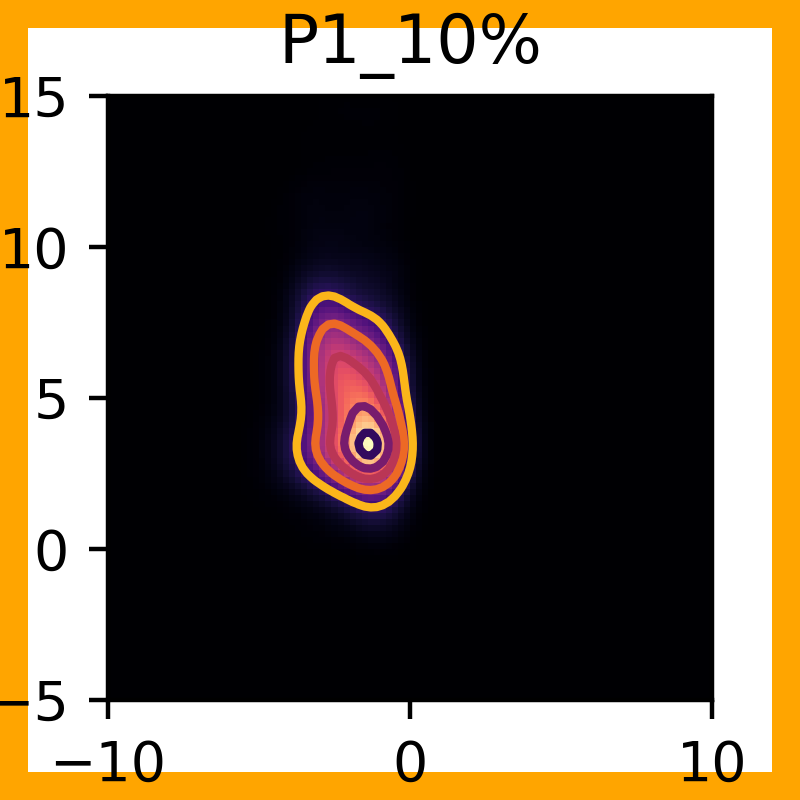} & 
\includegraphics[width=8em]{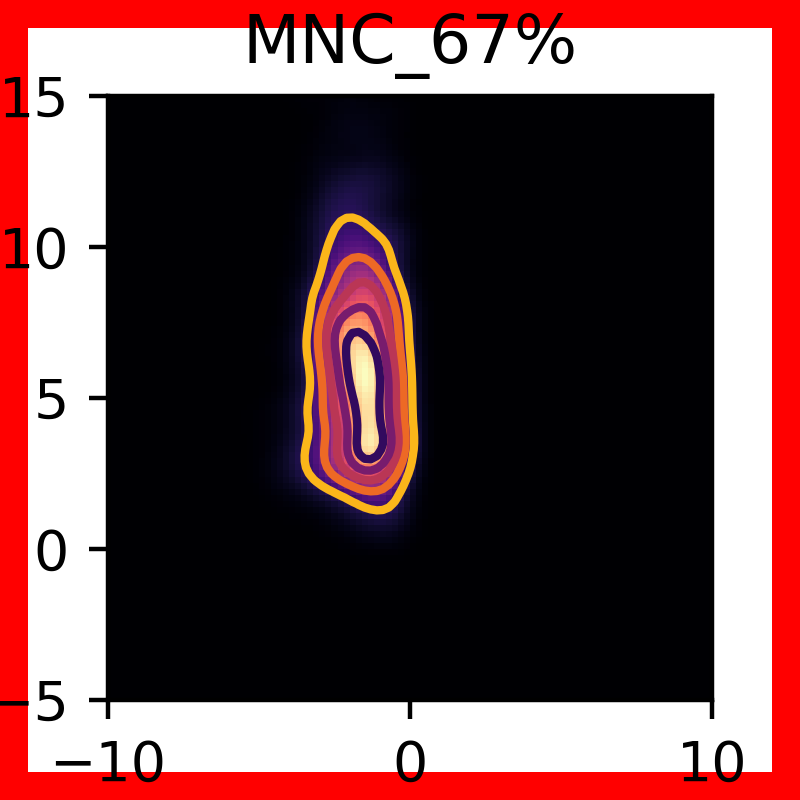} &
\includegraphics[width=8em]{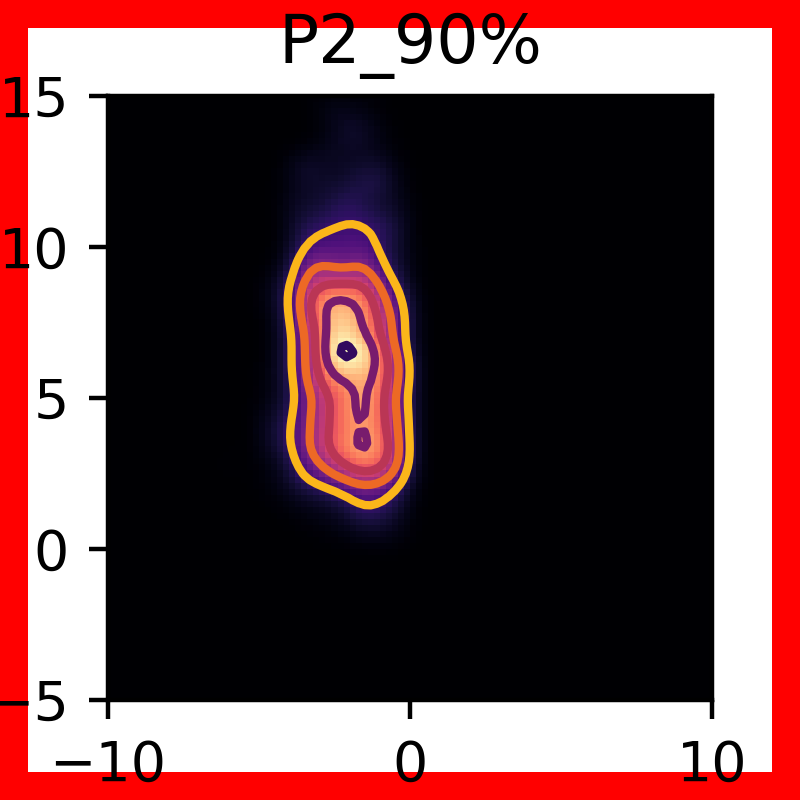}\\ 
\end{tabular}
\end{adjustwidth}
\end{table}

\begin{table}
\caption{A selection of heatmaps of each phase with contouring showing the Gaussian kernel density estimates in the PH$_2$ NE quadrant of the persistence diagrams corresponding to the pairing of the $g_2$ and $g_3$ sizes.}
\label{tab:transitionsph2ne}
\begin{adjustwidth}{-2cm}{}
\centering
\begin{tabular}{cc|cc|cc}
\multicolumn{2}{c|}{Phase O} & \multicolumn{2}{c|}{Phase I} & \multicolumn{2}{c}{Phase II} \\ \hline
\includegraphics[width=8em]{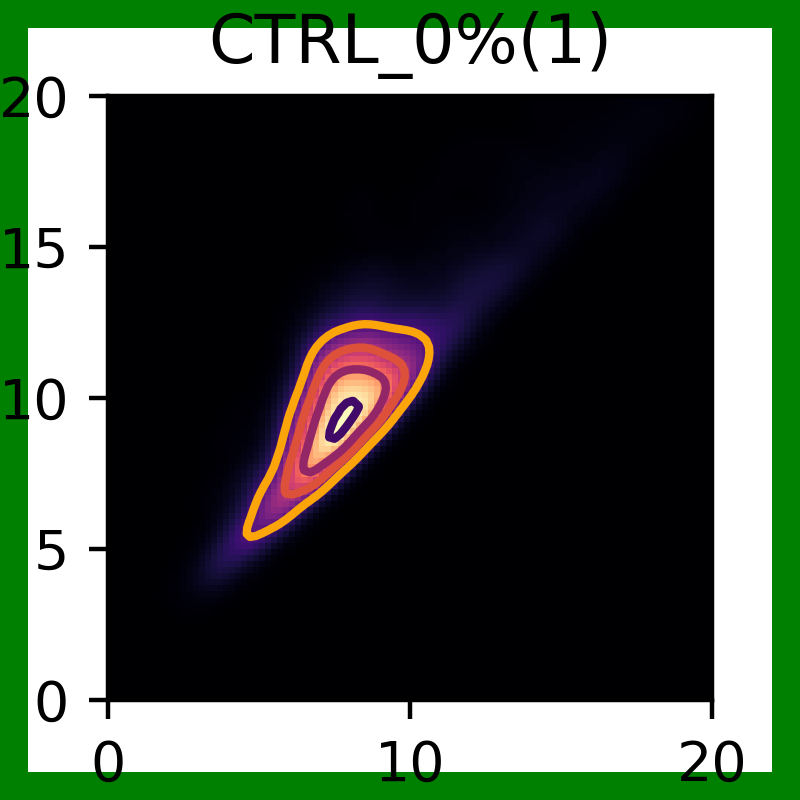} &    
\includegraphics[width=8em]{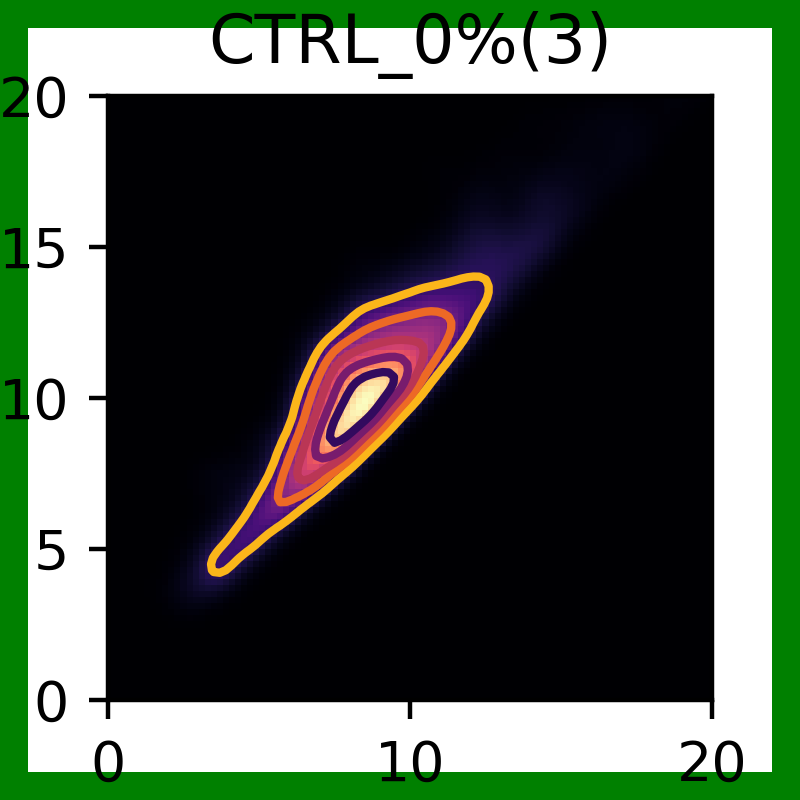} &
\includegraphics[width=8em]{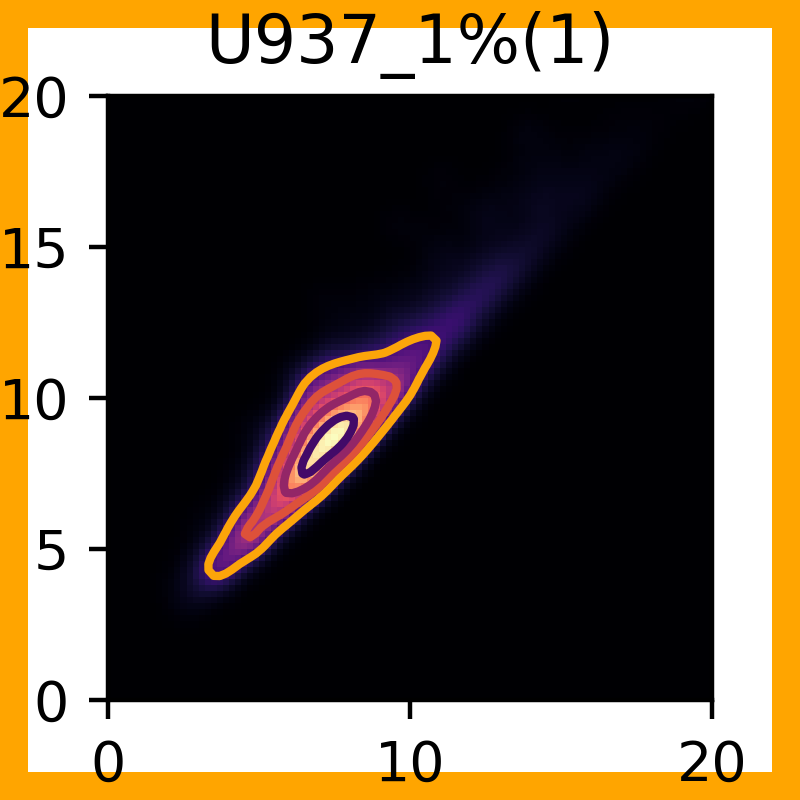} &
\includegraphics[width=8em]{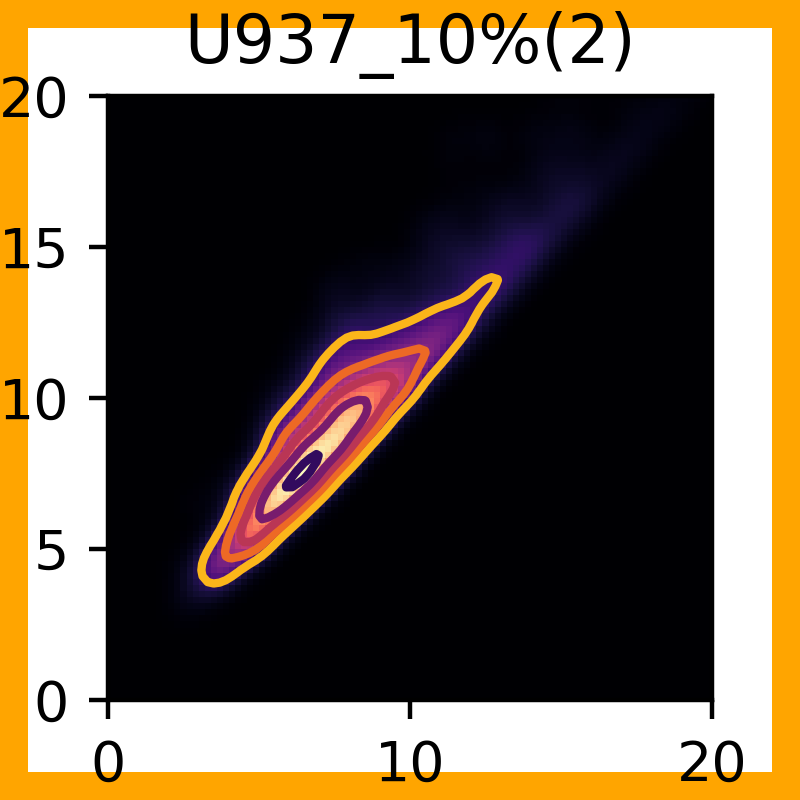} & 
\includegraphics[width=8em]{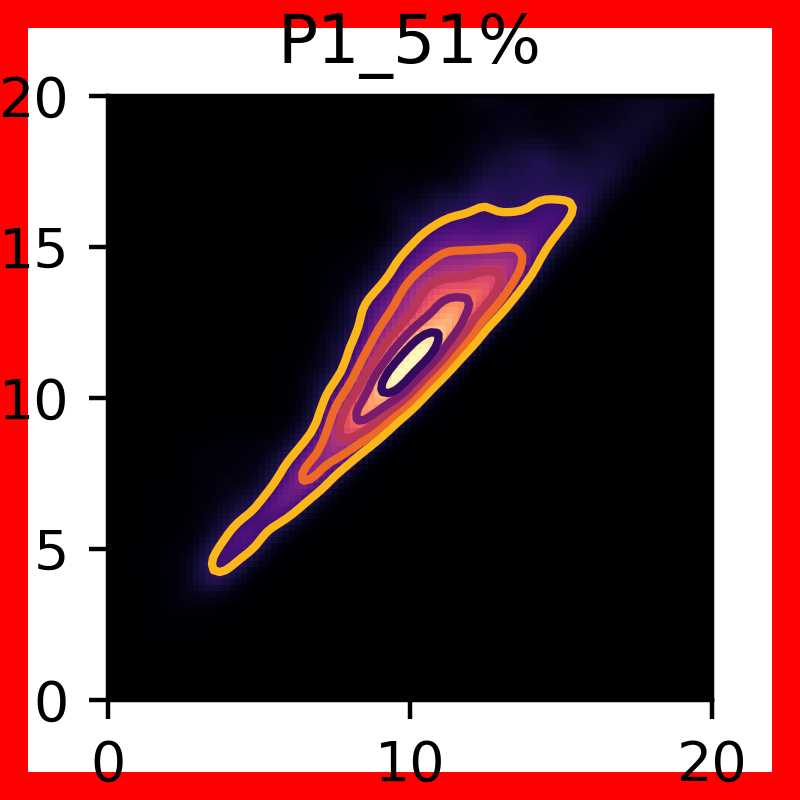} &
\includegraphics[width=8em]{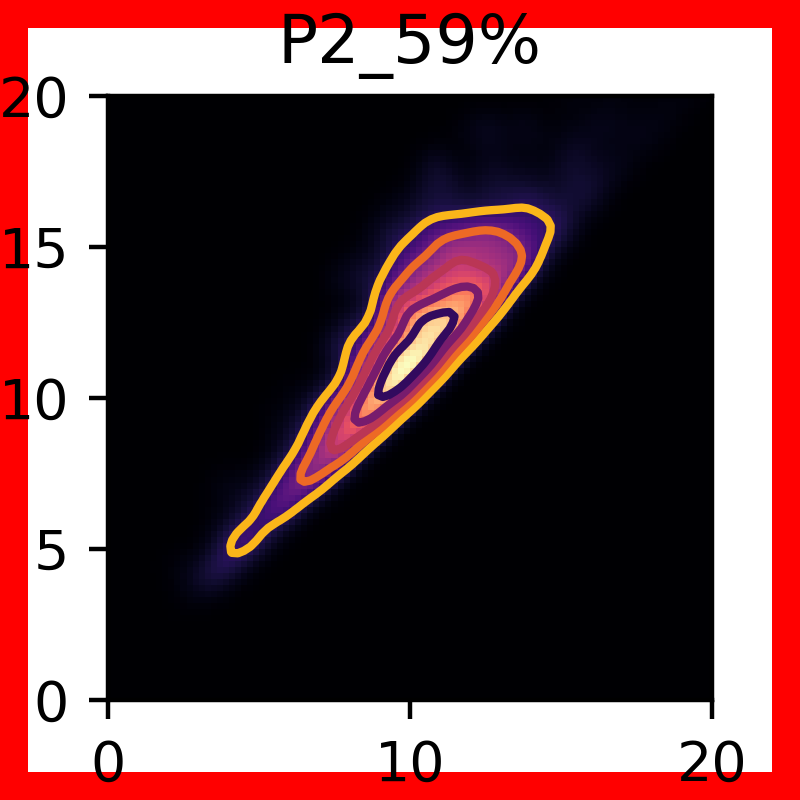}\\  
\includegraphics[width=8em]{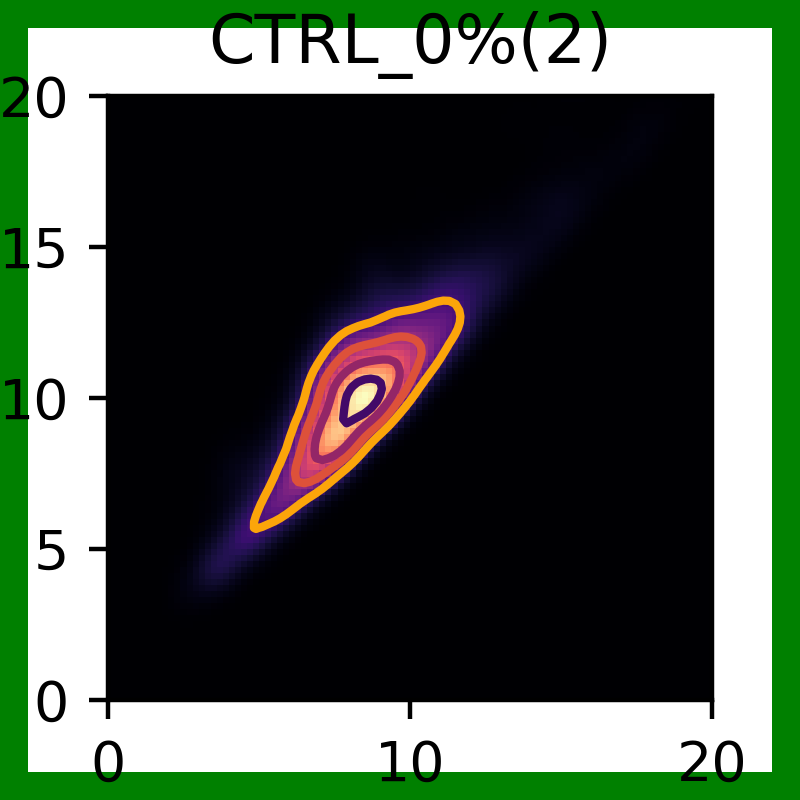} &
\includegraphics[width=8em]{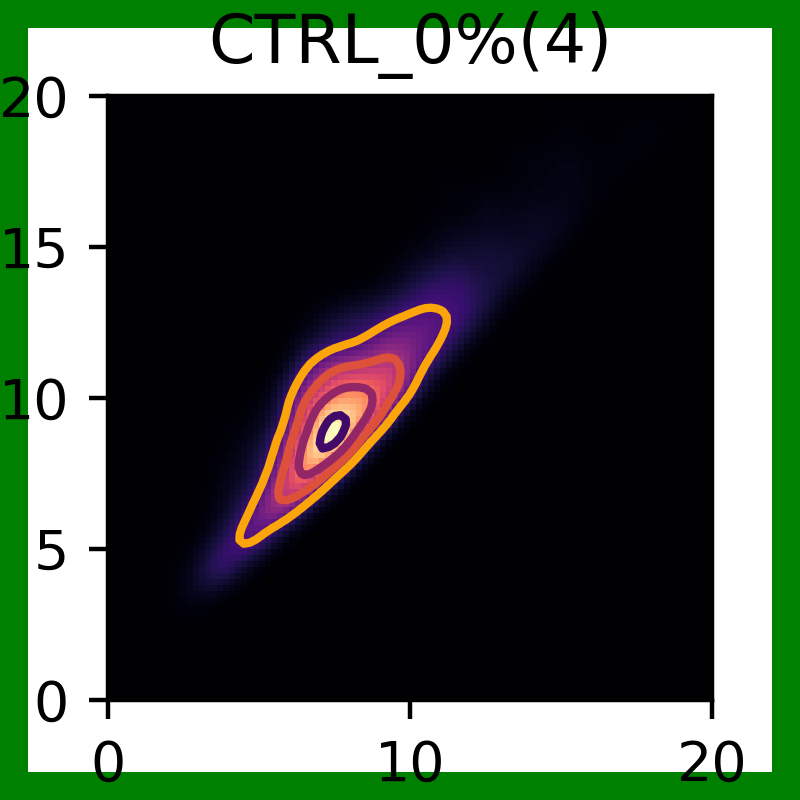} &
\includegraphics[width=8em]{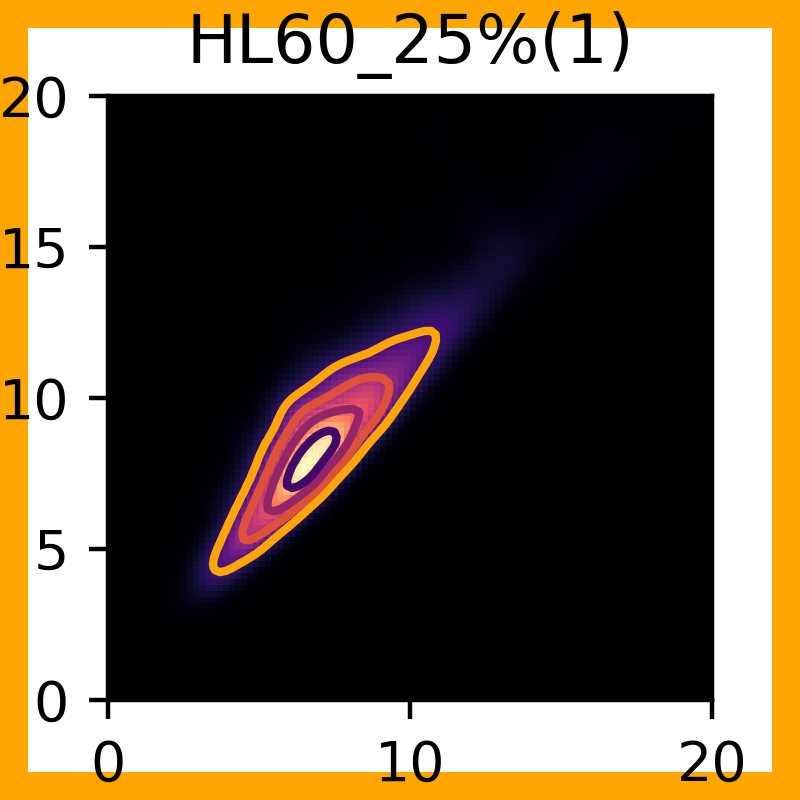} & 
\includegraphics[width=8em]{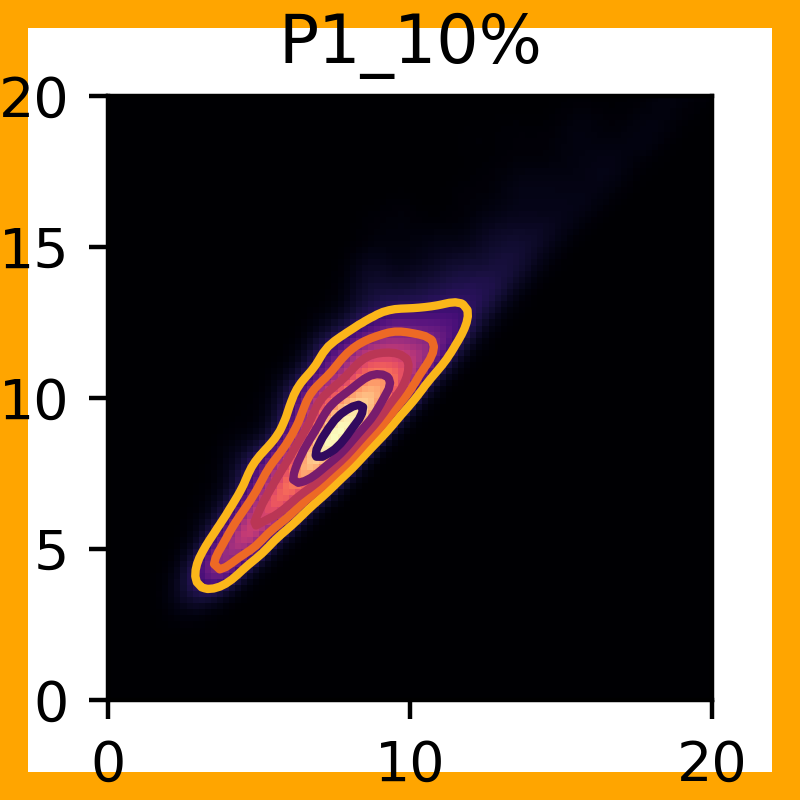} & 
\includegraphics[width=8em]{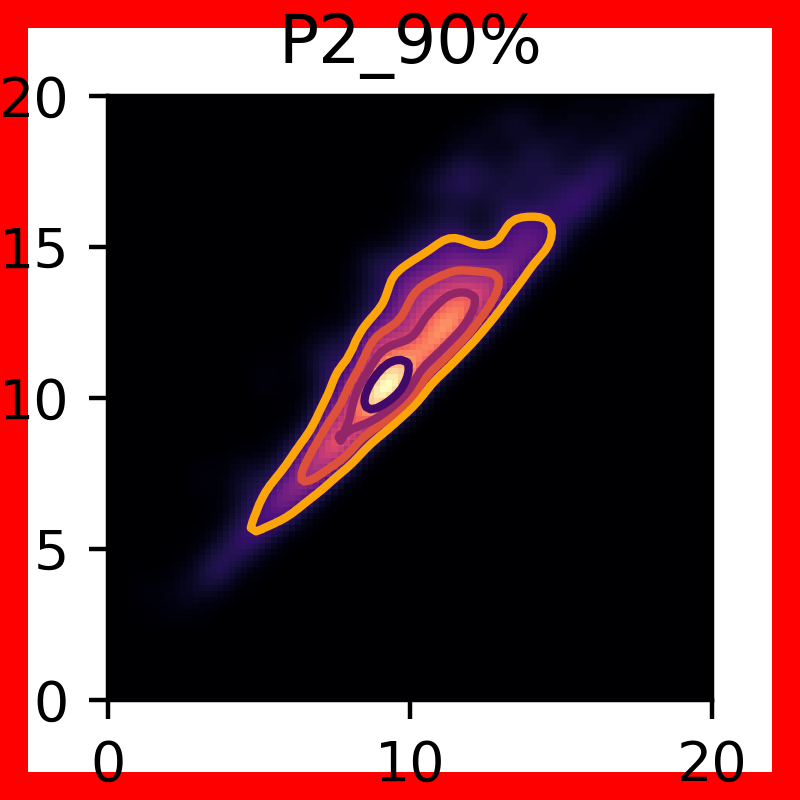} &
\includegraphics[width=8em]{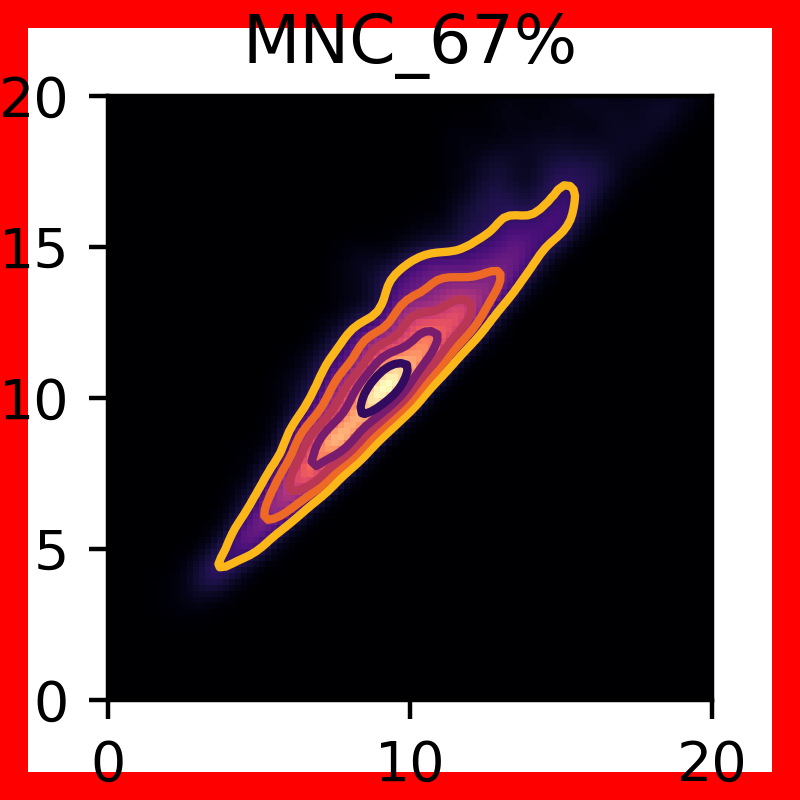}\\ 
\end{tabular}
\end{adjustwidth}
\end{table}

To explore whether the morphological distinctions identified through local texture analysis are also reflected at a global level, we perform hierarchical clustering on the Gaussian KDEs derived from the persistence diagrams. We apply standard agglomerative hierarchical clustering using 4 different linkage strategies---complete, average, single, and Ward---and compute a consensus over both linkage methods and bootstrap resamples. This helps mitigate the sensitivity of hierarchical clustering to specific algorithmic choices and yields a more robust global structure across samples. The resulting consensus dendrograms provide global, density-based visualizations of the morphological similarity across samples. The optimal number of clusters, set to 4, was determined via the elbow method based on the within-cluster sum of squares at each number of clusters.

\begin{figure}[htb]
    \centering
    \includegraphics[width=.7\linewidth]{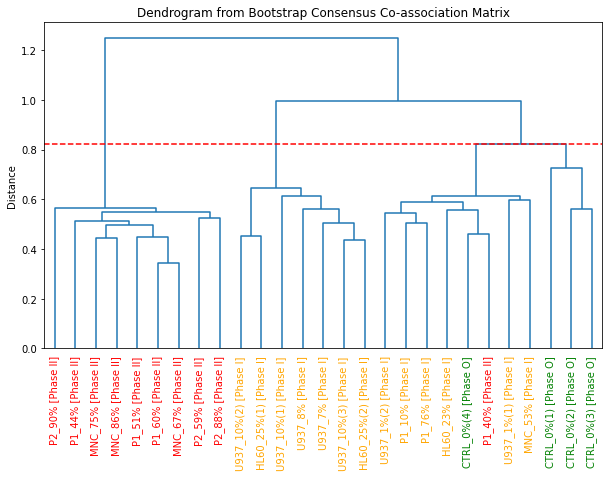}
    \caption{Hierarchical tree constructed from consensus over 100 bootstrap resamplings on PH$_1$ KDEs with Ward linkage.}
    \label{fig:bootstrap_tree}
\end{figure}

\begin{figure}[htb]
    \centering
    \includegraphics[width=.7\linewidth]{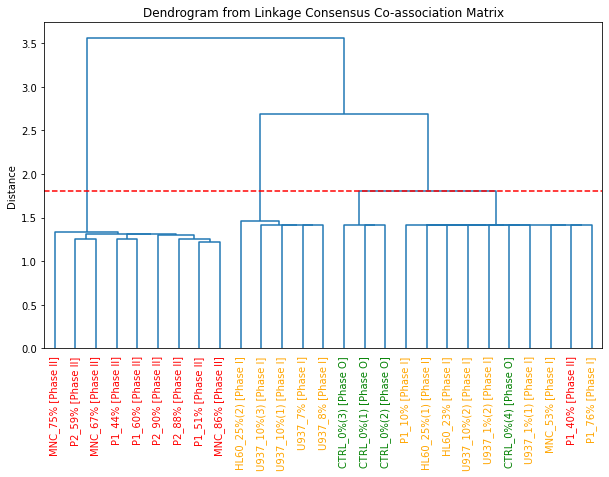}
    \caption{Hierarchical tree constructed from consensus over 4 linkage methods---complete, Ward, average, and single---on PH$_1$ KDEs }
    \label{fig:linkage_tree}
\end{figure}

The hierarchical tree computed from PH$_1$ KDEs with bootstrap consensus is shown in Figure~\ref{fig:bootstrap_tree}, and the tree computed using consensus over linkage methods is shown in Figure~\ref{fig:linkage_tree}. Given the optimal cluster count of 4, we cut the trees at approximate heights of 0.8 and 1.75, respectively, resulting in 4 major branches. These clusters largely align with the local phase classifications: 3 clusters consisting uniquely of solely Phase O, I and II samples, respectively, and a fourth cluster consisting of mostly Phase I samples, with two exceptions---CTRL sample at 0\% (Phase O) and P1 sample at 40\% (Phase II)---suggesting that these samples share intermediate global texture patterns. Similar consensus hierarchical tree for PH$_2$ can be found in Section \ref{app:consensus_tree} of the Supplementary Materials and supports a similar finding.

To quantify the alignment between the global clustering and local phase groupings, we compute several external clustering evaluation metrics: Adjusted Rand Index (ARI), Normalized Mutual Information (NMI), Normalized Variation of Information (VI), and normalized split/join distance. These results, summarized in Table~\ref{tab:clustering_agreement}, indicate moderate to strong agreement, particularly for the Phase I and Phase II samples.

\begin{table}[htb]
\centering
\caption{Agreement between hierarchical clustering (by linkage method) and local phase labels, measured using Adjusted Rand Index (ARI), NormaliZed Mutual Information (NMI), Variation of Information (VI), and split/join distance across regions.}
\label{tab:clustering_agreement}
\begin{tabular}{llcccc}
\toprule
\textbf{Metric} & \textbf{PH Region} & \textbf{Ward} & \textbf{Average} & \textbf{Complete} & \textbf{Single} \\
\midrule
\multirow{3}{*}{ARI} 
  & PH$_0$   & -0.015 & 0.023 & -0.009 & 0.071 \\
  & PH$_1$   & 0.626  & 0.626 & 0.502  & 0.673 \\
  & PH$_2$   & 0.463  & 0.529 & 0.502  & 0.498 \\
\midrule
\multirow{3}{*}{NMI} 
  & PH$_0$   & 0.131  & 0.157 & 0.137  & 0.325 \\
  & PH$_1$   & 0.761  & 0.761 & 0.638  & 0.716 \\
  & PH$_2$   & 0.552  & 0.614 & 0.638  & 0.628 \\
\midrule
\multirow{3}{*}{VI} 
  & PH$_0$   & 0.900 & 0.812 & 0.902 & 0.729 \\
  & PH$_1$   & 0.244 & 0.244 & 0.374 & 0.303 \\
  & PH$_2$   & 0.464 & 0.399 & 0.374 & 0.431 \\
\midrule
\multirow{3}{*}{Split/Join} 
  & PH$_0$   & 0.083 & 0.074 & 0.085 & 0.066 \\
  & PH$_1$   & 0.020 & 0.020 & 0.031 & 0.017 \\
  & PH$_2$   & 0.040 & 0.031 & 0.031 & 0.034 \\
\bottomrule
\end{tabular}
\end{table}

\section{Modeling Textural Changes with Gaussian Mixture Models on SDPH Diagrams}
\label{sec:modelling}

In Section \ref{sec:analysis}, we have shown that associations can be made between the SDPH diagrams and morphological features in the bone samples. This indicates that SDPH diagrams can quantify structural changes in bone marrow vasculature in an interpretable manner that is consistent with biological reasoning. Using localized analysis of the images, we assigned the samples into one of 3 morphologically-derived phases---O, I, and II---which approximately correspond to the control, early, and late stages of engraftment. Further, by computing the global SDPH over the knee regions and approximating the point distributions using a Gaussian kernel density estimate, we confirmed concordance with the phase allocations. We interpreted the transitions in the density patterns within the SDPH diagrams across phases as reflecting biological progression consistent with increasing levels of engraftment. To capture this behavior, we turn to our second contribution of this paper and propose a family of phase-dependent Gaussian mixture models (GMMs) to represent the common distributions observed for samples in each phase. This approach aligns with prior work on the statistical modeling of persistence diagrams, notably the Bayesian framework of \citet{maroulas_bayesian}, but differs in its focus on interpretable, low-dimensional mixtures tailored to biologically relevant, high-persistence features.

\subsection{Gaussian Mixture Models}
\label{sec:gmm}

Gaussian mixture models have been widely used for unsupervised clustering and density estimation \citep{gmm}; the models assume the data to be generated from a weighted mixture of Gaussian distributions which are parameterized by some unknown means and variances. Where the $k$-means algorithm is considered a hard clustering technique, GMMs allow for \emph{soft} clustering, where cluster assignments are associated with probabilities, allowing the model to capture more complex patterns in the data distribution. It is also common for the parameters of the GMM to depend on external parameters, such as time for applications in finance \citep{time_varying_mixture} and engineering \citep{sensors}, allowing for context to be integrated into the model. 

Due to the discrete nature of our samples and uncertainty in terms of the levels of engraftment, we introduce instead a family of phase-dependent GMMs to approximate our samples' SDPH diagrams. For each point in the SDPH diagram for each sample, we relabel the indices, so that points from samples of the same phase are viewed as a collective: Let $\mathbf{y_i}=\{y_{ij}\}_{j=1,\dots,n_i}$ for $i=0,1,2$ denote the collection of birth--death points from the SDPH diagrams of samples within phase $i$, where $n_i$ denotes the number of points in this phase and $j$ references the individual points in this collection. In addition, each point $y_{ij} = (b_{ij}, d_{ij})$ is weighted by its persistence computed as $w_{ij}=d_{ij}- b_{ij}$. Let $c_i\in \mathbb{N}$ be the size of the mixture for each phase, then the models are given by
$$
f_i(y_{ij}) = \sum^{c_i}_{m=1} \alpha_{im} \Phi(y_{ij}|\mu_{im}, \Sigma_{im}/w_{ij}),
$$
where $\sum^{c_i}_{m=1}\alpha_{im} =1$, and
$$
\Phi(y|\mu, \Sigma) = \frac{1}{2\pi|\Sigma|^{1/2}} \exp \left\{ -\frac{1}{2}(x-\mu)^\top\Sigma^{-1}(x-\mu)\right\}.
$$
Note that the notation for the weight follows from $\Phi(y_{ij}|\mu_{im}, \Sigma_{im})^{w_{ij}} \propto \Phi(y_{ij}|\mu_{im}, \Sigma_{im}/w_{ij})$.\\

\noindent
\textbf{Expectation--Maximization (EM) Algorithm.} To estimate the parameters $\{ a_{im}, \mu_{im}, \Sigma_{im}\}$, we use an iterative expectation--maximization (EM) algorithm, which can be broken down into two steps: An E-step, computing the probability of $y_{ij}$ belonging to each Gaussian distribution, 
$$ r_{ijm} = \frac{a_{im}\Phi(y_{ij}|\mu_{im}, \Sigma_{im}/w_{ij})}{\sum_{m'\neq m} a_{im'}\Phi(y_{ij}|\mu_{im'}, \Sigma_{im'}/w_{ij})};$$
and an M-step of maximizing the expected complete log-likelihood, 
$$\log(f_i(\mathbf{y_i})) = \sum^{n_i}_{j=1} \sum^{c_i}_{m=1} r_{ijm}\{\log(a_{im})+ \log(\Phi(y_{ij}|\mu_{im}, \Sigma_{im}/w_{ij}))\}.$$

We determine the size of the mixture model $c_i$ through trade-offs between the model performance evaluated using the Bayesian information criterion (BIC), the computational time, and the generalizability of the model across SDPH diagrams of all samples in the same phase. BIC is chosen for the application since it poses heavier penalties on the model complexity compared to other criteria, such as the Akaike information criterion (AIC), and is ideal for our application.\\

\noindent
\textbf{Nonparametric Bootstrap.}
Since we have no further way of generating more images from each engrafted sample or obtaining more biological samples, we use nonparametric bootstrap, a random subsampling method with replacement, to mimic the process of generating more data under the same conditions, i.e., the same level of cell engraftment.\\

\noindent
\textbf{A Continuous Regression-Based Model.} While our approach models disease progression using 3 discrete phases, a continuous regression-based framework could, in principle, offer a more detailed view of structural change. A natural candidate is a ``time''-varying Gaussian Mixture Model, where the component weights $\alpha_k(t)$, means $\mu_k(t)$, and covariances $\Sigma_k(t)$ vary with engraftment level $t$:
$$ f(y_t) = \sum_{k=1}^K \alpha_k(t) \, \Phi(y_t \mid \mu_k(t), \Sigma_k(t)). $$

However, this approach faces both biological and modeling limitations. Biologically, engraftment levels are measured at whole-body scale and may not reliably reflect local bone marrow states. In addition, inter-animal variability and different AML subtypes introduces further heterogeneity that complicates modeling. From a statistical perspective, the sparsity and repetition of time values, along with label noise and numerical instability, limit predictive performance. These challenges make the regression model suboptimal for our  dataset under study. We retain the phase-dependent formulation for its robustness and interpretability, and refer to Section \ref{app:regression} of the Supplementary Material for additional details and results for the proposed continuous model.

\subsection{Model Interpretation}\label{sec:model_interpretation} In this section, we fit our models to the full dataset to interpret the morphological structure characteristics of each phase. By examining the fitted distributions within each quadrant individually, we  are able to identify the dominant structural features and highlight how these evolve across phases. This enables us to characterize the progression of morphological changes from Phase O through to Phase II, offering a global summary of the vascular remodeling associated with AML progression.\\

\begin{figure}[htb]
\centering
\includegraphics[width=.3\linewidth]{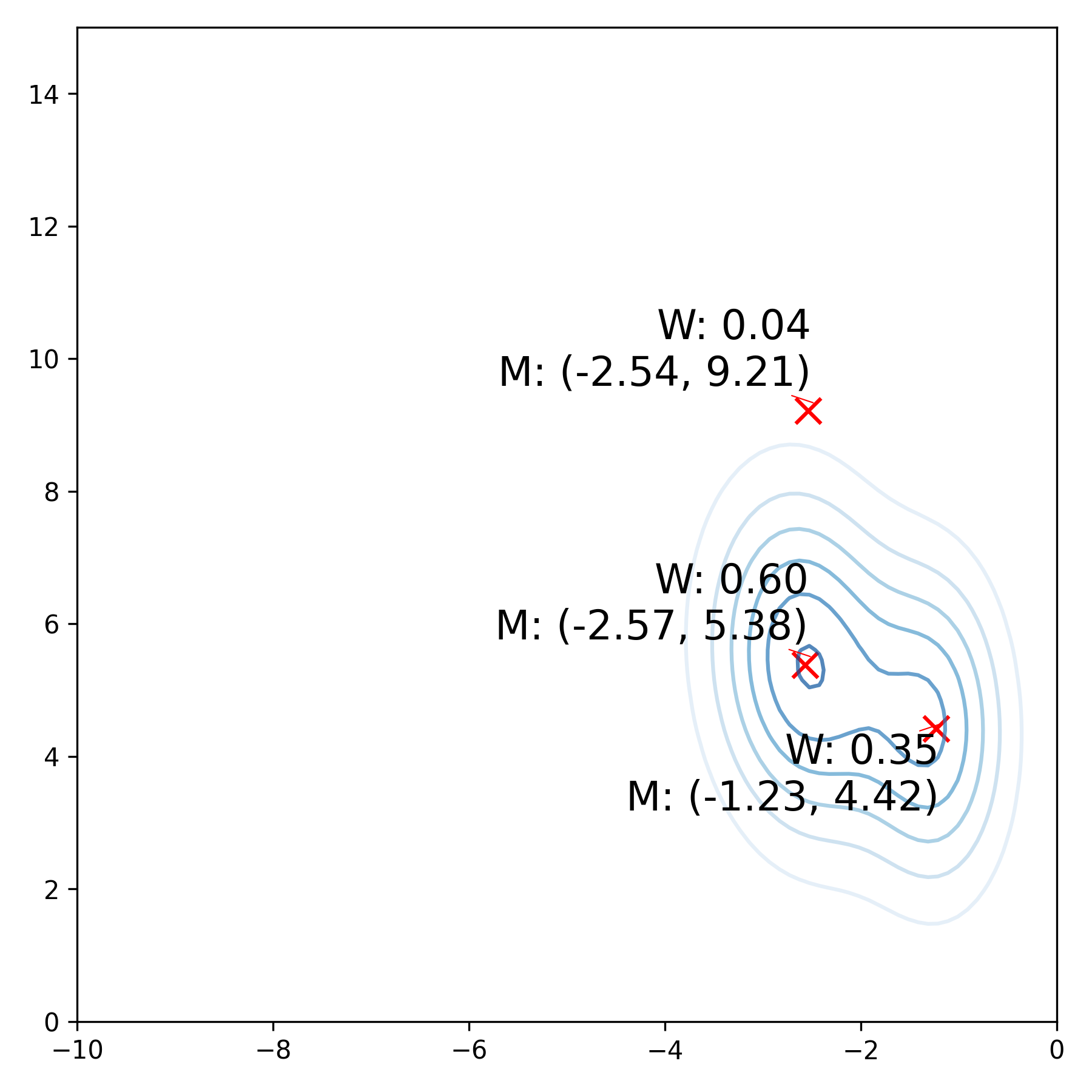}\hfill
\includegraphics[width=.3\linewidth]{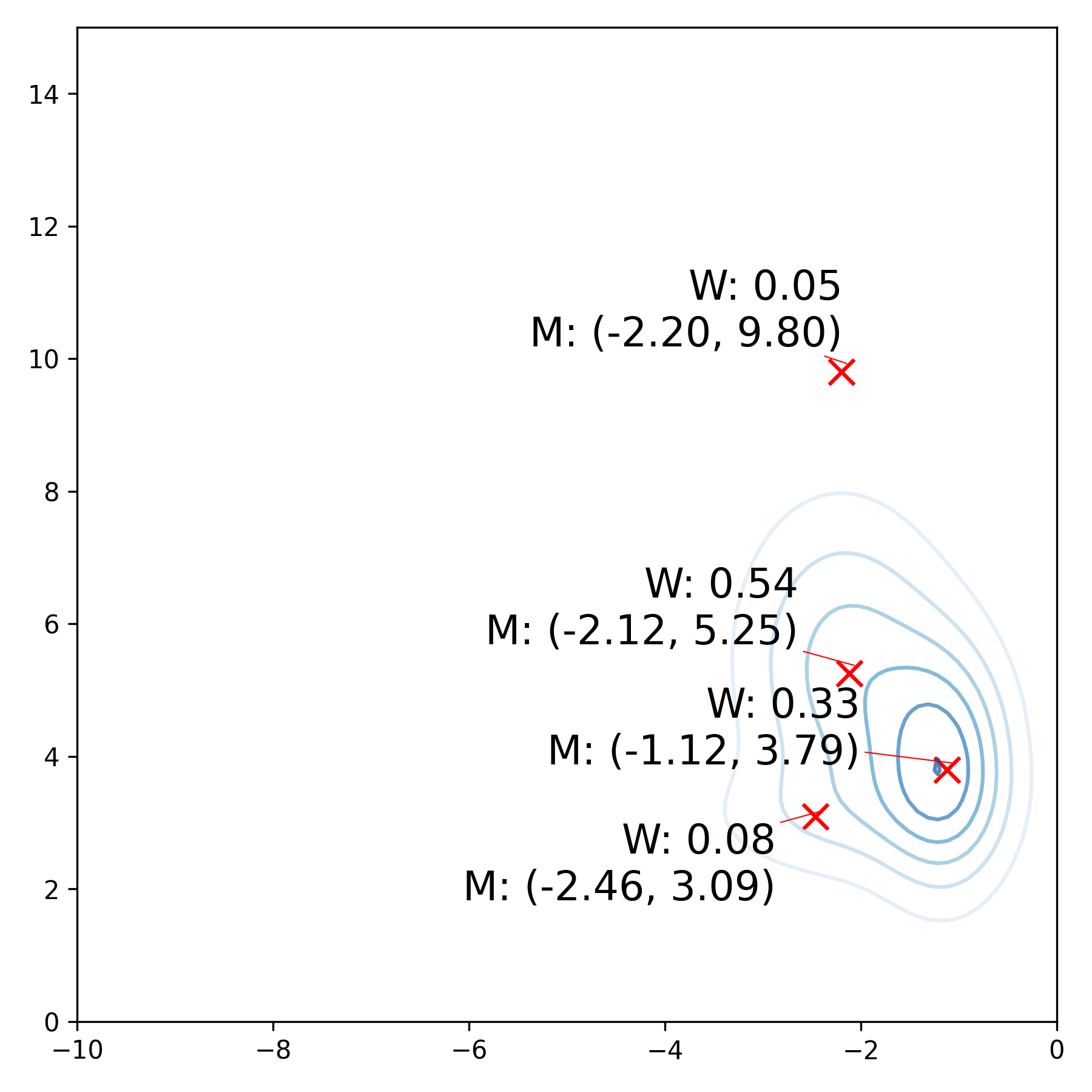}\hfill
\includegraphics[width=.3\linewidth]{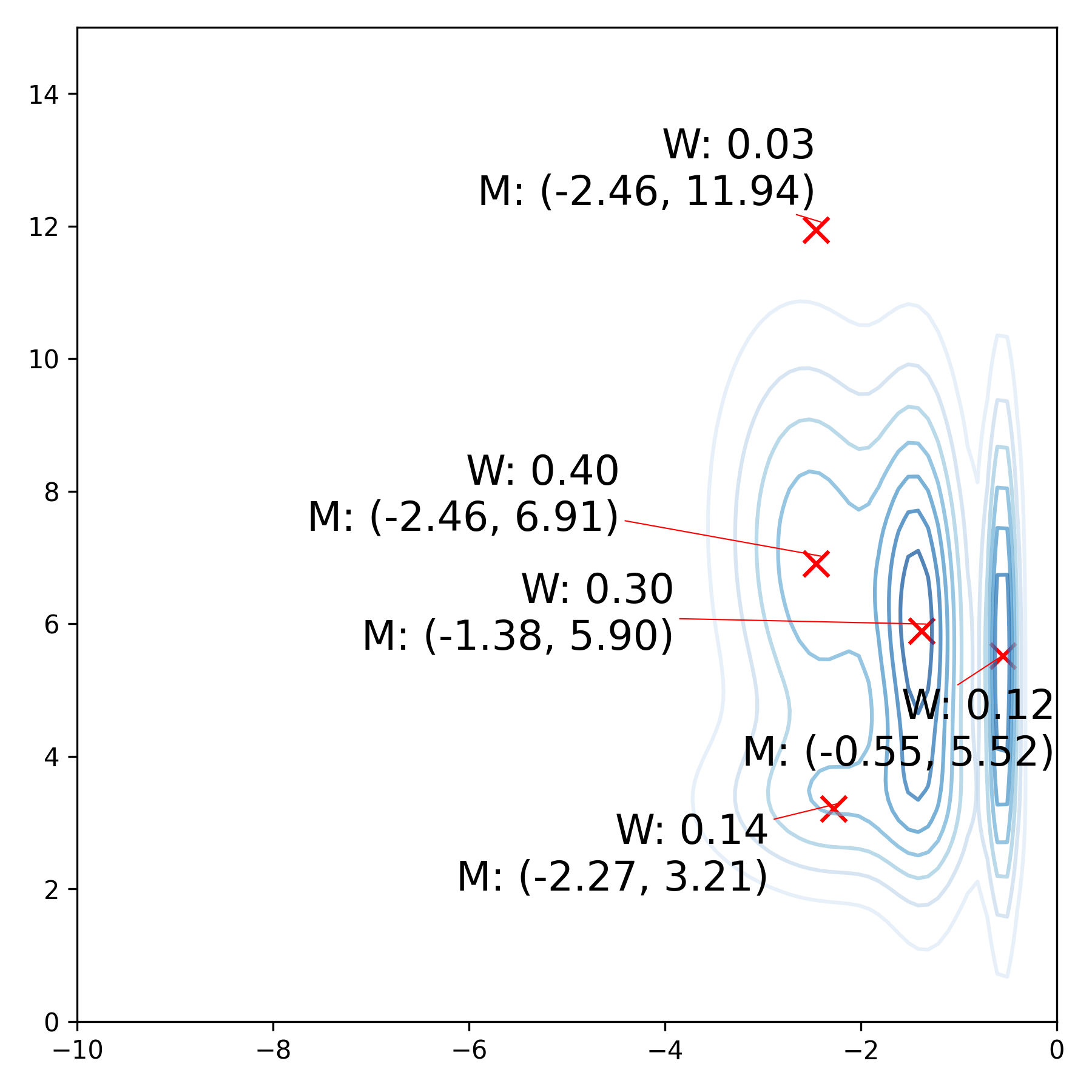}\\
\includegraphics[width=.3\linewidth]{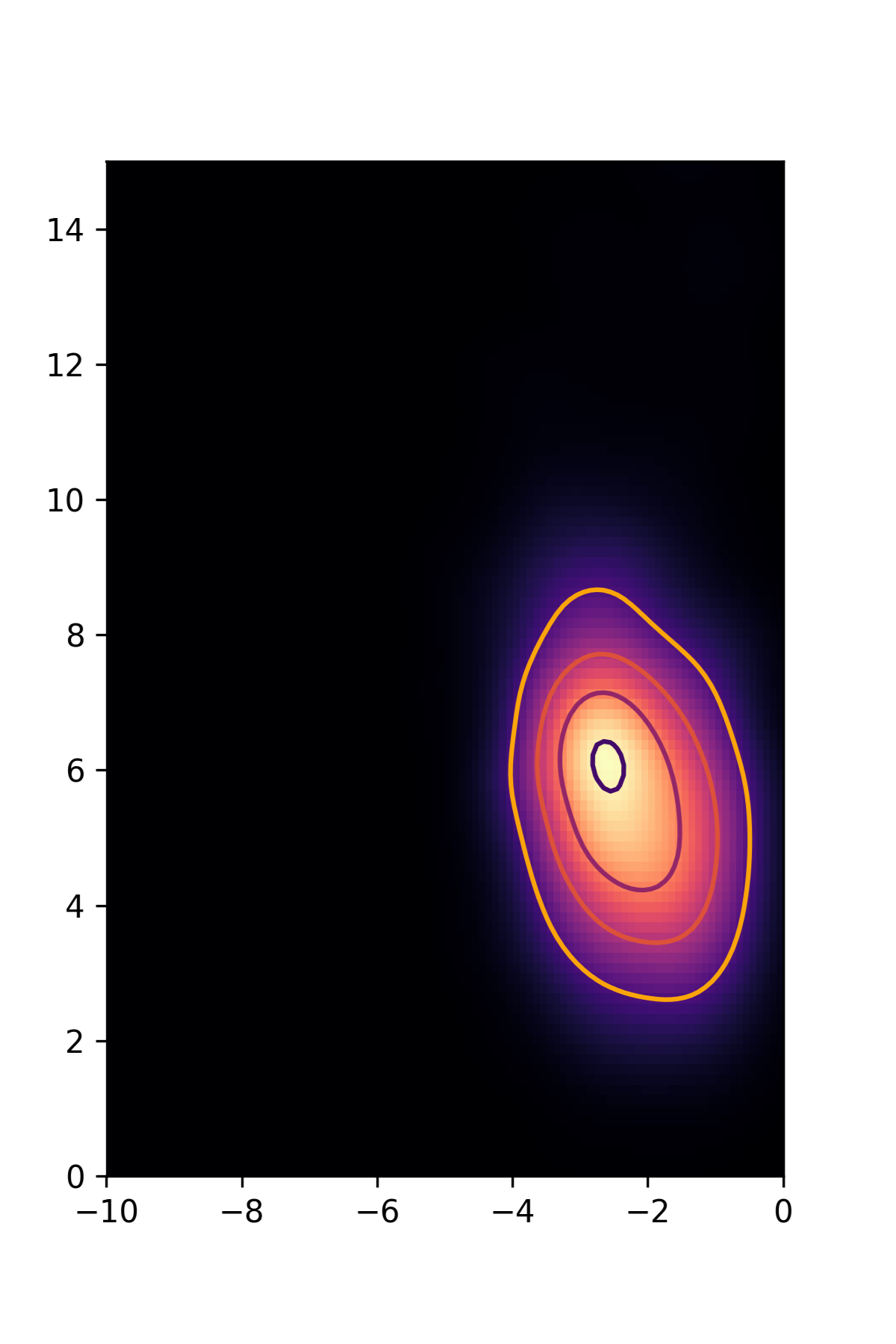}\hfill
\includegraphics[width=.3\linewidth]{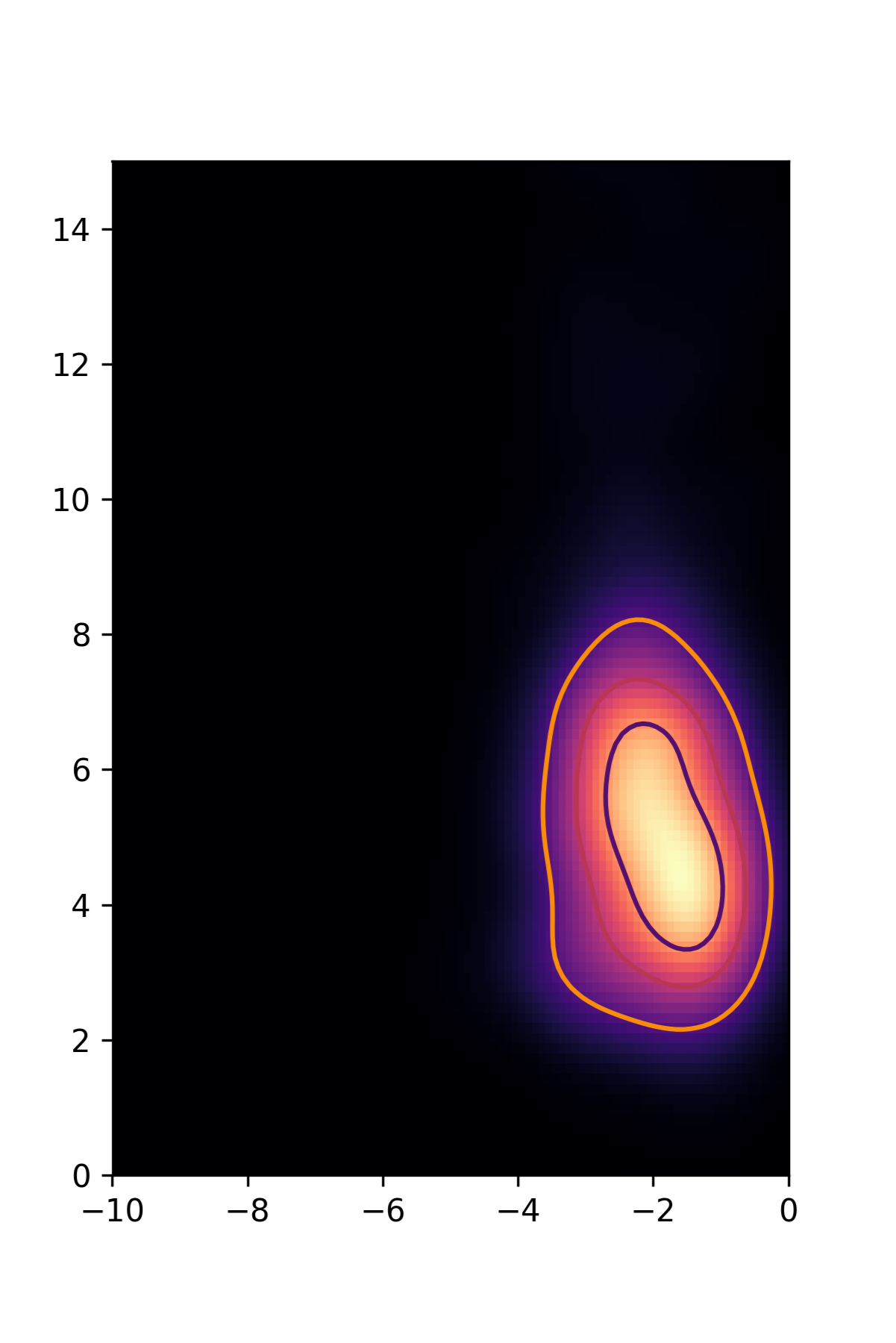}\hfill
\includegraphics[width=.3\linewidth]{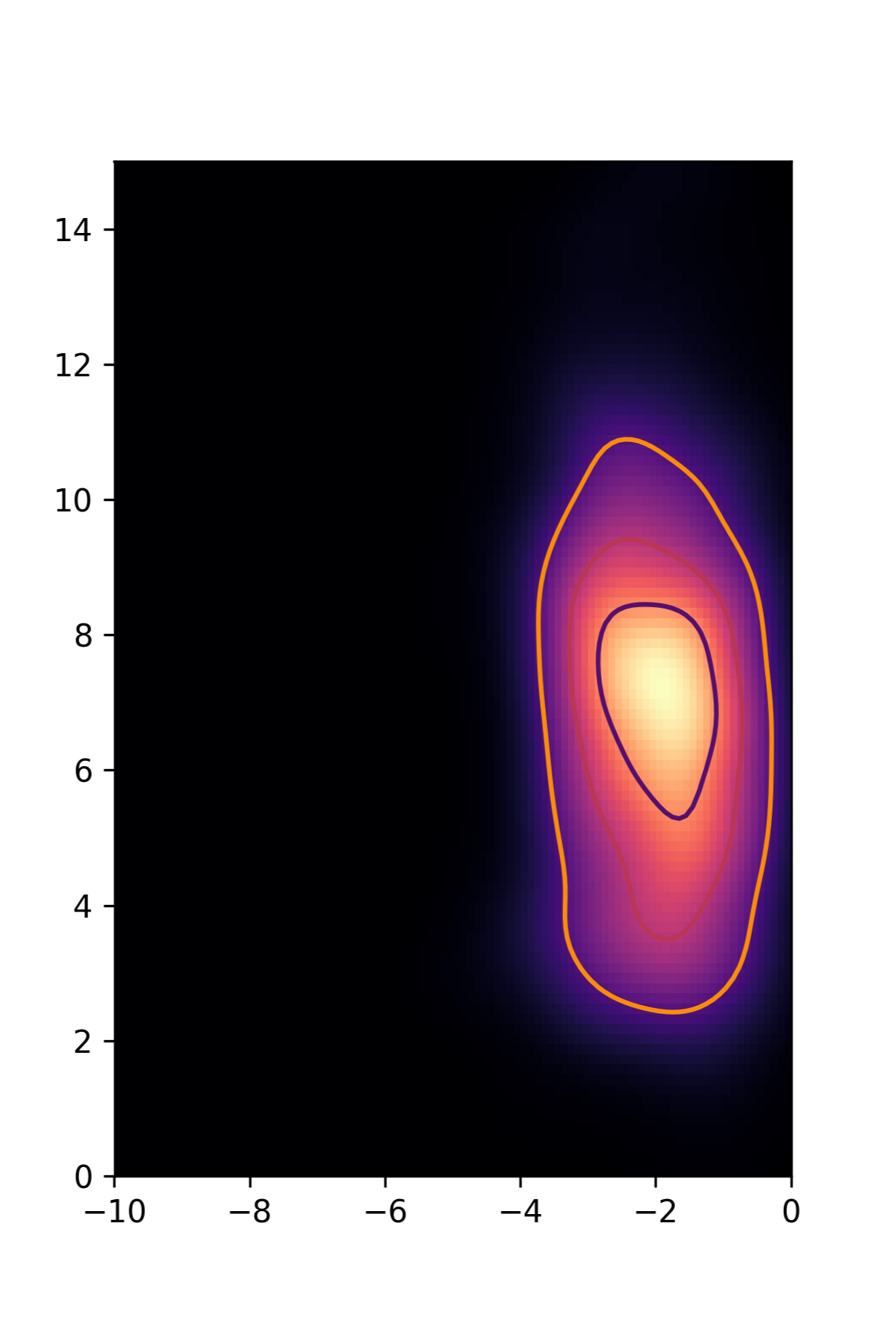}

\caption{Distributions of the weighted mixture of Gaussians for Phases O (left), I (center) and II (right) for $\text{PH}_1$ NW with the positions of the means and weights of Gaussians indicated in the top row of figures.}
\label{fig:ph1nw_dist}
\end{figure}

\noindent
\textbf{PH$_1$ NW.} We begin our evaluation by studying the $\text{PH}_1$ NW quadrant which characterizes the number of loops, their thickness, and sizes, and was seen to give rise to distinguishing features in the samples' estimated density between the different phases. On $\text{PH}_1$ NW samples, taking into account the trade-off between the model fit, computational time, and ensuring that the model is generic across all the samples within one phase, we found that the optimal number of components for the Phase O model to be 3, then increasing the number of components for each consecutive phase by 1 was found to distinguish the models sufficiently as well as allow good fits to the samples within each phase. Figure \ref{fig:ph1nw_dist} shows the GMMs for each phase and interpreting from the parameter of the GMMs, we found that compared to the CTRL samples, the Phase I samples tend to contain thinner and more compact vessels in the early stages of AML engraftment. Transitioning from Phase I to II, towards later stages of engraftment, we observe instead a shift in density towards the northwest direction in the diagram, in favor of higher values of $g_2$ and $r_1$ sizes, indicative of thicker and wider loops formed by vessels in the cavity, although not as much as those found in Phase O. \\

\begin{figure}[htb]

\centering
\includegraphics[width=.3\linewidth]{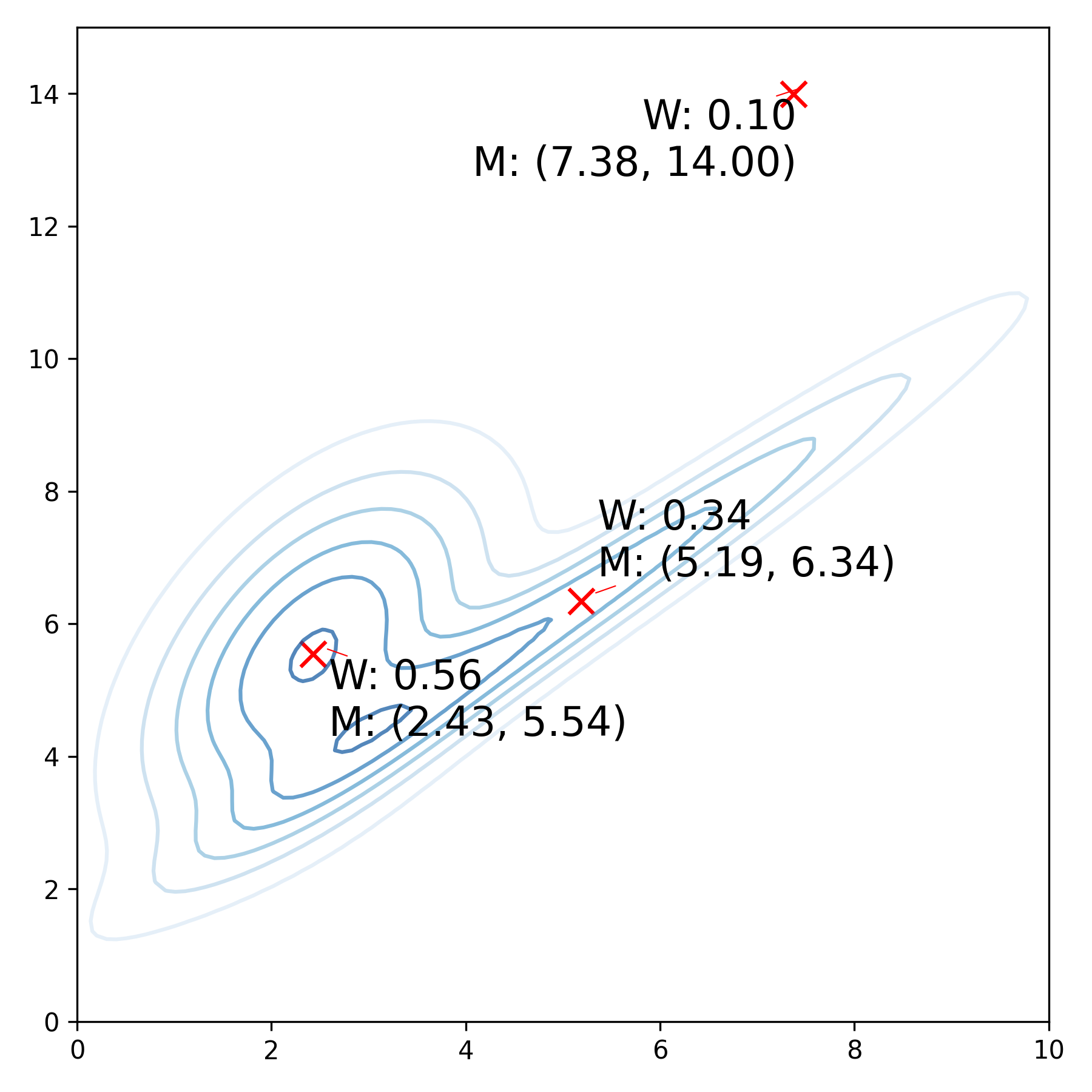}\hfill
\includegraphics[width=.3\linewidth]{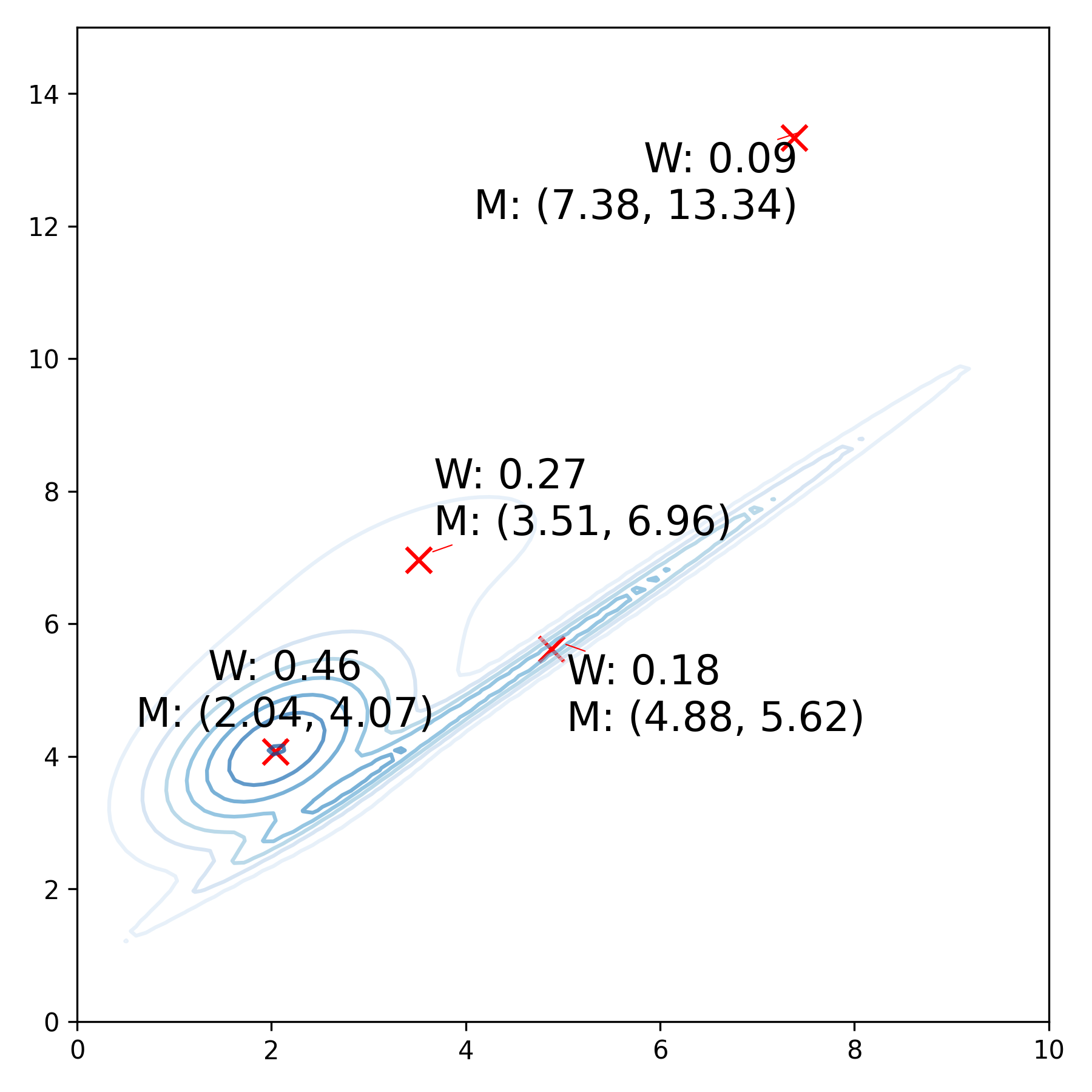}\hfill
\includegraphics[width=.3\linewidth]{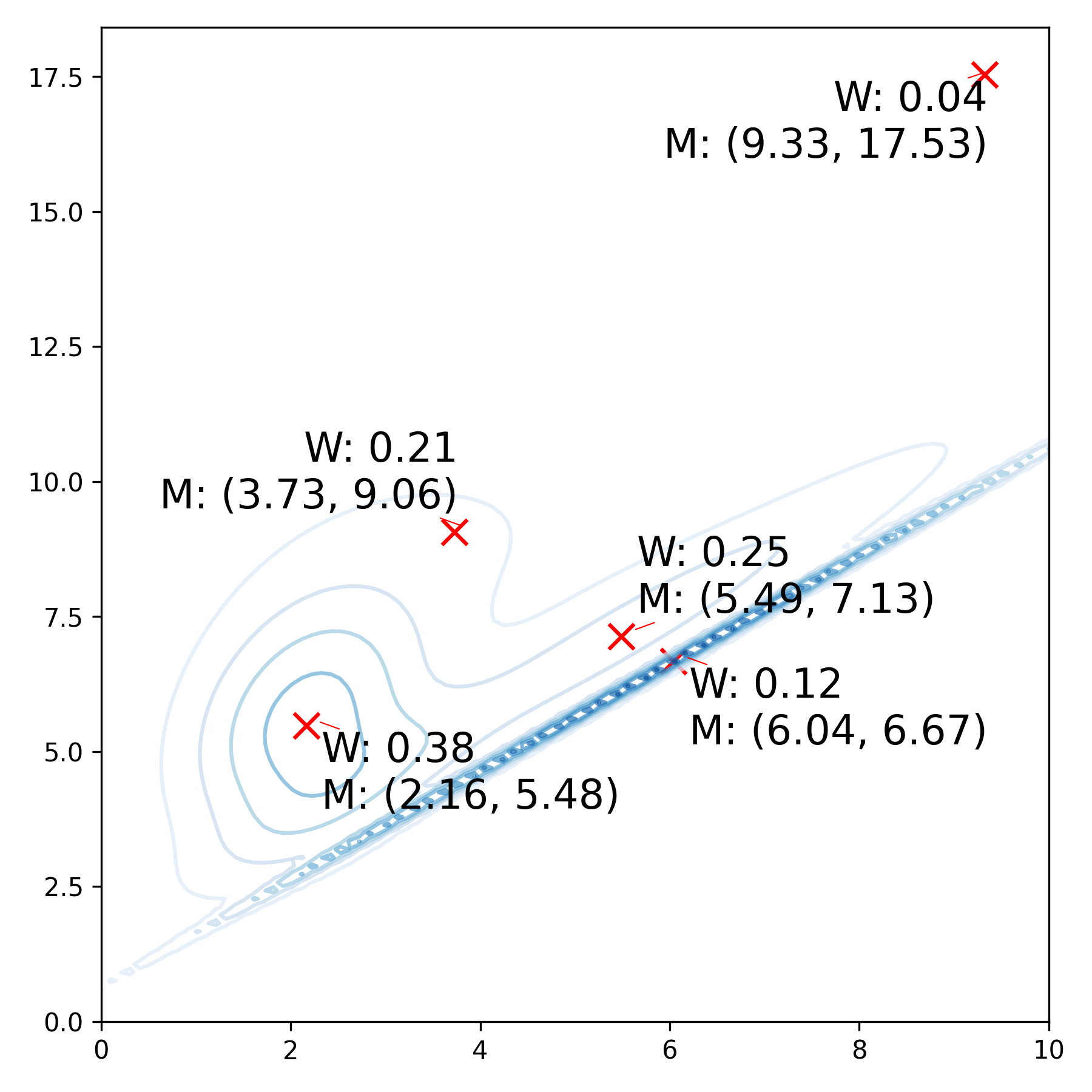}\\
\includegraphics[width=.3\linewidth]{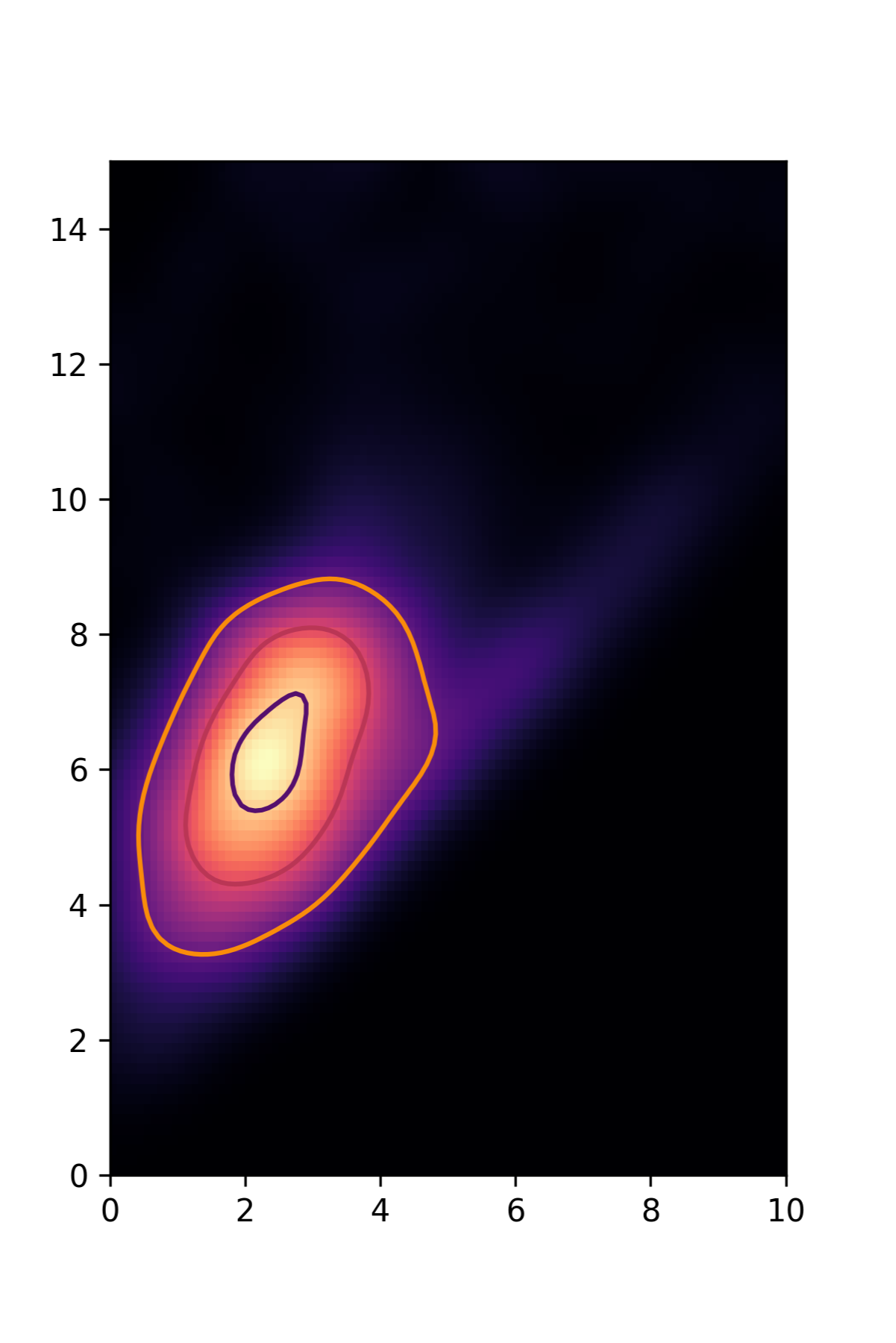}\hfill
\includegraphics[width=.3\linewidth]{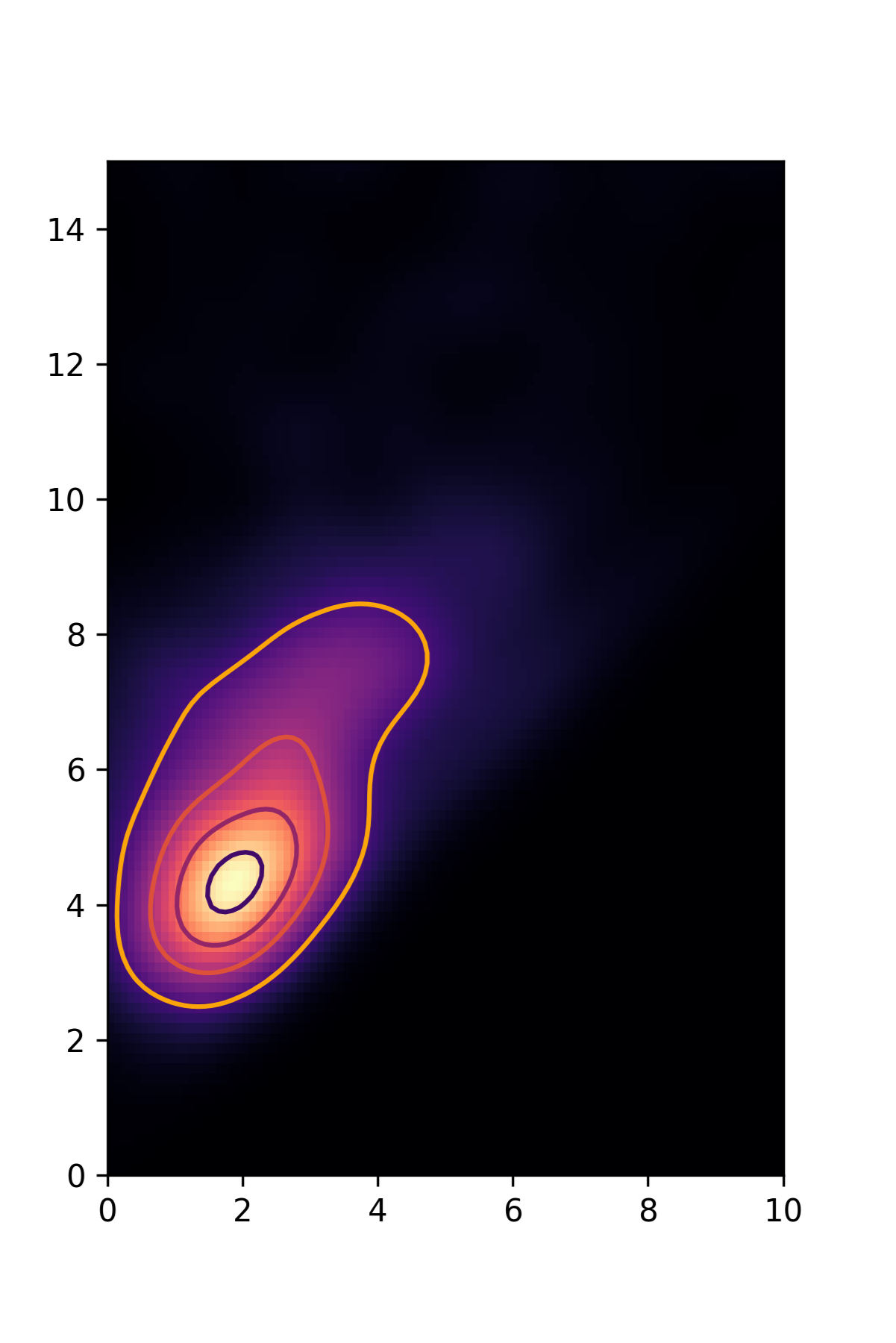}\hfill
\includegraphics[width=.3\linewidth]{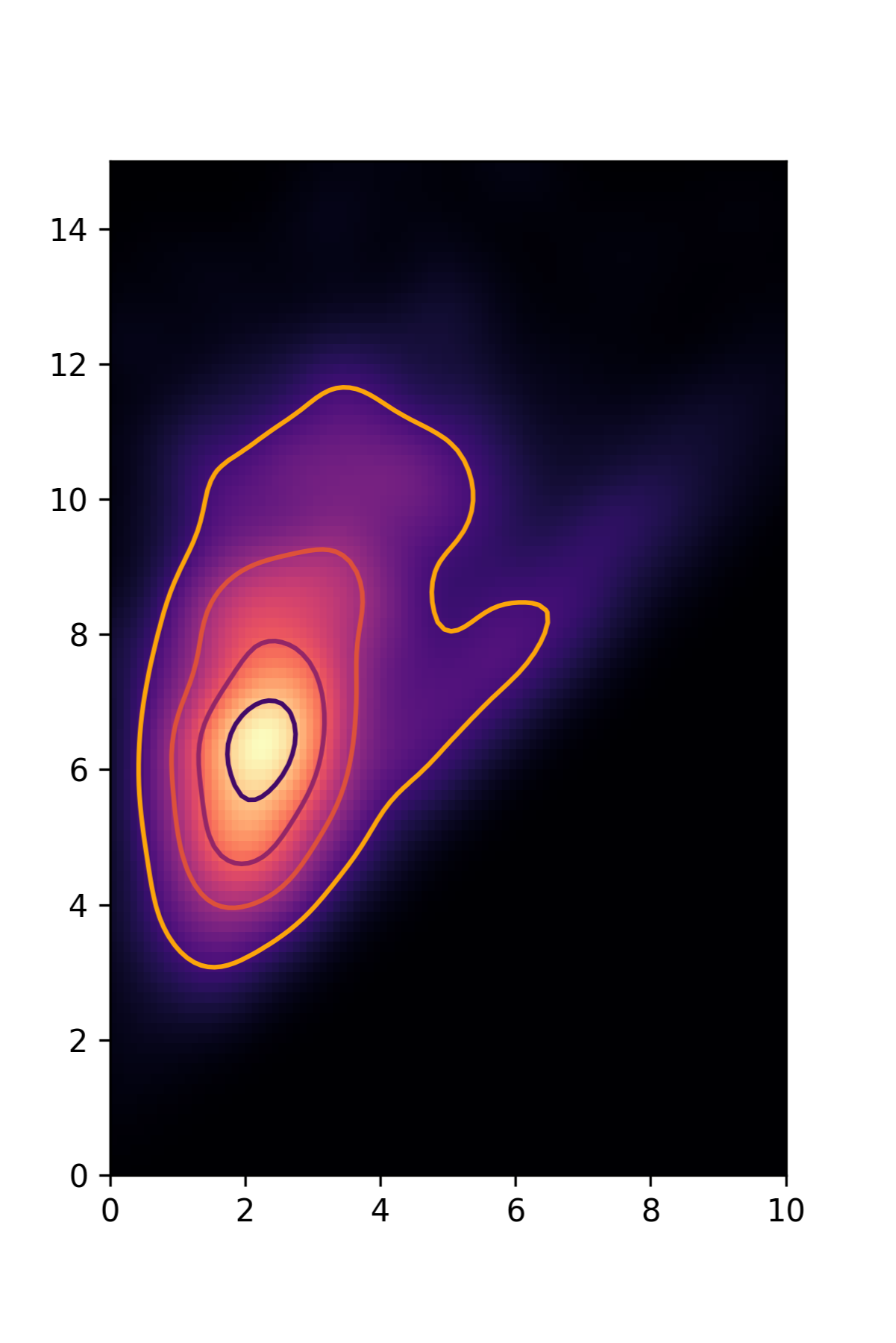}

\caption{Distributions of the weighted mixture of Gaussians for Phases O (left), I (center), and II (right) for $\text{PH}_1$ NE with the positions of the means and weights of Gaussians indicated in the top row of figures.}
\label{fig:ph1ne_dist}
\end{figure}

\noindent
\textbf{PH$_1$ NE.} We next study the neighboring quadrant PH$_1$ NE. From the transitions between the Phase O, I, and II models for $\text{PH}_1$ NE shown in Figure \ref{fig:ph1ne_dist}, we find that compared to the unaffected bone marrow vasculature in the CTRL samples, samples in the early stages of AML engraftment have more compact vasculature. We observe a higher probability around lower values of both $g_1$ and $g_2$ critical sizes, suggesting tighter spacing between vessels in general, both in the narrowing between loops and loops formed by vessels. Then, transitioning from the Phase I model to the Phase II model, we observe the region of high probability moving towards higher $g_1$ and $g_2$ values, suggesting a dilation of the underlying vessel structure.\\ 

\begin{figure}[htb]

\centering
\includegraphics[width=.3\linewidth]{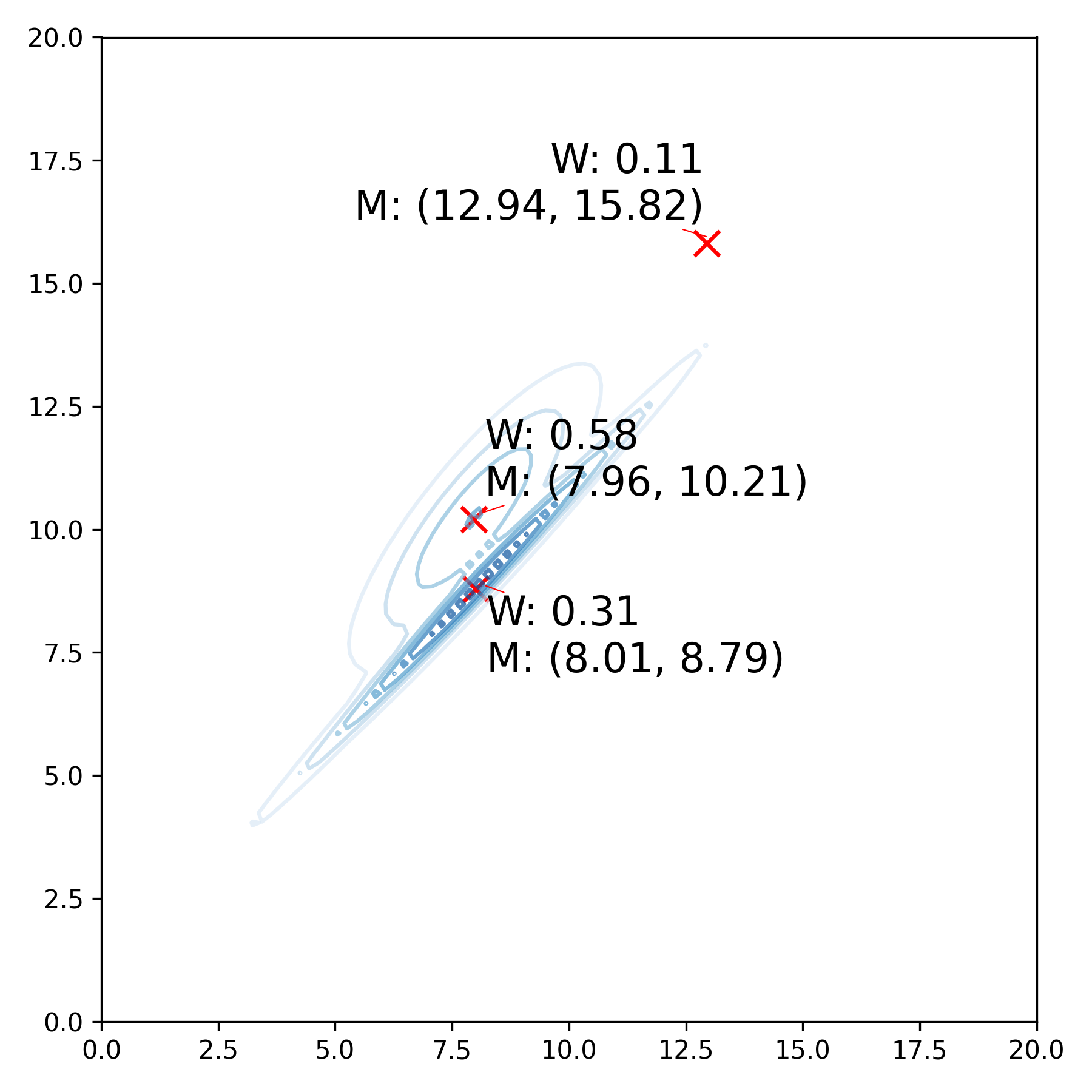}\hfill
\includegraphics[width=.3\linewidth]{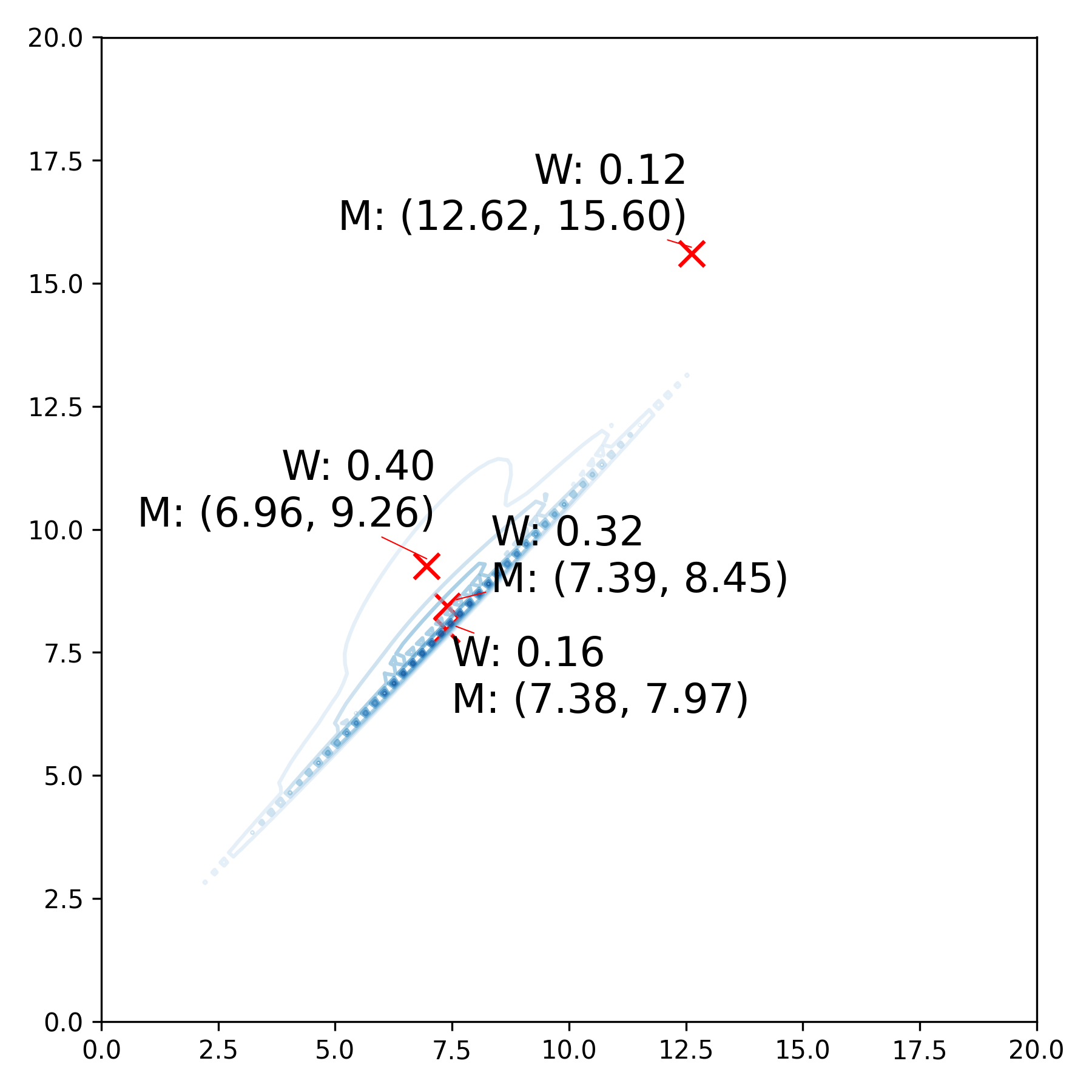}\hfill
\includegraphics[width=.3\linewidth]{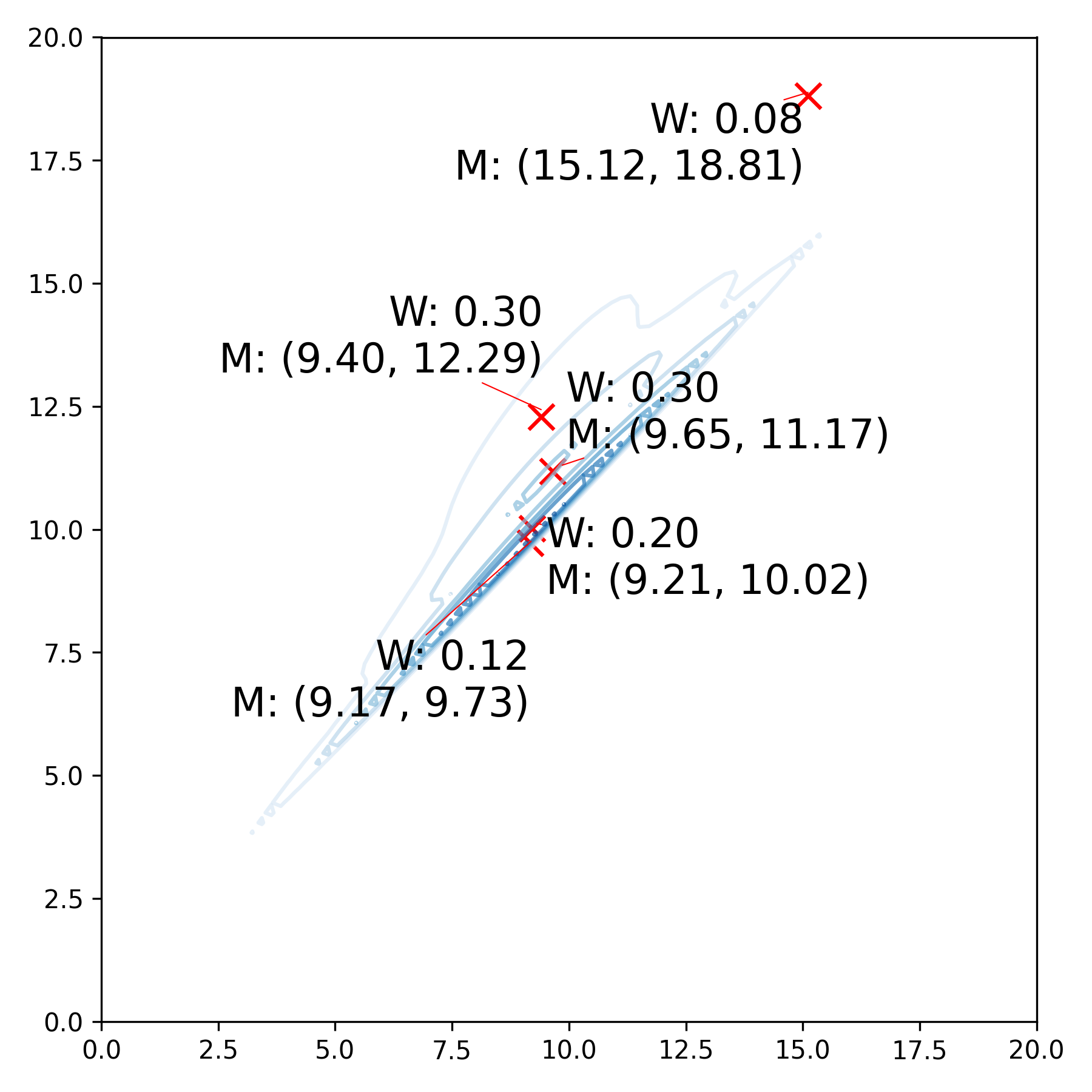}\\
\includegraphics[width=.33\linewidth]{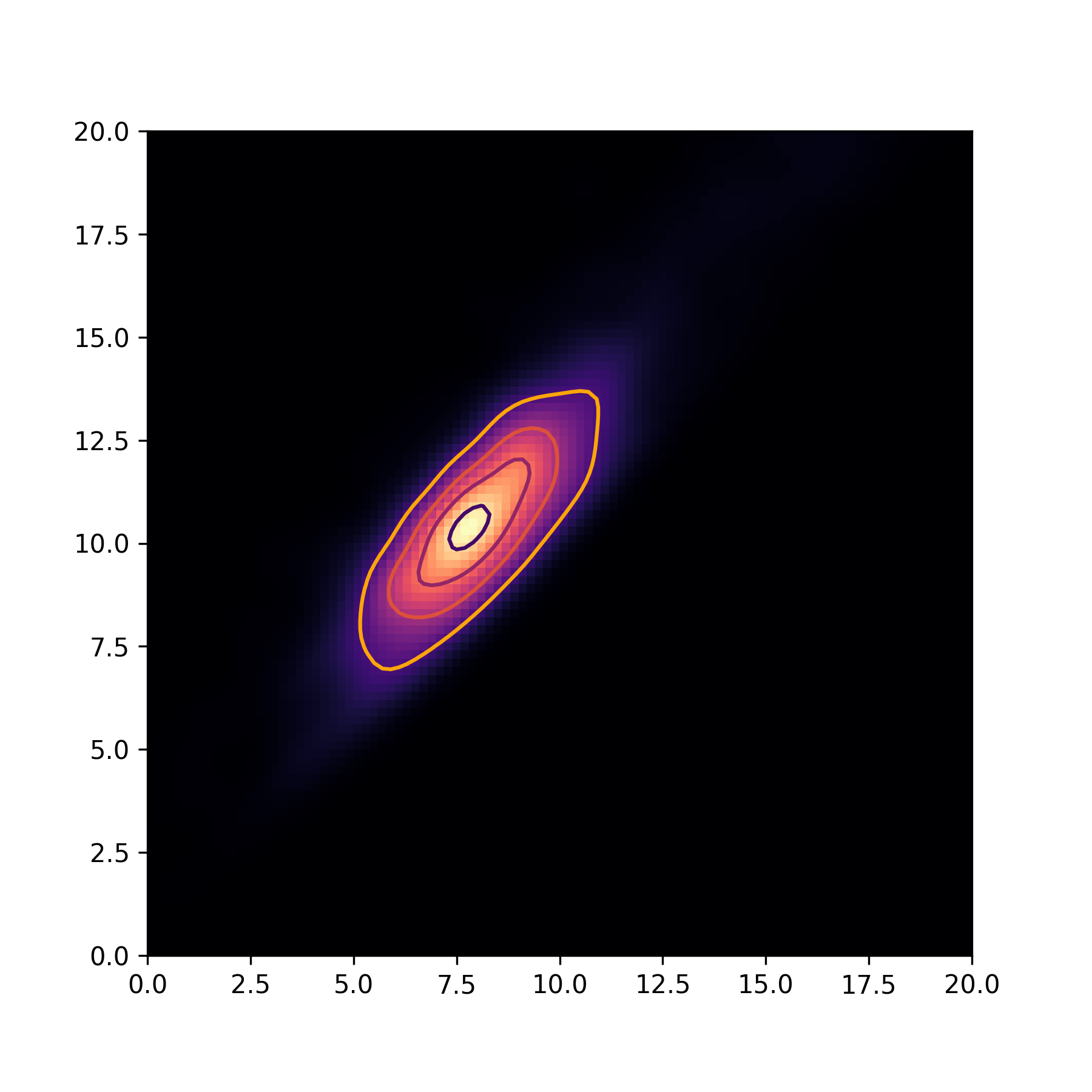}\hfill
\includegraphics[width=.33\linewidth]{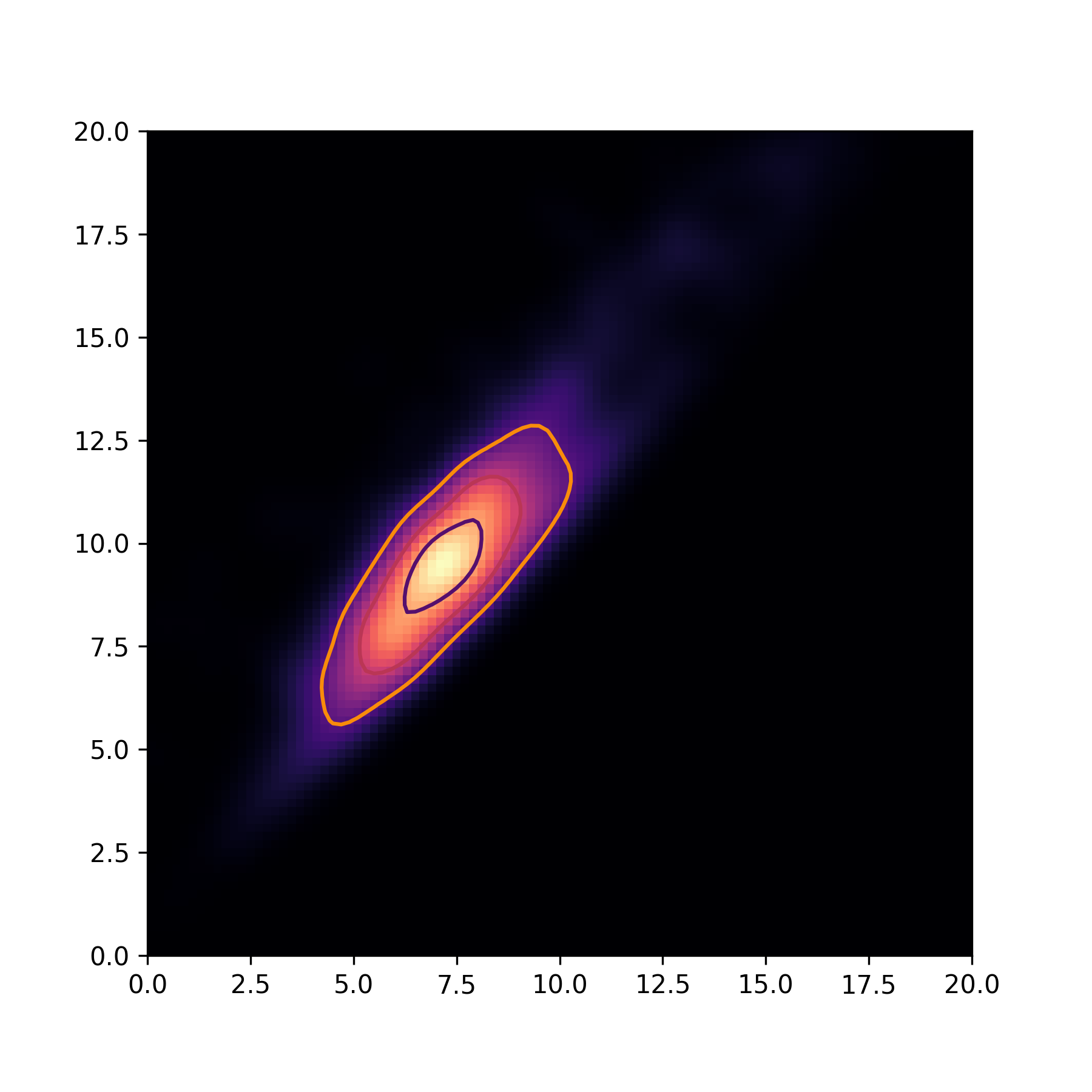}\hfill
\includegraphics[width=.33\linewidth]{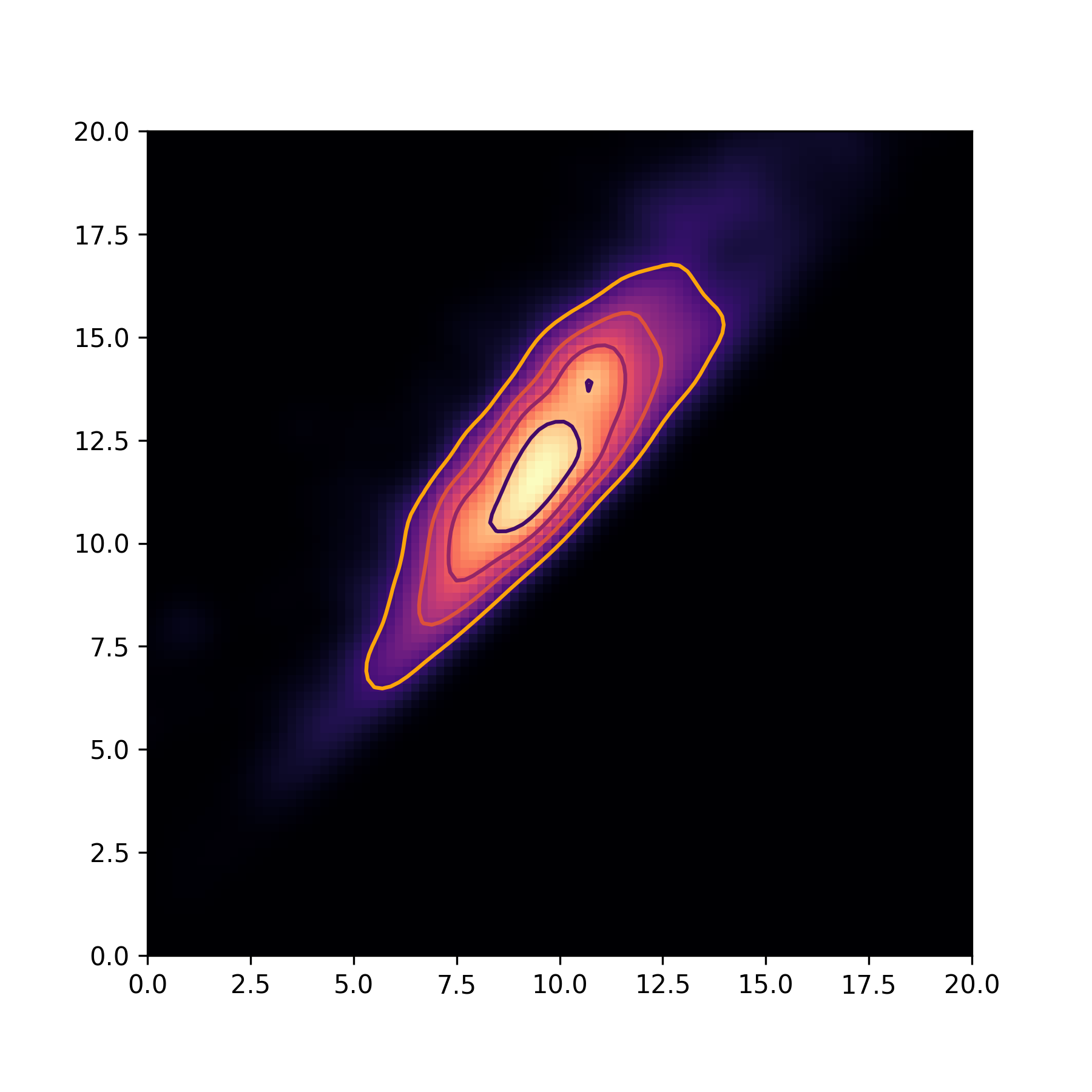}

\caption{Distributions of the weighted mixture of Gaussians for Phases O (left), I (center), and II (right) for $\text{PH}_2$ NE with the positions of the means and weights of Gaussians indicated in the top row of figures.}
\label{fig:ph2ne_dist}
\end{figure}

\noindent
\textbf{PH$_2$ NE.} Evaluating our model on the $\text{PH}_2$ NE quadrant. Between the Phase O and I models, we again find a shift in the region of high probability towards lower values of $g_2$ and $g_3$, indicating smaller loops and cavities formed by vessels. This suggests a more compact vasculature at the onset of angiogenesis. As we transition from the Phase I to the Phase II model, we find that the distribution shifts towards larger values of $g_2$ and $g_3$, indicating that the vasculature becomes more sparse. Figure \ref{fig:ph2ne_dist} shows the composition of the Gaussian mixture model and densities plotted as heatmaps. \\  

The collection of points in the $\text{PH}_0$ NW quadrant can be disregarded since they arise as a complication in the piecewise computation of SDPH along the bone images. In theory, there should only be one point with the quadrant for each sample representing the entire connected vessel structure. The points in the $\text{PH}_2$ NW can also be disregarded since it contains only a few points which should never, or only very rarely, occur since they correspond to bubbles of extravascular space trapped inside the lumen and may only appear due to noise. The evaluation for the $\text{PH}_0$ SW and $\text{PH}_1$ SW quadrant are available in Section \ref{app:model_eval} of the Supplementary Material as the type of characteristics described within these quadrants are not distinct between the phases.\\

\subsection{Model Evaluation}\label{sec:model_evaluation}

We evaluate the classification performance of our phase models using leave-one-out cross-validation (LOOCV) across all 27 samples under two regimes. In the first, standard LOOCV is performed directly on the original samples. Due to data imbalance, particularly the sparsity of Phase O samples, this setup can lead to biased or unstable predictions. In the second regime, we augment the training set using bootstrap resampling: within each LOOCV iteration, we concatenate training samples within each phase and generate 50 bootstrap subsamples per phase. This augmentation ensures balance across phases samples and allows for more stable estimation of the phase-dependent GMM parameters.

For each test sample, we compute its similarity to the fitted Phase O, I, and II models using two measures: the Hellinger distance and the symmetrized Kullback--Leibler (KL) divergence. The test sample is then assigned to the phase corresponding to the model with the smallest distance or divergence to its density GMM approximation. The Hellinger distance \citep{Hellinger} between densities $f$ and $g$ is defined as  
 $$
 H^2(f,g) = \frac{1}{2} \int (\sqrt{f(x)}-\sqrt{g(x)})^2 dx,
 $$ while the Kullback--Leibler (KL) divergence \citep{Kullback1951OnIA} is given by
 $$D_{KL}(f\|g) = \int_{x\in\mathcal{X}} f(x)\log \frac{f(x)}{g(x)}dx.$$
 As the KL divergence is asymmetric, we report the symmetrized version by averaging $D_{KL}(f\|g)$ and $D_{KL}(g\|f)$.

\begin{table}[htb]
\centering
\caption{Phase classification accuracy using Hellinger distance and KL divergence, with and without phase-wise bootstrap subsampling.}
\begin{tabular}{lccccc}
\toprule
\textbf{Method} & \textbf{PH1\_NE} & \textbf{PH0\_SW} & \textbf{PH1\_NW} & \textbf{PH1\_SW} & \textbf{PH2\_NE} \\
\midrule
\multicolumn{6}{l}{\textit{Hellinger-based classification}} \\
LOOCV & 0.704 & 0.482 & 0.852 & 0.370 & 0.889 \\
LOOCV \& bootstrap & 0.852 & 0.556 & 0.852 & 0.296 & 0.889 \\
\midrule
\multicolumn{6}{l}{\textit{KL-divergence-based classification}} \\
LOOCV & 0.778 & 0.482 & 0.852 & 0.185 & 0.889 \\
LOOCV \& bootstrap & 0.852 & 0.556 & 0.852 & 0.269 & 0.889 \\
\bottomrule
\end{tabular}
\label{tab:classification_results}
\end{table}

Table~\ref{tab:classification_results} summarizes the classification accuracies under both LOOCV regimes and distance metrics, evaluated over 5 representative quadrants. We observe high performance in the PH$_1$ NE, PH$_1$ NW, and PH$_2$ NE quadrants, with consistent improvements under bootstrap. In particular, classification accuracies under LOOCV with bootstrap reach 85.2\%, 85.2\%, and 88.9\% for these 3 regions respectively, with PH$_2$ NE yielding the highest overall accuracy. These results suggest that quadrant-specific models effectively capture structural differences in bone marrow morphology across the main stages of AML engraftment and are sufficient for reliable phase discrimination.

In contrast, classification performance is lower in the PH$_0$ SW and PH$_1$ SW regions. This aligns with earlier model interpretations where it was noted that these quadrants displayed greater structural ambiguity and less distinct separation between phases. A recurring source of misclassification is the P1 sample at 40\% engraftment, which was also flagged in the local and global analysis as a potential boundary case. This reflects the biological variability and overlap between phases, particularly at transitional stages of engraftment.

\subsection{Method Comparison}

We now compare the performance of our phase-dependent GMM applied to SDPH diagrams with several alternative approaches for phase classification, in order to assess the added value of our proposed framework over existing topological and morphological methods. Specifically, we consider: (i) a phase-dependent GMM applied to \emph{distance persistent homology} (DPH) diagrams; (ii) classification algorithms such as $k$-nearest neighbors ($k$-NN) and support vector machines (SVM/SVC) trained on \emph{SDPH landscapes} \citep{bubenik2015statisticaltopologicaldataanalysis}; (iii) classification on \emph{persistence images} \citep{persistence_images} computed from SDPH diagrams; and (iv) a baseline based on traditional \emph{morphometric features}.\\

\noindent
\textbf{Gaussian Mixture Models on Distance Persistent Homology.} We compute persistent homology based on distance function \eqref{eq:dist} from Section \ref{sec:sdph}, which tracks the distance of a point $x$ to the surface boundary of a set $A$, without any sign indicating whether the point $x$ lies inside or outside the set $A$ as in \eqref{eq:signed_dist}; we refer to this as \emph{distance persistent homology} (DPH). The resulting DPH persistence diagrams consist of the absolute values of the birth and death points in the SDPH diagrams. Note that this approach corresponds to computing classical distance-based persistent homology, which is the typical strategy in most application areas seeking to apply TDA tools.

To fit the phase-dependent GMM onto the DPH diagrams, we optimized the mixture by balancing computational efficiency with the BIC, also ensuring that no Gaussian component had zero weight. The optimal mixture sizes for the $\text{PH}_0$ and $\text{PH}_1$ diagrams for Phases O, I, and II were 3, 4, and 5, respectively. For the $\text{PH}_2$ diagrams, the optimal mixture sizes for Phases O, I, and II were 4, 2, and 2, respectively. For a quantitative comparison of model performance, we evaluated the classification performance of these models using LOOCV with phase-level bootstrap augmentation. Under this regime, the predictive accuracies achieved using the Hellinger distance were 66.7\% for PH$_0$, 70.4\% for PH$_1$, and 81.5\% for PH$_2$ diagrams. The corresponding results using the KL-divergence were 63.0\%, 70.4\%, and 85.2\%. The PH$_0$ and PH$_1$ results are considerably lower than those achieved by the corresponding SDPH-GMM models, reinforcing the added value of incorporating signed distance information. While standard DPH retains the topological summary of the data, it fails to capture the nuanced directional and morphological differences that SDPH encodes. These findings highlight the advantage of our model over a direct application of standard persistent homology. Note, the high performance of the PH$_2$ DPH--GMM models is to be expected. In a prior analysis, the PH$_2$ NE quadrant exhibited strong separation between phases, and the full PH$_2$ diagram differs from it only by the inclusion of the PH$_2$ NW region, which introduces mild noise. Nonetheless, our SDPH--GMM still achieves highest accuracy (88.9\%), demonstrating the overall superiority of our framework.
\\

\noindent
\textbf{Classification on SDPH Landscapes.} The algebraic construction of persistence diagrams makes them difficult to integrate into standard statistical and machine learning methodology, due to the complicated geometry that they induce \citep{turner_frechet}.  A popular approach to bypassing this problem is to apply a vectorization method; one of the most popular vectorizations is the \emph{persistence landscape} \citep{bubenik2015statisticaltopologicaldataanalysis}. Briefly, landscapes consider the convex hull of persistence points over the diagonal rotated to the $x$-axis, giving a functional summary of the persistence diagram which is feasible to work with in statistics and machine learning.  This comparison experiment studies the performance of another standard approach to using persistent homology.

We evaluate this again under LOOCV with bootstrap. We converted the resultant PH$_0$, PH$_1$, and PH$_2$ diagrams as well as the PH$_1$ NW, PH$_1$ NE, and PH$_2$ NE quadrants in each subsample into persistence landscapes. After training and testing using 4 classifiers---logistic regression, SVM, random forest, and $k$-NN---we observed varied accuracies depending on the specific diagram subset and classifier as shown in Table \ref{tab:PL+PI}. The highest accuracy of 63.0\% was achieved by random forest when using all dimensions of the landscapes, while accuracies for other combinations ranged from as low as 22.2\% (logistic regression on PH$_1$ diagrams and $k$-NN using all dimensions) to around 59.3\% (random forest on PH$_2$ NE quadrant). These results indicate that our proposed signed distance method clearly outperforms this standard vectorization approach and that the persistence landscape construction procedure applied to SDPH diagrams results in a significant loss of important information for accurate inference.\\

\noindent
\textbf{Classification on SDPH Images.} As an alternative vectorization scheme to persistence landscapes, persistence images preserve the importance of feature persistence while enabling the use of standard machine learning algorithms \citep{persistence_images}. Each point in the persistence diagram is mapped to the birth-persistence plane where it contributes a localized Gaussian kernel typically weighted by persistence allowing more persistent features to exert greater influence. The resulting continuous function is discretized over a grid to form a pixel-based image. 

Similar to the landscapes, we compute the persistence images for the PH$_0$, PH$_1$, and PH$_2$ diagrams as well as the PH$_1$ NW, PH$_1$ NE, and PH$_2$ NE quadrants. Under the same LOOCV and bootstrap regime, with logistic regression, SVM, random forest, and $k$-NN classifiers, persistence images consistently outperformed landscapes. As shown in Table \ref{tab:PL+PI}, the highest accuracy achieved was 81.5\% using logistic regression on PH$_1$ (NE) diagrams. Overall, accuracies ranged from 48.1\% to above 70\%, indicating improved discriminatory power relative to persistence landscapes. While this performance is competitive with our proposed model, which achieved a highest accuracy of 88.9\%, our goal is not only to attain high classification accuracy but also build an interpretable, parameterized model that captures and explains the progression of AML. This is achieved through our phase-dependent GMM framework grounded in the theoretical results of \cite{song2023generalized}, which lend interpretability to the possible shapes and sizes of vessels and gaps within the bone marrow vasculature over the course of AML progression.\\

\begin{table}[htb]
    \centering
    \caption{Classification accuracy using persistence landscapes and persistence images on different homological dimensions and quadrants of SDPH diagram with logistic, SVM, random forest and $k$-NN classifiers.}
    \label{tab:PL+PI}
    \begin{tabular}{lccccccc}
        \toprule
        \textbf{Classifier} & \textbf{All Dims} & \textbf{PH0} & \textbf{PH1} & \textbf{PH2} & \textbf{PH1 NW} & \textbf{PH1 NE} & \textbf{PH2 NE} \\
        \midrule
        \addlinespace[0.5em]
        \multicolumn{8}{l}{\textbf{Persistence Landscapes}} \\
        \addlinespace[0.2em]
        Logistic Regression & 0.556 & 0.556 & 0.222 & 0.519 & 0.296 & 0.333 & 0.481 \\
        SVM                 & 0.370 & 0.444 & 0.407 & 0.556 & 0.296 & 0.296 & 0.481 \\
        Random Forest       & 0.630 & 0.407 & 0.222 & 0.556 & 0.222 & 0.444 & 0.593 \\
        $k$-NN                 & 0.222 & 0.481 & 0.259 & 0.519 & 0.296 & 0.296 & 0.481 \\
        \addlinespace[0.8em]
        \multicolumn{8}{l}{\textbf{Persistence Images}} \\
        \addlinespace[0.2em]
        Logistic Regression & 0.704 & 0.630 & 0.815 & 0.741 & 0.815 & 0.778 & 0.741 \\
        SVM                 & 0.667 & 0.556 & 0.630 & 0.630 & 0.667 & 0.630 & 0.630 \\
        Random Forest       & 0.741 & 0.481 & 0.704 & 0.741 & 0.630 & 0.741 & 0.704 \\
        $k$-NN                 & 0.667 & 0.556 & 0.630 & 0.630 & 0.667 & 0.630 & 0.630 \\
        \bottomrule
    \end{tabular}
\end{table}

\noindent
\textbf{Classification on Morphometrics.} To contextualize the difficulty of phase classification using morphometric measures, we present a classification task on a simplified dataset. Specifically, we selected a representative sample from each of the Phase O, I, and II groups of samples and extracted 30 subregions from each for classification. For each subregion, we computed a comprehensive set of morphometric features derived from the literature \citep{BROWN2022100357,lafage,kvas}, encompassing vessel quantity, quality, network architecture, and spatial relationships. These included measures such as total vessel volume, vessel number per unit area, mean vessel diameter, curvature, tortuosity, total vessel length, junction counts, and inter-vessel distances.

We then trained 3 standard classifiers---random forest, logistic regression, and SVM---on these features to discriminate between the 3 phases. The resulting classification accuracies were 66.7\% (random forest), 64.4\% (logistic regression), and 54.4\% (SVM), indicating limited discriminative power. These results highlight the intrinsic difficulty of the task, especially when relying solely on conventional morphometric descriptors. Singular morphometric measures often fail to capture the complex, large-scale heterogeneity of bone marrow architecture, as evidenced by the poor classification performance in this setting. The overlapping and variable nature of vascular structures across phases likely contributes to the limited separability, underscoring the need for more expressive and topology-aware representations such as those employed in our framework.

\section{Discussion}
\label{sec:disucussion}

In this paper, we presented a new topological method for classifying and modeling the architectural changes in the bone marrow vasculature associated with the progression of AML. We quantified these changes using persistent homology applied to proprietary confocal microscopy images where the fineness of the resolution is a novelty in AML imaging. Specifically, we implemented SDPH to quantify and analyze the data, enabling us to categorize data samples into morphologically-derived phases corresponding to different stages of disease progression. Our analysis reveals that clustering based on SDPH summaries aligns closely with biological intuition regarding the stages of AML progression, indicating that SDPH clustering applied to images is is an effective method to study AML. We subsequently developed a class of phase-dependent GMMs that effectively approximated the spatial patterns captured within the SDPH diagrams, reflecting transitions in vascular morphology. We have shown that these models can be employed to classify new observations into phases or stages of AML progression.  Through a comparison with straightforward approaches that compute classical persistent homology or apply a standard vectorization, we show that on fine resolution confocal microscopy images, the SDPH framework is required for more meaningful inference results.

Our findings indicated that the $\text{PH}_2$ NE and PH$_1$ NW quadrants of the SDPH diagrams are notably more informative and discriminative across stages of engraftment compared to other regions. This is reflected in both the accuracy of our models and their predictive performance. This observation aligns with the understanding that points in this quadrant correspond to ``real'' textures, representing static cycles in the vascular network, rather than ``virtual'' textures (noise arising from the construction of the filtration of the signed distance function). Thus, modeling on the $\text{PH}_2$ NE or PH$_1$ NW regions alone provides sufficient information about the vessel structure to allow for accurate prediction of the disease progression.\\

\noindent
\textbf{Biological Implications.} While most previous studies have primarily focused on investigations of late-stage leukaemia morphology \citep{passaro, DUARTE20181507}, our work successfully distinguishes between the early and late stages of AML progression, highlighting clear morphological divergence between them quantified using SDPH.  In addition to our SDPH approach combined with GMMs for inference, an additional biological novelty of our work is the study of early vs.~late stage in a comparative context, while previous significant AML research interests have focused on studying primarily the late stage.  Our study was  possible due to the quality of our proprietary confocal microscopy imaging dataset at such fine resolutions.

The morphological features identified by our models and analyses correspond to known biological processes: in the early stage, we identified higher number of thinner, more compact vessels as a result of angiogenesis; in the late stage, vessels are sparser apart due to clusters of AML cells deforming the bone marrow causing spatial constraints. Given these distinct differences between early and late stages, our work provides a solid, quantitative foundation for further research into the biological mechanisms driving these patterns, particularly during the early stages of AML progression, and the development to later stages.\\

\noindent
\textbf{Limitations and Future Outlook.} Our study was subject to notable constraints, given the nature of the problem of interest and the data available.  Our data are limited, given the high time and equipment costs required to generate them, which necessitated the use of subsampling methods for classification.  Another limitation is the inherent uncertainty due to the levels of engraftment having been derived as global averages of bones other than the imaged femur, which arises in the reported percentage, and thus prevents their use as exogenous variables. 

There are important and interesting future directions of research in the statistical and computational components of our study, in addition to the previously-discussed biological component.  Given a larger collection with more continuous measures of engraftment levels or time after engraftment, we envision extending our model to a ``time''-dependent Gaussian mixture model for the progression, where the parameters of the GMM vary with time or engraftment level, rather than fixed for each phase. This enhancement could lead to more accurate predictions of AML progression. We also envision that our framework could be adapted for application to other biological systems with similar vascular structures. Moreover, our model naturally lends itself to extensions for uncertainty quantification, for example, by constructing confidence regions around GMM components to assess compatibility with a given phase model. This, in turn, may help with identifying atypical or borderline sample behaviors arising from biological variation---patterns we have already begun to observe in our analysis.

\section*{Software and Data Availability}
The raw imaging data from this study is proprietary data and cannot be openly shared, however, the summaries of the analysis such as the vectorized summaries, SDPH diagrams, and code to replicate steps of the methodology can be found at \href{https://github.com/Qiquan-Wang/AML_SDPH}{this repository}.

We utilized the following packages in Python for the computation of SDPH: giotto-tda \citep{giotto}; GUDHI \citep{gudhi}; and CubicalRipser \citep{kaji2020cubical}. Clustering was performed using the scikit-learn package in Python \citep{scikit-learn}. The KeOps package \citep{keops} was used for the efficient computation of Gaussian kernel estimates. The pomegranate package \citep{schreiber2018pomegranate} was used for the training and evaluation of the models. A comprehensive list of packages used can be found in the above repository.

\section*{Acknowledgments}

This work was supported by the Francis Crick Institute, which receives its core funding from Cancer Research UK (CC1045), the UK Medical Research Council (CC1045), and the Wellcome Trust (CC1045). 

Q.W.~is funded by a CRUK--Imperial College London Convergence Science PhD studentship at the UK EPSRC Centre for Doctoral Training in Modern Statistics and Machine Learning (2021 cohort, PIs Monod/Williams), which is supported by Cancer Research UK under grant reference [CANTAC721\textbackslash10021].
A.S.~was funded by the Francis Crick Institute and Imperial College London.
A.B.~was funded by an junior EHA fellowship. 
A.M.~is supported by the Engineering and Physical Sciences Research Council under grant reference [EP/Y028872/1].

\bibliographystyle{authordate3}
\bibliography{main}

\vfill\eject

\appendix

\section{Tri-Cluster Texture Decomposition of the Long Regions of Bone Samples}
\label{App:long_region_texture_decomposition}

\begin{figure}[htb]
    \centering
    \includegraphics[width=\linewidth]{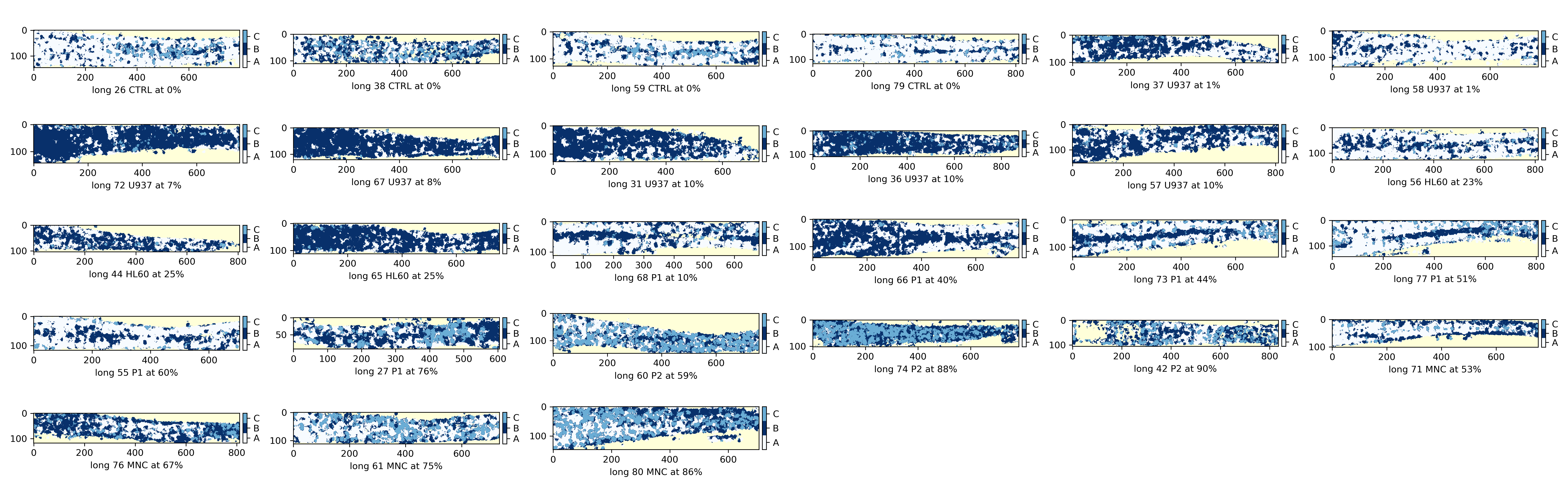}
    \caption{Spatial mapping of the 3 textures onto 2D slices of the long regions of bone samples \citep{annathesis} (the textures are derived from the same clustering that produced Figures 7 and 8 from the article for the knee region via $k$-means clustering).}
    \label{fig:long_texture}
\end{figure}

In Section 4.2 of the main paper, we presented the texture decomposition of the knee region of the bone sample with 3 clusters. We note that applying the same technique to the long region of the femur bones may not yield comparable results, given the greater structural variability and complexity observed along the bone’s length. Here we present the exact comparison on the long region of the bone when we apply the same texture decomposition with 3 clusters. Figure \ref{fig:long_texture} shows the 3 clusters projected onto 2D slices of the long bone samples. We label the white, light blue, and dark blue colored textures as textures A, B, and C. We then see that, similar to the knee regions of the sample, the CTRL samples are mainly composed of texture A, however, this behavior is also seen in other samples such as the P1 samples and U937 sample at 1\% engraftment. The introduction of a large proportion of the darker texture C distinguishes the samples with low levels of engraftment from the CTRL samples. Transitioning to the samples with higher levels of engraftment, we observe diminishing proportions of texture C replaced by largely texture B and texture A. Yet, we see visible similarities between the CTRL samples and samples with medium levels of engraftment ($\sim $55\%), which cannot be explained by the patterns identified in the knee region and may be due to the highly nonlinear spread of the AML cells in the bone.

\begin{figure}[htb]
    \centering
    \begin{subfigure}{0.45\textwidth}
        \includegraphics[width=\linewidth]{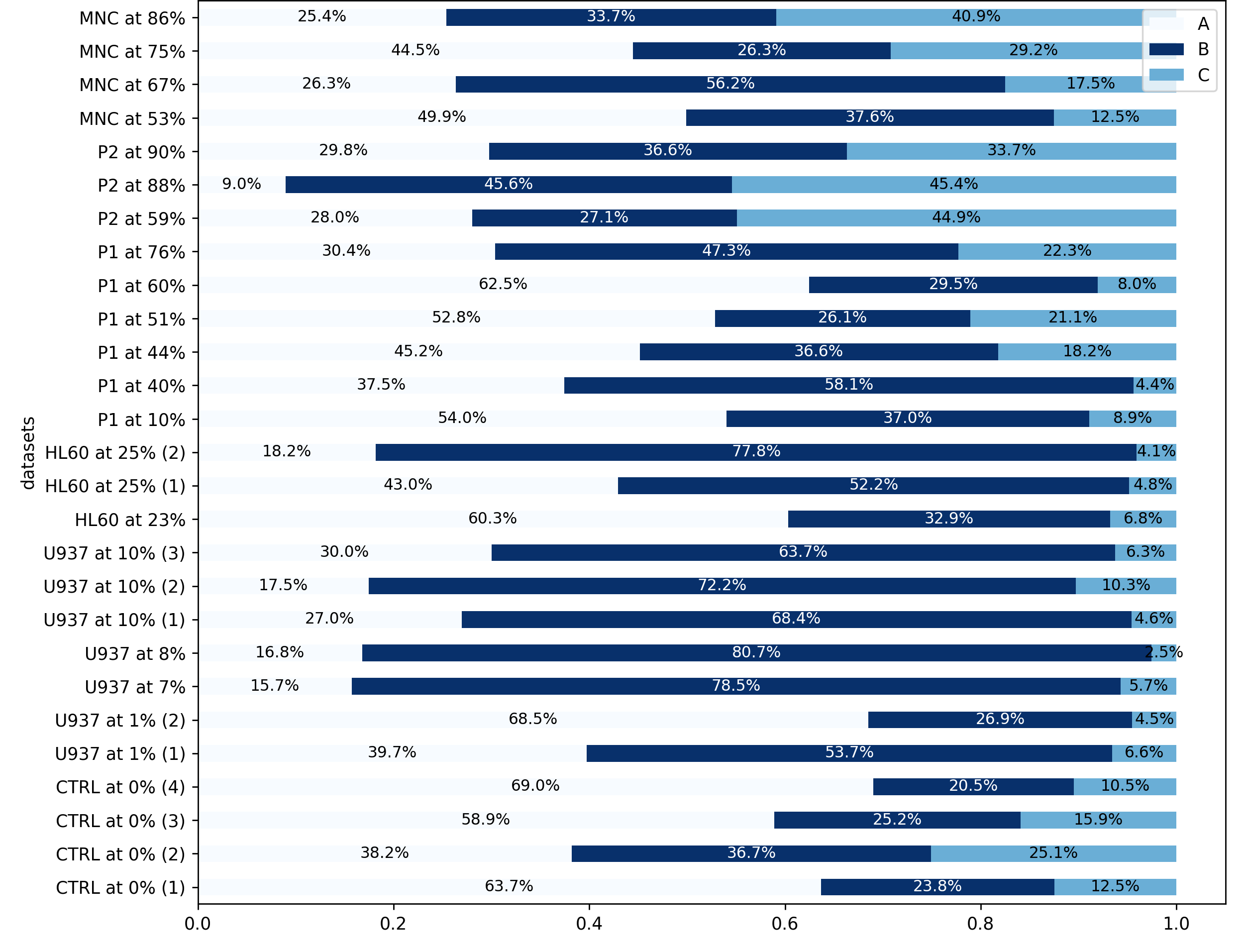}
        \caption{Texture decomposition of samples into 3 textural clusters represented with percentages.}
        \label{fig:long_percentages}
    \end{subfigure}
    \begin{subfigure}{0.53\textwidth}
        \includegraphics[width=\linewidth]{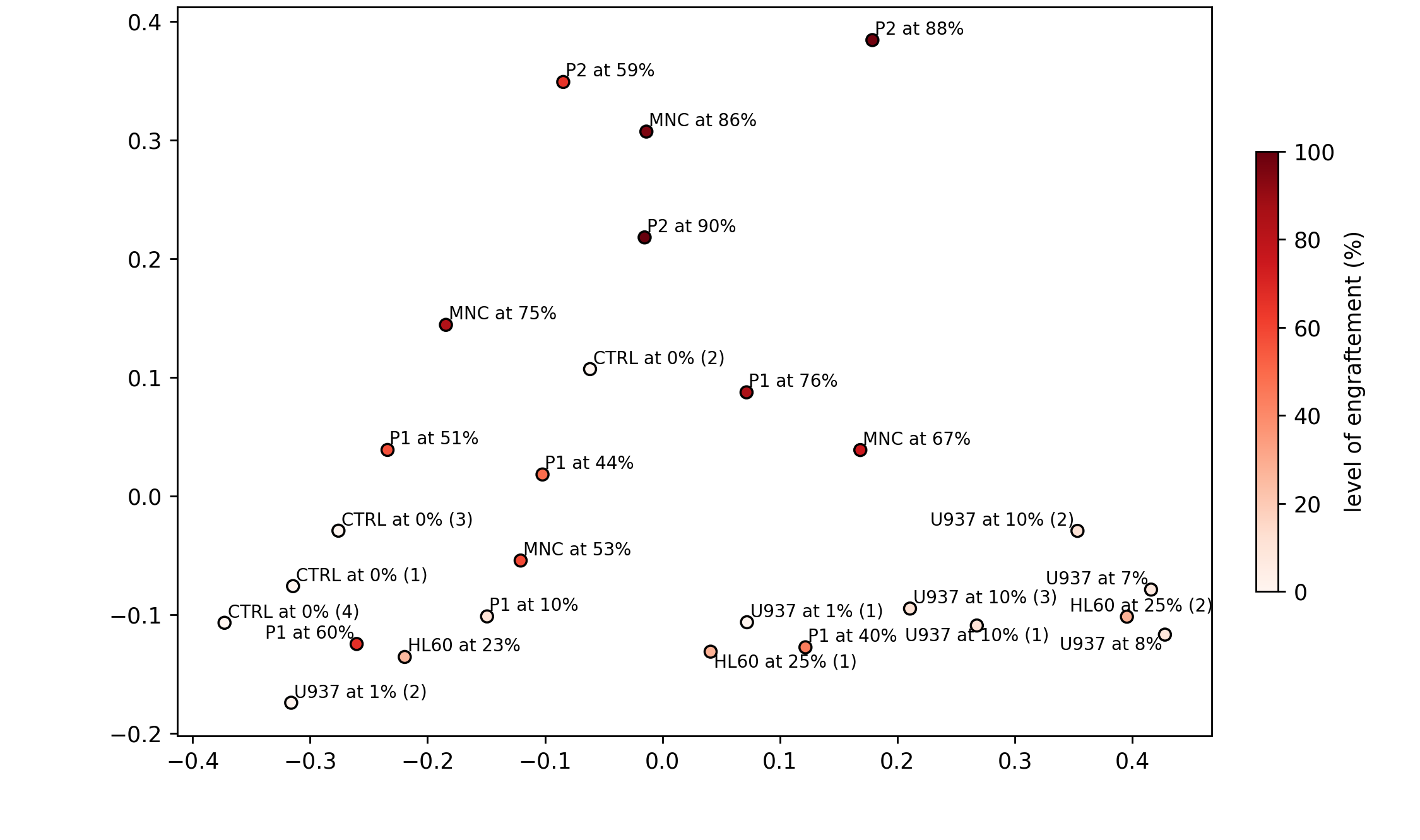}
        \caption{Projection of texture decomposition into first 2 principal components via PCA for 3 textural clusters.}
        \label{fig:long_pca}
    \end{subfigure}
    \caption{Local texture analysis of long region of bone samples using clustering with 3 clusters.}
    \label{fig:mainfig}
\end{figure}
Figure \ref{fig:long_percentages} shows the proportions of each texture present in each sample, where we apply PCA to the decomposition, embedding the texture decomposition of each sample into the plane spanned by the first 2 principal components of the reduction. Figure \ref{fig:long_pca} displays this embedding. Here, the boundaries are less discernible; in particular, the second CTRL sample presents itself as more similar to the samples with high levels of engraftment and the P1 60\% sample is found to be ``closer'' to the CTRL samples and samples with much lower levels of engraftment, despite some similarities to those behaviors identified in the knee region. Furthermore, this finding is consistent with higher numbers of clusters experimented.

\section{Exploring Alternative Numbers of Clusters and Clustering Algorithms in Texture Decomposition}
\label{app:clustering_exploration}

In Section 4.2 of the article, we noted that alternative clustering algorithms, such as GMM and CLARA, and with different numbers of textural clusters, produce similar groupings of the samples into the 3 phases. In this section, we illustrate the agreement in group allocations with further examples and figures.

\subsection{Experimenting with Different Number of Clusters using the $k$-Means Algorithm in the Cluster Decomposition}
\label{app:kmeans_different_k}

In Figures 7 and 9 in Section 4.2 of the article, we showed that the texture decomposition with 3 clusters gave rise to 3 distinct clusters of samples in the PCA-embedded space corresponding to the different phases. We show with the following Figures \ref{fig:kmeans10}, \ref{fig:kmeans15}, and \ref{fig:kmeans25}, that using different and larger numbers of clusters (10, 15, and 25 respectively) in the analysis nevertheless leads to the same groupings. The phases are colored using green, orange, and red to represent Phases O, I, and II. We observe that the different colored samples can be grouped distinctly on the PCA-embedded plane.

\begin{figure}[h]
    \centering
    \begin{subfigure}{0.45\textwidth}
    \captionsetup{width=0.95\textwidth}
        \includegraphics[width=\linewidth]{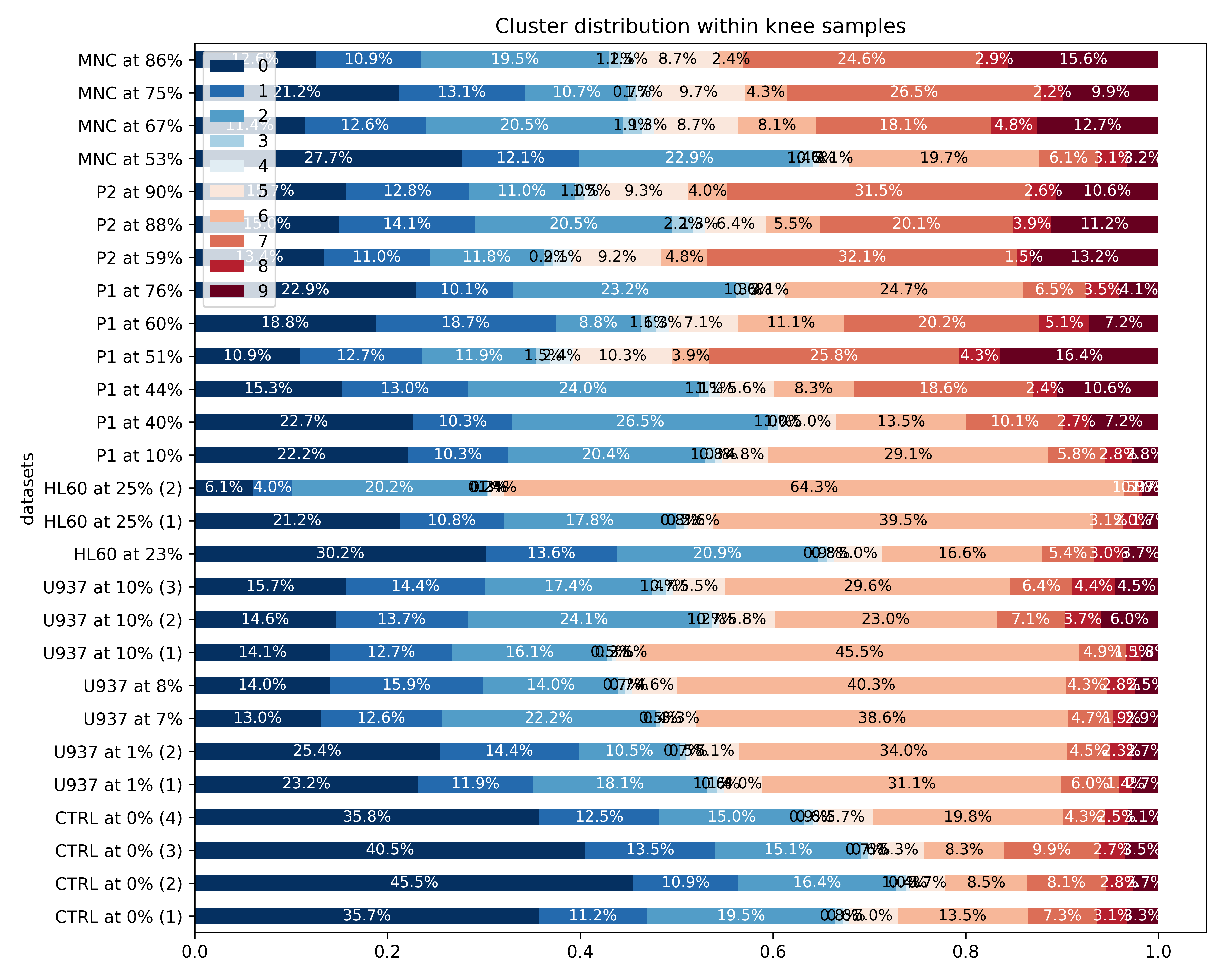}
        \caption{Texture decomposition of samples into 10 textural clusters represented with percentages.}
        \label{fig:10kmeanspercent}
    \end{subfigure}
    \begin{subfigure}{0.53\textwidth}
    \captionsetup{width=0.95\textwidth}
        \includegraphics[width=\linewidth]{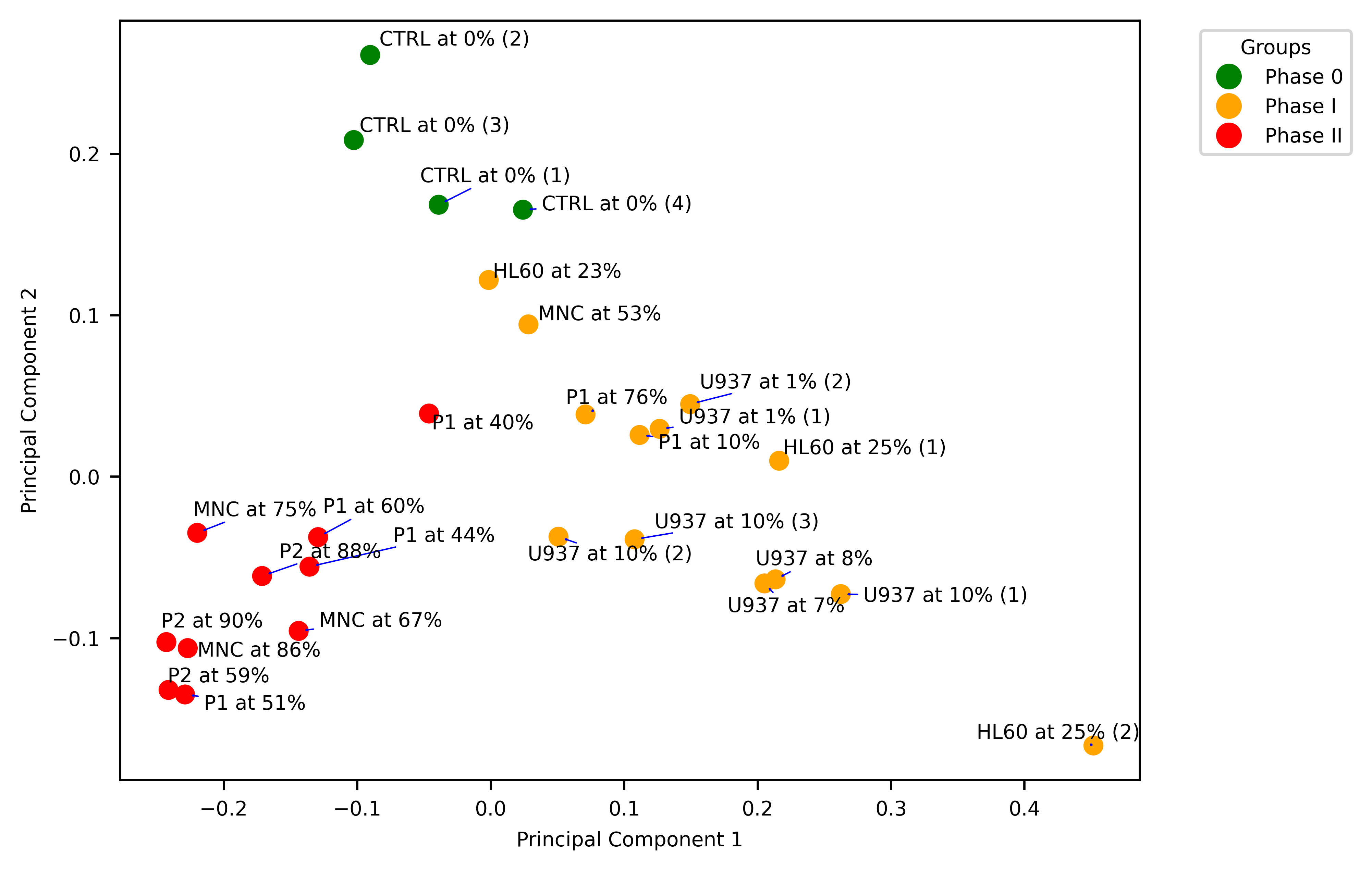}
        \caption{Projection of texture decomposition into first 2 principal components via PCA for 10 textural clusters,}
        \label{fig:10kmeanspca}
    \end{subfigure}
    \caption{Local texture analysis of knee region of bone samples using clustering with 10 clusters.}
    \label{fig:kmeans10}
\end{figure}

\begin{figure}[htb]
    \centering
    \begin{subfigure}{0.45\textwidth}
    \captionsetup{width=0.95\textwidth}
        \includegraphics[width=\linewidth]{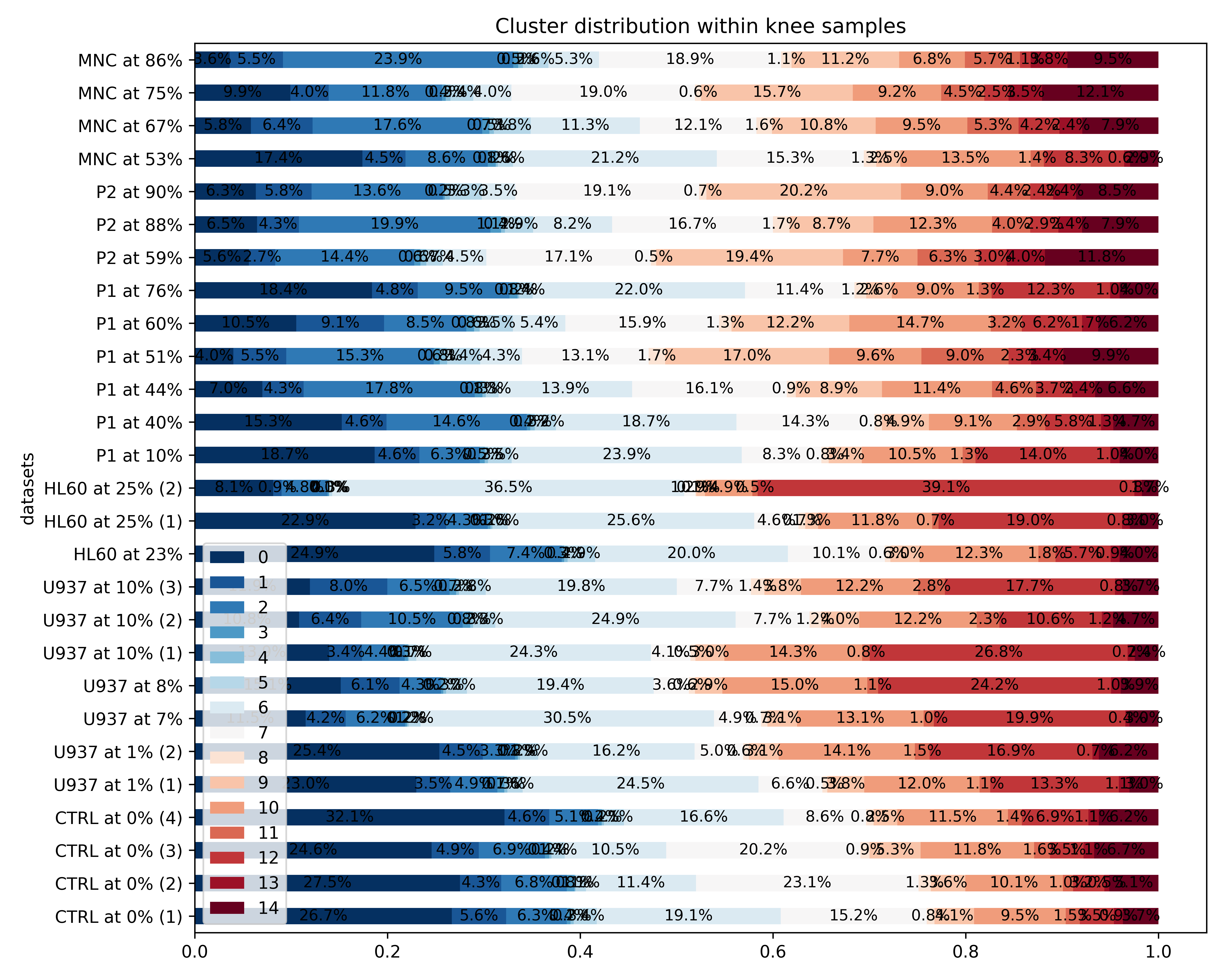}
        \caption{Texture decomposition of samples into 15 textural clusters represented with percentages.}
        \label{fig:15kmeanspercent}
    \end{subfigure}
    \begin{subfigure}{0.53\textwidth}
    \captionsetup{width=0.95\textwidth}\includegraphics[width=\linewidth]{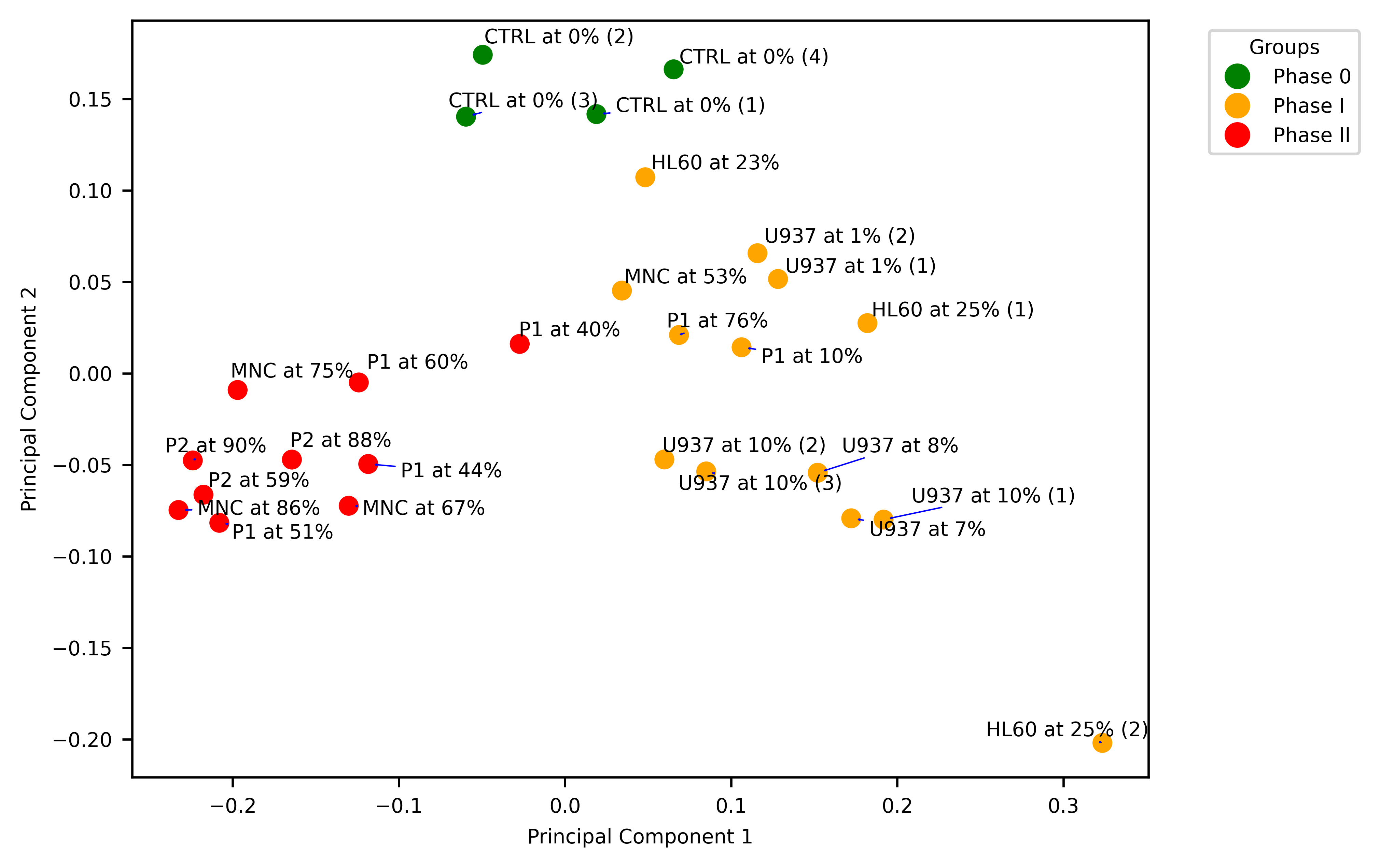}
        \caption{Projection of texture decomposition into first 2 principal components via PCA for 15 textural clusters.}
        \label{fig:15kmeanspca}
    \end{subfigure}
    \caption{Local texture analysis of knee region of bone samples using clustering with 15 clusters.}
    \label{fig:kmeans15}
\end{figure}

\begin{figure}[H]
    \centering
    \begin{subfigure}{0.45\textwidth}
    \captionsetup{width=0.95\textwidth}
    \includegraphics[width=\linewidth]{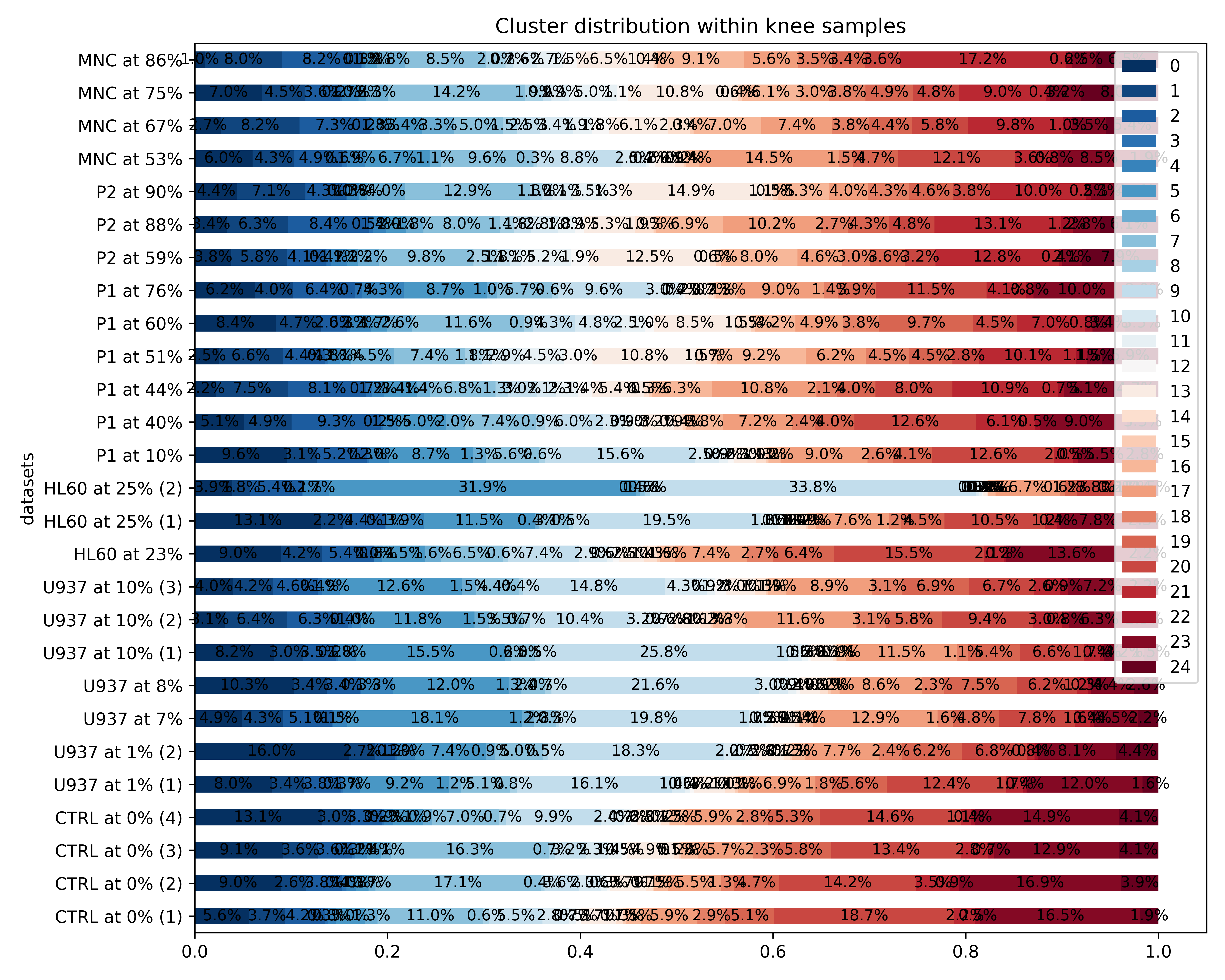}
        \caption{Texture decomposition of samples into 25 textural clusters represented with percentages.}
        \label{fig:25kmeanspercent}
    \end{subfigure}
    \begin{subfigure}{0.53\textwidth}
    \captionsetup{width=0.95\textwidth}
        \includegraphics[width=\linewidth]{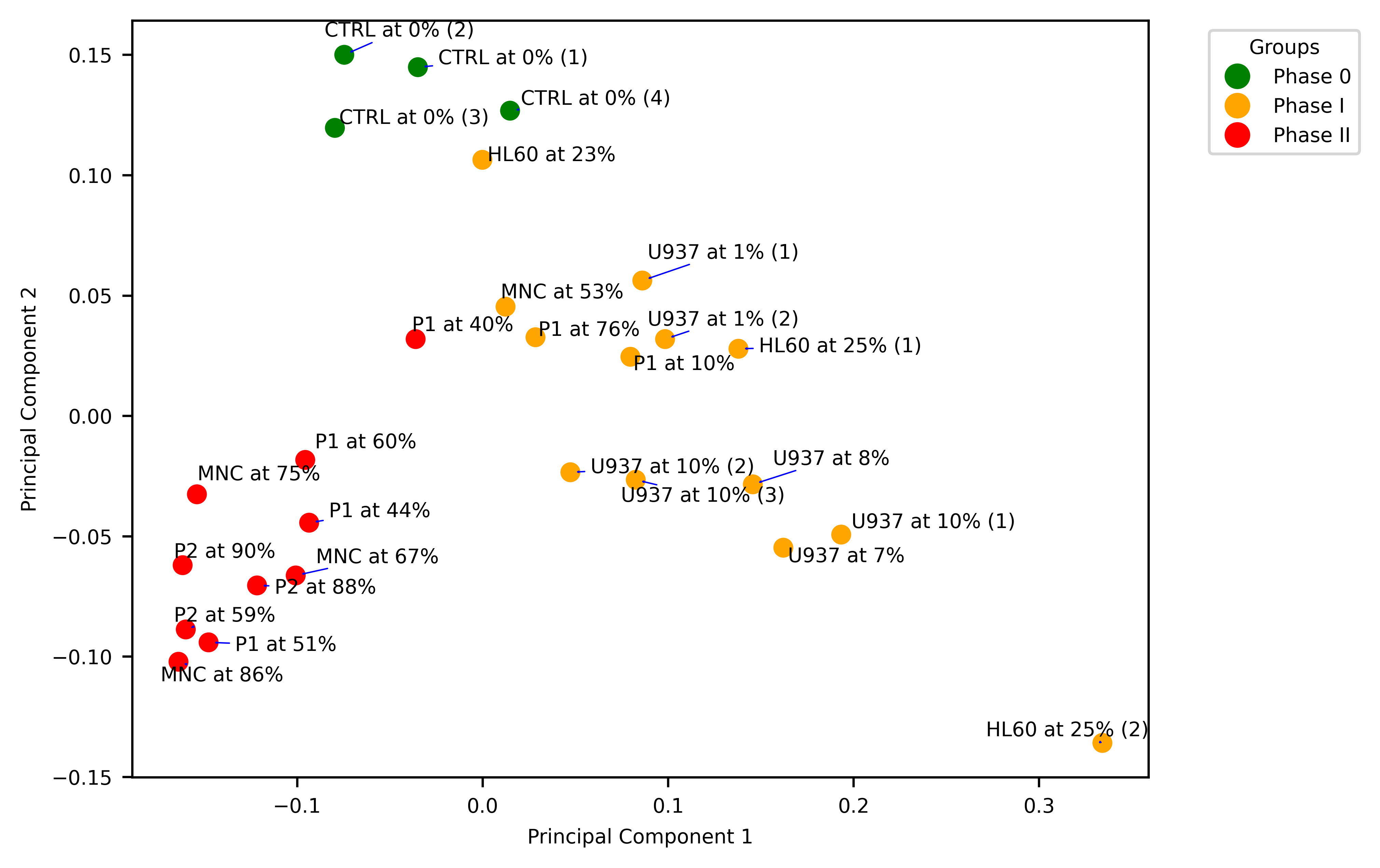}
        \caption{Projection of texture decomposition into first 2 principal components via PCA for 25 textural clusters.}
        \label{fig:25kmeanspca}
    \end{subfigure}
    \caption{Local texture analysis of knee region of bone samples using clustering with 25 clusters.}
    \label{fig:kmeans25}
\end{figure}

\subsection{Alternative Clustering Methods for Texture Decomposition}
\label{app:alternative_alg}

To further verify the consistency of our clustering results, we experimented with alternative clustering methods to identify similar textures in the bone samples and indeed the following results show coherence in the phase groupings. The methods chosen were the Gaussian mixture model (GMM, \cite{gmm}) clustering and the Clustering Large Application (CLARA) algorithm, which is an extension of the Partition around medoids (PAM) method \citep{clara}, due to their computational simplicity. Other methods such as hierarchical clustering, DBSCAN (Density-Based Spatial Clustering of Applications with Noise), or PAM, entail computational complexity of $O(n^2)$ or greater, where $n$ is the number of features in total across all the samples, which means that the memory required to complete the clustering would be in the order of petabytes.

We now briefly outline the methods. Partition around medoids (PAM), also known as $k$-medoids, is a clustering algorithm that is similar to the $k$-means algorithm with the modification that the centroids now instead can only be points in the dataset. This makes the algorithm more robust to outliers and the computed means are more meaningful, however it is more computationally expensive, especially for large datasets as distance matrices need to be computed for the entire dataset, making the complexity $O(n^2)$. CLARA is an extension of PAM which deals with the computational complexity of large datasets through subsampling. CLARA subsamples multiple subsets of the original data and within iterations retains the medoids of the subset with the lowest average or total within-cluster dissimilarity. GMM is described in Section 5.1 of the article and the data points are assigned to clusters according to probabilities of belonging to each Gaussian distribution, denoted above by $r_{ijk}$.

\subsubsection{Results using GMM Clustering}
\label{app:gmm}

\begin{figure}[htb]
    \centering
    \begin{subfigure}{0.45\textwidth}
    \captionsetup{width=0.95\textwidth}
        \includegraphics[width=\linewidth]{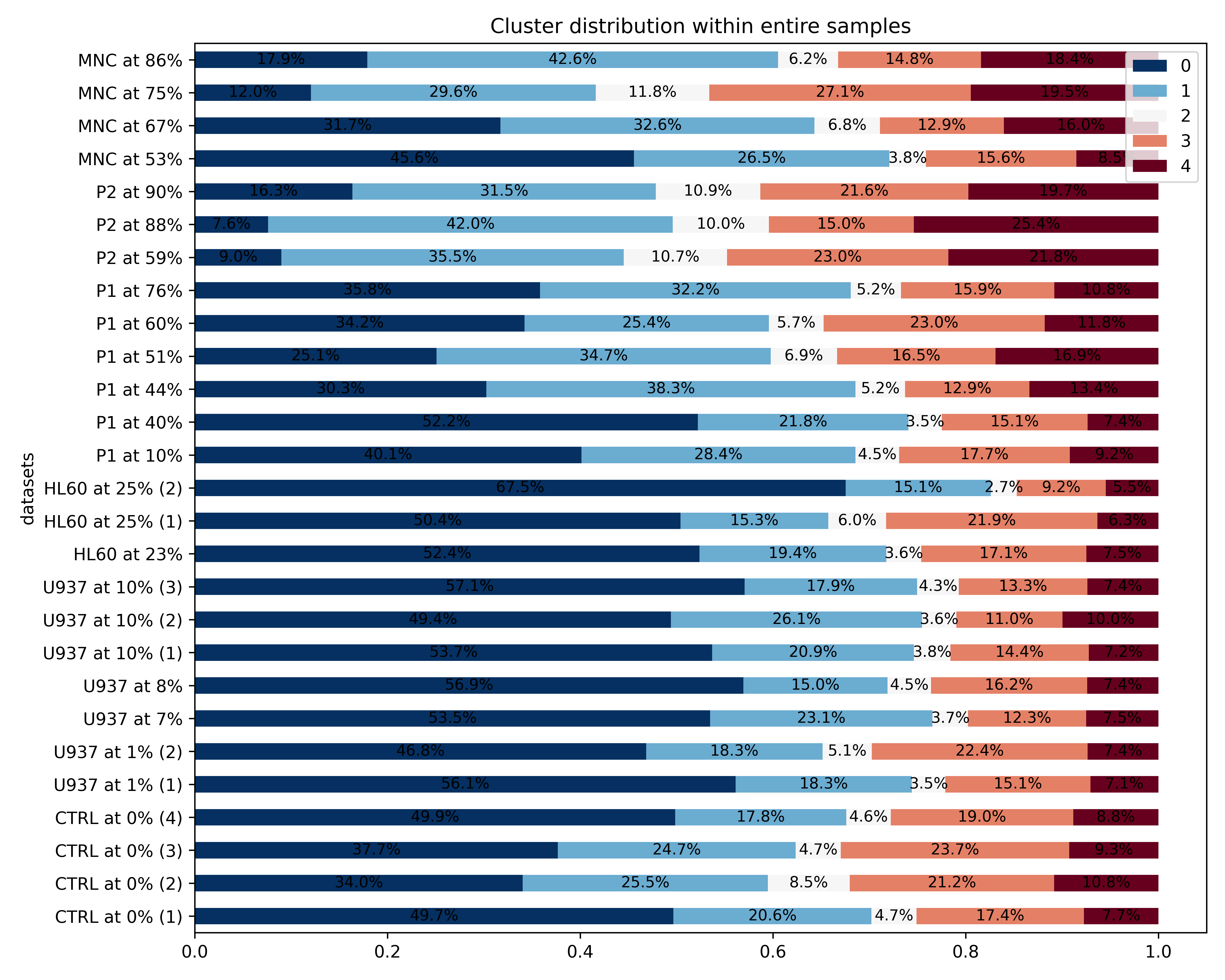}
        \caption{Texture decomposition of samples into five textural clusters using GMM clustering }
        \label{fig:5gmmpercent}
    \end{subfigure}
    \begin{subfigure}{0.53\textwidth}
    \captionsetup{width=0.95\textwidth}
        \includegraphics[width=\linewidth]{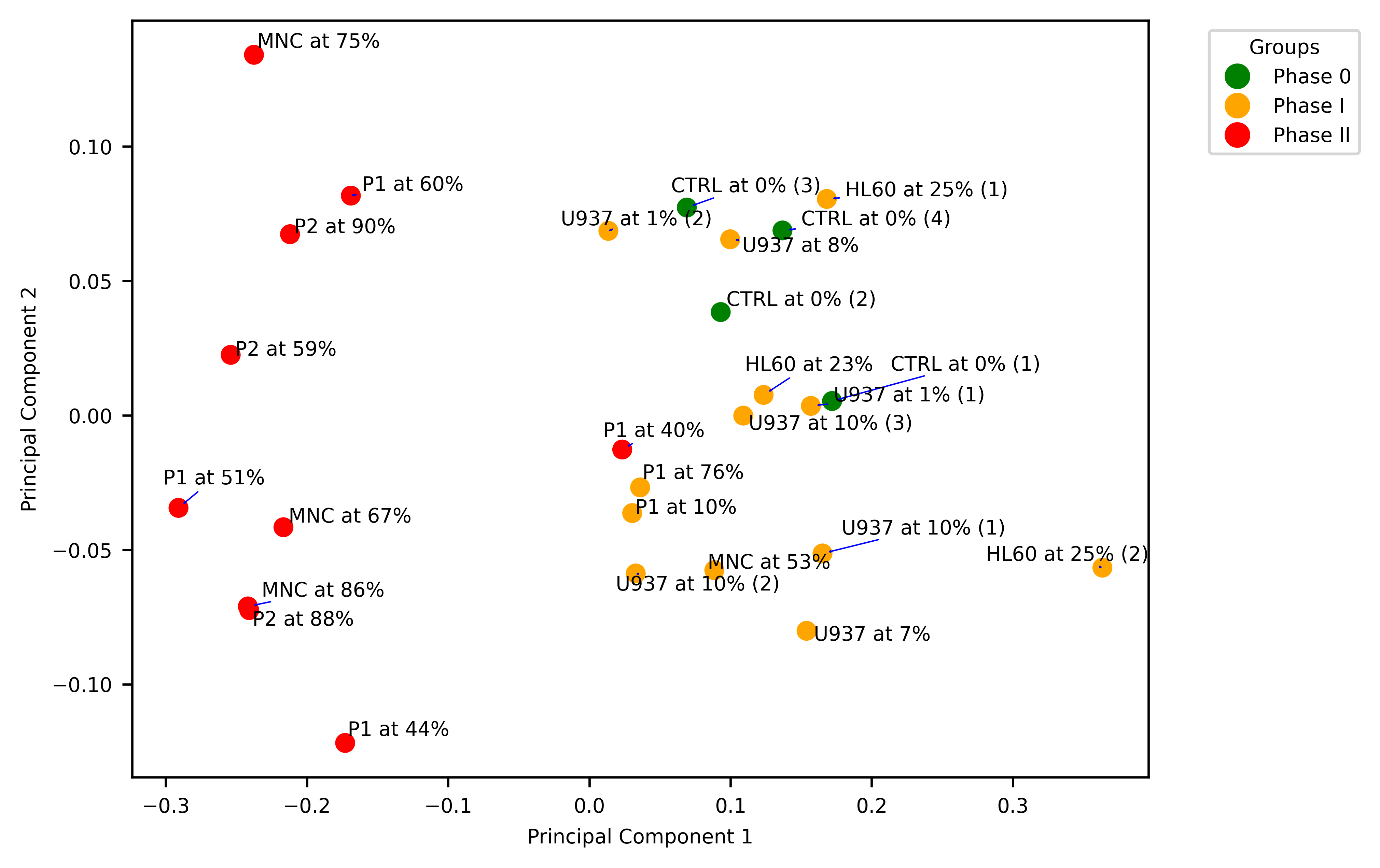}
        \caption{Projection of texture decomposition into first 2 principal components via PCA for five textural clusters.}
        \label{fig:5gmmpca}
    \end{subfigure}
    \caption{Local texture analysis of knee region of bone samples using GMM clustering with 5 clusters.}
    \label{fig:gmm5}
\end{figure}

\begin{figure}[H]
    \centering
    \begin{subfigure}{0.45\textwidth}
    \captionsetup{width=0.95\textwidth}
        \includegraphics[width=\linewidth]{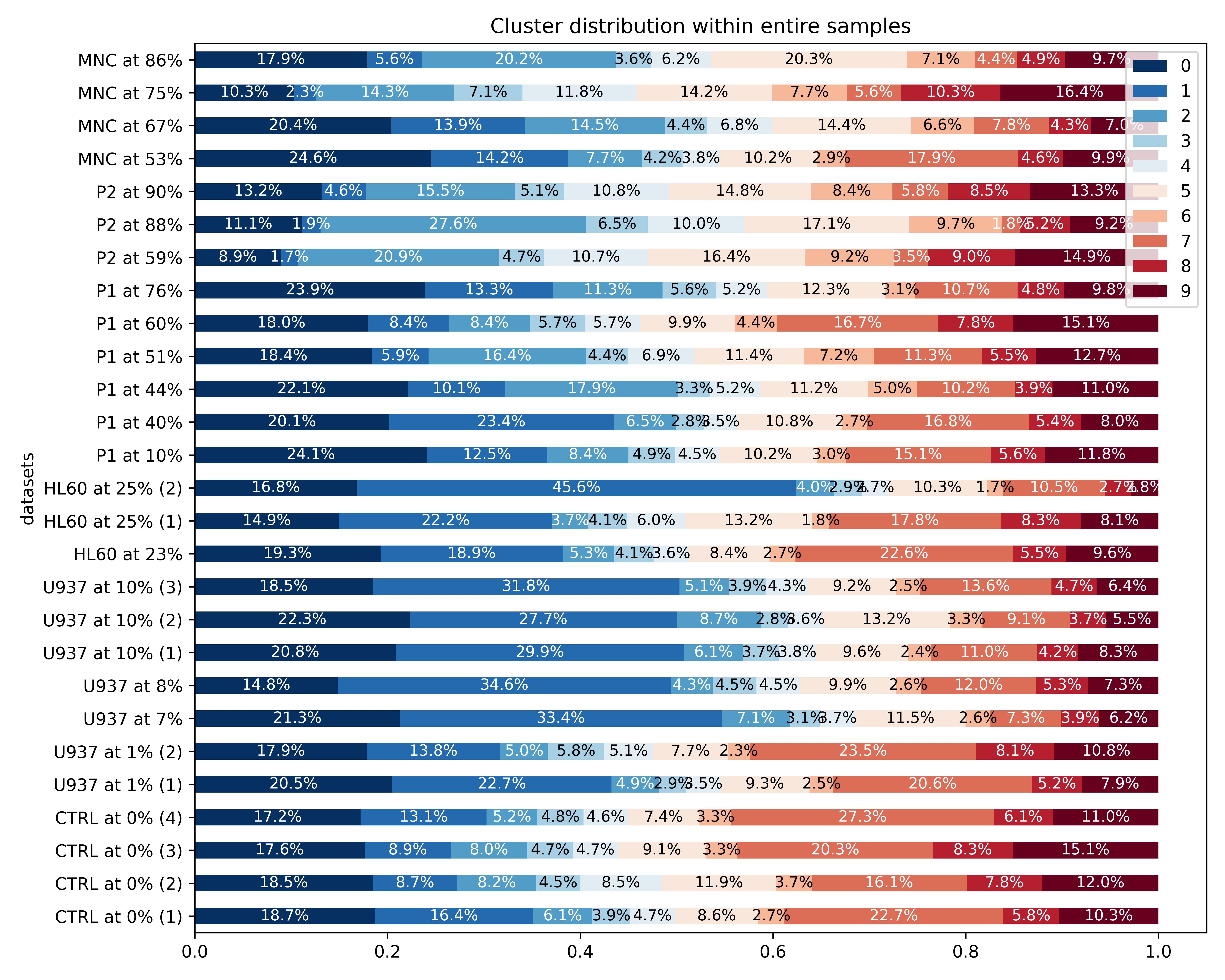}
        \caption{Texture decomposition of samples into ten textural clusters using GMM clustering }
        \label{fig:10gmmpercent}
    \end{subfigure}
    \begin{subfigure}{0.53\textwidth}
    \captionsetup{width=0.95\textwidth}
        \includegraphics[width=\linewidth]{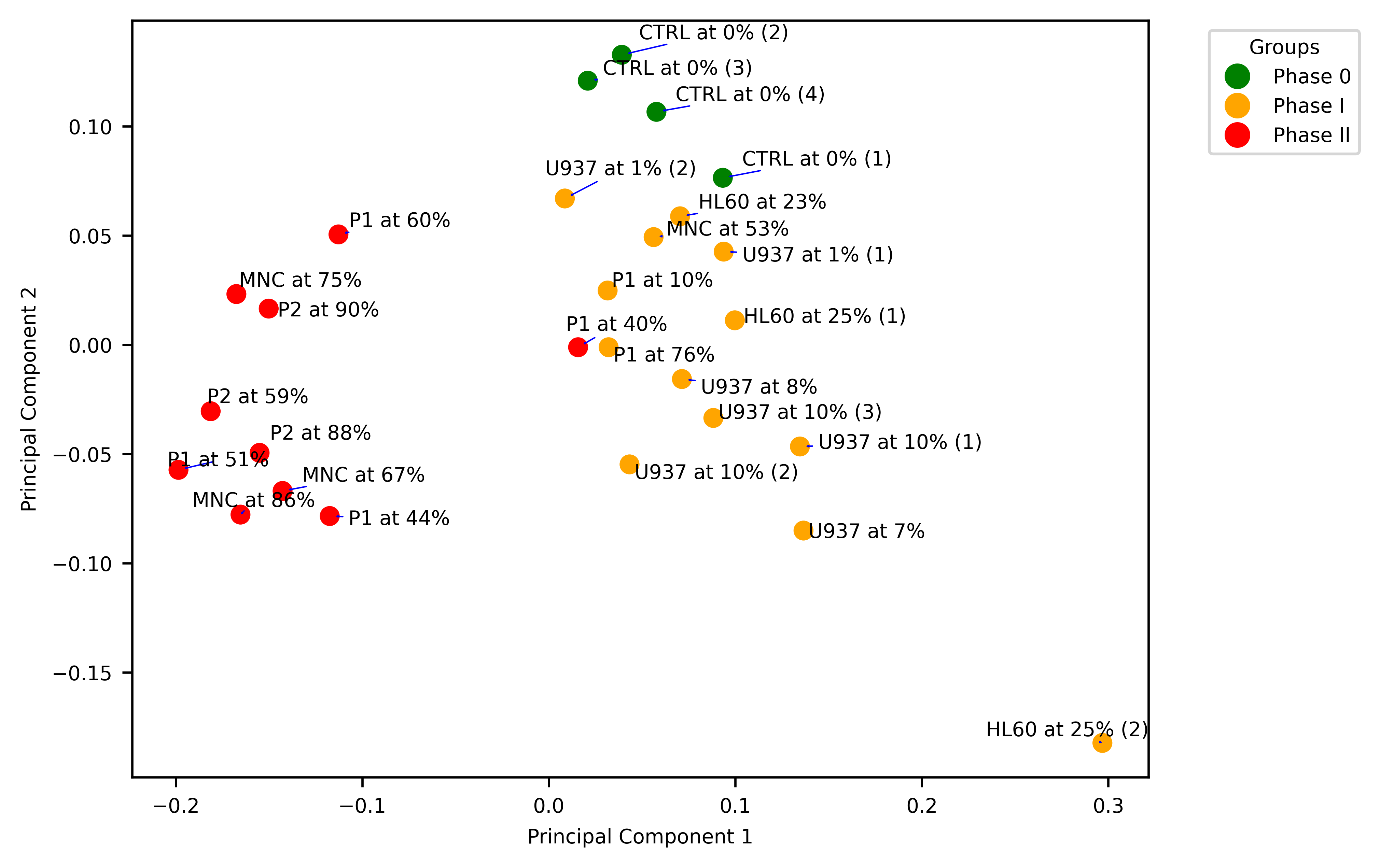}
        \caption{Projection of texture decomposition into first 2 principal components via PCA for 10 textural clusters.}
        \label{fig:10gmmpca}
    \end{subfigure}
    \caption{Local texture analysis of knee region of bone samples using GMM clustering with 10 clusters.}
    \label{fig:gmm10}
\end{figure}

Using GMM clustering, we obtain results similar to those shown in Figures \ref{fig:gmm5} and \ref{fig:gmm10} for 5 and 10 clusters. For some number of clusters, the CTRL samples can be seen as distinct from the samples of the other two phase groups as in Figure \ref{fig:5gmmpca}, while for others, the CTRL samples are clustered with other phase groups as in Figure \ref{fig:10gmmpca}. However, from the experimental design, we understand the CTRL samples are distinct from the other samples as they are the unmodified bone samples and so we still maintain a control grouping for these samples. 

\subsubsection{Results using CLARA}
\label{app:clara}

\begin{figure}[htb]
    \centering
    
    \begin{subfigure}{0.48\textwidth}
        \includegraphics[width=\linewidth]{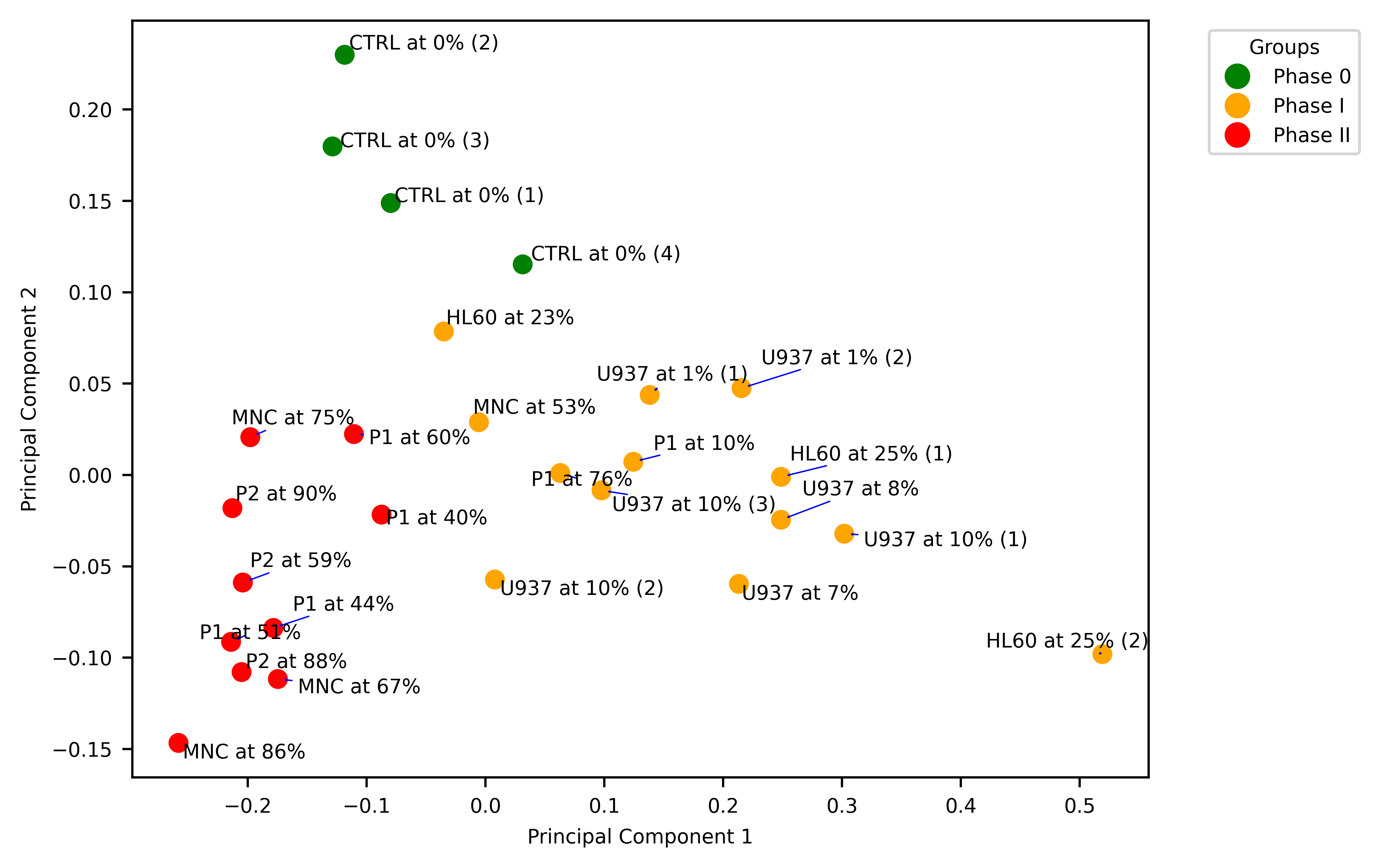}
        \caption{PCA projection for 6 textures.}
        \label{fig:clara6}
    \end{subfigure}
    \hfill
    \begin{subfigure}{0.48\textwidth}
        \includegraphics[width=\linewidth]{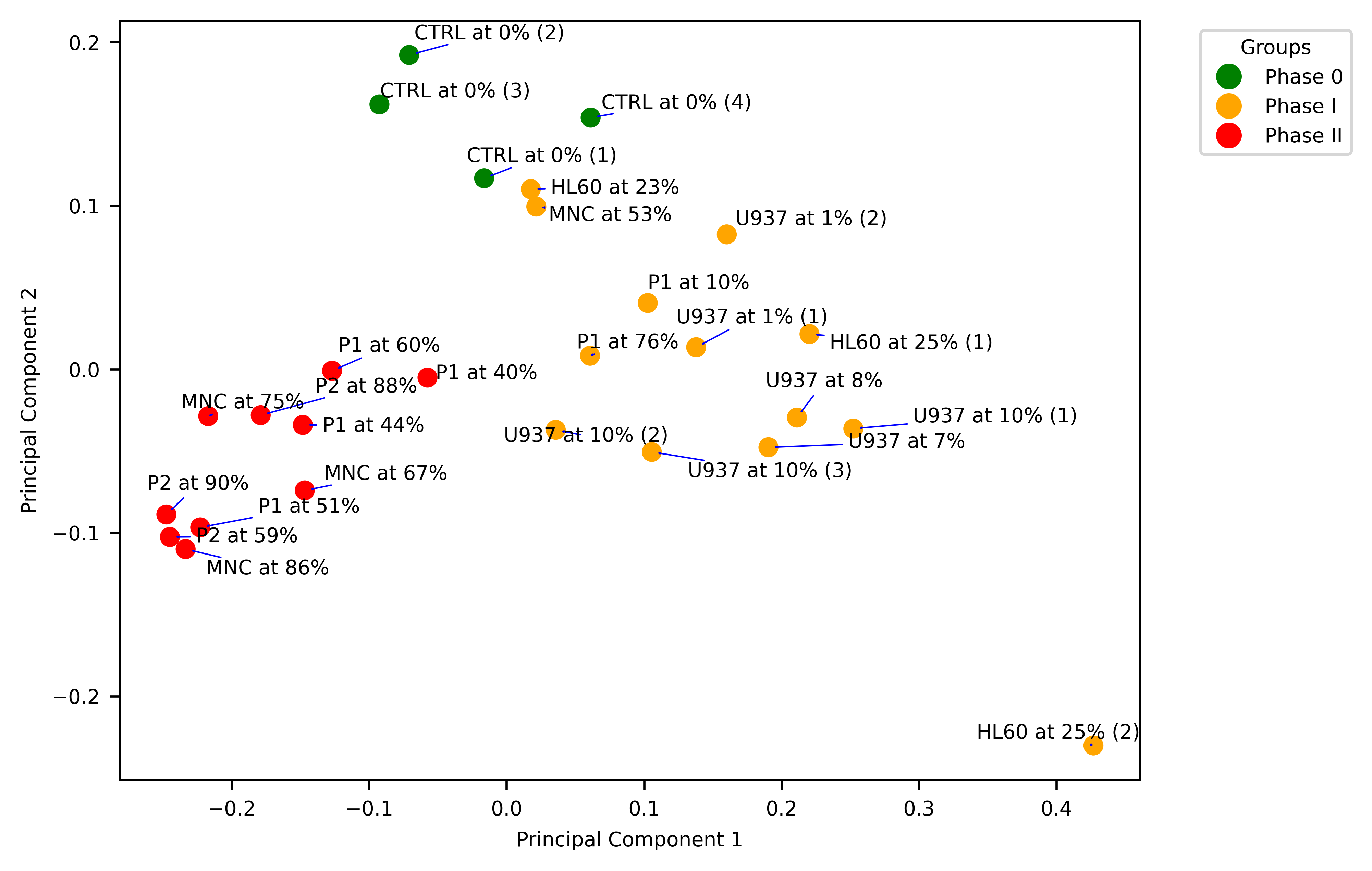}
        \caption{PCA projection for 10 textures.}
        \label{fig:clara10}
    \end{subfigure}
    \vspace{\baselineskip}
    \begin{subfigure}{0.48\textwidth}
        \includegraphics[width=\linewidth]{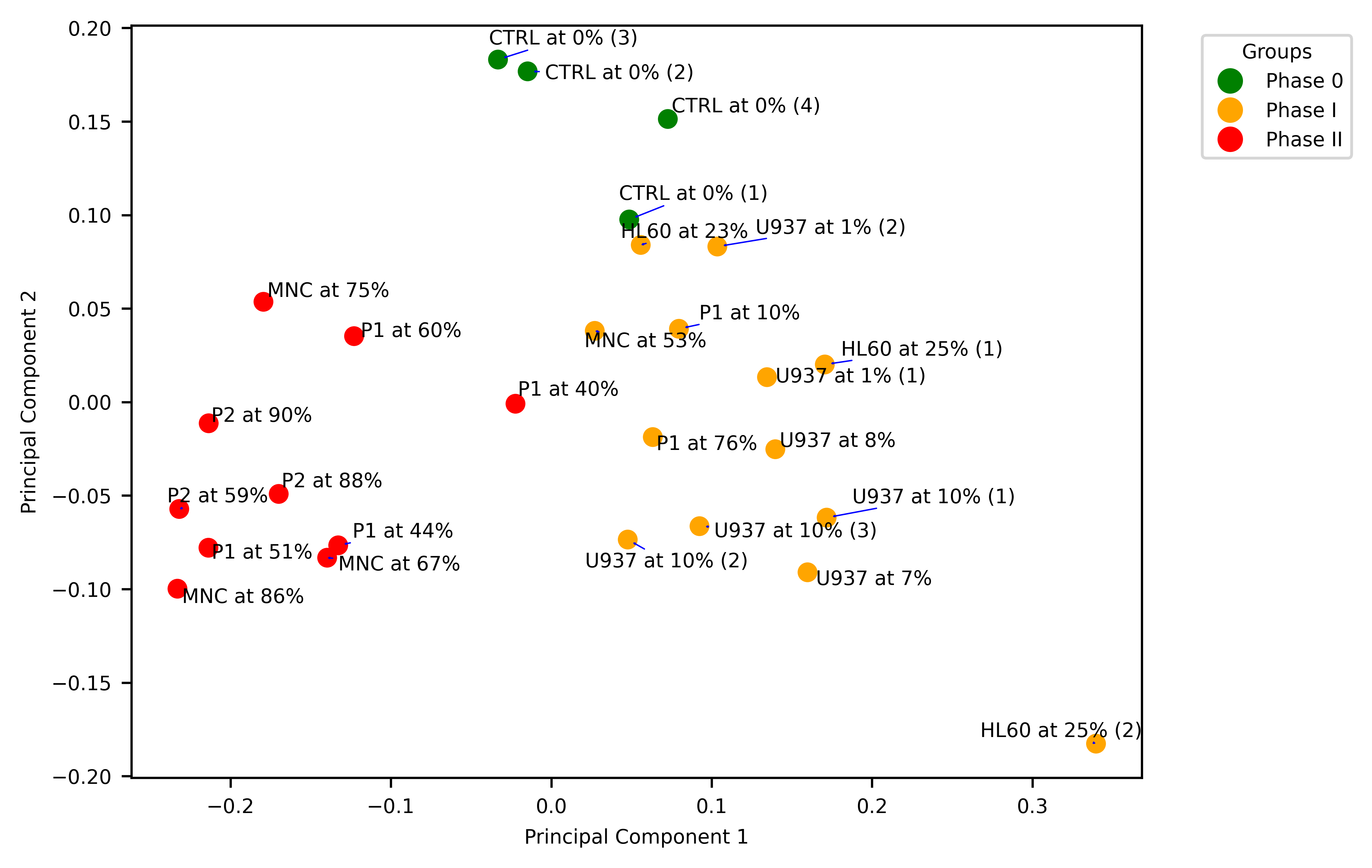}
        \caption{PCA projection for 15 textures.}
        \label{fig:clara15}
    \end{subfigure}
    \hfill
    \begin{subfigure}{0.48\textwidth}
        \includegraphics[width=\linewidth]{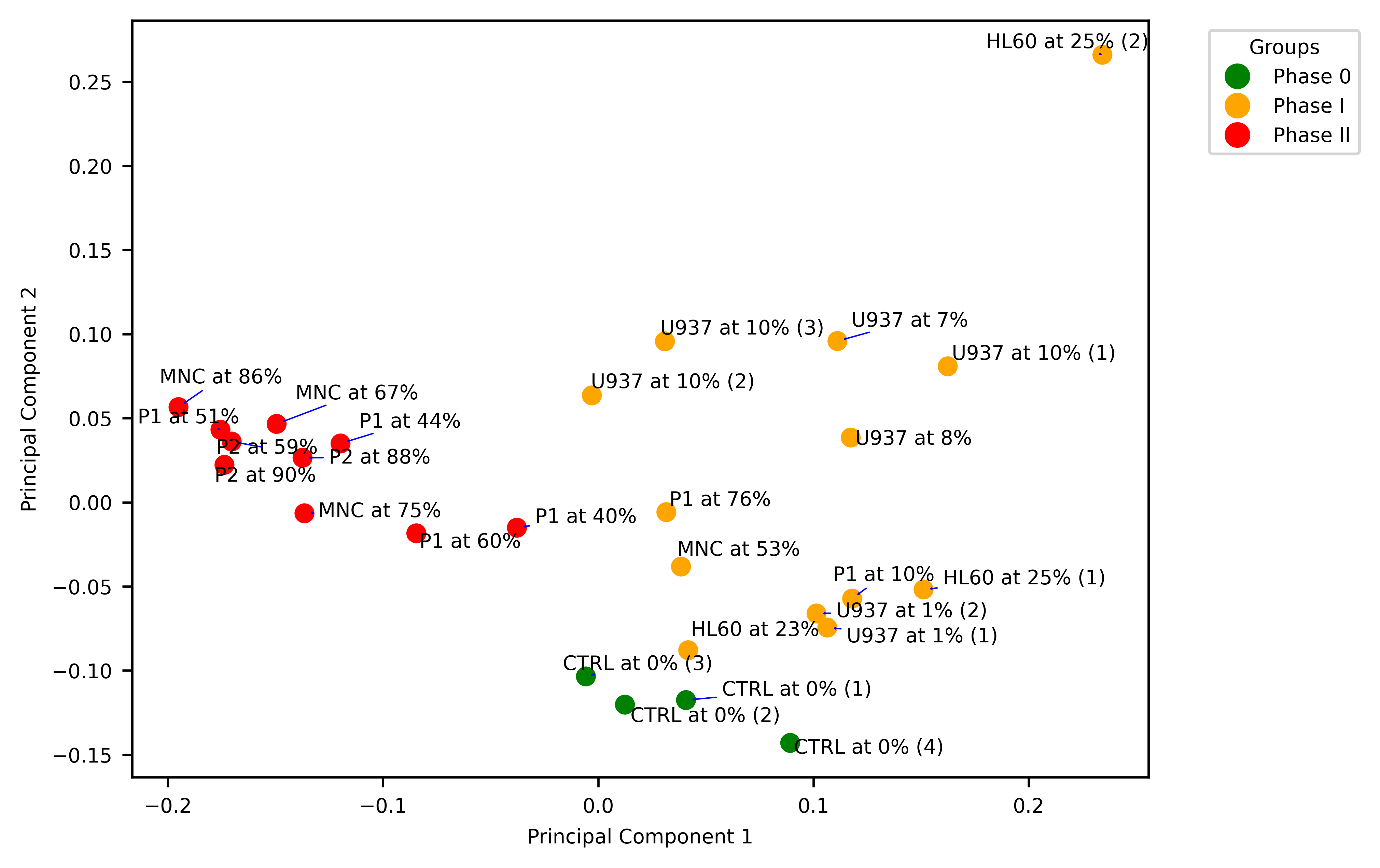}
        \caption{PCA projection for 25 textures.}
        \label{fig:clara25}
    \end{subfigure}
    \caption{Projection of texture decompositions into first 2 principal
components via PCA for 6, 10, 15, and 25 textural clusters obtained using CLARA.}
    \label{fig:clara}
\end{figure}

Following the same methodology, Figure \ref{fig:clara} shows the distribution of samples in the first 2 principal components according to PCA on the texture decomposition for 6, 10, 15, and 25 texture clusters. We see that for each of these numbers of clusters, the samples of different phases can be seen to segregate into distinct groups as presented in the main study. We also note that in general, the HL60 sample with 25\% engraftment appears anomalous, far from the group of all samples.

\section{Persistence-Weighted Diagrams}\label{app:heatmap}

In this section, we include the persistence-weighted diagrams for all the samples grouped by quadrants of the persistence diagram.

\begin{table}[htb]\sffamily
    \caption{Persistence-weighted diagrams show the $\text{PH}_1$ NW regions for different knee region samples. The colored boundaries indicate the phase for different samples. A green border indicates a Phase O sample, an orange border indicates a Phase I sample, and a red border indicates a Phase II sample.}
    \label{tbl:kneeph1ne}
    \begin{adjustwidth}{-1cm}{}
    \begin{tabular}{c|ccccccc}
    \toprule
    Type &  \multicolumn{7}{c}{Knee region samples} \\
    \midrule
    CTRL & \adjustbox{valign=c}{\includegraphics[width=6em]{figure/knee/CTRL_0_1_heatmap.png}} & \adjustbox{valign=c}{\includegraphics[width=6em]{figure/knee/CTRL_0_1_heatmap.png}} &\adjustbox{valign=c}{\includegraphics[width=6em]{figure/knee/CTRL_0_3_heatmap.png}} & \adjustbox{valign=c}{\includegraphics[width=6em]{figure/knee/CTRL_0_4_heatmap.png}} & & & \\
    MNC & \adjustbox{valign=c}{\includegraphics[width=6em]{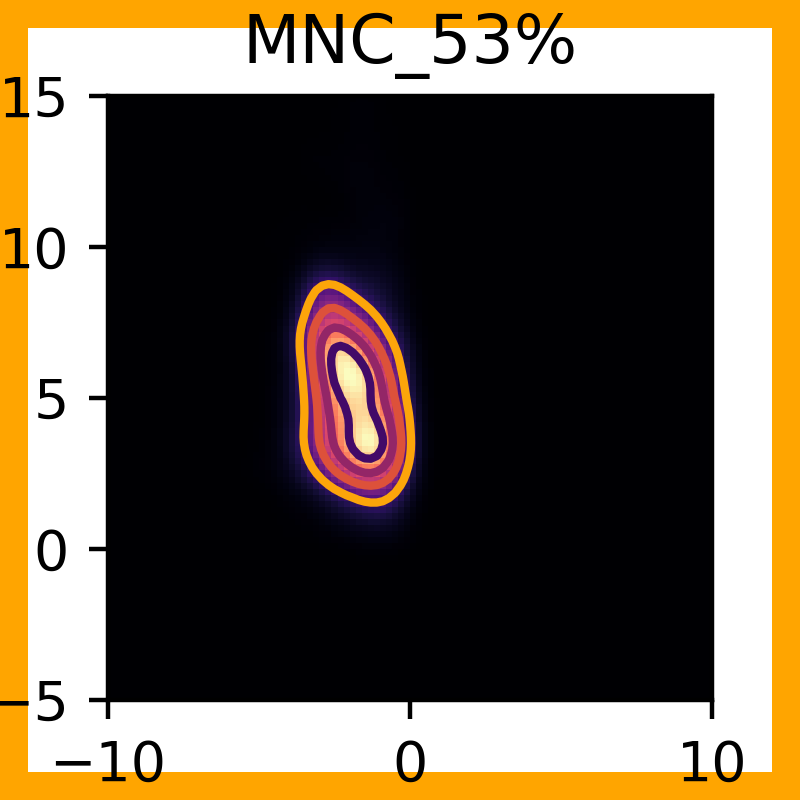}} & \adjustbox{valign=c}{\includegraphics[width=6em]{figure/knee/MNC_67_heatmap.png}} &\adjustbox{valign=c}{\includegraphics[width=6em]{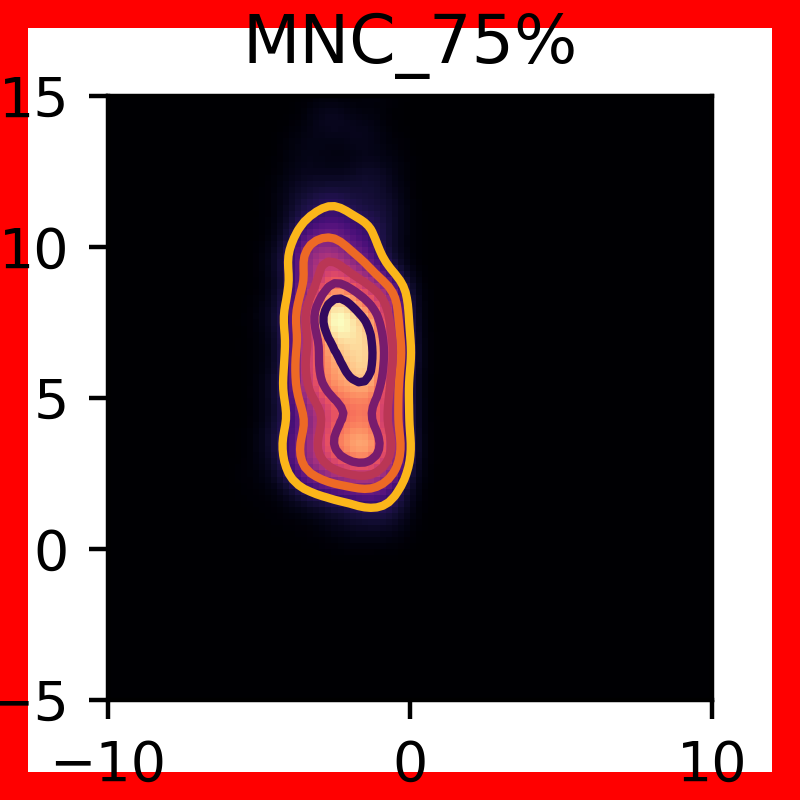}} & \adjustbox{valign=c}{\includegraphics[width=6em]{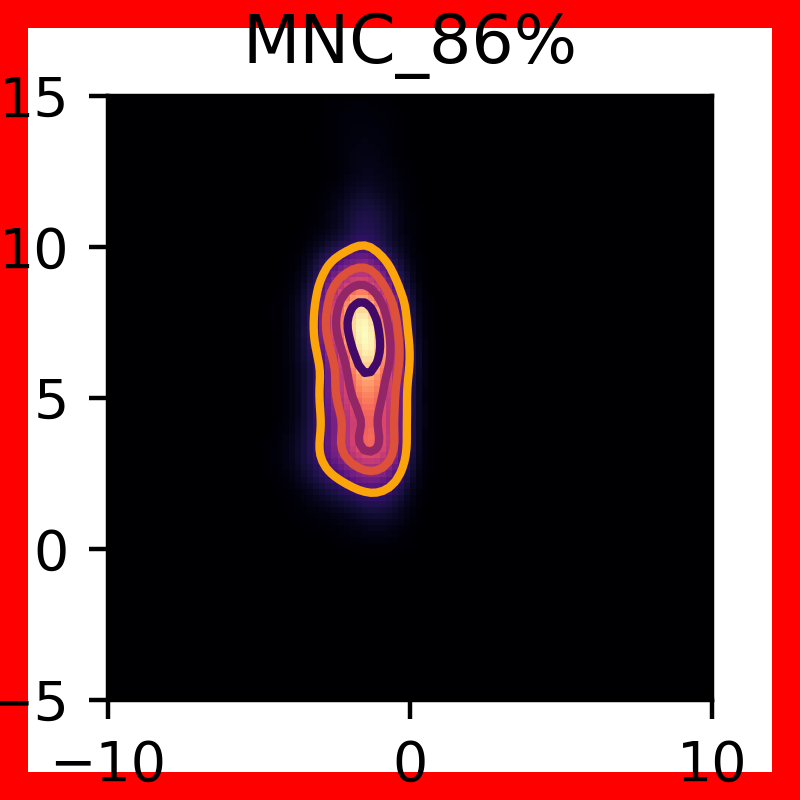}}& & & \\
    U937 & \adjustbox{valign=c}{\includegraphics[width=6em]{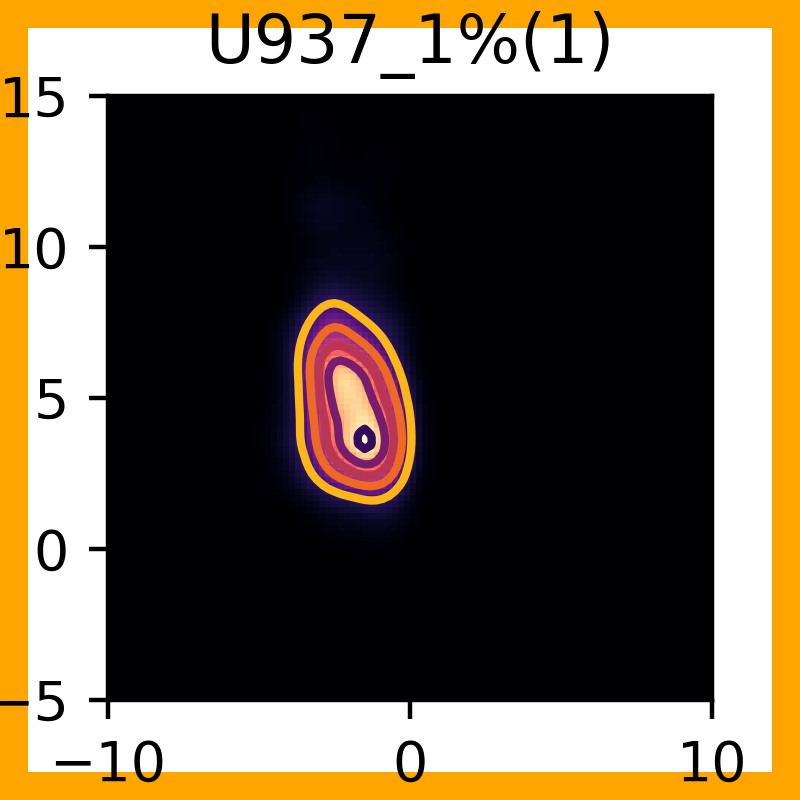}} &\adjustbox{valign=c}{\includegraphics[width=6em]{figure/knee/U937_1_2_heatmap.png}} &\adjustbox{valign=c}{\includegraphics[width=6em]{figure/knee/U937_7_heatmap.png}} & \adjustbox{valign=c}{\includegraphics[width=6em]{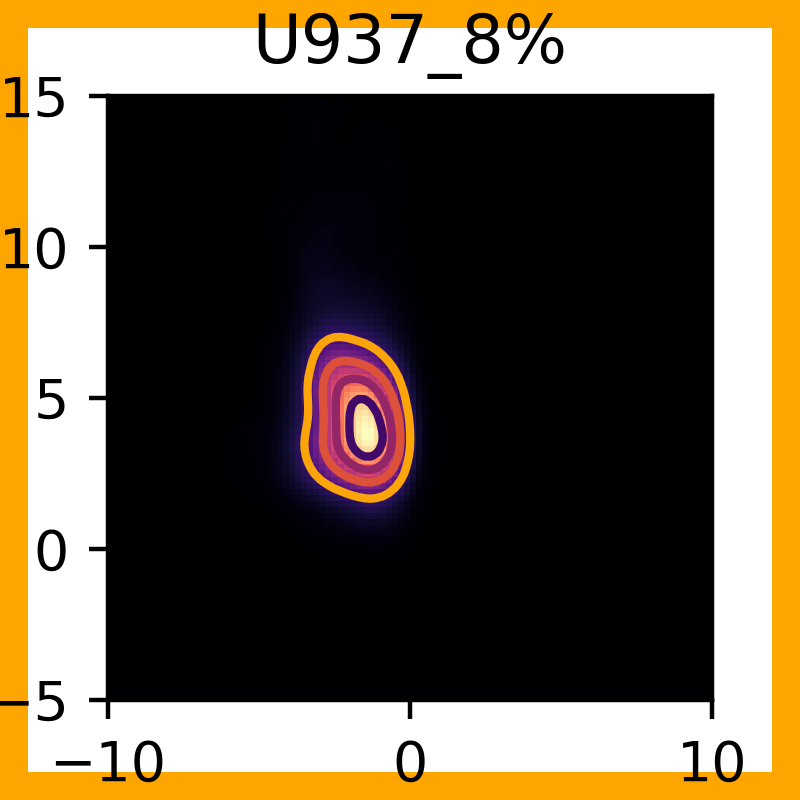}}& \adjustbox{valign=c}{\includegraphics[width=6em]{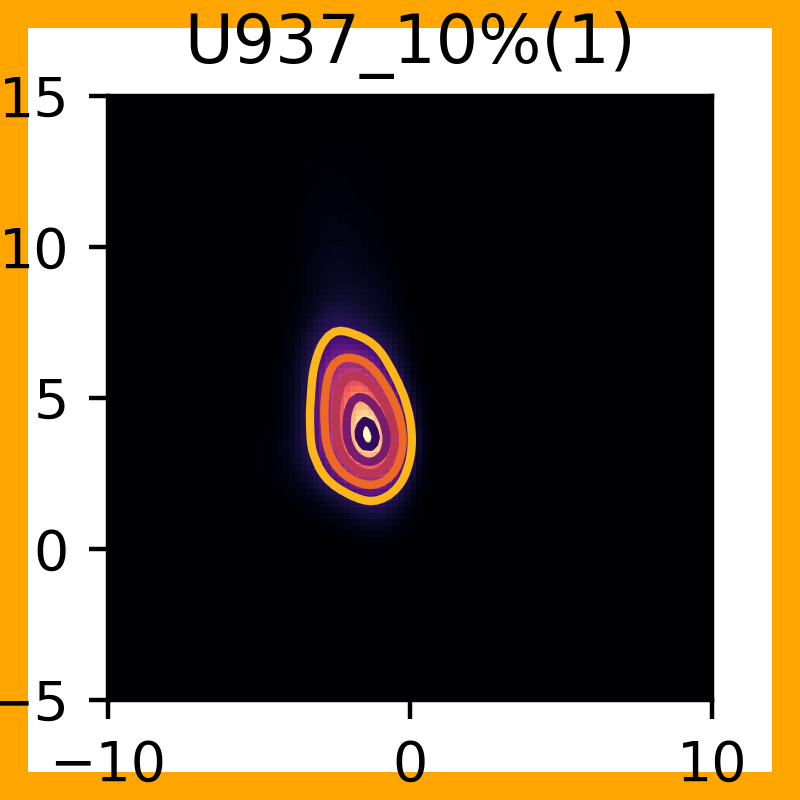}}&\adjustbox{valign=c}{\includegraphics[width=6em]{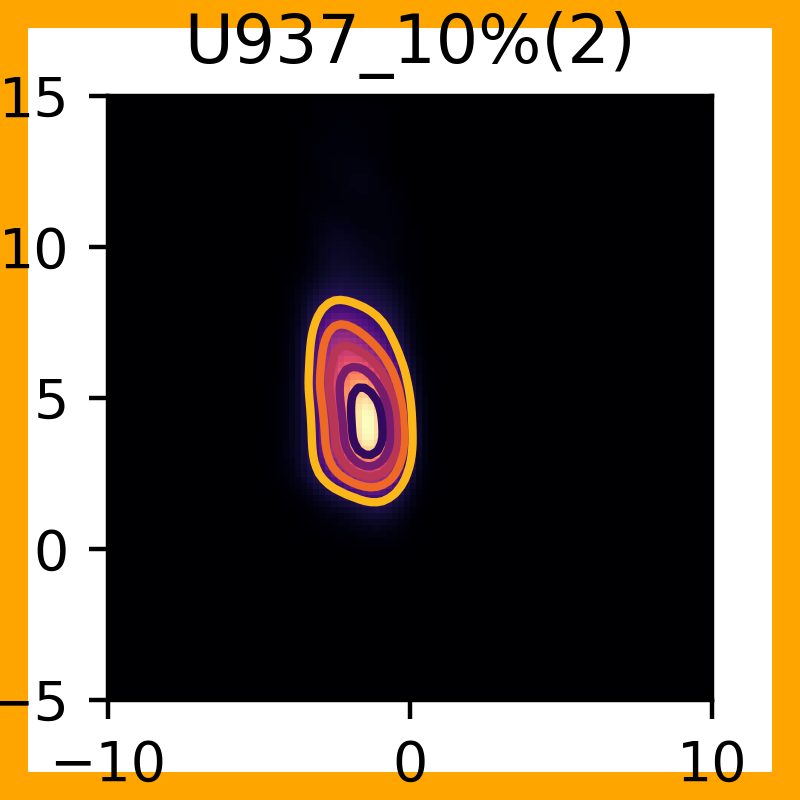}} &\adjustbox{valign=c}{\includegraphics[width=6em]{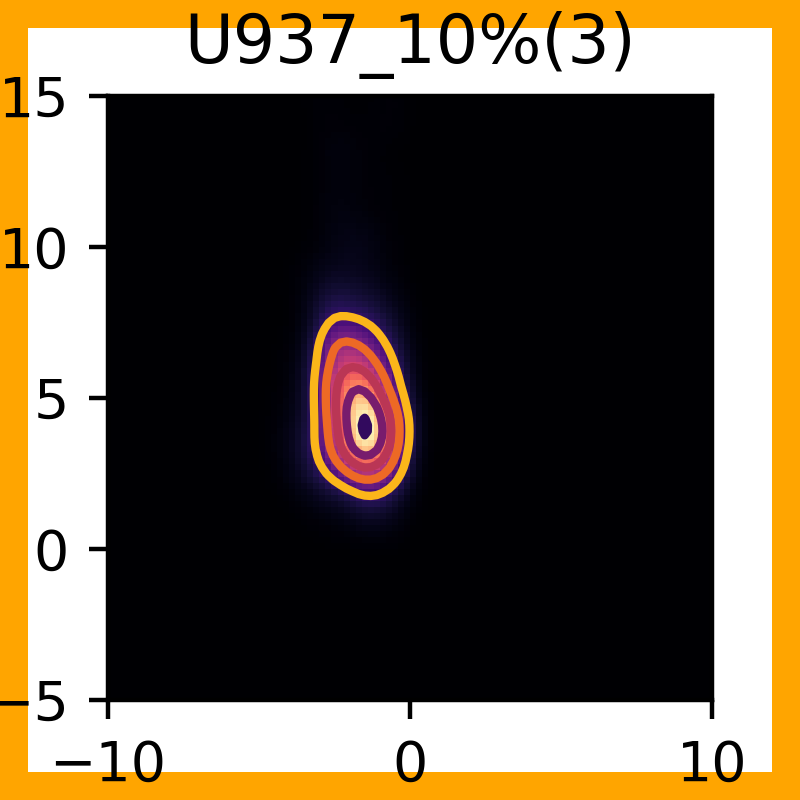}}  \\
    HL60 & \adjustbox{valign=c}{\includegraphics[width=6em]{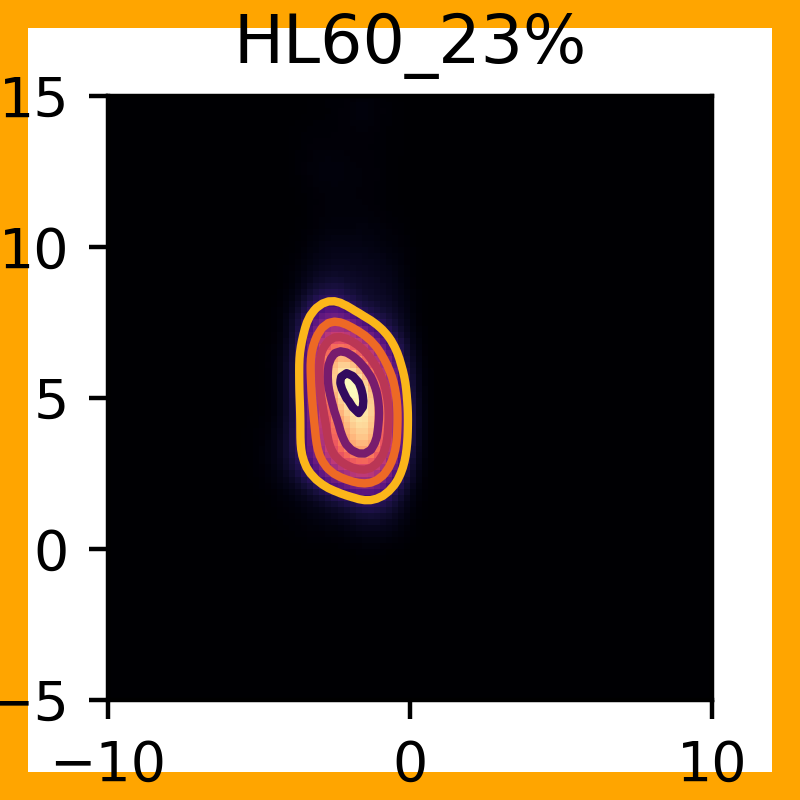}}& \adjustbox{valign=c}{\includegraphics[width=6em]{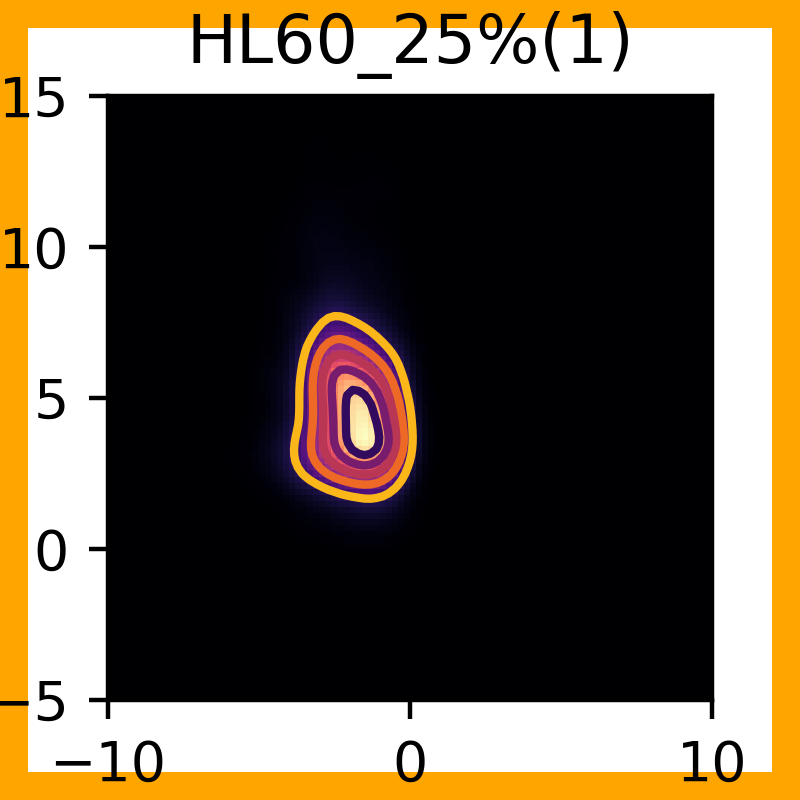}}& \adjustbox{valign=c}{\includegraphics[width=6em]{figure/knee/HL60_25_2_heatmap.png}}& & & &  \\
    P1 &\adjustbox{valign=c}{\includegraphics[width=6em]{figure/knee/P1_10_heatmap.png}} &\adjustbox{valign=c}{\includegraphics[width=6em]{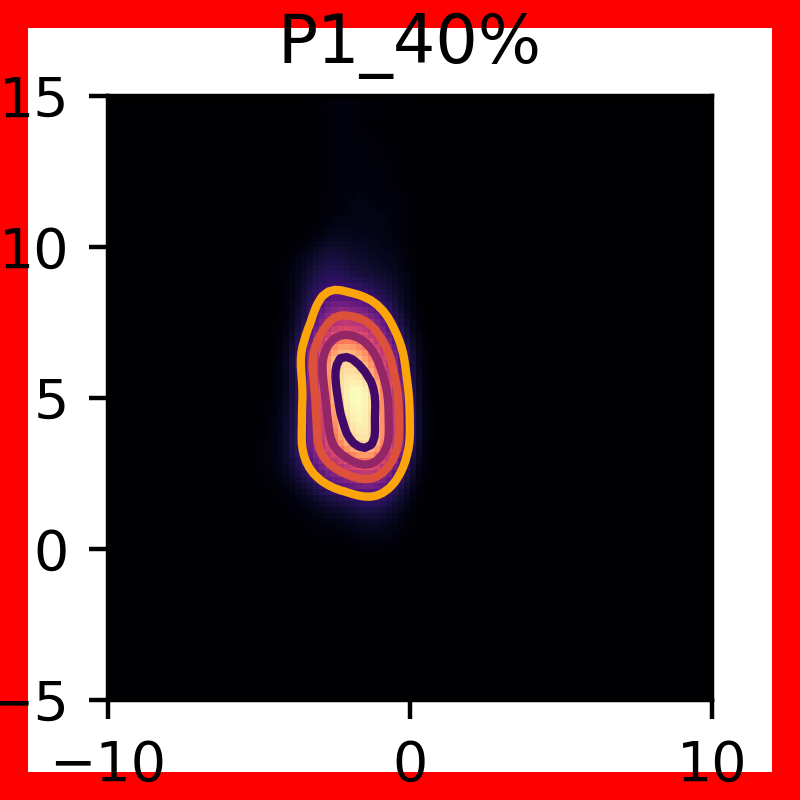}} &\adjustbox{valign=c}{\includegraphics[width=6em]{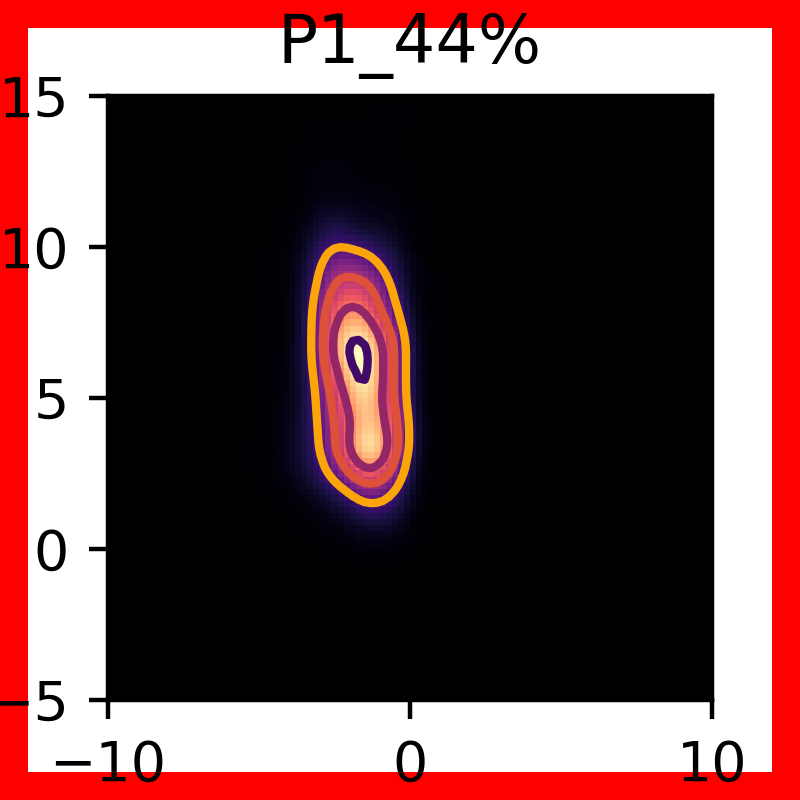}} & \adjustbox{valign=c}{\includegraphics[width=6em]{figure/knee/P1_51_heatmap.png}}& \adjustbox{valign=c}{\includegraphics[width=6em]{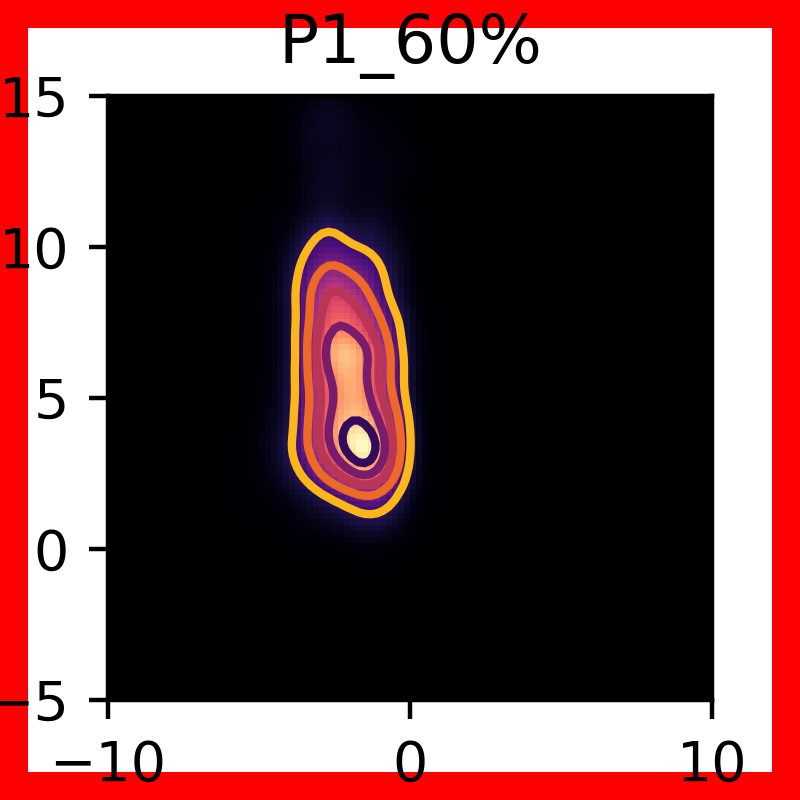}}&\adjustbox{valign=c}{\includegraphics[width=6em]{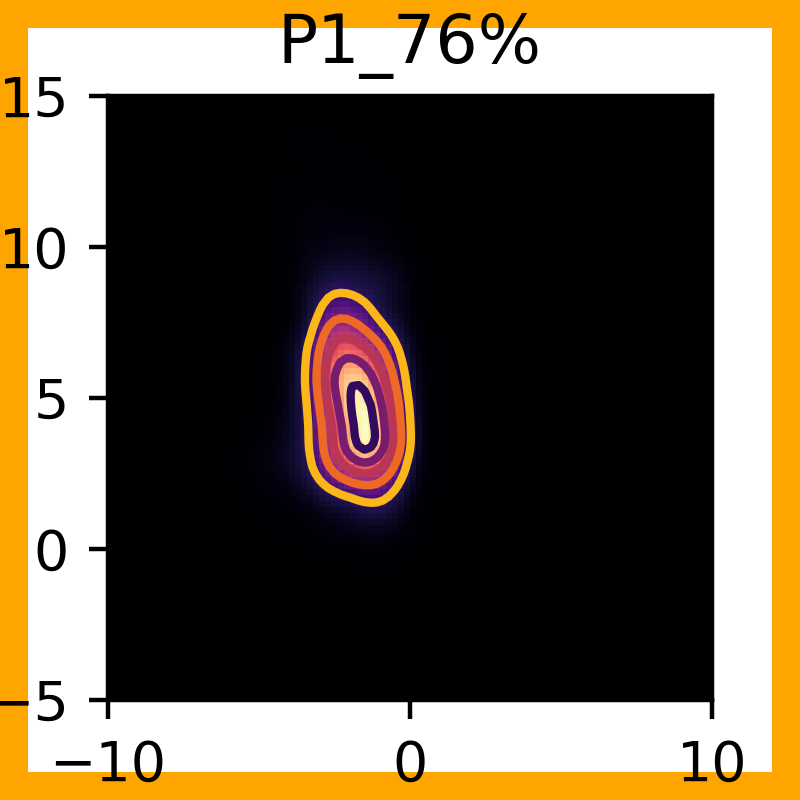}} & \\
    P2 &\adjustbox{valign=c}{\includegraphics[width=6em]{figure/knee/P2_59_heatmap.png}} & \adjustbox{valign=c}{\includegraphics[width=6em]{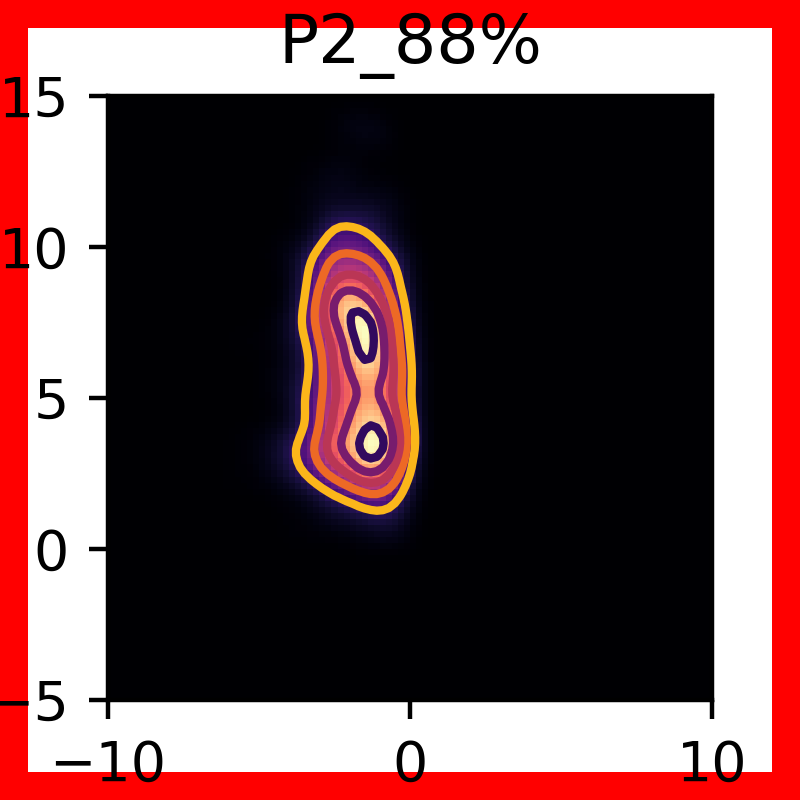}}&\adjustbox{valign=c}{\includegraphics[width=6em]{figure/knee/P2_90_heatmap.png}} & & & & \\
    \bottomrule
    \end{tabular}
    \end{adjustwidth}
\end{table}

\begin{table}[htb]\sffamily 
    \caption{Table showing $\text{PH}_1$ NW regions of persistence-weighted Gaussian kernel density estimates for long regions of all bone samples. Diagrams are outlined in different colors showing their respective phase determined from the local texture analysis: Phase O (green); Phase I (orange); Phase II (red). }
    \label{tbl:longph1ne} 
    \begin{adjustwidth}{-1cm}{}
    \begin{tabular}{c|ccccccc} 
    \toprule 
    Type &  \multicolumn{7}{c}{Long region samples} \\ 

    \midrule 

    CTRL & \adjustbox{valign=c}{\includegraphics[width=6em]{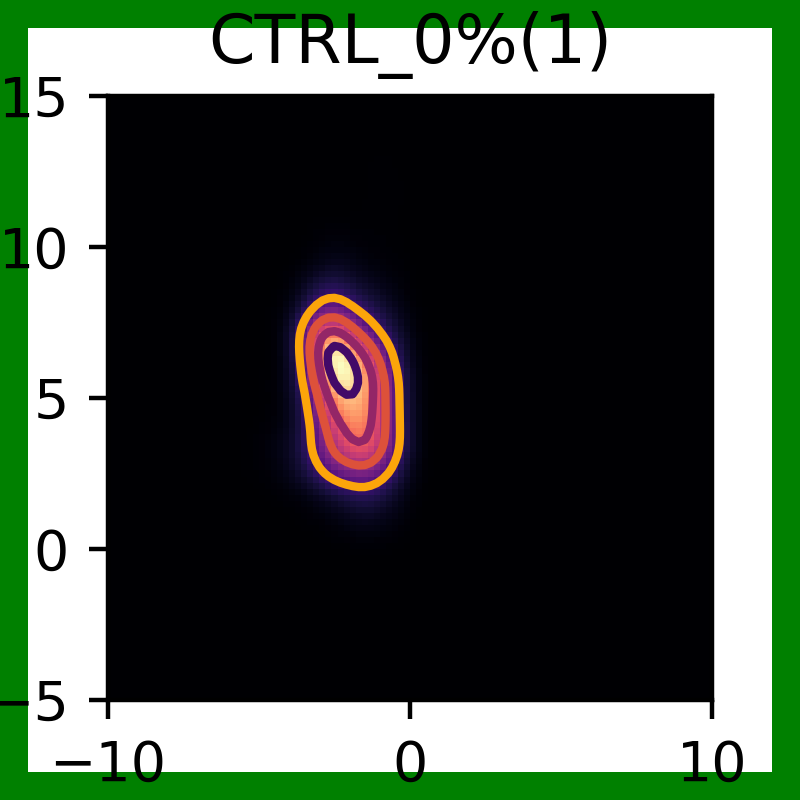}} & \adjustbox{valign=c}{\includegraphics[width=6em]{figure/long/long_CTRL_0_1_heatmap.png}} &\adjustbox{valign=c}{\includegraphics[width=6em]{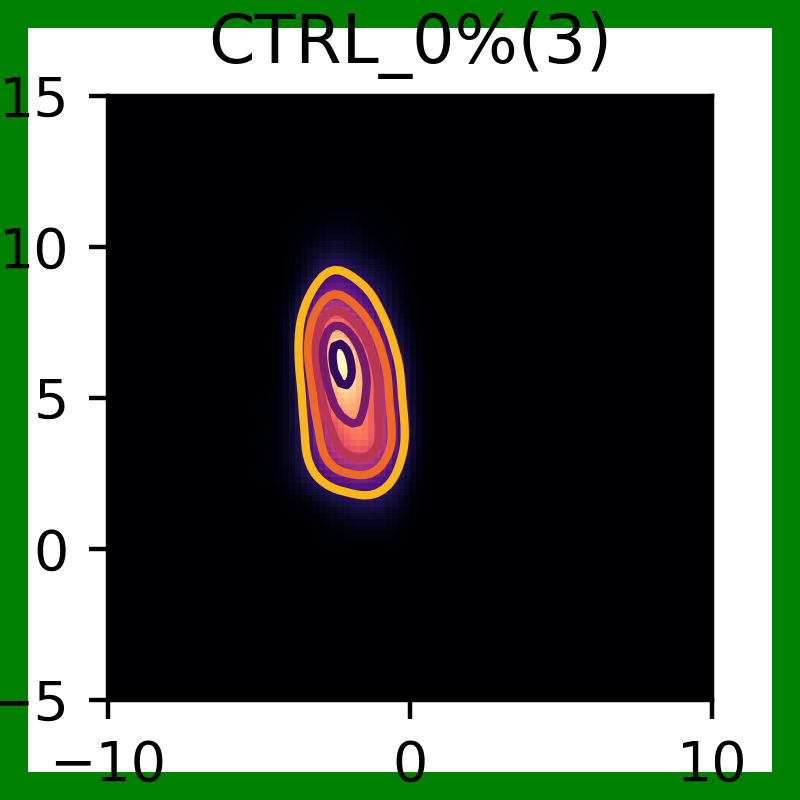}} & \adjustbox{valign=c}{\includegraphics[width=6em]{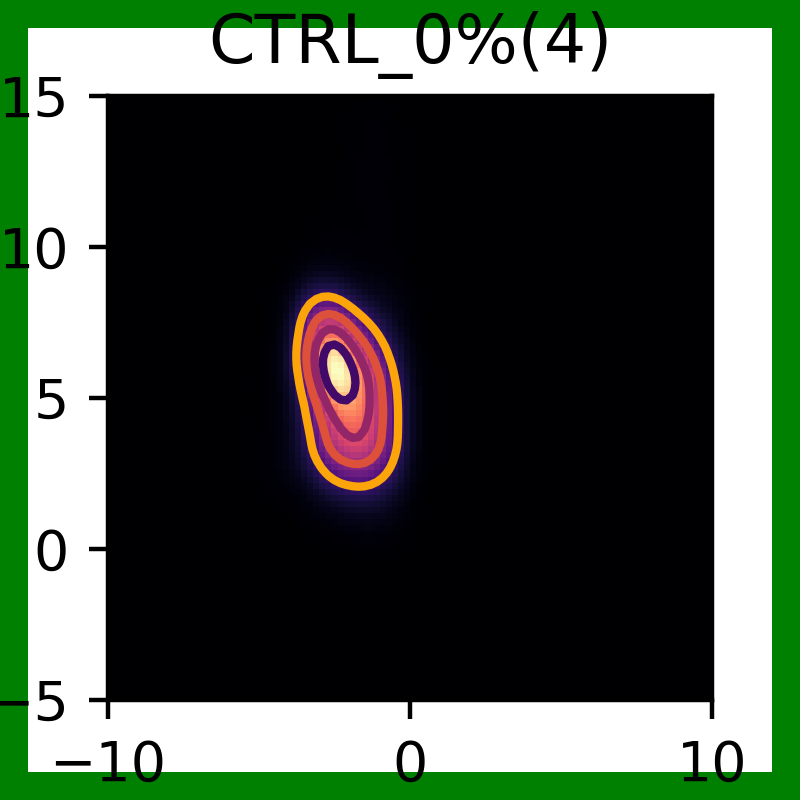}} & & & \\ 

    MNC & \adjustbox{valign=c}{\includegraphics[width=6em]{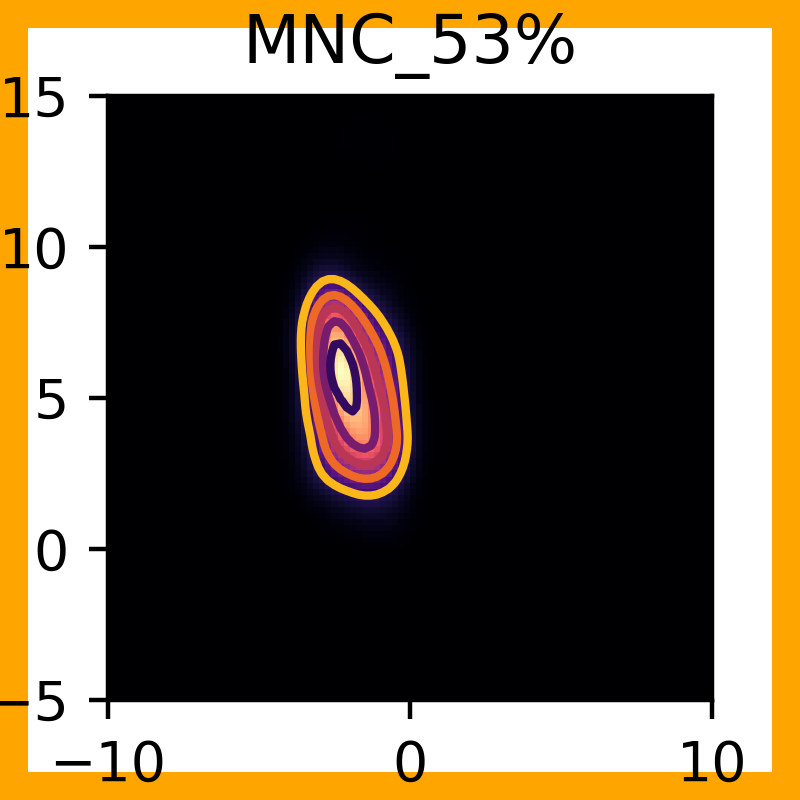}} & \adjustbox{valign=c}{\includegraphics[width=6em]{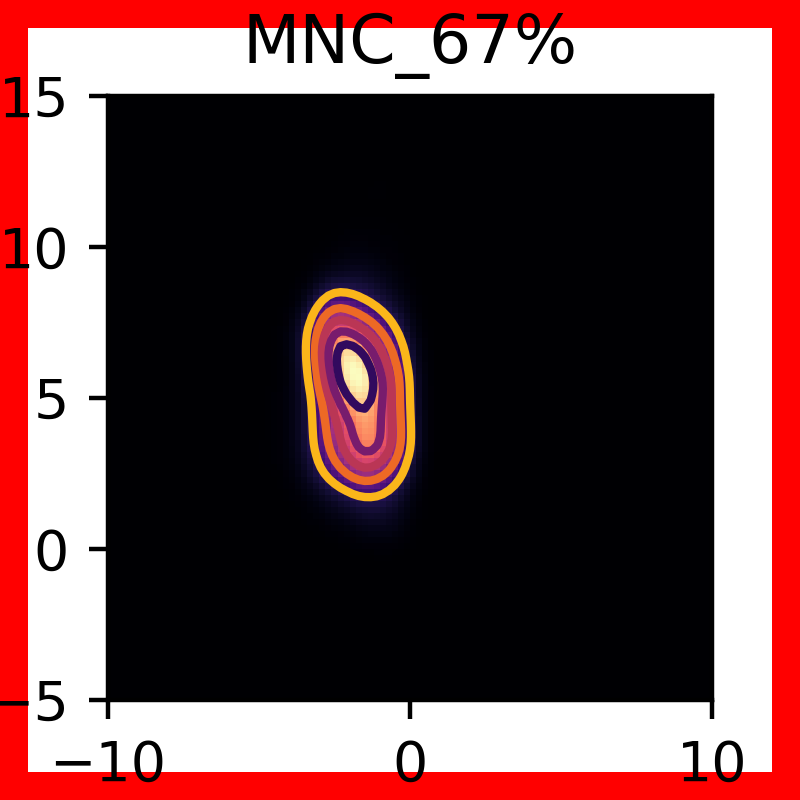}} &\adjustbox{valign=c}{\includegraphics[width=6em]{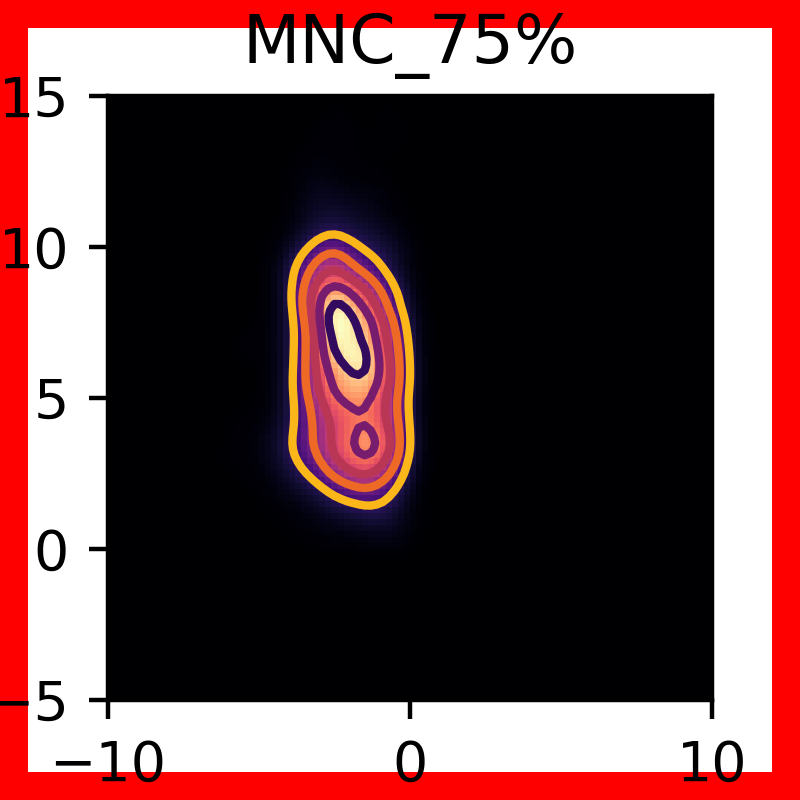}} & \adjustbox{valign=c}{\includegraphics[width=6em]{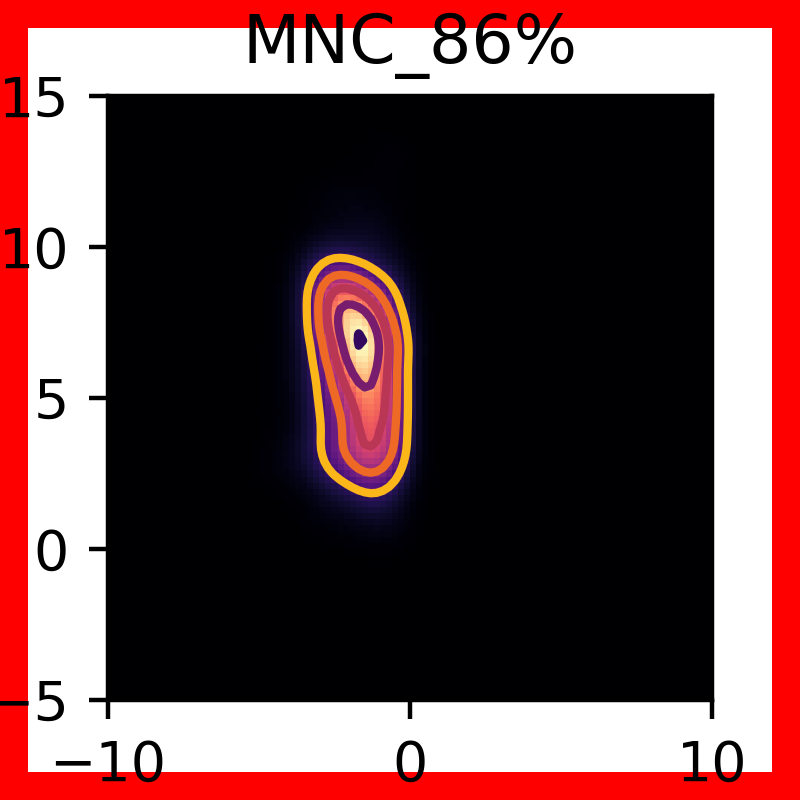}}& & & \\ 

    U937 & \adjustbox{valign=c}{\includegraphics[width=6em]{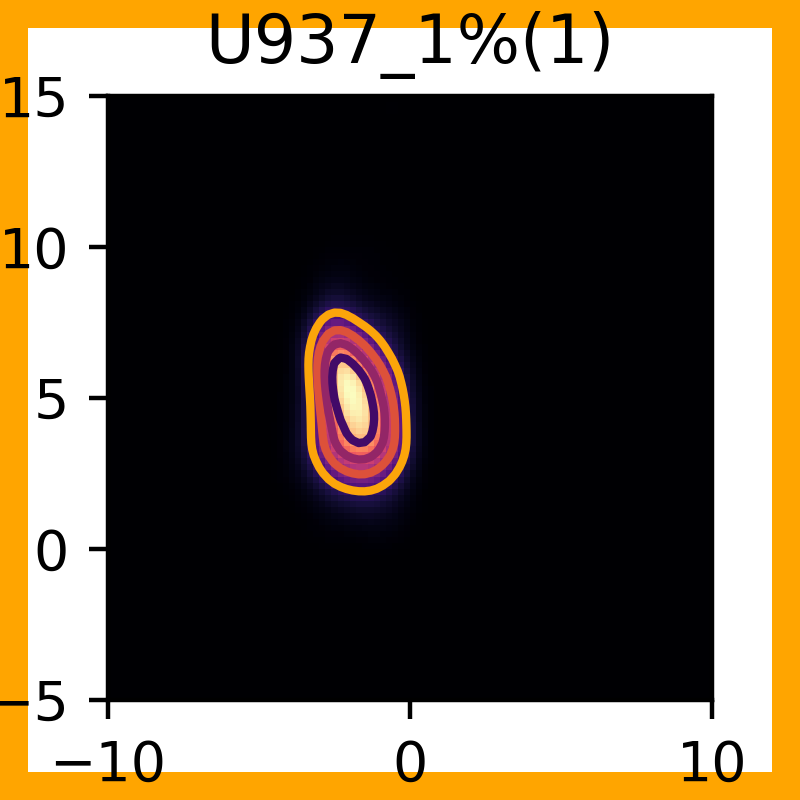}} &\adjustbox{valign=c}{\includegraphics[width=6em]{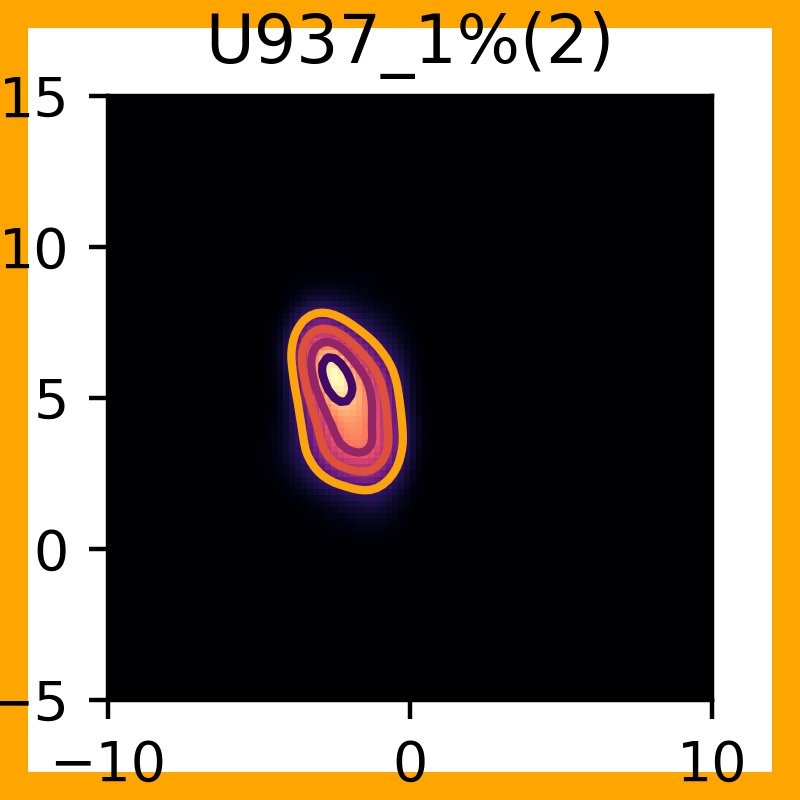}} &\adjustbox{valign=c}{\includegraphics[width=6em]{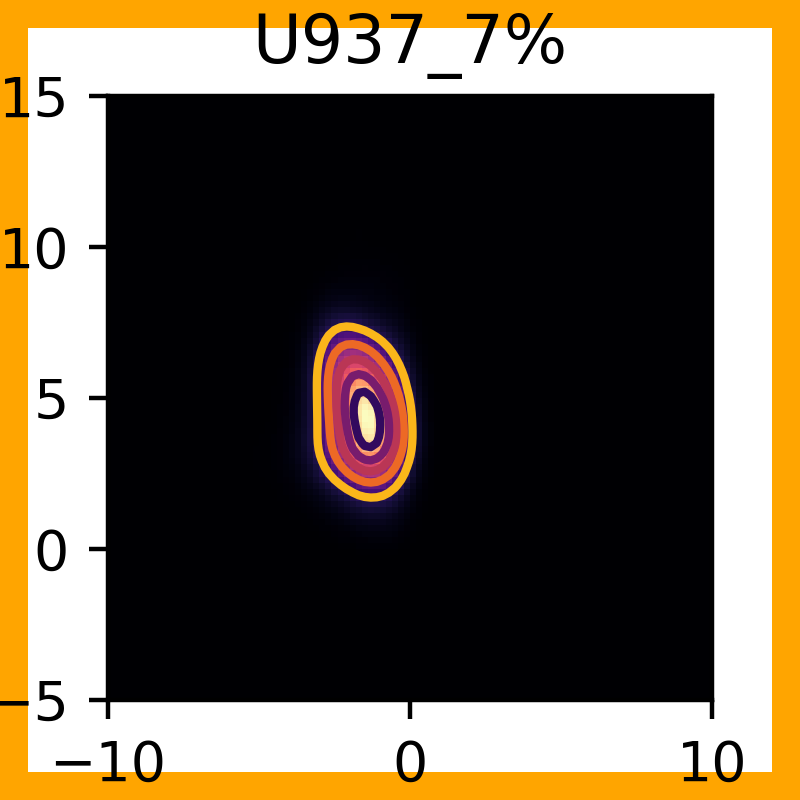}} & \adjustbox{valign=c}{\includegraphics[width=6em]{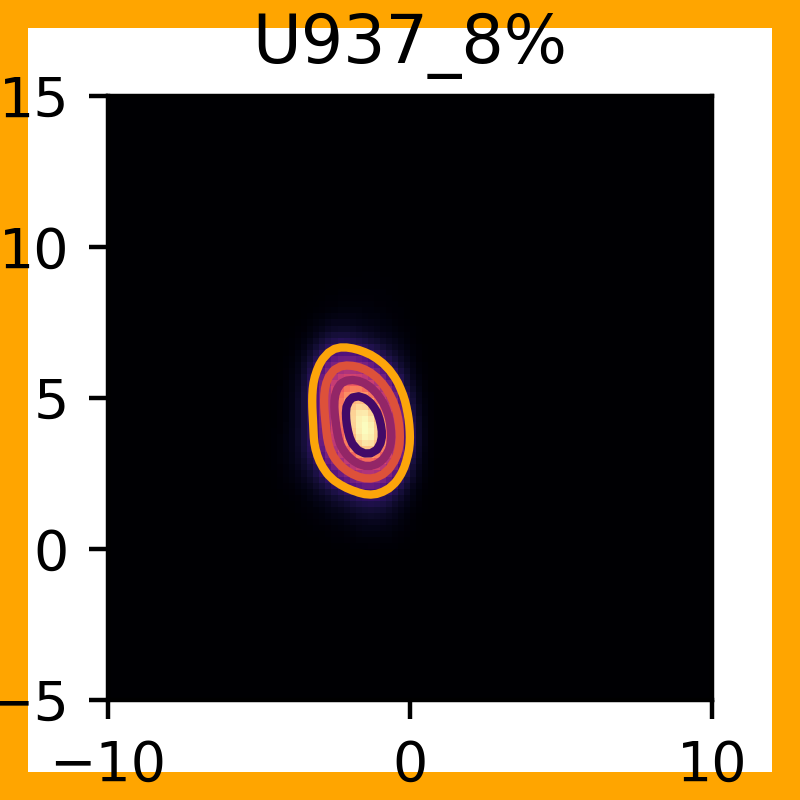}}& \adjustbox{valign=c}{\includegraphics[width=6em]{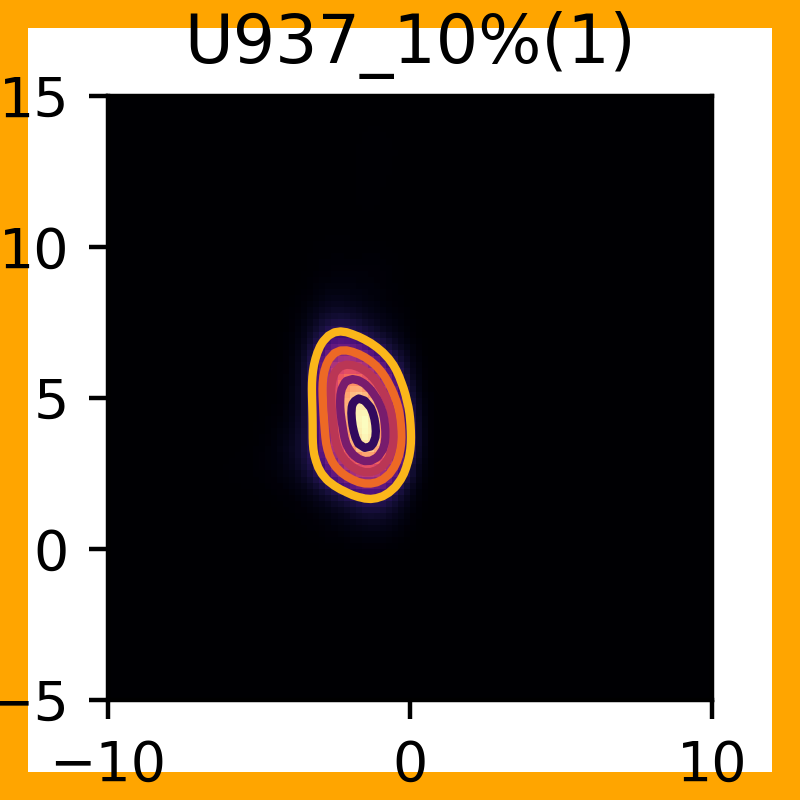}}&\adjustbox{valign=c}{\includegraphics[width=6em]{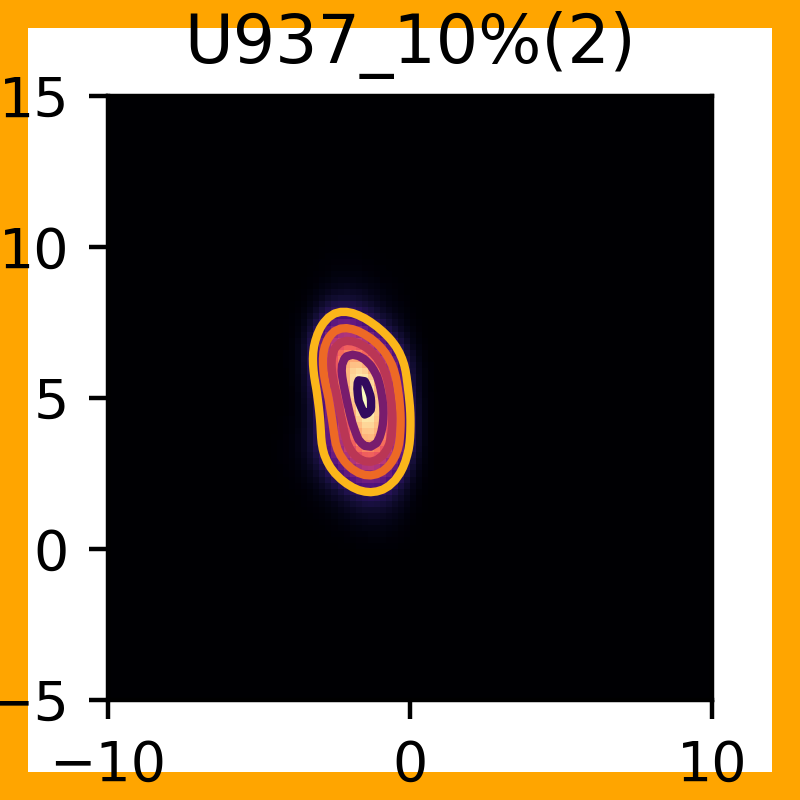}} &\adjustbox{valign=c}{\includegraphics[width=6em]{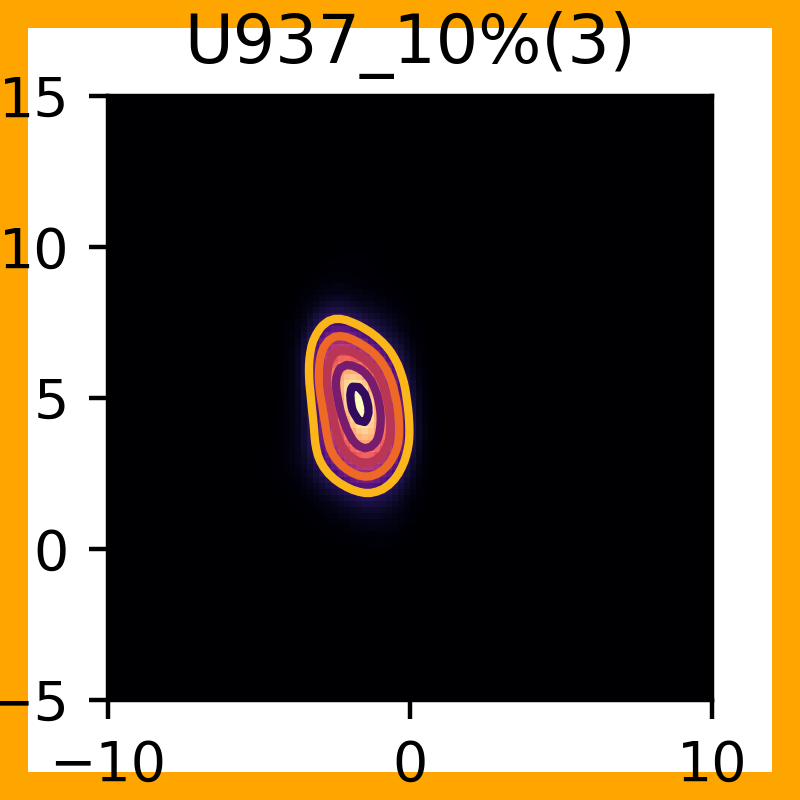}}  \\ 

    HL60 & \adjustbox{valign=c}{\includegraphics[width=6em]{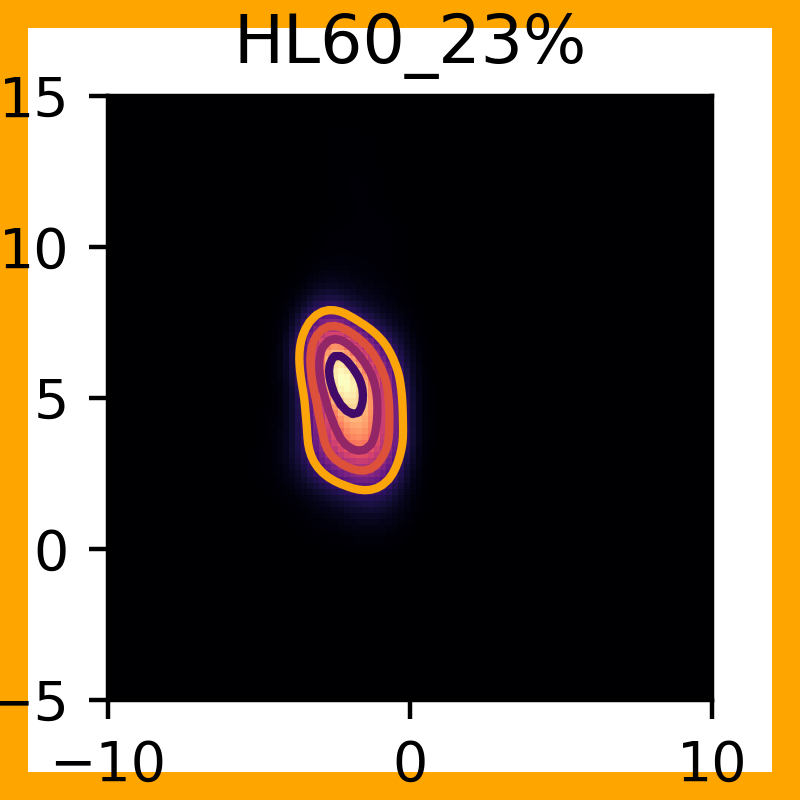}}& \adjustbox{valign=c}{\includegraphics[width=6em]{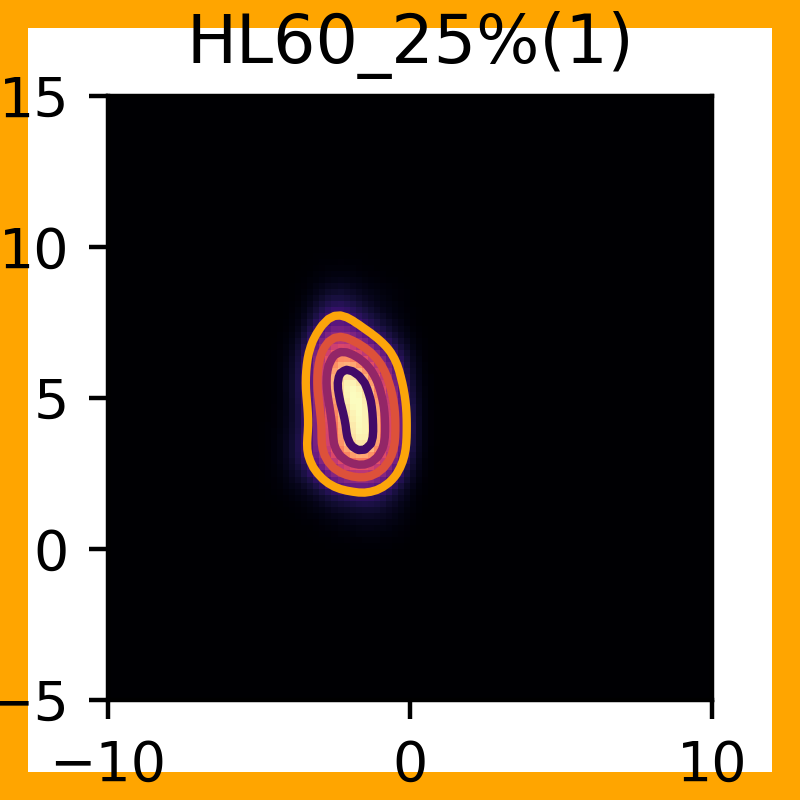}}& \adjustbox{valign=c}{\includegraphics[width=6em]{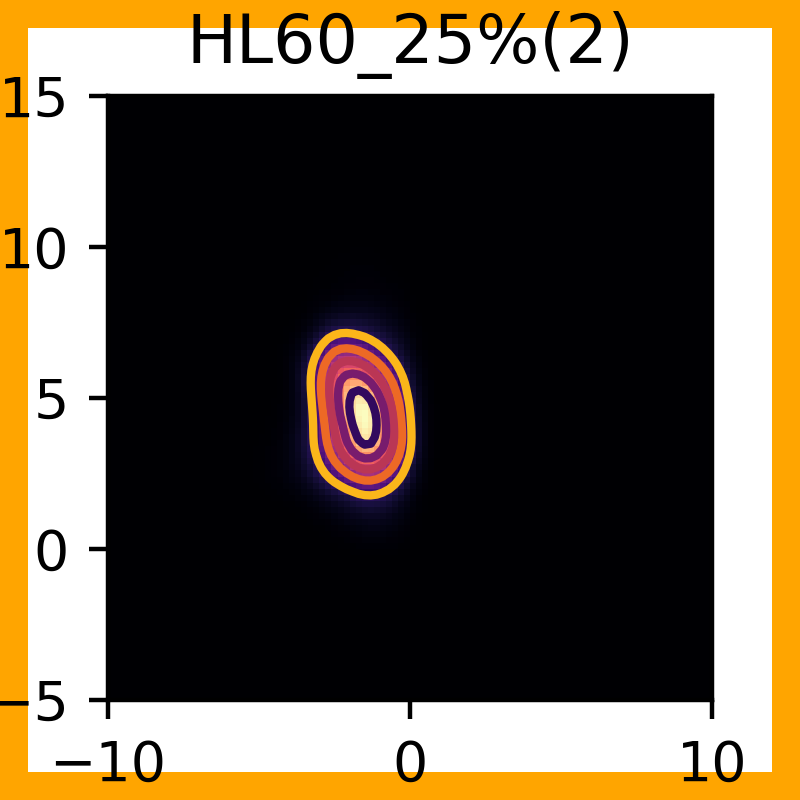}}& & & &  \\ 

    P1 &\adjustbox{valign=c}{\includegraphics[width=6em]{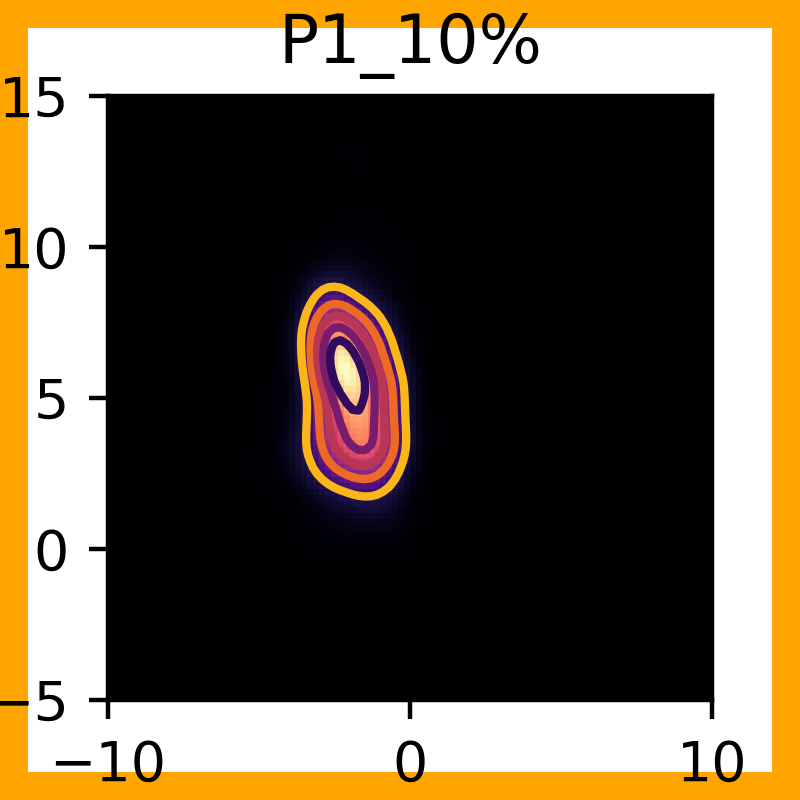}} &\adjustbox{valign=c}{\includegraphics[width=6em]{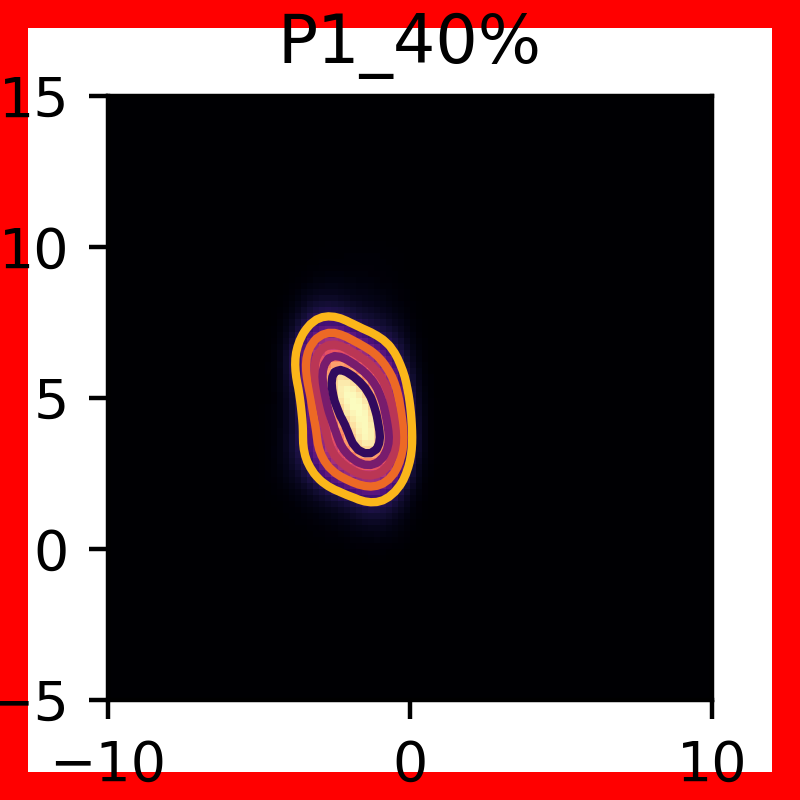}} &\adjustbox{valign=c}{\includegraphics[width=6em]{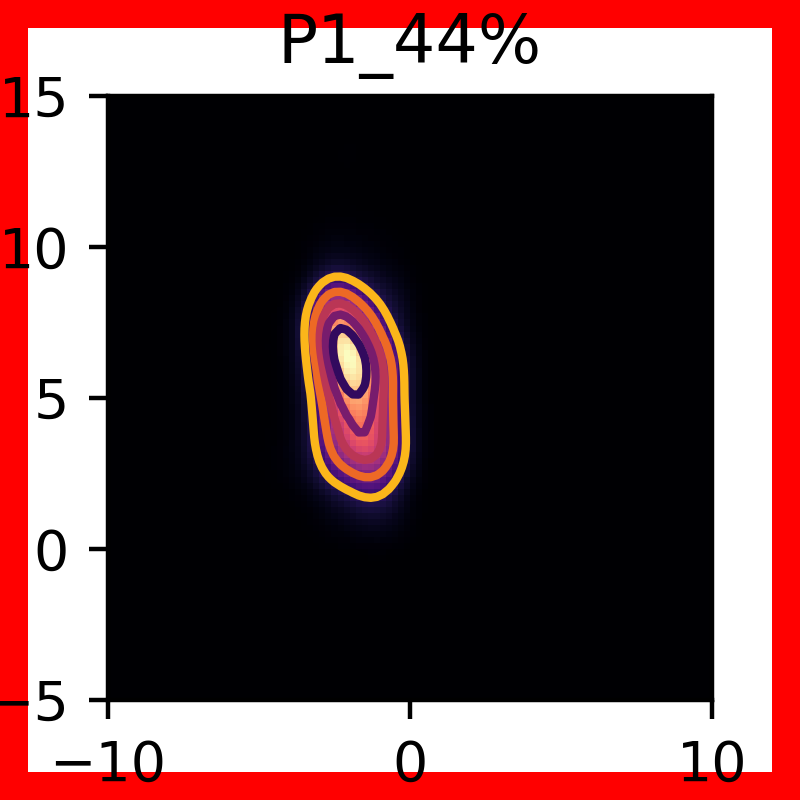}} & \adjustbox{valign=c}{\includegraphics[width=6em]{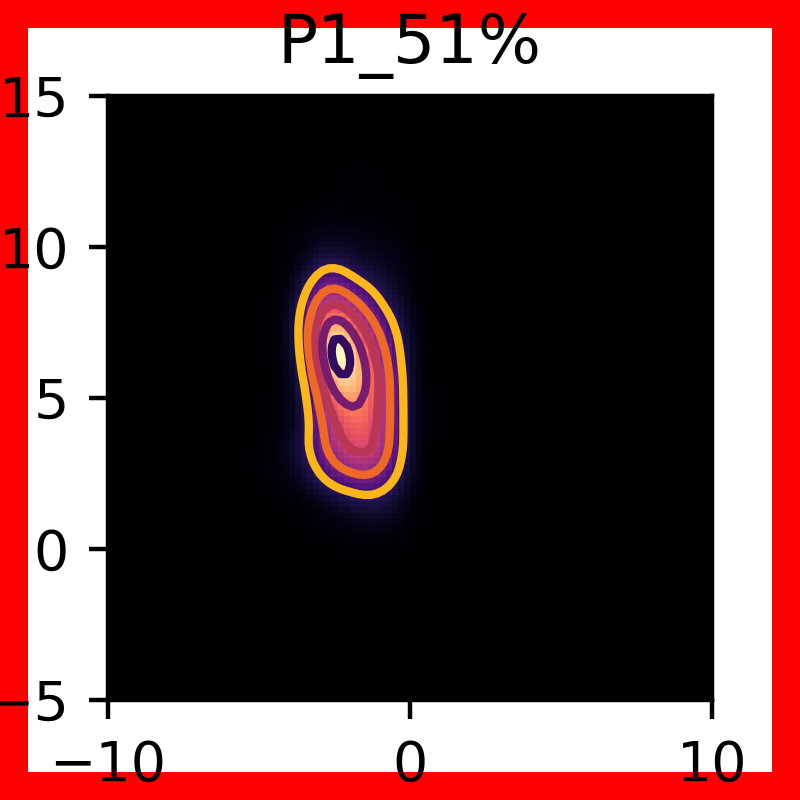}}& \adjustbox{valign=c}{\includegraphics[width=6em]{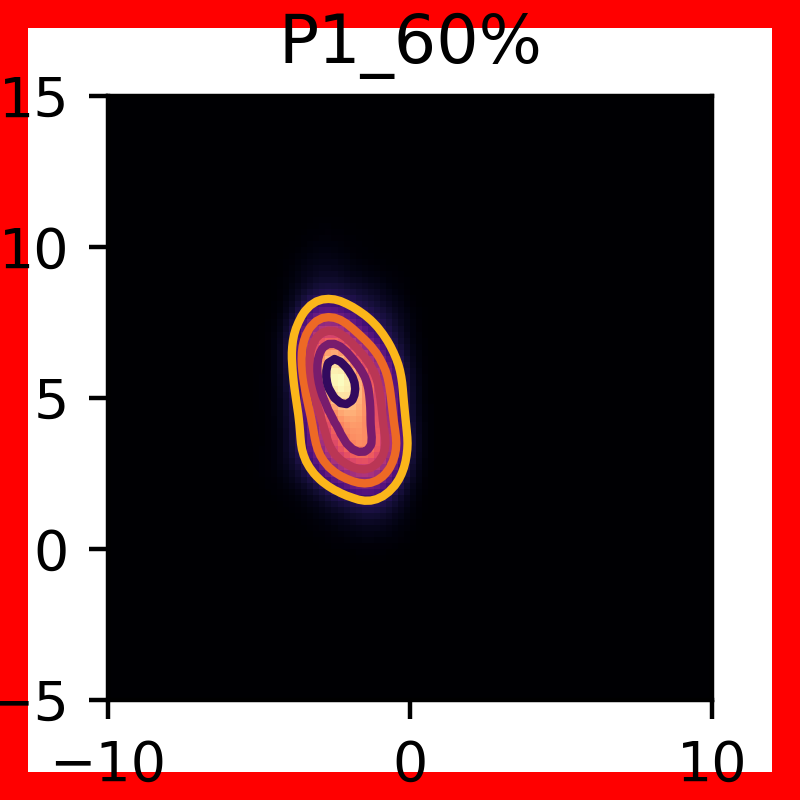}}&\adjustbox{valign=c}{\includegraphics[width=6em]{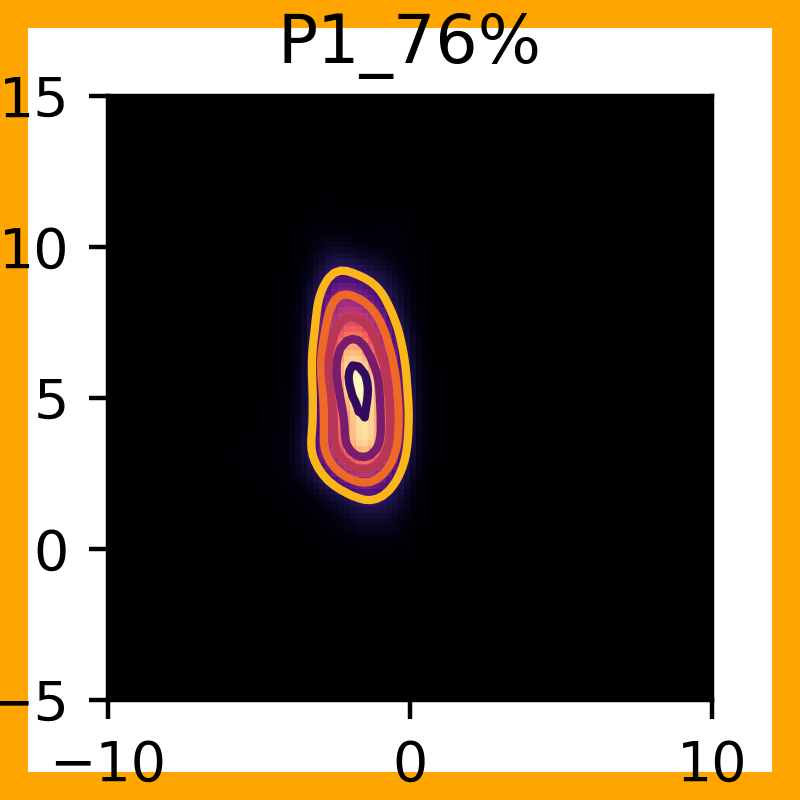}} & \\ 

    P2 &\adjustbox{valign=c}{\includegraphics[width=6em]{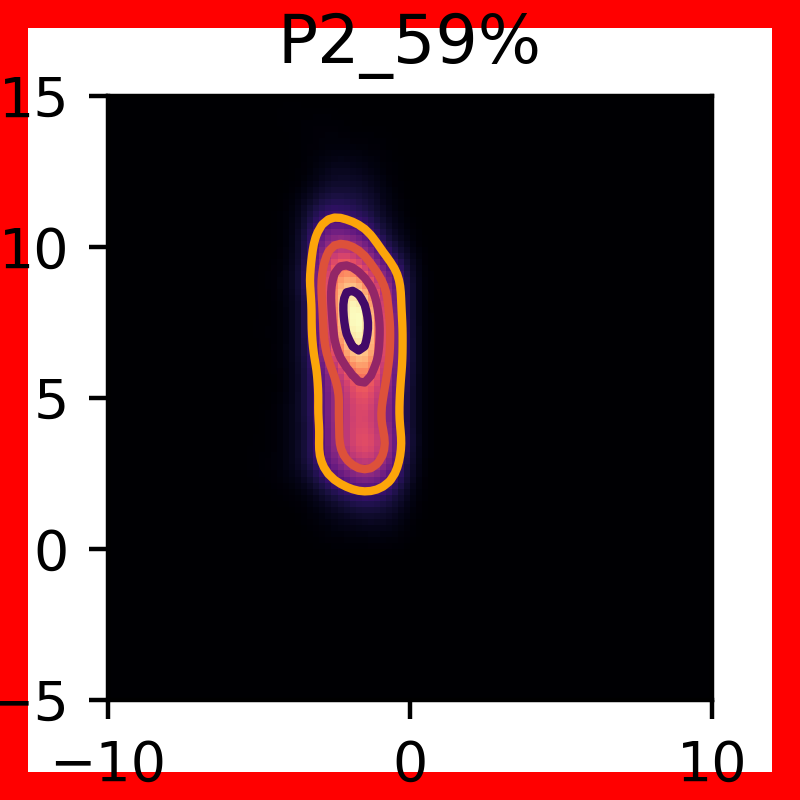}} & \adjustbox{valign=c}{\includegraphics[width=6em]{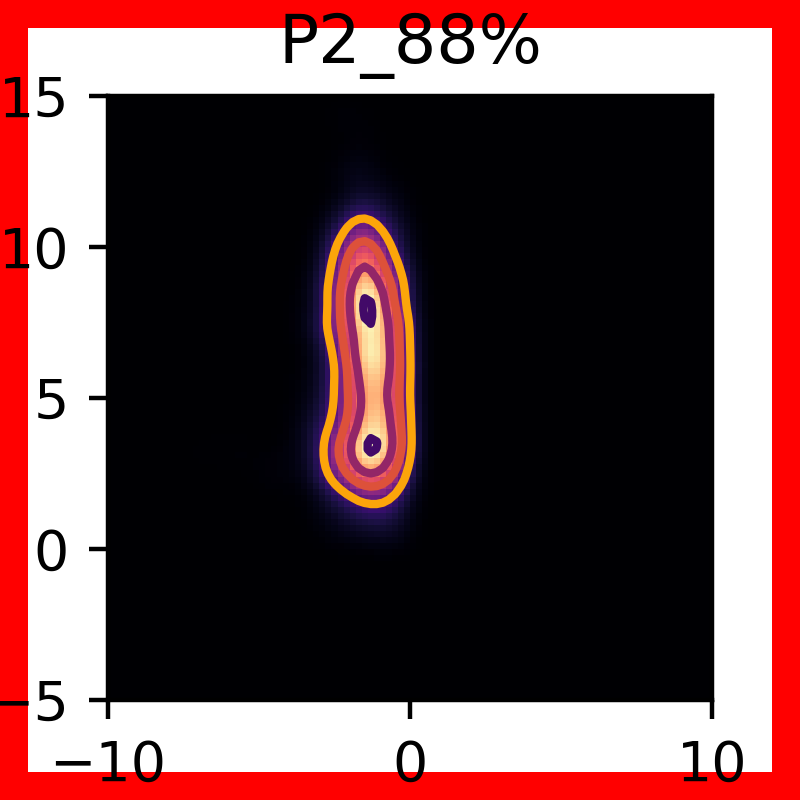}}&\adjustbox{valign=c}{\includegraphics[width=6em]{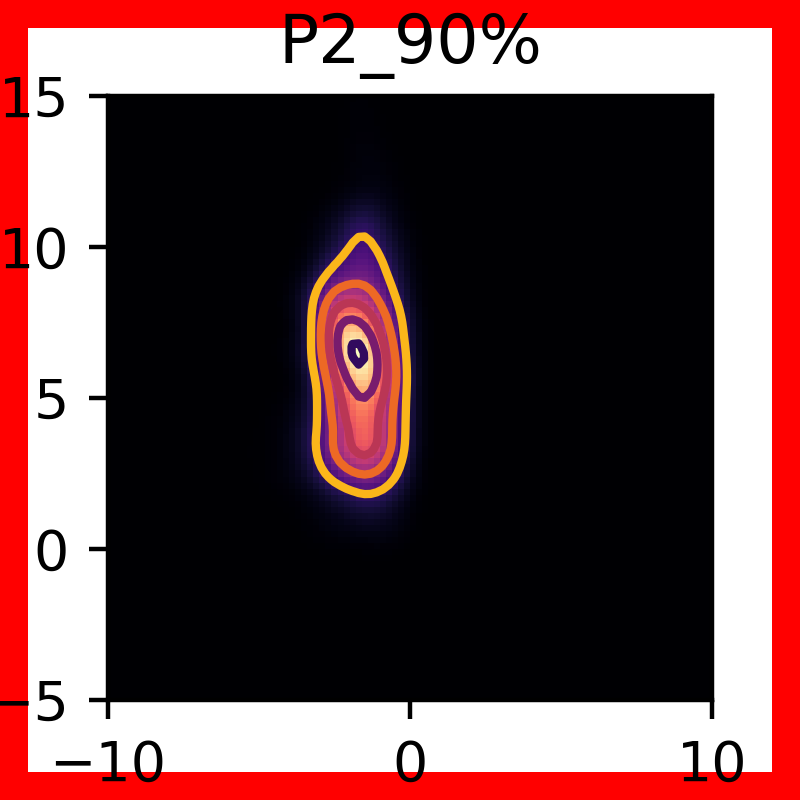}} & & & & \\ 

    \bottomrule 

    \end{tabular} 
    \end{adjustwidth}
\end{table} 

\begin{table}[htb]
\caption{Table showing $\text{PH}_2$ NE regions of persistence-weighted Gaussian kernel density estimates for long regions of all bone samples. Diagrams are outlined in different colors showing their respective phase determined from the local texture analysis: Phase O (green); Phase I (orange); Phase II (red). }
\label{tab:ph2ne}
\begin{adjustwidth}{-1cm}{}
\begin{tabular}{c|ccccccc}
\toprule
Type                      & \multicolumn{6}{c}{Samples} \\ \hline
Phase O                   &\adjustbox{valign=c}{\includegraphics[width=6em]{figure/knee_ph2ne/ph2ne_CTRL_0_1_heatmap.png}}    &\adjustbox{valign=c}{\includegraphics[width=6em]{figure/knee_ph2ne/ph2ne_CTRL_0_2_heatmap.png}}    & \adjustbox{valign=c}{\includegraphics[width=6em]{figure/knee_ph2ne/ph2ne_CTRL_0_3_heatmap.png}}   &\adjustbox{valign=c}{\includegraphics[width=6em]{figure/knee_ph2ne/ph2ne_CTRL_0_4_heatmap.png}}    &    &   & \\ \hline
\multirow{2}{*}{Phase I}  & \adjustbox{valign=c}{\includegraphics[width=6em]{figure/knee_ph2ne/ph2ne_U937_1_1_heatmap.png}}   & \adjustbox{valign=c}{\includegraphics[width=6em]{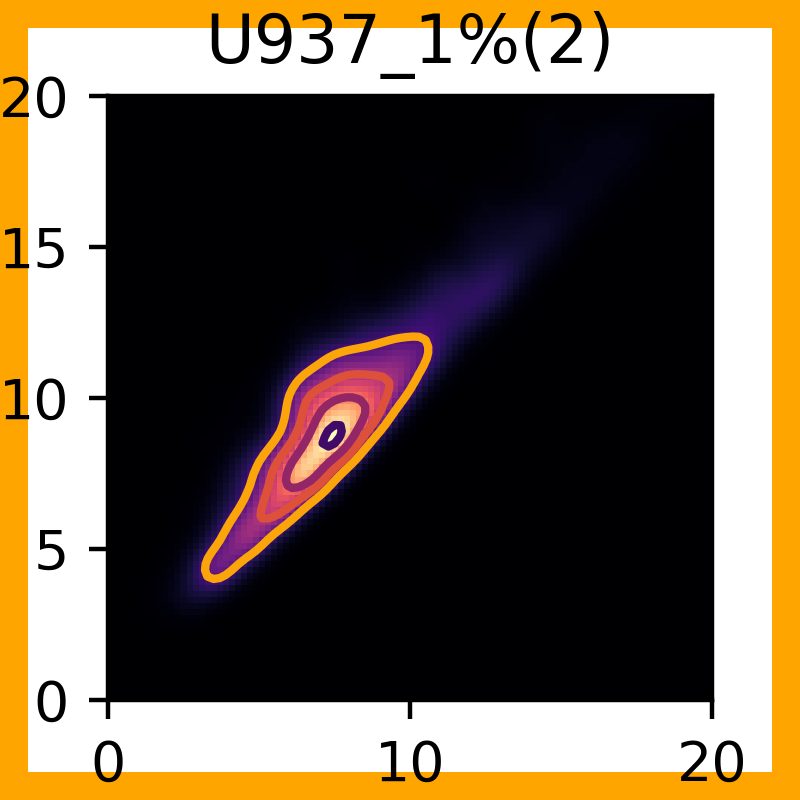}}   & \adjustbox{valign=c}{\includegraphics[width=6em]{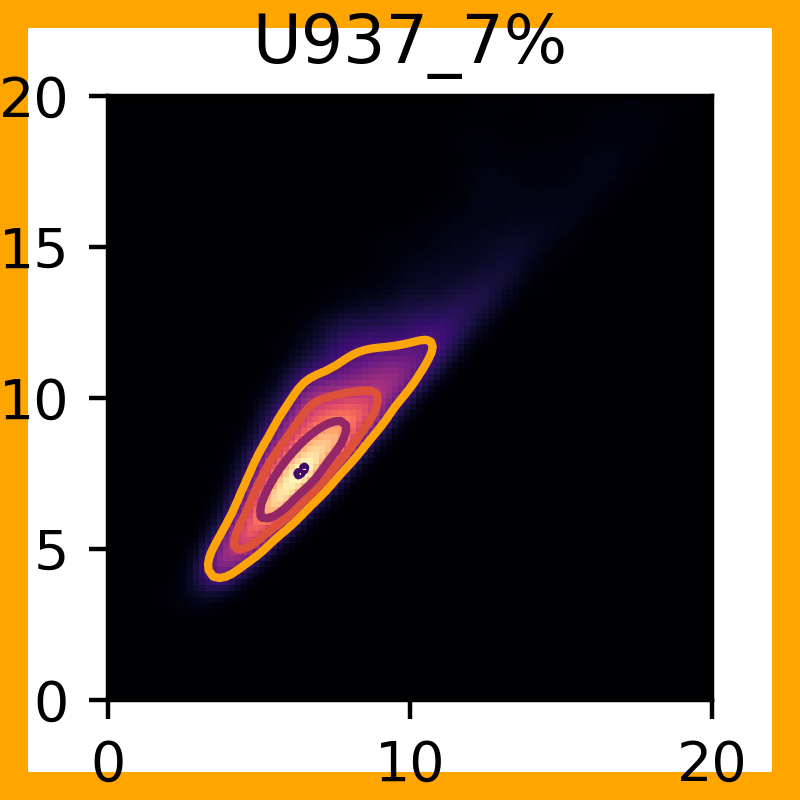}}   & \adjustbox{valign=c}{\includegraphics[width=6em]{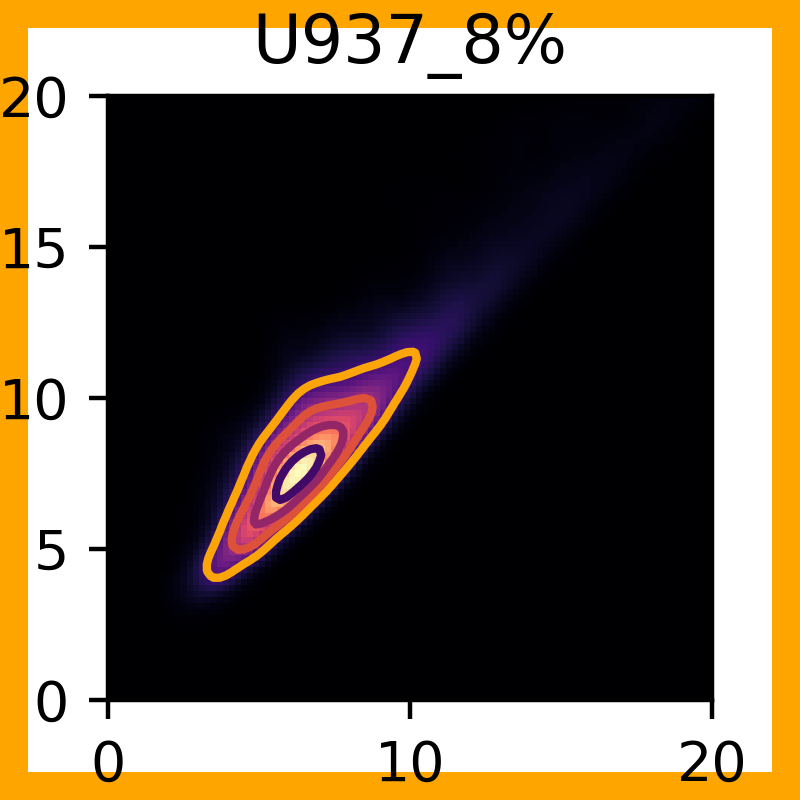}}   & \adjustbox{valign=c}{\includegraphics[width=6em]{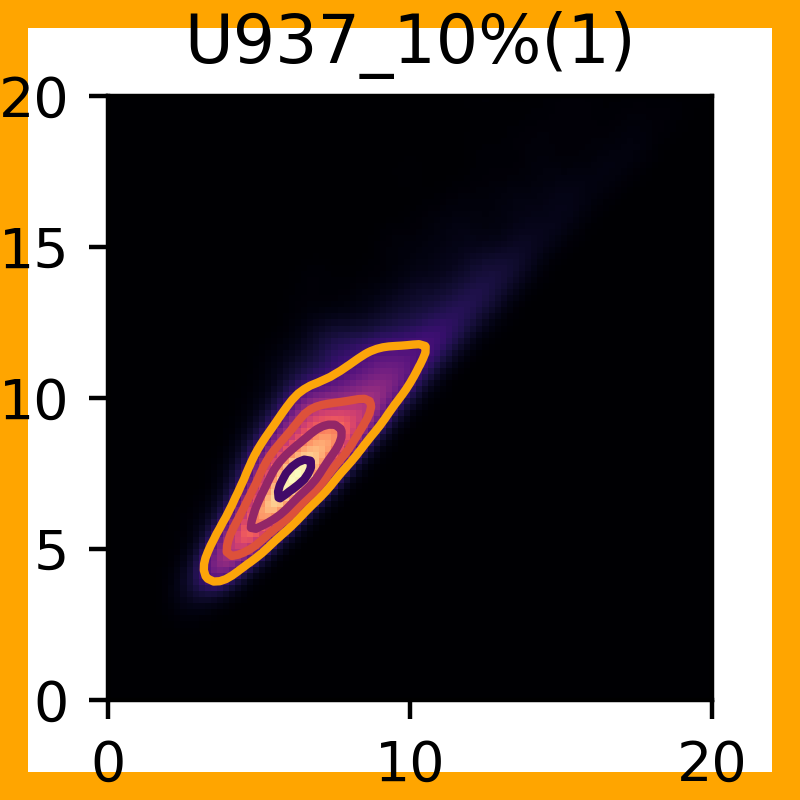}}   &\adjustbox{valign=c}{\includegraphics[width=6em]{figure/knee_ph2ne/ph2ne_U937_10_2_heatmap.png}}  
 & \adjustbox{valign=c}{\includegraphics[width=6em]{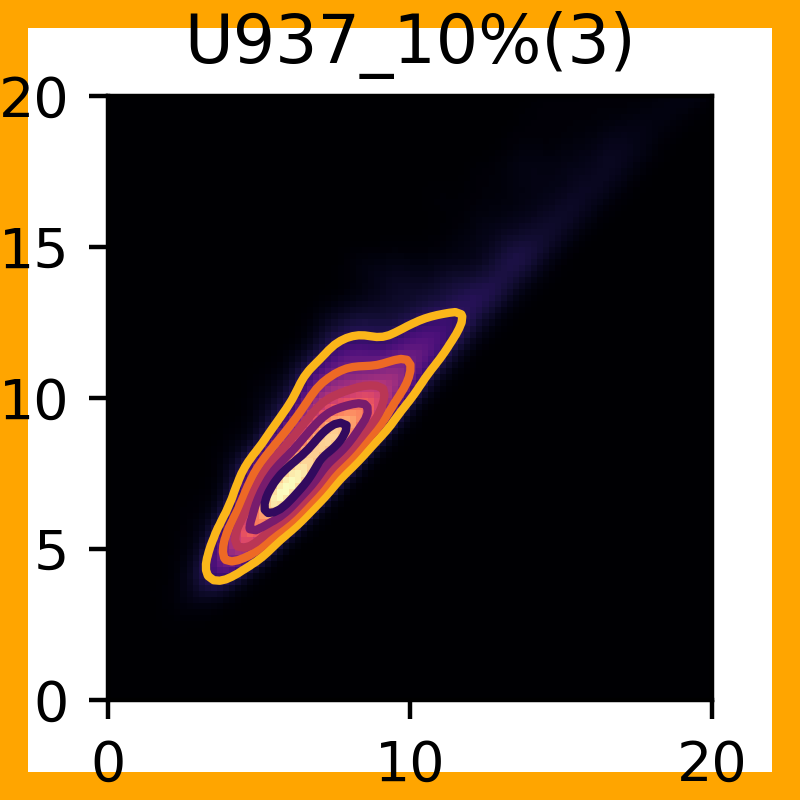}}  \\ 
 & \adjustbox{valign=c}{\includegraphics[width=6em]{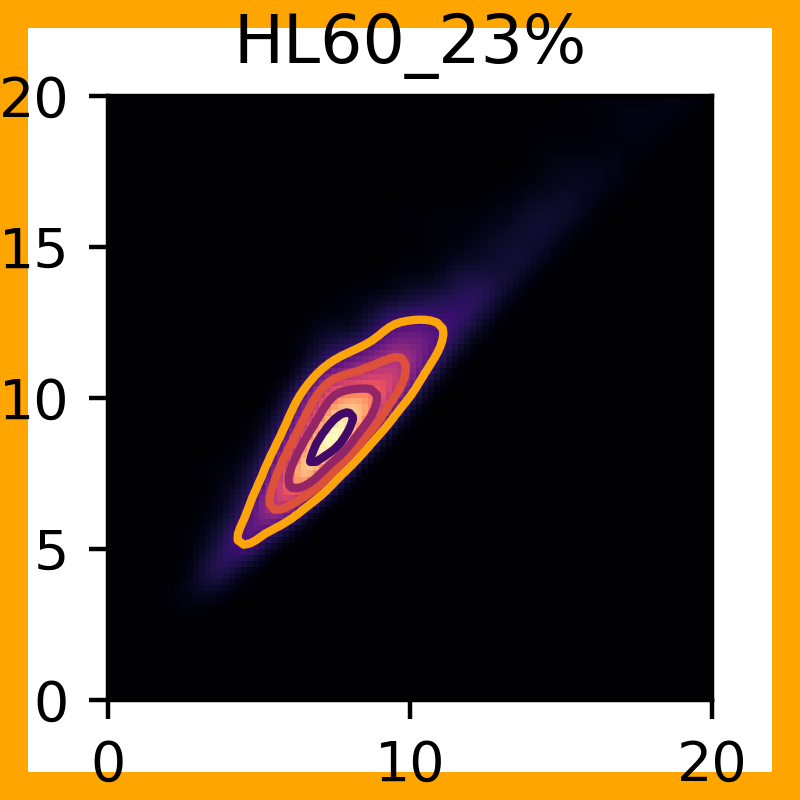}}    &
 \adjustbox{valign=c}{\includegraphics[width=6em]{figure/knee_ph2ne/ph2ne_HL60_25_1_heatmap.png}} &\adjustbox{valign=c}{\includegraphics[width=6em]{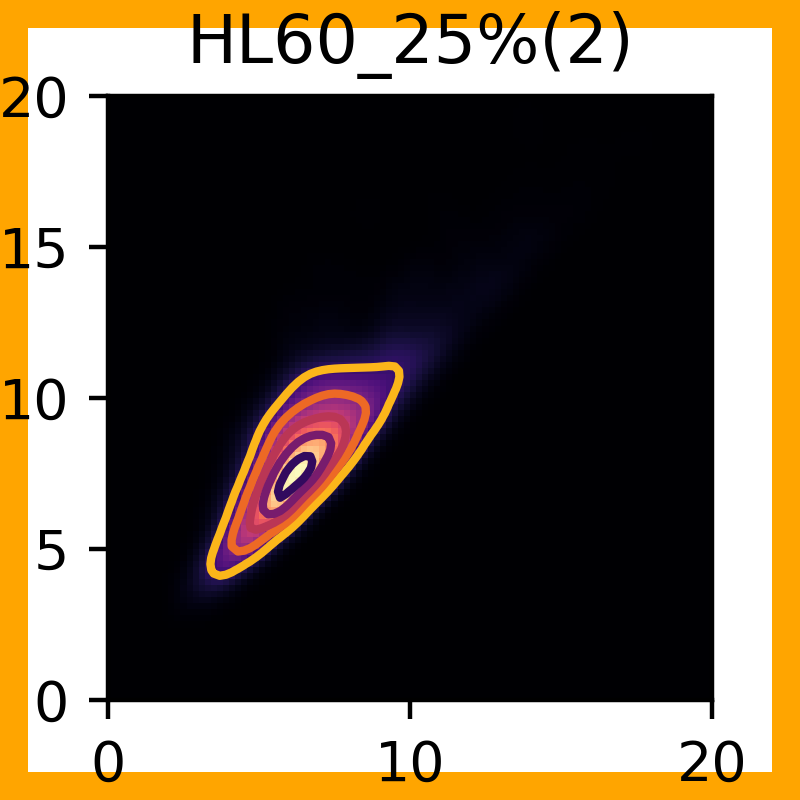}}     &\adjustbox{valign=c}{\includegraphics[width=6em]{figure/knee_ph2ne/ph2ne_P1_10_heatmap.png}}     &\adjustbox{valign=c}{\includegraphics[width=6em]{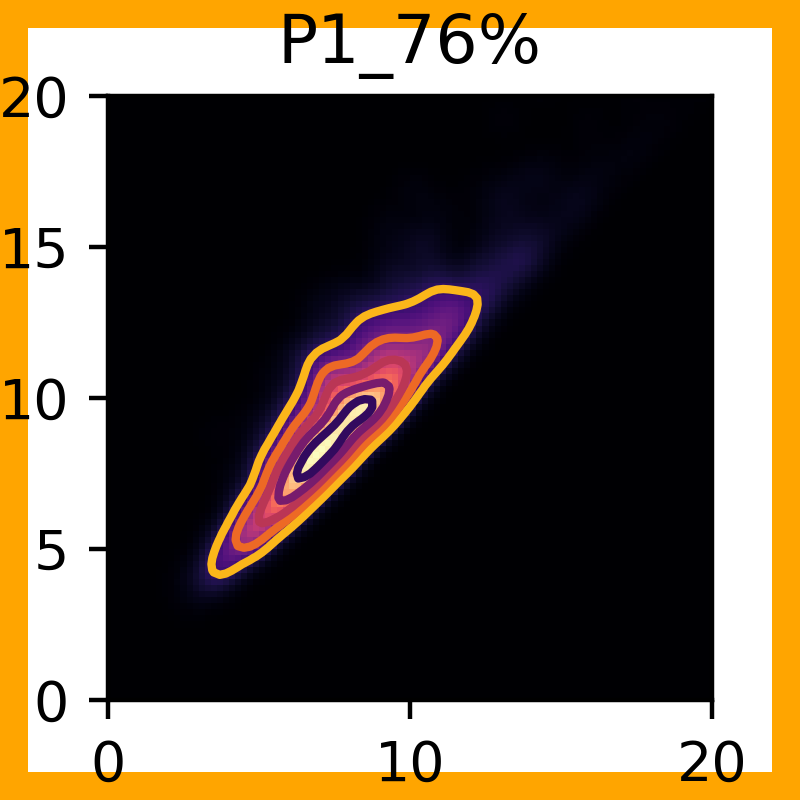}}  & \adjustbox{valign=c}{\includegraphics[width=6em]{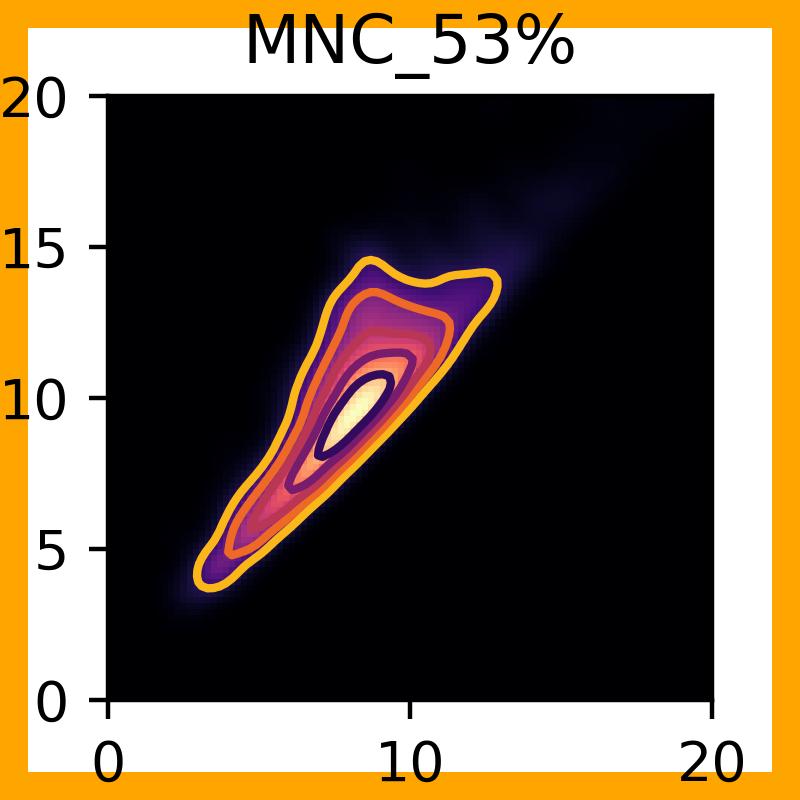}} &    \\ \hline
\multirow{2}{*}{Phase II} &\adjustbox{valign=c}{\includegraphics[width=6em]{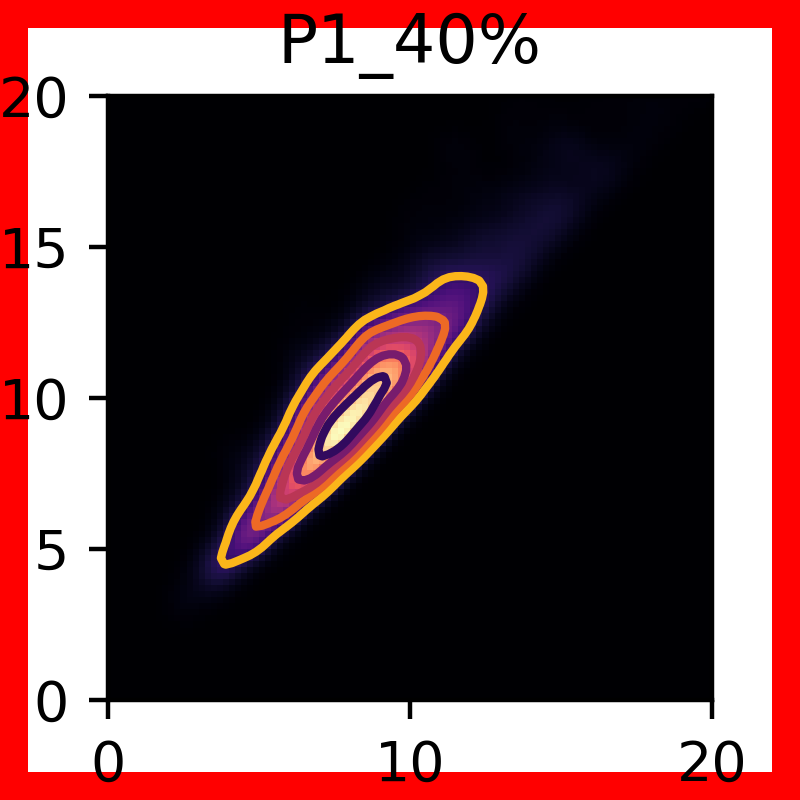}}     &
\adjustbox{valign=c}{\includegraphics[width=6em]{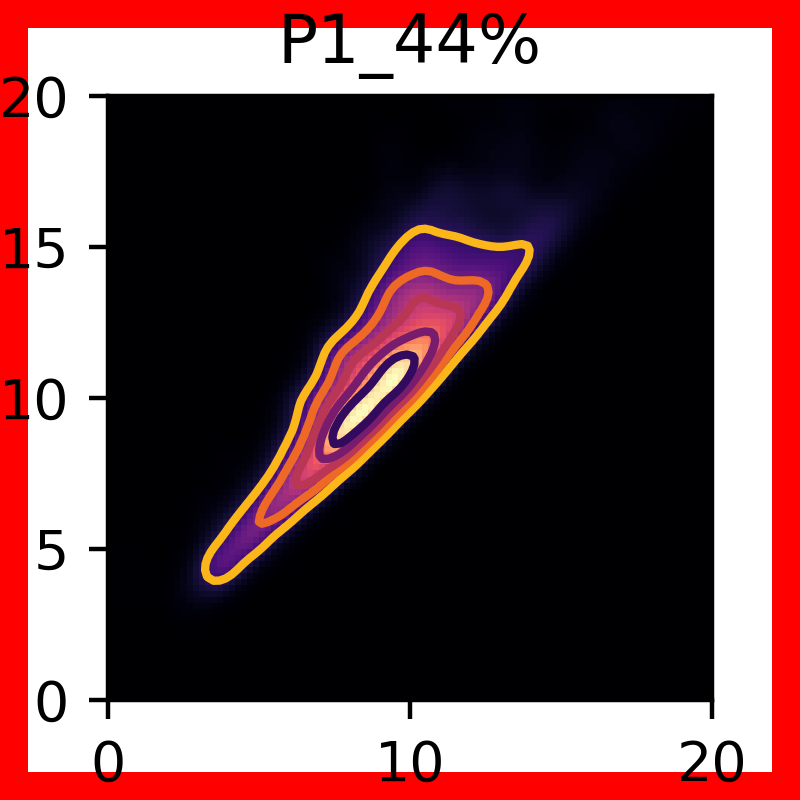}}& 
\adjustbox{valign=c}{\includegraphics[width=6em]{figure/knee_ph2ne/ph2ne_P1_51_heatmap.png}}& 
\adjustbox{valign=c}{\includegraphics[width=6em]{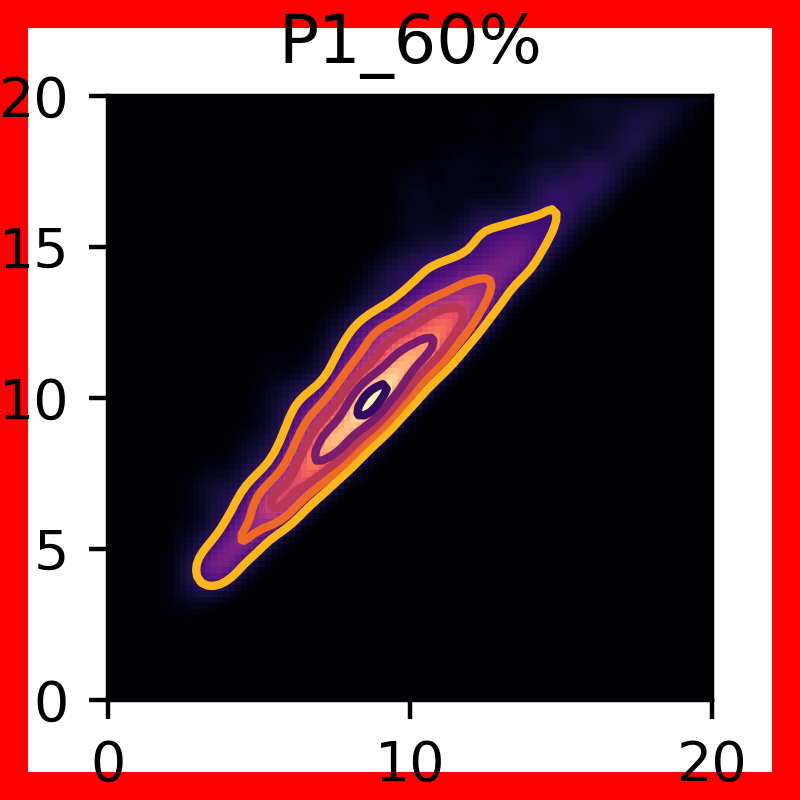}}&
\adjustbox{valign=c}{\includegraphics[width=6em]{figure/knee_ph2ne/ph2ne_P2_59_heatmap.png}}&
\adjustbox{valign=c}{\includegraphics[width=6em]{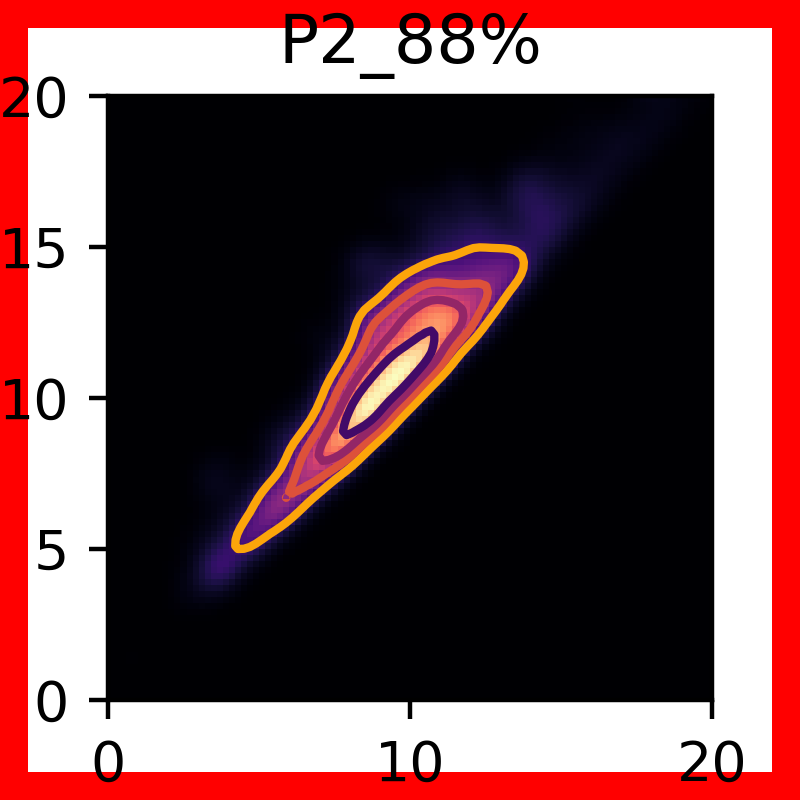}}& 
\adjustbox{valign=c}{\includegraphics[width=6em]{figure/knee_ph2ne/ph2ne_P2_90_heatmap.png}}\\
&
\adjustbox{valign=c}{\includegraphics[width=6em]{figure/knee_ph2ne/ph2ne_MNC_67_heatmap.png}} & 
\adjustbox{valign=c}{\includegraphics[width=6em]{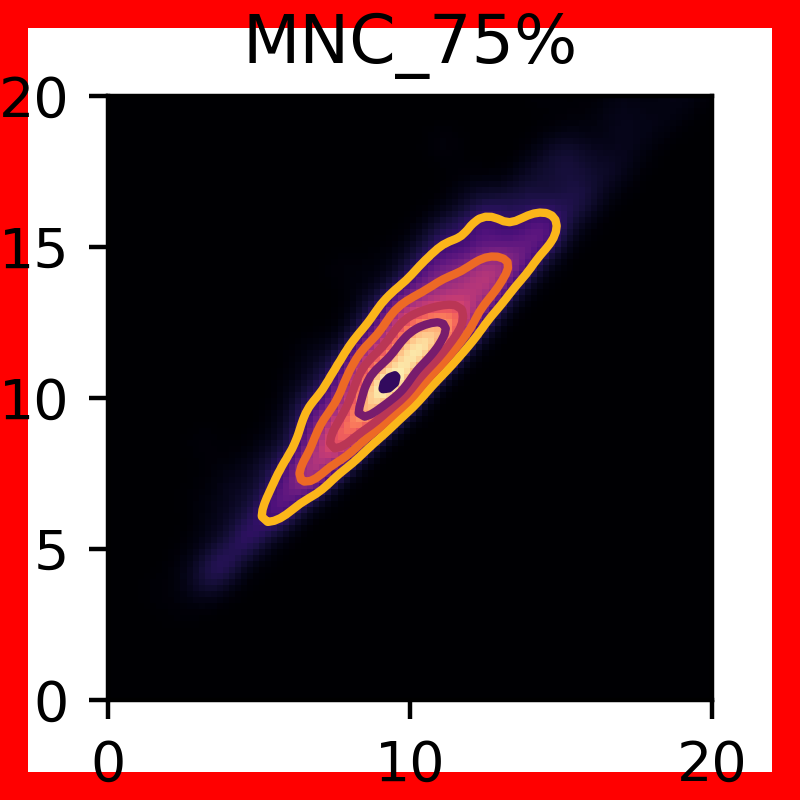}} & 
\adjustbox{valign=c}{\includegraphics[width=6em]{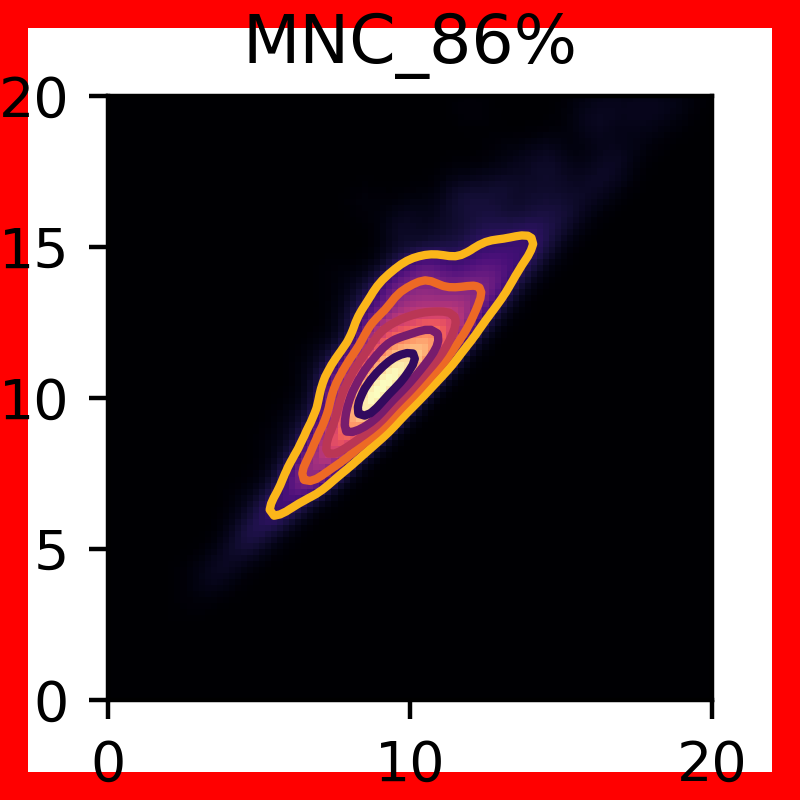}} &    &    &    &   \\
\bottomrule
\end{tabular}
\end{adjustwidth}
\end{table}

\section{Consensus Hierarchical Clustering}\label{app:consensus_tree}

\begin{figure}[htb]
    \centering
    \includegraphics[width=.7\linewidth]{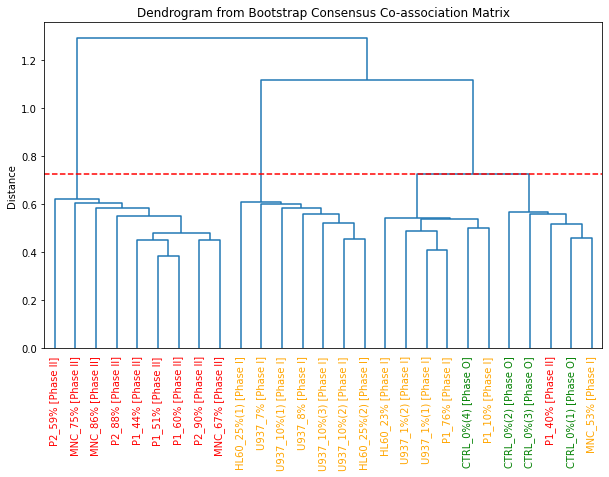}
    \caption{Hierarchical tree constructed from consensus over 100 bootstrap resamplings on PH2 KDEs with Ward linkage}
    \label{fig:ph2_bootstrap_tree}
\end{figure}

\begin{figure}[htb]
    \centering
    \includegraphics[width=.7\linewidth]{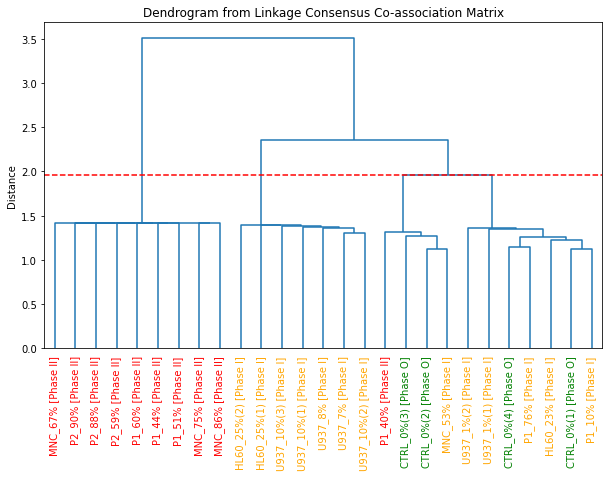}
    \caption{Hierarchical tree constructed from consensus over 4 linakge methods---complete, Ward, average, and single---on PH2 KDEs.}
    \label{fig:ph2_linkage_tree}
\end{figure}

In this section, we present the additional consensus hierarchical trees for PH$_2$ KDEs. Figure \ref{fig:ph2_bootstrap_tree} shows the consensus hierarchical clusters obtained across bootstrap resamplings, while Figure \ref{fig:ph2_linkage_tree} depicts consensus across different linkage methods. In both cases, we extract the 4 clusters by cutting the dendrograms at appropriate heights. Among the resultant clusters, 2 clusters correspond exclusively to Phase II and Phase I samples, respectively, with the P1 40\% sample once again appearing outside the sole Phase II group as an outlier. The remaining 2 clusters are composed primarily of Phase O and Phase I samples, reflecting the agreement to the Phase labels.

\section{Continuous Modeling of Disease Progression}\label{app:regression}

While our primary analyses rely on 3 inferred discrete disease phases, we recognize that AML progression is inherently continuous. A regression-based framework could, in principle, offer a more fine-grained representation of disease state. However, several biological and experimental factors present nontrivial challenges to implementing such an approach robustly in our setting. Firstly, each data sample in our study corresponds to a single terminally sacrificed mouse, eliminating the possibility of within-subject longitudinal modeling. Moreover, the biological heterogeneity across mice---including differences in immune response, genetic background, and AML subtype---complicates the assumption of a shared continuous timeline across animals. Furthermore, engraftment levels, often used as a proxy for disease burden, were measured from skeletal sites distinct from the imaged femur and are thus subject to both biological and spatial noise. These values should be interpreted as imprecise, interval-censored indicators of progression, without clearly defined temporal bounds.

Despite these limitations, we investigated a prototype continuous-time model based on a spline-interpolated time-varying Gaussian Mixture Model (GMM). In this framework, each time point $t$ is associated with a density
$$f_t(x) = \sum^K_{k=1} \alpha_k(t) \Phi (x| \mu_k(t), \Sigma_k(t)$$
where $\alpha_k(t)$, $\mu_k(t)$, and $\Sigma_k(t)$ are the mixture weights, means, and covariances, respectively, and the weights and means are then smoothed using spline functions for interpolation. While this provides a flexible framework for capturing smooth temporal evolution in the data, it is also affected by the volatility of the estimated Gaussian parameters and by the limited number of temporal time points, which can undermine stability and reliability in downstream inference.

\begin{figure}[htb]
    \centering
    \includegraphics[width=.6\linewidth]{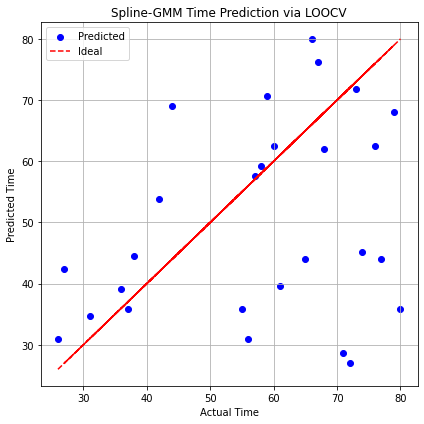}
    \caption{Plot of predicted and true engraftment level from prototype regression model with leave-one-out cross-validation (with engraftment levels as time).}
    \label{fig:pred_reg}
\end{figure}

We evaluated predictive performance using leave-one-out cross-validation (LOOCV) and the results are presented in Figure \ref{fig:pred_reg}. Although the model exhibited some alignment with the engraftment levels, its overall accuracy was limited. Performance was affected by several factors: the sparsity and repetition of time points complicated spline fitting; numerical instability arose during inversion of the spline mappings for prediction; and noise in the engraftment labels further impeded reliable inference. More sophisticated strategies---such as incorporating stochastic interpolation, treating time as a latent variable in a Bayesian framework, or modeling measurement error explicitly---may offer improvements. However, these approaches substantially increase model complexity and risk overfitting given the limited sample size. For these reasons, we adopt a more interpretable and statistically stable discretization into 3 phases, which preserves meaningful biological structure without overburdening the data.

\section{Model Interpretation: Additional Figures and Results}\label{app:model_eval}

In this section, we provide additional results on our model evaluation discussed in Section 5.2 of the article.

\subsection{Mixture Size Selection}

\begin{figure}[htbp]
    \centering
    \begin{subfigure}[t]{0.45\textwidth}
    \captionsetup{width=0.95\textwidth}
        \centering
        \includegraphics[width = \textwidth]{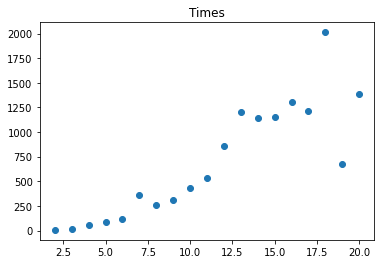}
        \caption{Plot of computation time in seconds required for model fitting against model size.}
        \label{fig:ph0sw_time}
    \end{subfigure}%
    \begin{subfigure}[t]{0.45\textwidth}
    \captionsetup{width=0.95\textwidth}
        \centering
        \includegraphics[width = \textwidth]{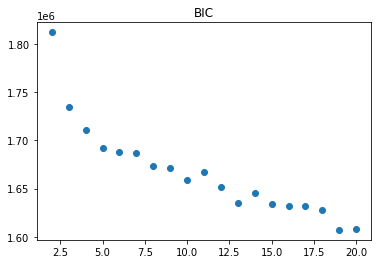}
        \caption{Plot of Bayesian information criterion (BIC) of model fit against model size.}
        \label{fig:ph0sw_bic}        
    \end{subfigure}%
    \caption{Plots showing the computation time required and model fit across a range of model sizes for the $\text{PH}_0$ SW quadrant.}
    \label{fig:ph0_sw_time_bic}
\end{figure}

\begin{figure}[htbp]
    \centering
    \begin{subfigure}[t]{0.45\textwidth}
    \captionsetup{width=0.95\textwidth}
        \centering
        \includegraphics[width = \textwidth]{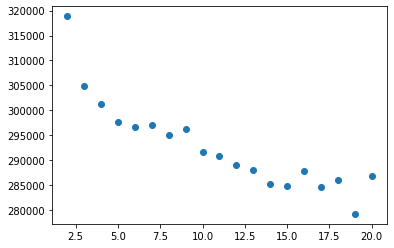}
        \caption{Plot of BIC of model fit against model size for Phase O data.}
        \label{fig:ph0sw_bic0}
    \end{subfigure}
    \begin{subfigure}[t]{0.45\textwidth}
    \captionsetup{width=0.95\textwidth}
        \centering
        \includegraphics[width = \textwidth]{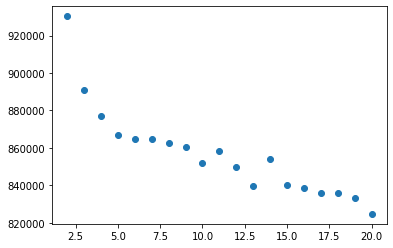}
        \caption{Plot of BIC of model fit against model size for Phase I data.}
        \label{fig:ph0sw_bic1}        
    \end{subfigure}%
    \\
    \begin{subfigure}[t]{0.45\textwidth}
    \captionsetup{width=0.95\textwidth}
        \centering
        \includegraphics[width = \textwidth]{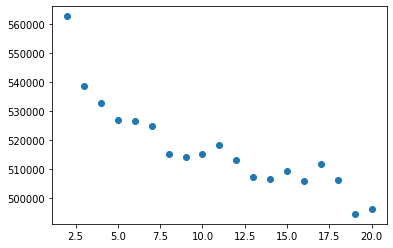}
        \caption{Plot of BIC of model fit against model size for Phase II data.}
        \label{fig:ph0sw_bic2}        
    \end{subfigure}%
    \caption{Plots showing the model fit measured by BIC across a range of model sizes for each phase of the $\text{PH}_0$ SW quadrant.}
    \label{fig:ph0_sw_bics}
\end{figure}

The number of components of our mixture model is determined by a trade-off between the goodness of fit of the models and the computation time. We explored a range of sizes of models from 2 to 20 components and we cap the maximum size to 20 for generalizability of the models across a range of engraftment levels within the same phase, with larger models being more restrictive. Below, as an example, we present the choice of model size for the $\text{PH}_0$ SW quadrant. Figure \ref{fig:ph0_sw_time_bic} shows the comparison of computation time and model fit measured by the Bayesian information criterion (BIC) with increasing model sizes. We selected BIC as the criterion since it demonstrates greater penalization on the complexity of the model and converges asymptotically to the true model if present. Balancing the need for the model to approximate the data well, and for the algorithm to run in appropriate time with bootstrapping, we found a small model with around 5 components to be optimal. Figure \ref{fig:ph0_sw_bics} further studies the model fit within each of the phases; we observe ``elbows'' when the model contains 3 to 5 components, after which the decrease in BIC is noticeably smaller. Transitioning from the Phase O model to the Phase I, we also found that incrementing the model size by one led to better model fit.

\subsection{Model Evaluation for PH$_0$ SW and PH$_1$ SW Quadrants}
\begin{figure}[htbp]

\centering
\includegraphics[width=.3\textwidth]{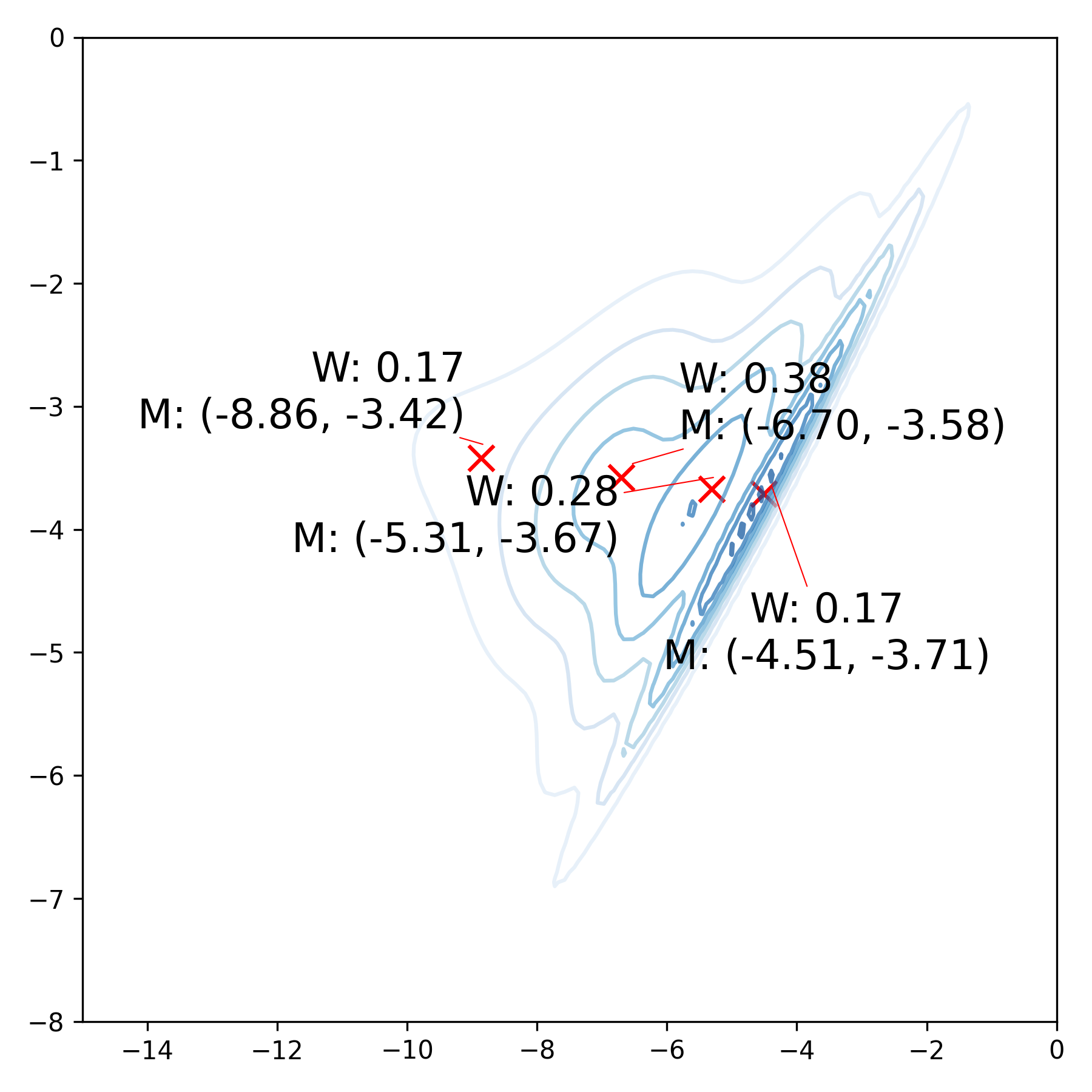}\hfill
\includegraphics[width=.3\textwidth]{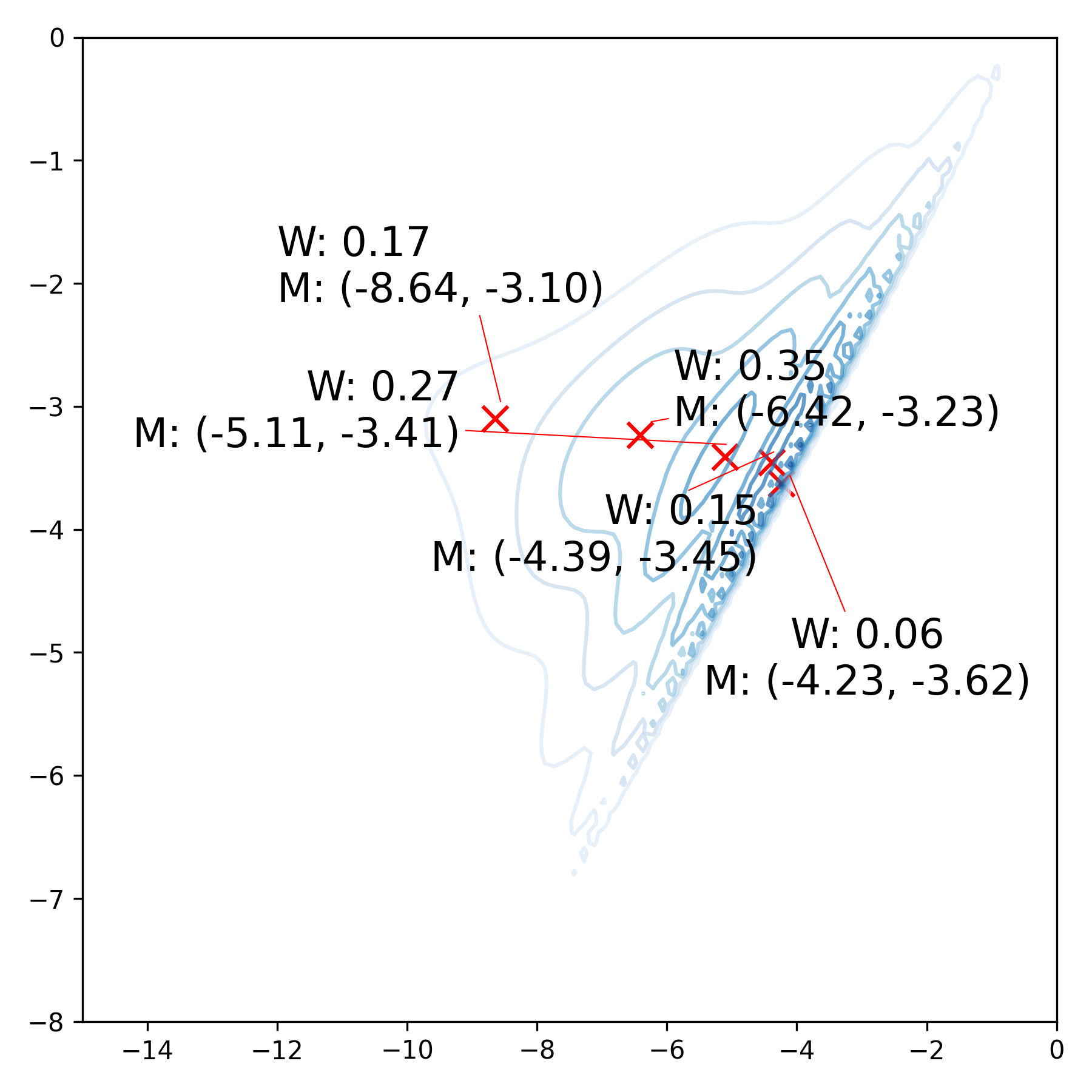}\hfill
\includegraphics[width=.3\textwidth]{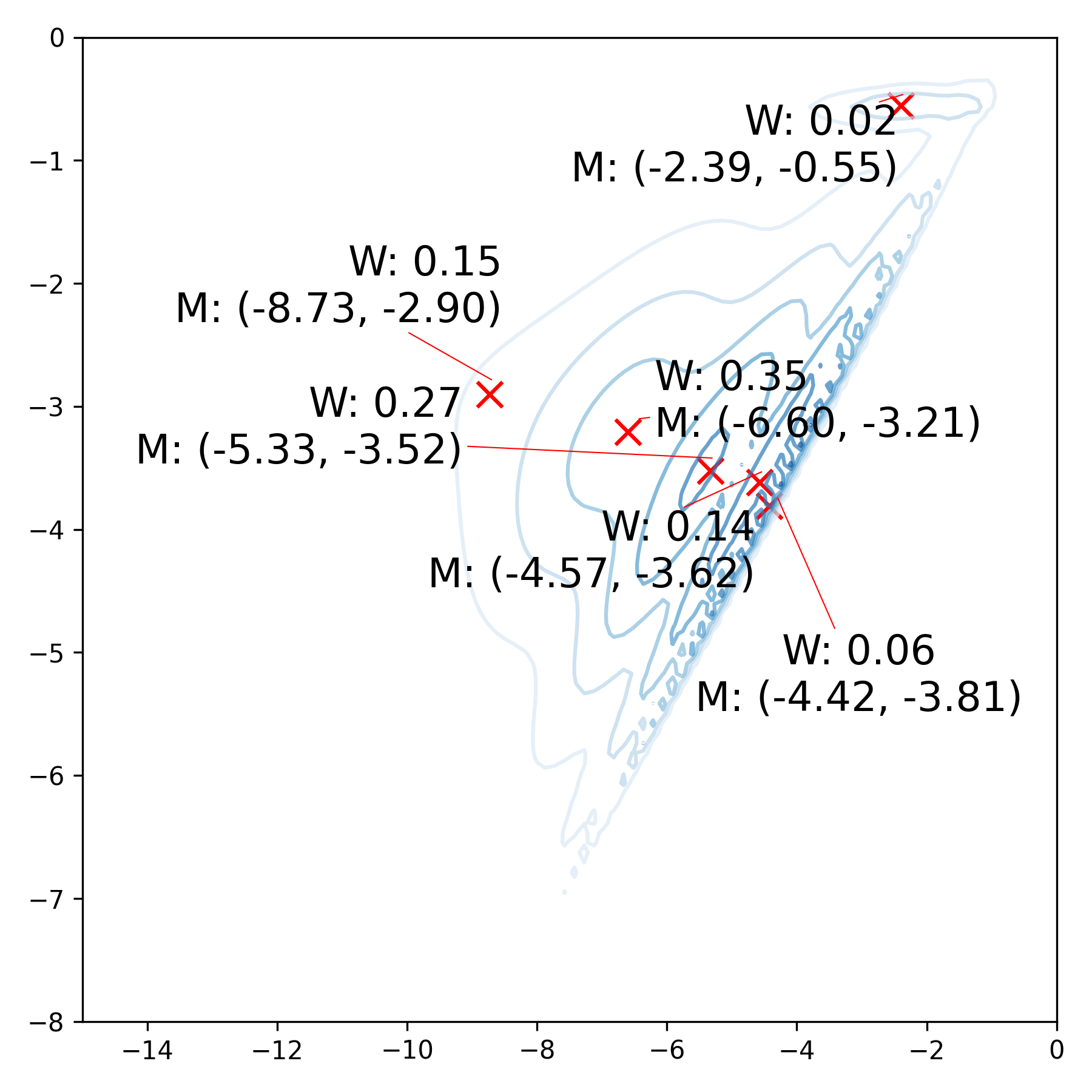}\\
\includegraphics[width=.33\textwidth]{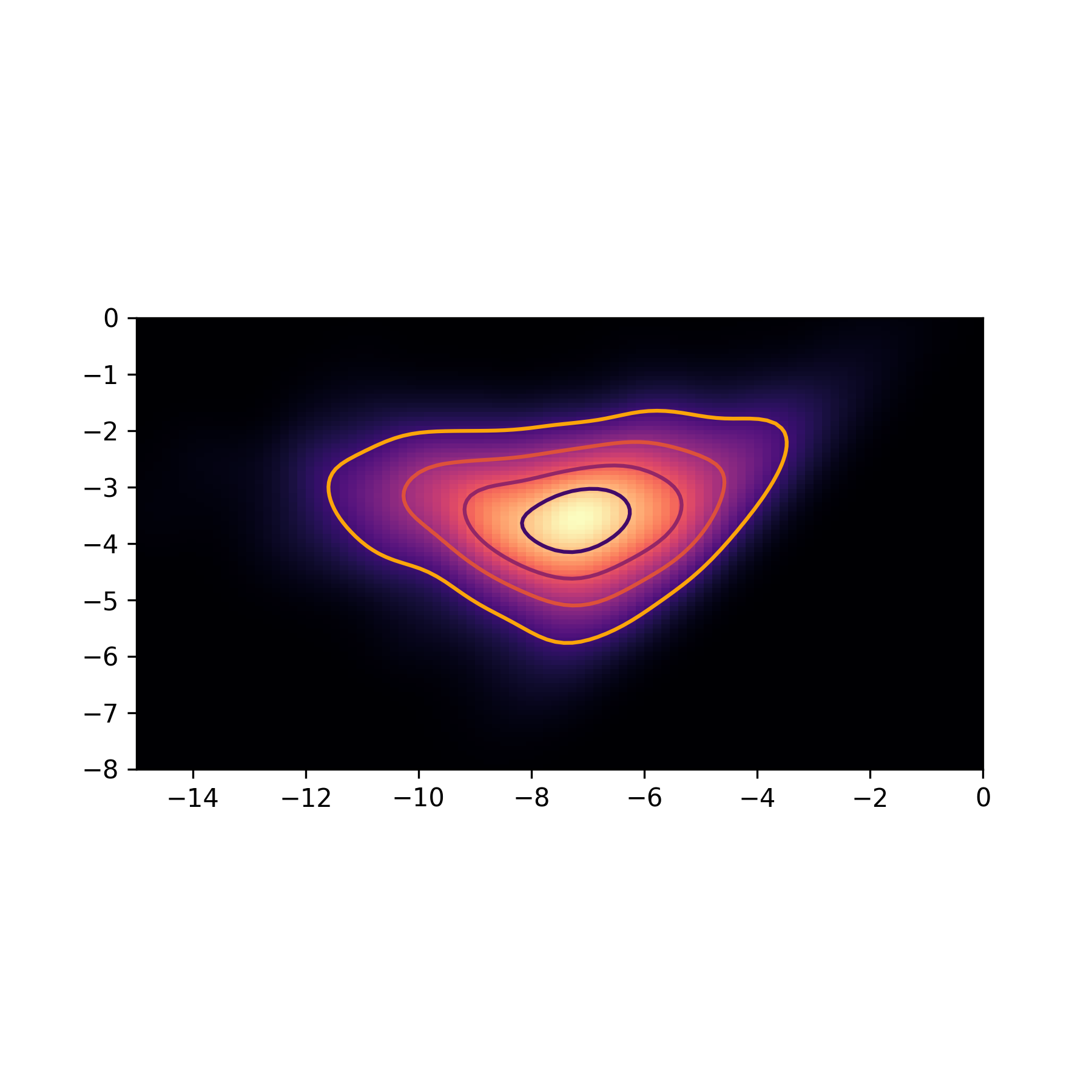}\hfill
\includegraphics[width=.33\textwidth]{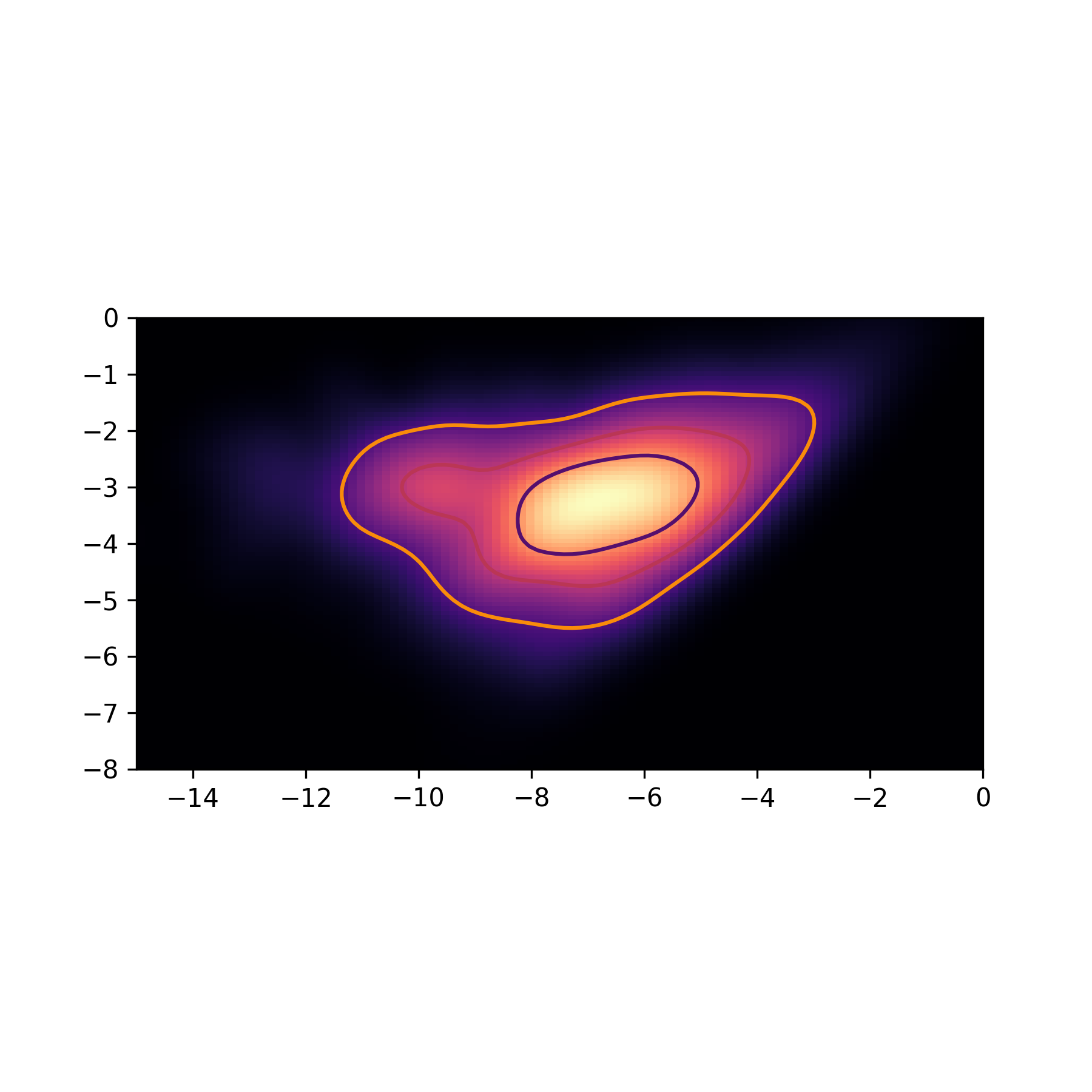}\hfill
\includegraphics[width=.33\textwidth]{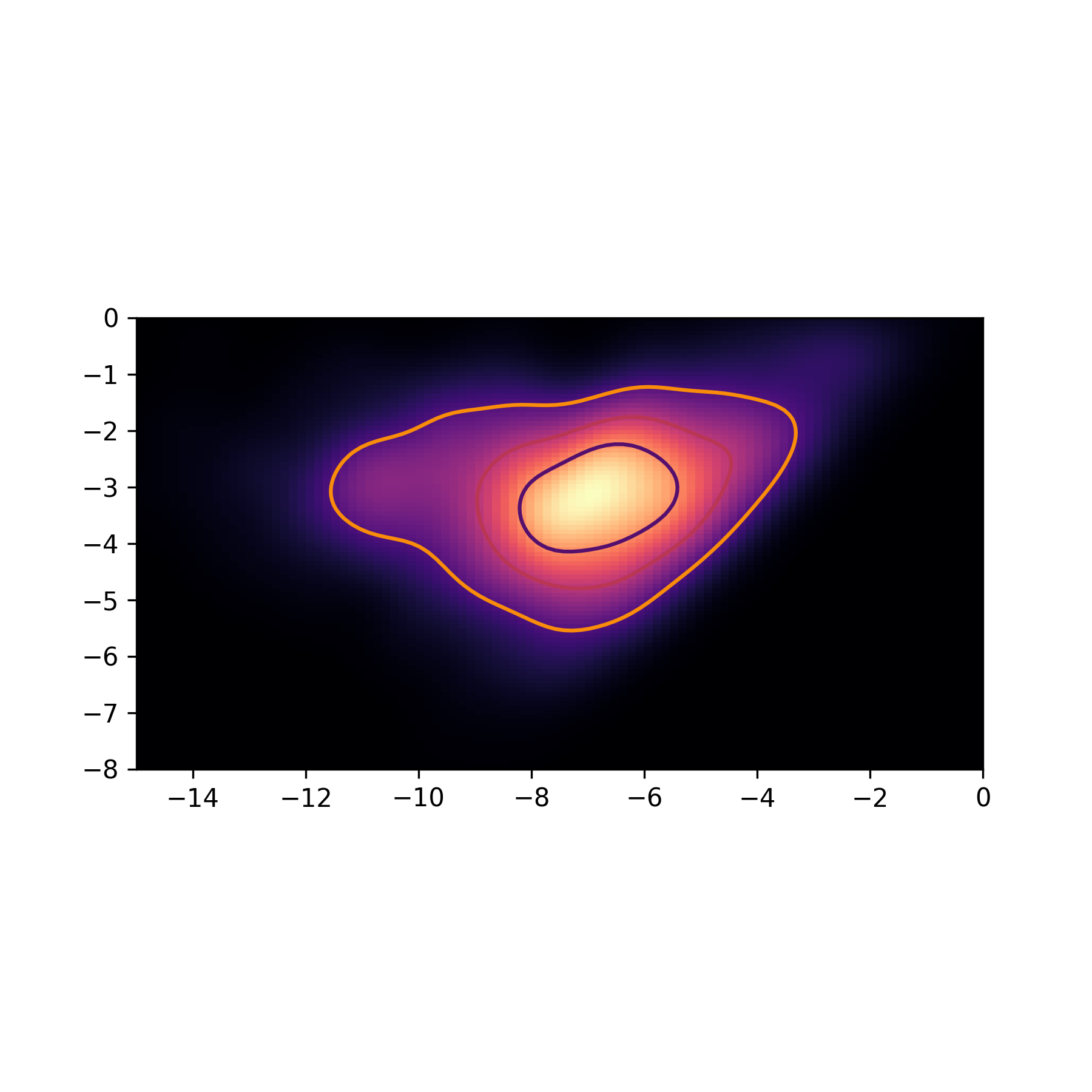}

\caption{Figures showing distributions of the weighted mixture of Gaussians for Phases O (left), I (center), and II (right) for $\text{PH}_0$ SW.}
\label{fig:ph0sw_dist}
\end{figure}

In this section, we show our model fitted to the PH$_0$ SW and PH$_1$ SW quadrants. In the PH$_0$ SW quadrant, the $x, \,y$ axes correspond to values of $r_0$ and $r_1$, representing vessel thickness at bumps and narrowings along the vessel. Across the Phase O, I, and II models as shown in Figure \ref{fig:ph0sw_dist}, we observe relatively stable distributions with minimal shift in density across the phases. This suggests that the morphological features captured in this quadrant remain largely unaffected during AML progression.

\begin{figure}[htbp]

\centering
\includegraphics[width=.3\textwidth]{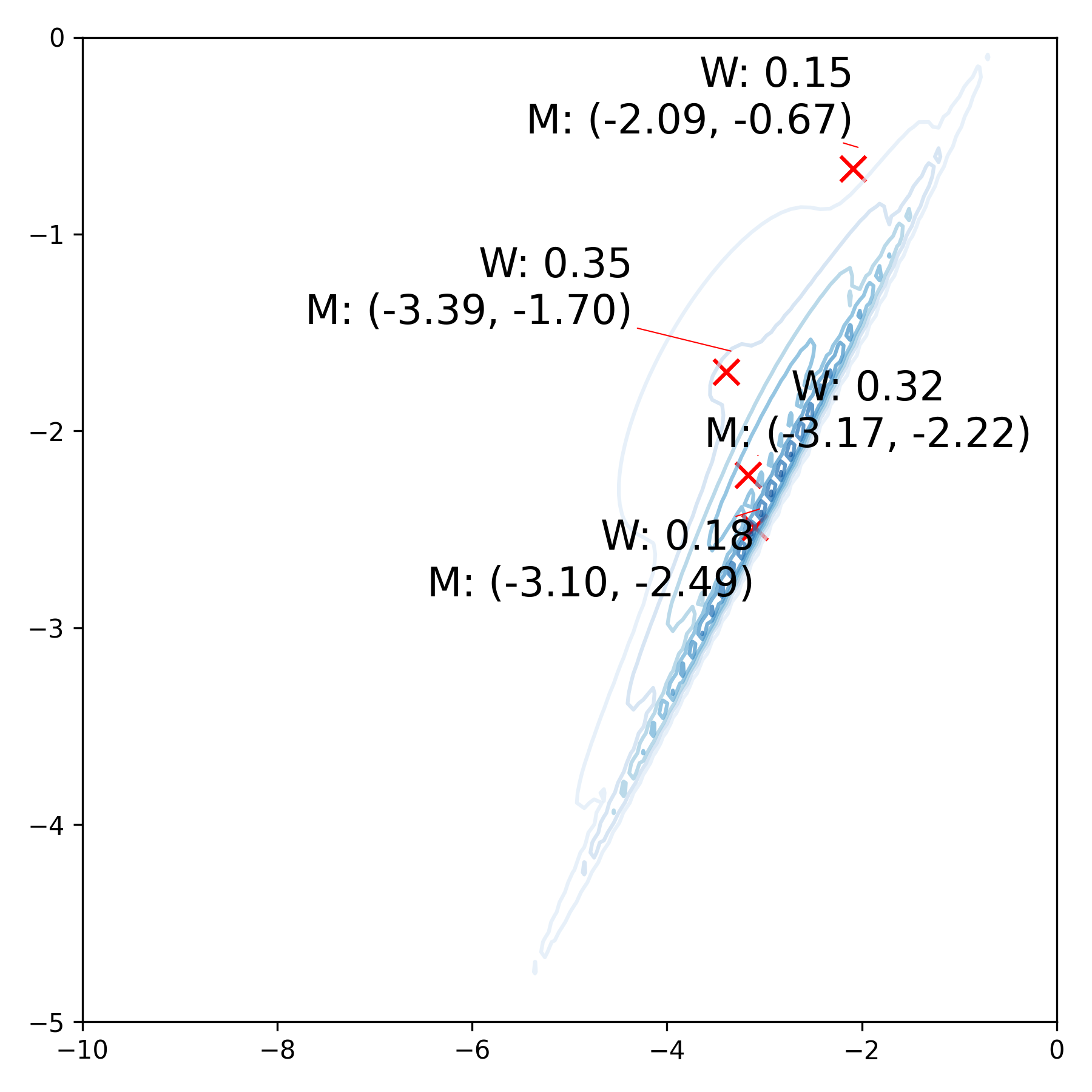}\hfill
\includegraphics[width=.3\textwidth]{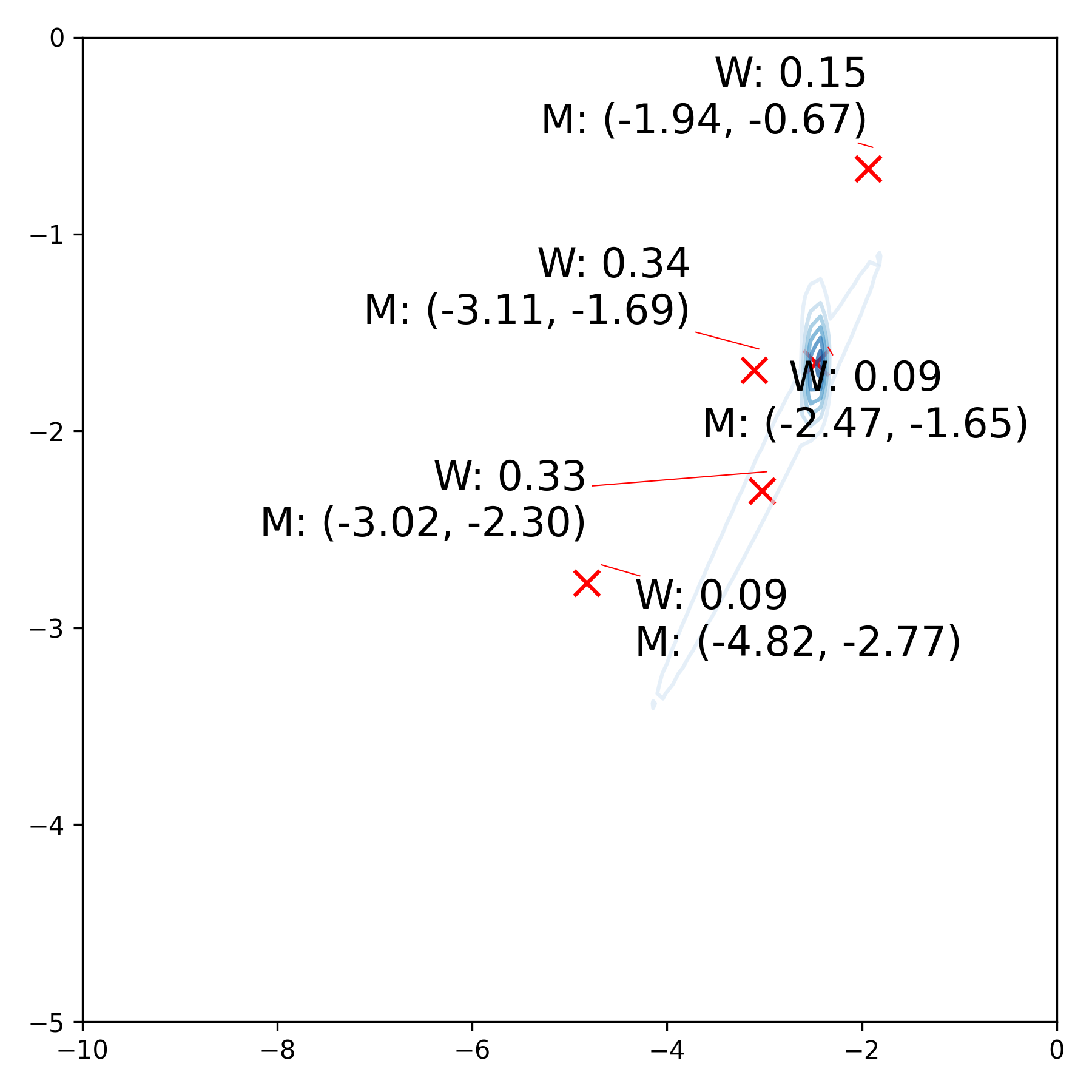}\hfill
\includegraphics[width=.3\textwidth]{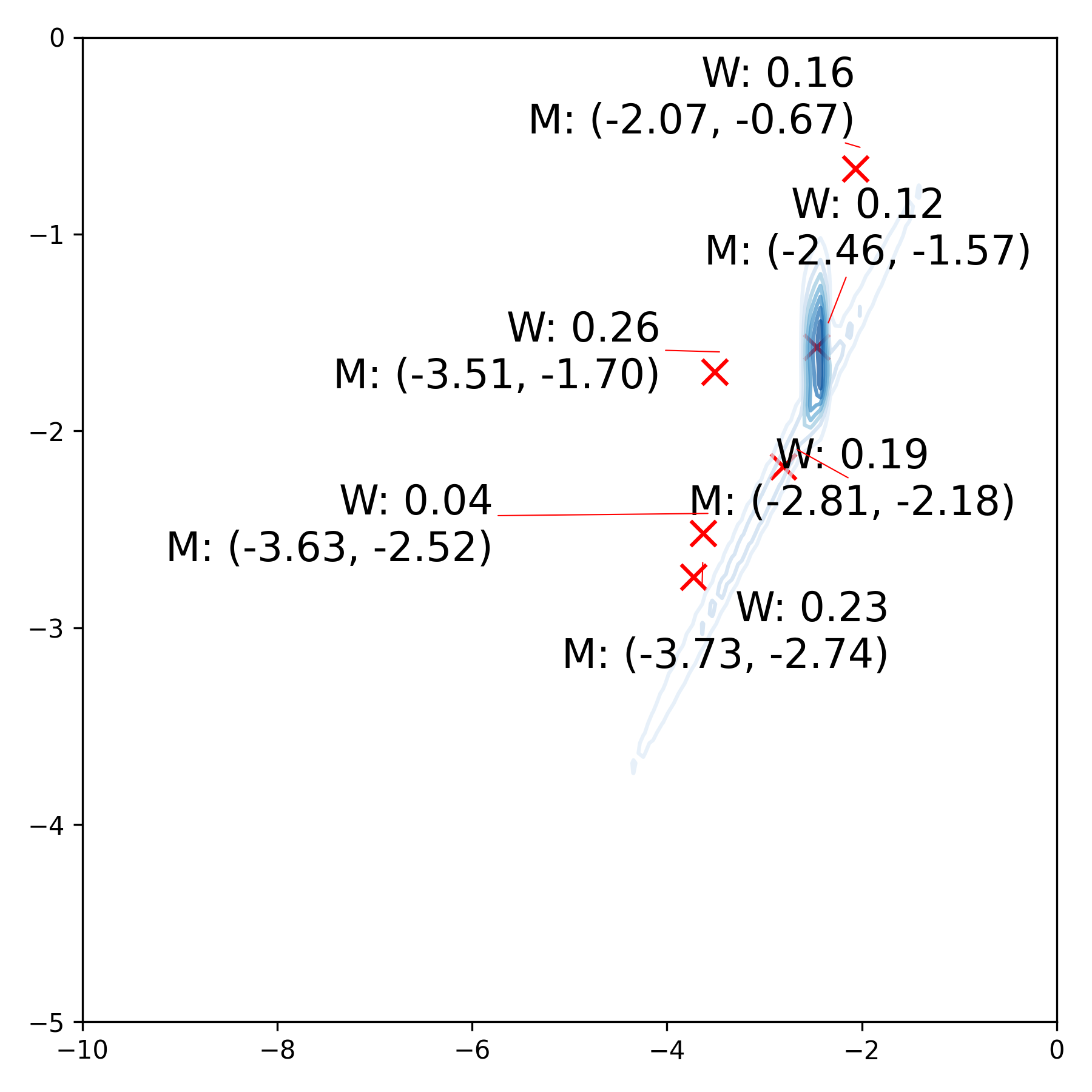}\\
\includegraphics[width=.33\textwidth]{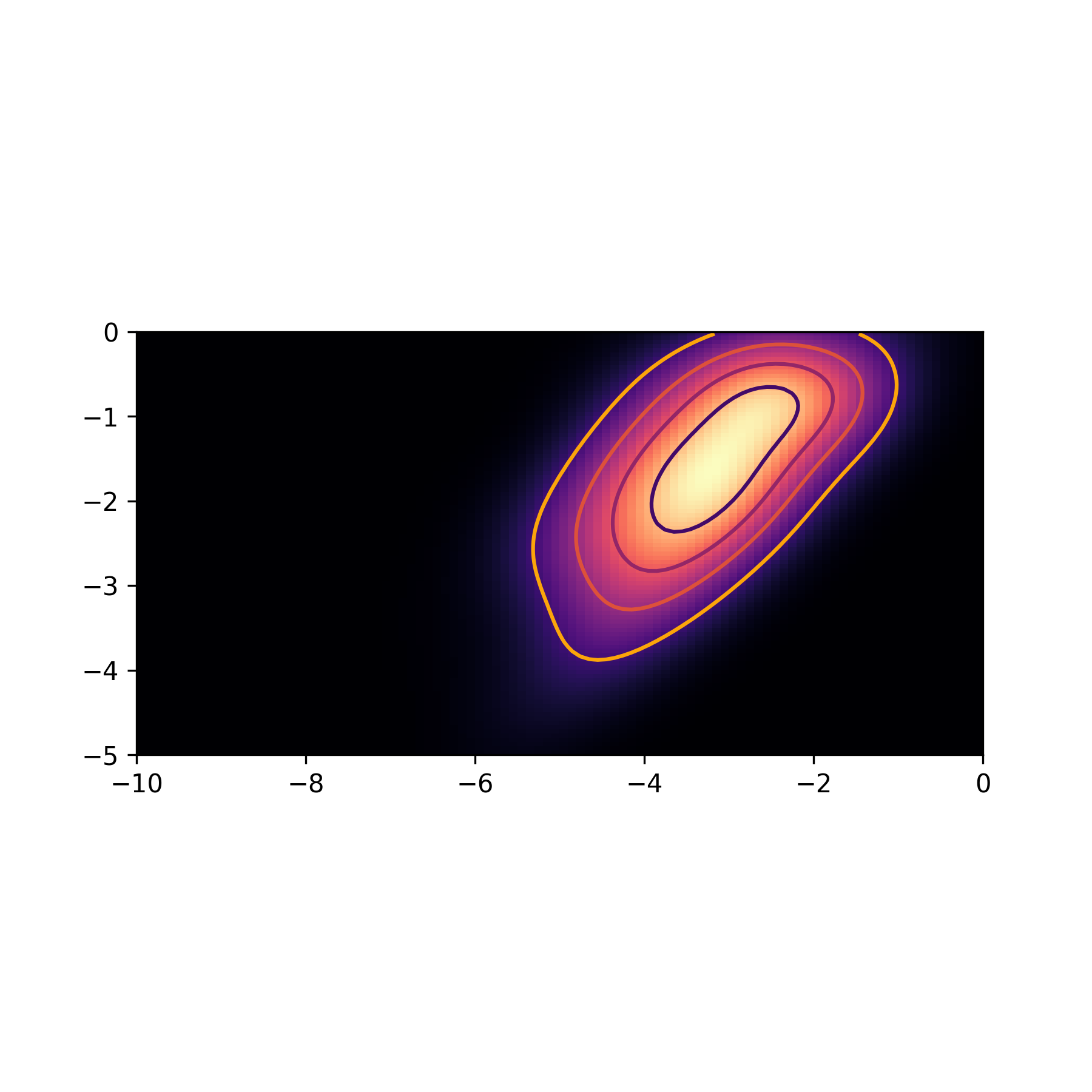}\hfill
\includegraphics[width=.33\textwidth]{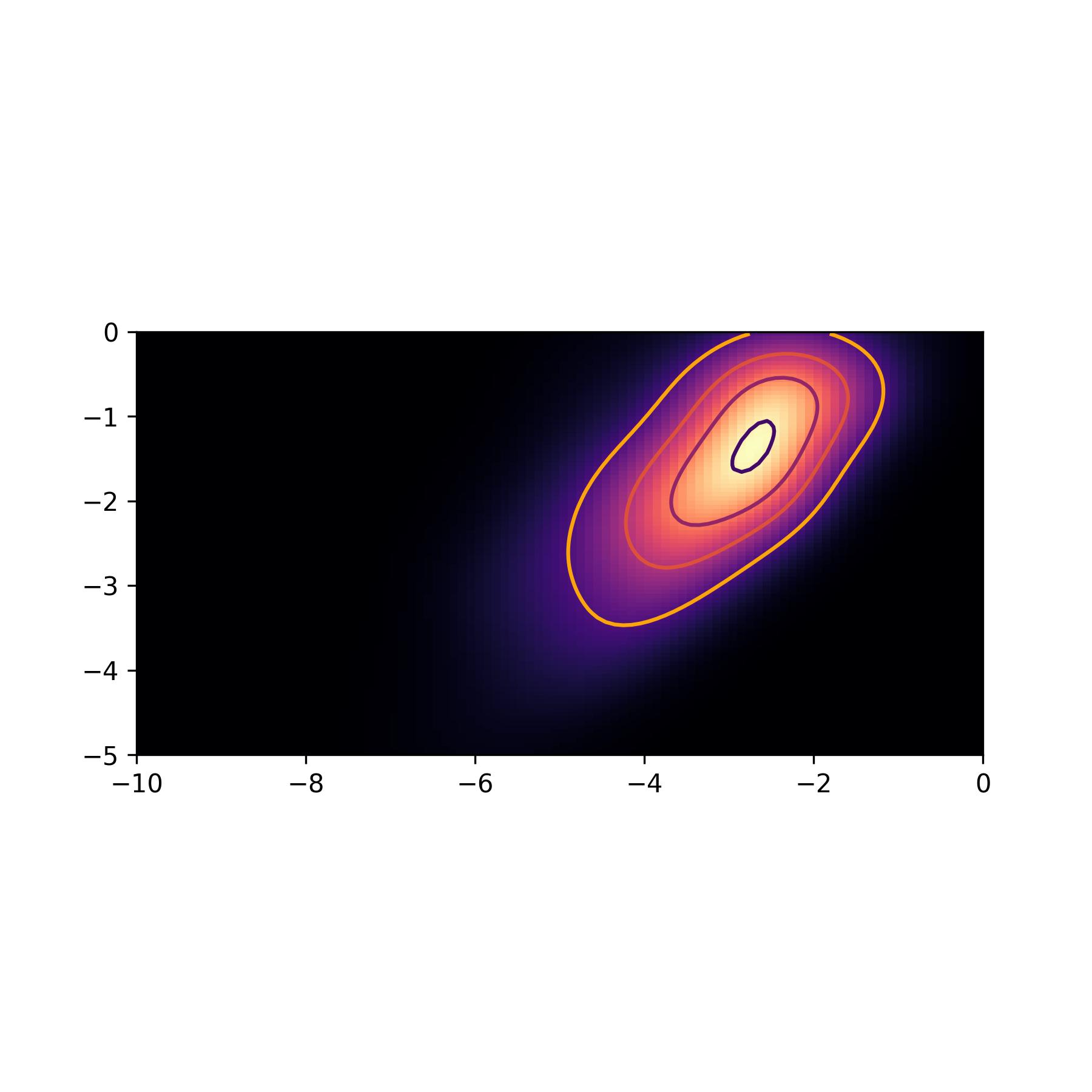}\hfill
\includegraphics[width=.33\textwidth]{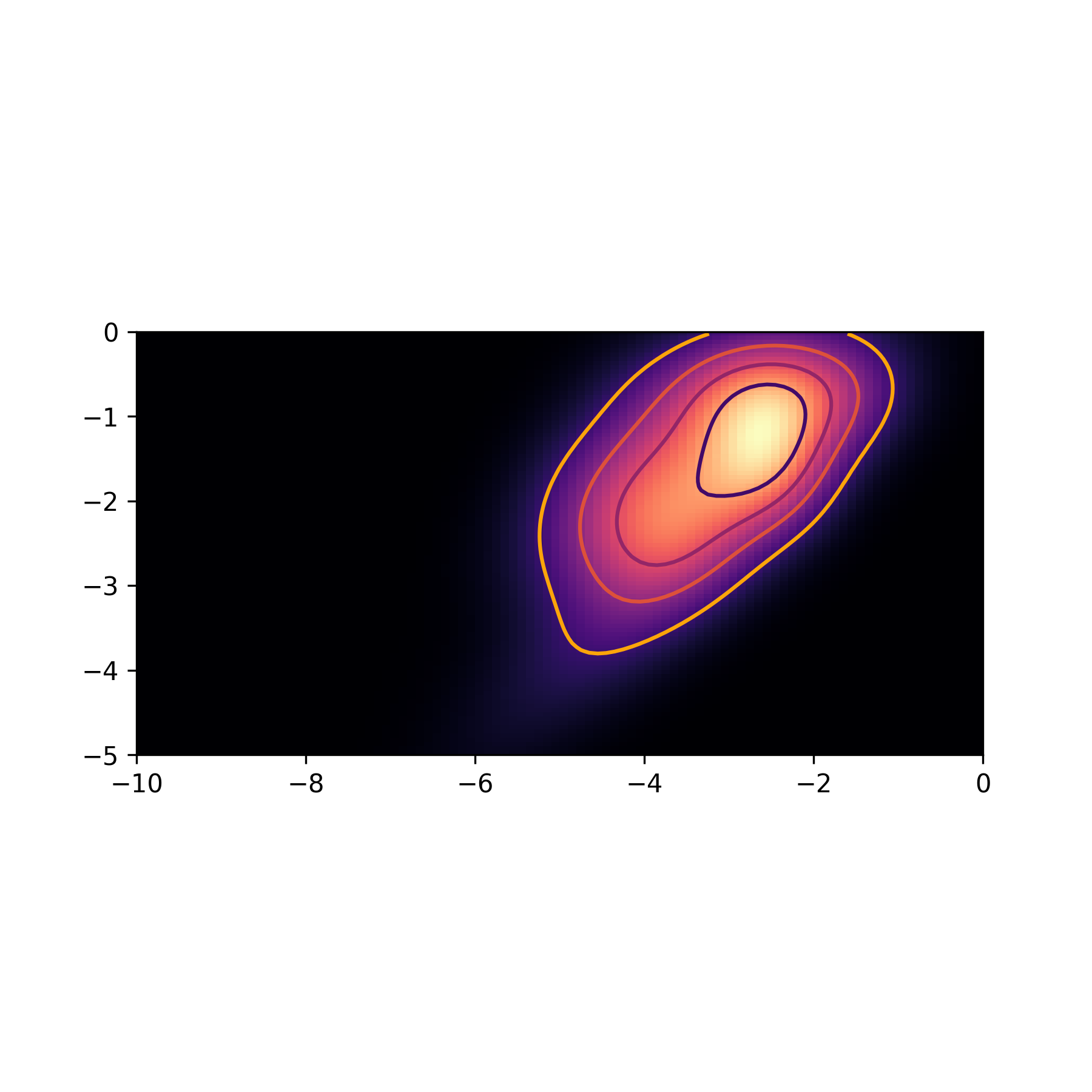}

\caption{Figures showing distributions of the weighted mixture of Gaussians for Phases O (left), I (center), and II (right) for $\text{PH}_1$ SW.}
\label{fig:ph1sw_dist}
\end{figure}

Within the PH$_1$ SW quadrant, the axes correspond to the values of $r_1$ and $r_2$, quantifying vessel thicknesses at narrowings and dimples, we observe a subtle but consistent trend. As shown in Figure \ref{fig:ph1sw_dist}, compared to Phase O, the Phase I and Phase II models exhibit a slight shift of the high-probability region toward lower values of both $r_1$ and $r_2$. This suggests a modest reduction in vessel size at these critical points, indicating that as AML progresses, vessels may become more constricted or exhibit more prominent surface depressions in the vessel structure.

\end{document}